\documentclass[12pt,twoside]{article}

\usepackage{amsmath}
\usepackage{graphicx}
\usepackage{epsfig}
\usepackage{rotating}
\usepackage{pstcol}
\usepackage{cite}
\newcommand{\mysection}[1]{\section{\boldmath #1}}
\newcommand{\mysubsection}[1]{\subsection[#1]{\boldmath #1}}
\newcommand{\mysubsubsection}[1]{\subsubsection[#1]{\boldmath #1}}
\newcommand{\mysubsubsubsection}[1]{\subsubsubsection{\boldmath #1}}

\newcommand{\lesssim}{\ensuremath{\raise-.5ex\hbox{$\buildrel<\over\sim$}\,}} 

\def\dof{{\rm dof}}

\newcommand\VCKM{{V}}
\newcommand\etacpf{{\eta_f}}
\newcommand\etacp{{\eta}}

\renewcommand\Im{{\rm Im}}

\newcommand\Abar{\kern 0.18em\overline{\kern -0.18em A}{}}
\newcommand\Af{A_f}
\newcommand\Abarf{\Abar_f}
\newcommand\Afbar{A_{\bar f}}
\newcommand\Abarfbar{\Abar_{\bar f}}
\newcommand\Acp{{\cal A}}
\newcommand\Adirnoncp{\ensuremath{\langle{\cal A}_{f\bar f}\rangle}\xspace}
\newcommand\mc{\multicolumn}
\newcommand\ph{\phantom}
\newcommand {\cbf}{\ensuremath{{\cal B}}}
\newcommand {\qq}{\ensuremath{q^2}}

\newcommand {\vcb}{\ensuremath{|V_{cb}|}}
\newcommand {\vub}{\ensuremath{|V_{ub}|}}

\newcommand {\Bxclnu}{\ensuremath{\Bb\to X_c\ell\nub}}

\newcommand {\breco}{\ensuremath{B_{reco}}}

\def\Bp      {\ensuremath{B^{+}}}
\def\Bm      {\ensuremath{B^{-}}}
\def\Bz      {\ensuremath{B^{0}}}
\def\Bs      {\ensuremath{B_{s}}}

\newcommand{\BzbDplnu}    {\ensuremath{\bar{B}^{0}\to D^{+}\ell^{-}\nub}}
\newcommand{\BzbDstarlnu} {\ensuremath{\bar{B}^{0}\to D^{*+}\ell^{-}\nub}}

\newcommand {\Bxulnu}{\ensuremath{B \to X_u \ell \bar{\nu}}\hbox{ }}

\usepackage{relsize}

\def\beq{\begin{equation}}
\def\eeq#1{\label{#1}\end{equation}}
\def\eeqn{\end{equation}}

\def\beqa{\begin{eqnarray}}
\def\eeqa#1{\label{#1}\end{eqnarray}}
\def\eeqan{\end{eqnarray}}

\let\bar=\overbar

\def\etal{{\it et al.}}
\def\ie{{\it i.e.}}
\def\eg{{\it e.g.}}
\def\etc{{\it etc.}}
\def\cf{{\it cf.}}

\def\Dslash{\ensuremath{\not{\hbox{\kern-4pt $D$}}}\xspace}
\def\dslash{\not{\hbox{\kern-2pt $\del$}}}

\def\BR{\mbox{\rm BR}}
\def\ee{e^+e^-}

\def\msb{{\bar{\ssstyle M \kern -1pt S}}}

\RequirePackage{xspace}

\usepackage{relsize}
\def\babar{\mbox{\slshape B\kern-0.1em{\smaller A}\kern-0.1em
    B\kern-0.1em{\smaller A\kern-0.2em R}}\xspace}
\def\belle{\mbox{\normalfont Belle}\xspace}
\newcommand{\dzero}{D\O\xspace}

\def\ee         {\ensuremath{e^-e^-}\xspace}

\def\nub        {\ensuremath{\overline{\nu}}\xspace}

\def\nub        {\ensuremath{\overline{\nu}}\xspace}

\def\ubar  {\ensuremath{\overline u}\xspace}

\def\sbar  {\ensuremath{\overline s}\xspace}

\def\b  {\ensuremath{b}\xspace}

\def\piz   {\ensuremath{\pi^0}\xspace}

\def\pip   {\ensuremath{\pi^+}\xspace}
\def\pim   {\ensuremath{\pi^-}\xspace}
\def\pipi  {\ensuremath{\pi^+\pi^-}\xspace}

\def\etapr {\ensuremath{\eta^{\prime}}\xspace}

\def\Kbar  {\kern 0.2em\overline{\kern -0.2em K}{}\xspace}

\def\Kmp   {\ensuremath{K^\mp}\xspace}
\def\Kp    {\ensuremath{K^+}\xspace}
\def\Km    {\ensuremath{K^-}\xspace}
\def\KS    {\ensuremath{K^0_{\scriptscriptstyle S}}\xspace} 
\def\KL    {\ensuremath{K^0_{\scriptscriptstyle L}}\xspace}

\def\Kstar   {\ensuremath{K^*}\xspace}

\def\Kstarmp   {\ensuremath{K^{*\mp}}\xspace}
\def\Kz   {\ensuremath{K^0}\xspace}
\def\Kzb   {\ensuremath{\Kbar^0}\xspace}
\def\KzKzb {\ensuremath{K^0 \kern -0.16em \Kzb}\xspace}

\def\Dz    {\ensuremath{D^0}\xspace}
\def\Dbar  {\kern 0.2em\overline{\kern -0.2em D}{}\xspace}

\def\Dzb   {\ensuremath{\Dbar^0}\xspace}
\def\DzDzb {\ensuremath{D^0 {\kern -0.16em \Dzb}}\xspace}
\def\Dp    {\ensuremath{D^+}\xspace}

\def\Dstar   {\ensuremath{D^*}\xspace}

\def\Dstarz  {\ensuremath{D^{*0}}\xspace}

\def\Dstarp  {\ensuremath{D^{*+}}}

\def\DorDstar   {\ensuremath{D^{(*)}}\xspace}
\def\DorDstarz  {\ensuremath{D^{(*)0}}\xspace}
\def\DorDstarzb {\ensuremath{\Dbar^{(*)0}}\xspace}

\def\Ds    {\ensuremath{D^+_s}\xspace}

\def\Bz    {\ensuremath{B^0}\xspace}
\def\B     {\ensuremath{B}\xspace}
\def\Bbar  {\kern 0.18em\overline{\kern -0.18em B}{}\xspace}
\def\Bb    {\ensuremath{\Bbar}\xspace}
\def\Bzb   {\ensuremath{\Bbar^0}\xspace}
\def\Bu    {\ensuremath{B^+}\xspace}

\def\Bmp   {\ensuremath{B^\mp}\xspace}
\def\Bs    {\ensuremath{B_s}\xspace}
\def\Bsb   {\ensuremath{\Bbar_s}\xspace}
\def\BB    {\ensuremath{B\Bbar}\xspace} 
\def\BzBzb {\ensuremath{B^0 {\kern -0.16em \Bzb}}\xspace}

\def\jpsi  {\ensuremath{{J\mskip -3mu/\mskip -2mu\psi\mskip 2mu}}\xspace}

\mathchardef\Upsilon="7107
\def\Y#1S{\ensuremath{\Upsilon{(#1S)}}\xspace}

\def\FourS {\Y4S}

\mathchardef\Deltares="7101
\mathchardef\Xi="7104
\mathchardef\Lambda="7103
\mathchardef\Sigma="7106
\mathchardef\Omega="710A
\def\Deltabar   {\kern 0.25em\overline{\kern -0.25em \Deltares}{}\xspace}
\def\Lbar {\kern 0.2em\overline{\kern -0.2em\Lambda\kern 0.05em}\kern-0.05em{}\xspace}
\def\Sigbar{\kern 0.2em\overline{\kern -0.2em \Sigma}{}\xspace}
\def\Xibar{\kern 0.2em\overline{\kern -0.2em \Xi}{}\xspace}
\def\Obar{\kern 0.2em\overline{\kern -0.2em \Omega}{}\xspace}
\def\Nbar{\kern 0.2em\overline{\kern -0.2em N}{}\xspace}
\def\Xb{\kern 0.2em\overline{\kern -0.2em X}{}}

\def\BR{{\ensuremath{\cal B}}}

\newcommand{\tev}{\ensuremath{\mathrm{Te\kern -0.1em V}}\xspace}
\newcommand{\gev}{\ensuremath{\mathrm{Ge\kern -0.1em V}}\xspace}
\newcommand{\mev}{\ensuremath{\mathrm{Me\kern -0.1em V}}\xspace}
\newcommand{\kev}{\ensuremath{\mathrm{ke\kern -0.1em V}}\xspace}
\newcommand{\ev}{\ensuremath{\mathrm{e\kern -0.1em V}}\xspace}
\newcommand{\gevc}{\ensuremath{{\mathrm{Ge\kern -0.1em V\!/}c}}\xspace}
\newcommand{\mevc}{\ensuremath{{\mathrm{Me\kern -0.1em V\!/}c}}\xspace}
\newcommand{\gevcc}{\ensuremath{{\mathrm{Ge\kern -0.1em V\!/}c^2}}\xspace}
\newcommand{\mevcc}{\ensuremath{{\mathrm{Me\kern -0.1em V\!/}c^2}}\xspace}

\def\mus  {\ensuremath{\rm \,\mus}\xspace}

\def\ps   {\ensuremath{\rm \,ps}\xspace}

\def\mus        {\ensuremath{\,\mu{\rm s}}\xspace}    
\def\ps         {\ensuremath{{\rm \,ps}}\xspace}  
\def\gsim{{~\raise.15em\hbox{$>$}\kern-.85em
          \lower.35em\hbox{$\sim$}~}\xspace}
\def\lsim{{~\raise.15em\hbox{$<$}\kern-.85em
          \lower.35em\hbox{$\sim$}~}\xspace}

\def\CP                 {\ensuremath{C\!P}\xspace}
\def\CPT                {\ensuremath{C\!PT}\xspace}

\def\pep2{PEP-II}

\def\rhobar {\ensuremath{\overline{\rho}}\xspace}
\def\etabar {\ensuremath{\overline{\eta}}\xspace}

\def\Vub  {\ensuremath{|V_{ub}|}\xspace}

\def\stwob{\ensuremath{\sin\! 2 \beta   }\xspace}

\def\deltamd{\ensuremath{{\rm \Delta}m_d}\xspace}

\xspace

\newcommand{\hepph} [1]  {hepph #1}

\def\jetset74   {\mbox{\tt Jetset \hspace{-0.5em}7.\hspace{-0.2em}4}}

\newcommand{\aerr}[4]   {\mbox{${{#1}^{+ #2}_{- #3}\pm #4}$}}
\newcommand{\berr}[4]   {\mbox{${{#1}\pm #2^{+ #3}_{- #4}}$}}
\newcommand{\cerr}[3]   {\mbox{${{#1}^{+ #2}_{- #3}}$}}
\newcommand{\aerrsy}[5] {\mbox{${{#1}^{+ #2 + #4}_{- #3 - #5}}$}}

\newcommand{\berrsyt}[6] {\mbox{${{#1}\pm #2^{+ #3 + #5}_{- #4 - #6}}$}}
\newcommand{\err}[3]   {\mbox{${{#1}\pm{#2}\pm{#3}}$}}
\newcommand{\nodata}{$$}
\def\etapr{{\eta^{\prime}}}

\def\sgline{\noalign{\vskip 0.10truecm\hrule\vskip 0.10truecm}}
\def\sglinespt{\noalign{\vskip 0.05truecm\hrule}}
\def\sglinespb{\noalign{\hrule\vskip 0.05truecm}}

\newcommand{\kz}    {\mbox{$K^0$}}

\newcommand{\RPP}{}

\begin{document}

\setcounter{page}{1}

\title{\begin{flushleft}
\end{flushleft}
\vskip 20pt
Averages of $b$-hadron Properties \\
as of Winter 2005 }
\author{Heavy Flavor Averaging Group (HFAG)\footnote{
The members involved in the HFAG averages for winter 2005 updates 
in this document are:
  K.~Anikeev, 
  E.~Barberio, 
  P.~Chang, 
  T.~Gershon,    
  R.~Harr,       
  A.~H\"ocker,   
  T.~Iijima,
  D.~Kirkby,
  R.~Kowalewski,   
  F.~Lehner,      
  A.~Limosani,
  V.~Luth,
  Y.~Sakai, 
  O.~Schneider, 
  C.~Schwanda,
  J.~Smith, 
  A.~Stocchi,    
  K.~Trabelsi,   
  R.~Van~Kooten, 
 and
  C.~Weiser. 
  }
 }
\maketitle
\thispagestyle{empty}
\begin{abstract}
This article reports world averages for measurements on
$b$-hadron properties obtained by the
Heavy Flavor Averaging Group (HFAG) using the available results as of
winter 2005 conferences. 
In the averaging, the input parameters used in the various analyses are 
adjusted (rescaled) to common values, and all known correlations are 
taken into account.
The averages include lifetimes,
neutral meson mixing parameters, semileptonic decay parameters, rare
decay branching fractions, and \CP violation measurements.
\end{abstract}

\newpage
\tableofcontents
\newpage


\mysection{Introduction}
\label{sec:intro}

 The flavor dynamics is one of the important elements in understanding
the nature of particle physics.  The accurate knowledge of properties of
heavy flavor hadrons, especially $b$ hadrons, play an essential role for
determination of the Cabbibo-Kobayashi-Maskawa (CKM) matrix~\cite{ref_ckm}.
Since asymmetric $B$ factories started their operation, available amounts
of $B$ meson samples has been dramatically increased and the accuracies
of measurements have been improved.
Tevatron experiments also started to provide rich results on $B$ hadron
decays with increased Run-II data samples.
 
 The Heavy Flavor Averaging Group (HFAG) has been formed in 2002, continuing 
the activities of LEP Heavy Flavor Steering group~\cite{LEPHFS}, 
to provide the
averages for measurements dedicated to the $b$-flavor related quantities.
The HFAG consists of representatives and contacts from the experimental
groups: \babar, \belle, CDF, CLEO, \dzero, and LEP. 

 The HFAG is currently organized into four subgroups.\footnote{
``$B \to $ Charm decays'' group has been newly formed, but averages were
not provided for winter 2005 conferences.}
\begin{itemize}
\item the ``Lifetime and mixing'' group provides
averages for $b$-hadron lifetimes, $b$-hadron fractions in $\Upsilon(4S)$ decay
and high energy collisions, and various parameters in $B^0$ and $B_s^0$
oscillation (mixing).

\item the ``Semileptonic $B$ decays'' group provides averages
for inclusive and exclusive $B$-decay branching fractions, and best values
of the CKM matrix elements $|V_{cb}|$ and $|V_{ub}|$. 

\item the ``$\CP(t)$ and Unitarity Triangle angles'' group provides averages for
time-dependent $\CP$ asymmetry parameters and angles of the unitarity
triangles.

\item the ``Rare decays'' group provides averages of branching fractions and
their asymmetries between $B$ and $\bar B$ for charmless mesonic,
radiative, leptonic, and baryonic $B$ decays.
\end{itemize}

The first two subgroups continue the activities from LEP working
groups with some reorganization (merging four groups into two groups).
The latter two groups are newly formed to take care of new results 
which are available from asymmetric $B$ factory experiments.

This article is an update of the first HFAG document\cite{hfag_hepex}, 
and we report the world averages using the available results
as of winter 2005 conferences (Moriond and CKM05 etc.).  
All results that are publicly available, including 
recent preliminary results, are used in averages.  
We do not use preliminary results which remain unpublished for a long time
or for which no publication is planned. 
Close contacts have been established between representatives from
the experiments and members of different subgroups in charge of the
averages, to ensure that the data are prepared in a form suitable
for combinations.  

We do not scale 
the error of an average 
(as is presently done by the Particle Data Group~\cite{Eidelman:2004wy})
in case $\chi^2/\dof > 1$, where $\dof$ is the number of 
degrees of freedom in the average calculation.
In this case, we examine the systematics of each measurement and 
try to understand them.  
Unless we find possible systematic discrepancies between the measurements, 
we do not make any special treatment for the calculated error.  
We provide the confidence level of the fit so that
one can know the consistency of the measurements included in the average.
We attach a warning message in case that some special treatment is done
or the approximation used in the average calculation may not be good enough
(\eg, Gaussian error is used in averaging though the likelihood 
indicates non-Gaussian behavior).

Section~\ref{sec:method} describes the methodology for
averaging various quantities in the HFAG.  
In the averaging, the input parameters used in the various analyses are 
adjusted (rescaled) to common values, and, where possible, known 
correlations are taken into account. 
The general philosophy and tools for calculations of averages are presented.


Sections~\ref{sec:life_mix}--\ref{sec:rare} describe the averaging of 
the quantities from each subgroup mentioned above.

A summary of the averages described in this article is given in
Sec.~\ref{sec:summary}.   

 The complete listing of averages and plots described in this article
are also available on the HFAG Web page:
 
 {\tt http://www.slac.stanford.edu/xorg/hfag } and 

 {\tt http://belle.kek.jp/mirror/hfag } (KEK mirror site).

\section{Methodology } \label{sec:method} 
The general averaging problem that HFAG faces is to combine the
information provided by different measurements of the same parameter,
to obtain our best estimate of the parameter's value and
uncertainty. The methodology described here focuses on the problems of
combining measurements performed with different systematic assumptions
and with potentially-correlated systematic uncertainties. Our methodology
relies on the close involvement of the people performing the
measurements in the averaging process.

Consider two hypothetical measurements of a parameter $x$, which might
be summarized as
\begin{align*}
x &= x_1 \pm \delta x_1 \pm \Delta x_{1,1} \pm \Delta x_{2,1} \ldots \\
x &= x_2 \pm \delta x_2 \pm \Delta x_{1,2} \pm \Delta x_{2,2} \ldots
\; ,
\end{align*}
where the $\delta x_k$ are statistical uncertainties, and
the $\Delta x_{i,k}$ are contributions to the systematic
uncertainty. One popular approach is to combine statistical and
systematic uncertainties in quadrature
\begin{align*}
x &= x_1 \pm \left(\delta x_1 \oplus \Delta x_{1,1} \oplus \Delta
x_{2,1} \oplus \ldots\right) \\
x &= x_2 \pm \left(\delta x_2 \oplus \Delta x_{1,2} \oplus \Delta
x_{2,2} \oplus \ldots\right)
\end{align*}
and then perform a weighted average of $x_1$ and $x_2$, using their
combined uncertainties, as if they were independent. This approach
suffers from two potential problems that we attempt to address. First,
the values of the $x_k$ may have been obtained using different
systematic assumptions. For example, different values of the \Bz
lifetime may have been assumed in separate measurements of the
oscillation frequency $\deltamd$. The second potential problem is that
some contributions of the systematic uncertainty may be correlated
between experiments. For example, separate measurements of $\deltamd$
may both depend on an assumed Monte-Carlo branching fraction used to
model a common background.

The problems mentioned above are related since, ideally, any quantity $y_i$
that $x_k$ depends on has a corresponding contribution $\Delta x_{i,k}$ to the
systematic error which reflects the uncertainty $\Delta y_i$ on $y_i$
itself. We assume that this is the case, and use the values of $y_i$ and
$\Delta y_i$ assumed by each measurement explicitly in our
averaging (we refer to these values as $y_{i,k}$ and $\Delta y_{i,k}$
below). Furthermore, since we do not lump all the systematics
together,
we require that each measurement used in an average have a consistent
definition of the various contributions to the systematic uncertainty.
Different analyses often use different decompositions of their systematic
uncertainties, so achieving consistent definitions for any potentially
correlated contributions requires close coordination between HFAG and
the experiments. In some cases, a group of
systematic uncertainties must be lumped to obtain a coarser
description that is consistent between measurements. Systematic uncertainties
that are uncorrelated with any other sources of uncertainty appearing
in an average are lumped with the statistical error, so that the only
systematic uncertainties treated explicitly are those that are
correlated with at least one other measurement via a consistently-defined
external parameter $y_i$. When asymmetric statistical or systematic
uncertainties are quoted, we symmetrize them since our combination
method implicitly assumes parabolic likelihoods for each measurement.

\begin{figure}
\begin{center}
\includegraphics[width=6.0in]{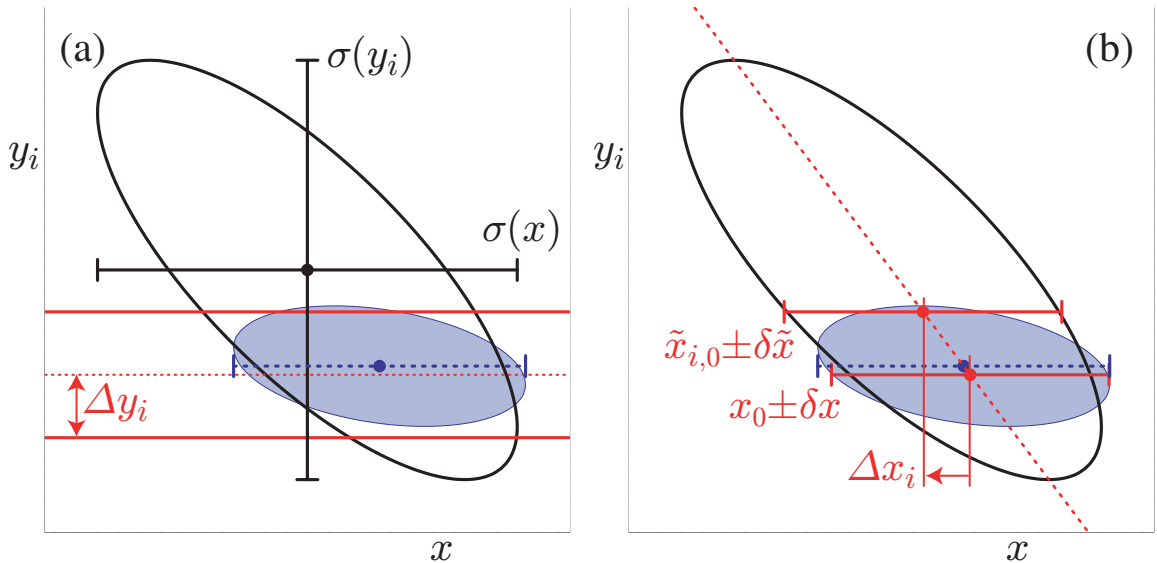}
\end{center}
\caption{The left-hand plot, (a), compares the 68\% confidence-level
  contours of a
  hypothetical measurement's unconstrained (large ellipse) and
  constrained (filled ellipse) likelihoods, using the Gaussian
  constraint on $y_i$ represented by the horizontal band. The solid
  error bars represent the statistical uncertainties, $\sigma(x)$ and
  $\sigma(y_i)$, of the unconstrained likelihood. The dashed
  error bar shows the statistical error on $x$ from a
  constrained simultaneous fit to $x$ and $y_i$. The right-hand plot,
  (b), illustrates the method described in the text of performing fits
  to $x$ only with $y_i$ fixed at different values. The dashed
  diagonal line between these fit results has the slope
  $\rho(x,y_i)\sigma(y_i)/\sigma(x)$ in the limit of a parabolic
  unconstrained likelihood. The result of the constrained simultaneous
  fit from (a) is shown as a dashed error bar on $x$.}
\label{fig:singlefit}
\end{figure}

The fact that a measurement of $x$ is sensitive to the value of $y_i$
indicates that, in principle, the data used to measure $x$ could
equally-well be used for a simultaneous measurement of $x$ and $y_i$, as
illustrated by the large contour in Fig.~\ref{fig:singlefit}(a) for a hypothetical
measurement. However, we often have an external constraint $\Delta
y_i$ on the value of $y_i$ (represented by the horizontal band in
Fig.~\ref{fig:singlefit}(a)) that is more precise than the constraint
$\sigma(y_i)$ from
our data alone. Ideally, in such cases we would perform a simultaneous
fit to $x$ and $y_i$, including the external constraint, obtaining the
filled $(x,y)$ contour and corresponding dashed one-dimensional estimate of
$x$ shown in Fig.~\ref{fig:singlefit}(a). Throughout, we assume that
the external constraint $\Delta y_i$ on $y_i$ is Gaussian.

In practice, the added technical complexity of a constrained fit with
extra free parameters is not justified by the small increase in
sensitivity, as long as the external constraints $\Delta y_i$ are
sufficiently precise when compared with the sensitivities $\sigma(y_i)$
to each $y_i$ of the data alone. Instead, the usual procedure adopted
by the experiments is to perform a baseline fit with all $y_i$ fixed
to nominal values $y_{i,0}$, obtaining $x = x_0 \pm \delta
x$. This baseline fit neglects the uncertainty due to $\Delta y_i$, but
this error can be mostly recovered by repeating the fit separately for
each external parameter $y_i$ with its value fixed at $y_i = y_{i,0} +
\Delta y_i$ to obtain $x = \tilde{x}_{i,0} \pm \delta\tilde{x}$, as
illustrated in Fig.~\ref{fig:singlefit}(b). The absolute shift,
$|\tilde{x}_{i,0} - x_0|$, in the central value of $x$ is what the
experiments usually quote as their systematic uncertainty $\Delta x_i$
on $x$ due to the unknown value of $y_i$. Our procedure requires that
we know not only the magnitude of this shift but also its sign. In the
limit that the unconstrained data is represented by a parabolic
likelihood, the signed shift is given by
\begin{equation}
\Delta x_i = \rho(x,y_i)\frac{\sigma(x)}{\sigma(y_i)}\,\Delta y_i \;,
\end{equation}
where $\sigma(x)$ and $\rho(x,y_i)$ are the statistical uncertainty on
$x$ and the correlation between $x$ and
$y_i$ in the unconstrained data.
While our procedure is not
equivalent to the constrained fit with extra parameters, it yields (in
the limit of a parabolic unconstrained likelihood) a central value
$x_0$ that agrees 
to ${\cal O}(\Delta y_i/\sigma(y_i))^2$ and an uncertainty $\delta x
\oplus \Delta x_i$ that agrees to ${\cal O}(\Delta y_i/\sigma(y_i))^4$.

In order to combine two or more measurements that share systematics
due to the same external parameters $y_i$, we would ideally perform a
constrained simultaneous fit of all data samples to obtain values of
$x$ and each $y_i$, being careful to only apply the constraint on each
$y_i$ once. This is not practical since we generally do not have
sufficient information to reconstruct the unconstrained likelihoods
corresponding to each measurement. Instead, we perform the two-step
approximate procedure described below.

\begin{figure}
\begin{center}
\includegraphics[width=6.0in]{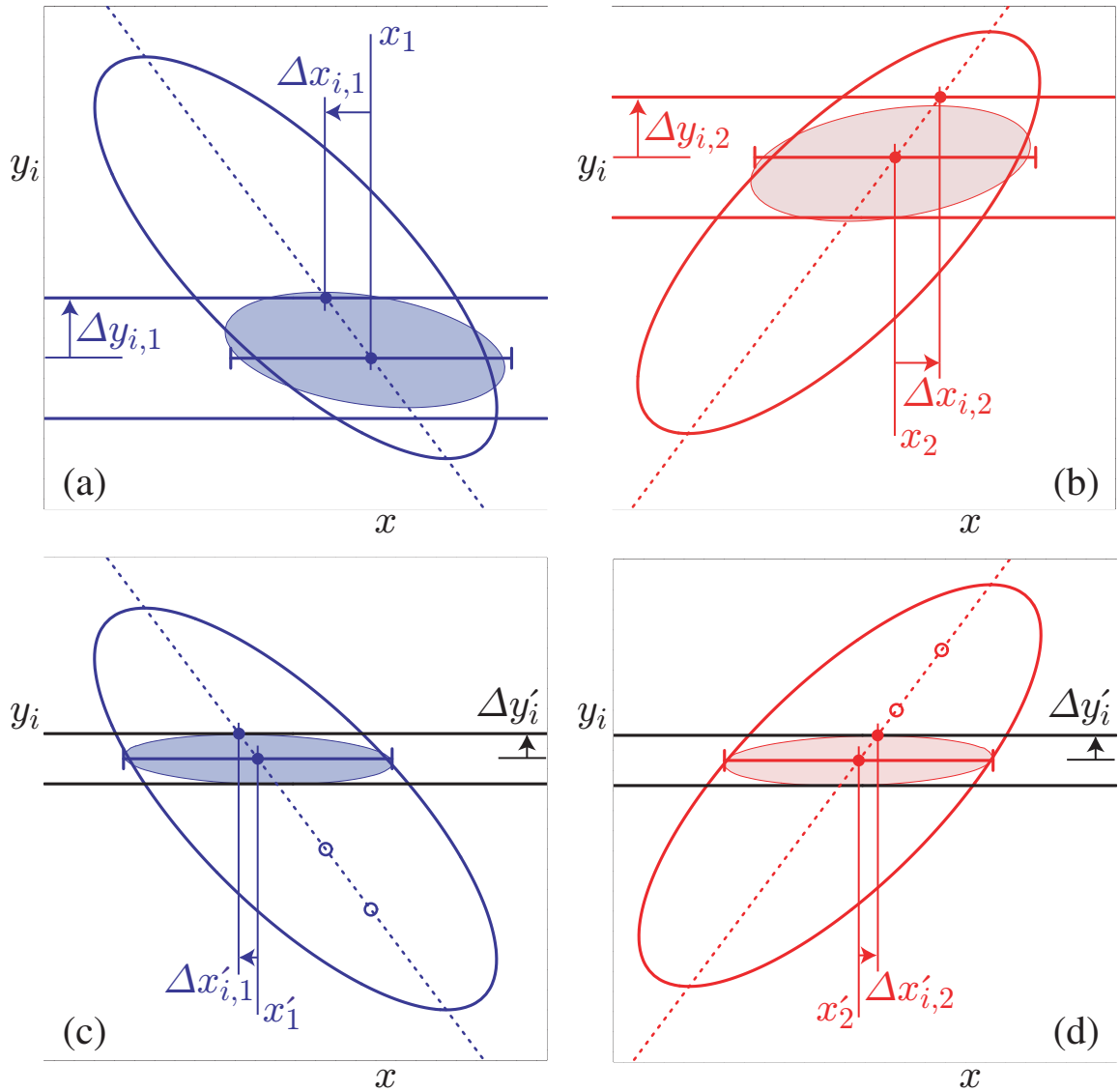}
\end{center}
\caption{The upper plots, (a) and (b), show examples of two individual
  measurements to be combined. The large ellipses represent their
  unconstrained likelihoods, and the filled ellipses represent their
  constrained likelihoods. Horizontal bands indicate the different
  assumptions about the value and uncertainty of $y_i$ used by each
  measurement. The error bars show the results of the approximate
  method described in the text for obtaining $x$ by performing fits
  with $y_i$ fixed to different values. The lower plots, (c) and (d),
  illustrate the adjustments to accommodate updated and consistent
  knowledge of $y_i$ described in the text. Hollow circles mark the
  central values of the unadjusted fits to $x$ with $y$ fixed, which
  determine the dashed line used to obtain the adjusted values. }
\label{fig:multifit}
\end{figure}

Figs.~\ref{fig:multifit}(a,b) illustrate two
statistically-independent measurements, $x_1 \pm (\delta x_1 \oplus
\Delta x_{i,1})$ and $x_2\pm(\delta x_i\oplus \Delta x_{i,2})$, of the same
hypothetical quantity $x$ (for simplicity, we only show the
contribution of a single correlated systematic due to an external
parameter $y_i$). As our knowledge of the external parameters $y_i$
evolves, it is natural that the different measurements of $x$ will
assume different nominal values and ranges for each $y_i$. The first
step of our procedure is to adjust the values of each measurement to
reflect the current best knowledge of the values $y_i'$ and ranges
$\Delta y_i'$ of the external parameters $y_i$, as illustrated in
Figs.~\ref{fig:multifit}(c,b). We adjust the
central values $x_k$ and correlated systematic uncertainties $\Delta
x_{i,k}$ linearly for each measurement (indexed by $k$) and each
external parameter (indexed by $i$):
\begin{align}
x_k' &= x_k + \sum_i\,\frac{\Delta x_{i,k}}{\Delta y_{i,k}}
\left(y_i'-y_{i,k}\right)\\
\Delta x_{i,k}'&= \Delta x_{i,k}\cdot \frac{\Delta y_i'}{\Delta
  y_{i,k}} \; .
\end{align}
This procedure is exact in the limit that the unconstrained
likelihoods of each measurement is parabolic.

\begin{figure}
\begin{center}
\includegraphics[width=3.5in]{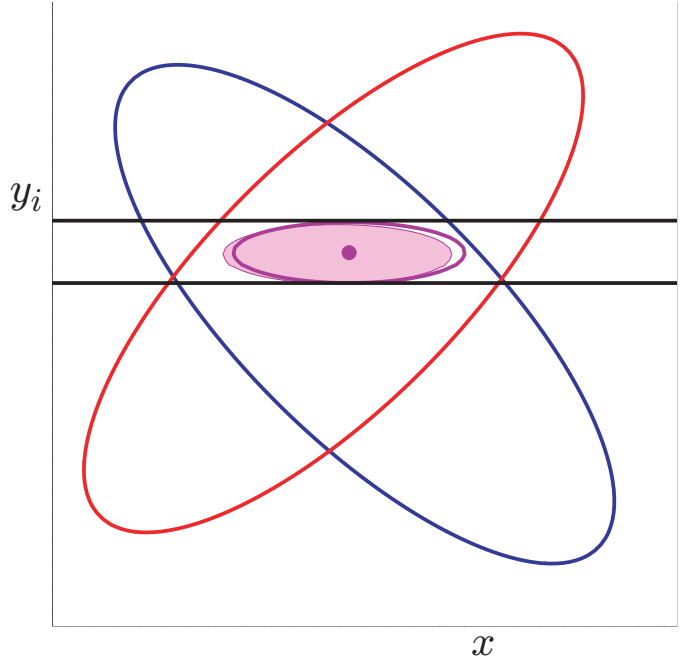}
\end{center}
\caption{An illustration of the combination of two hypothetical
  measurements of $x$ using the method described in the text. The
  ellipses represent the unconstrained likelihoods of each measurement
  and the horizontal band represents the latest knowledge about $y_i$ that
  is used to adjust the individual measurements. The filled small
  ellipse shows the result of the exact method using ${\cal
  L}_{\text{comb}}$ and the hollow small ellipse and dot show the
  result of the approximate method using $\chi^2_{\text{comb}}$.}
\label{fig:fit12}
\end{figure}

The second step of our procedure is to combine the adjusted
measurements, $x_k'\pm (\delta x_k\oplus \Delta x_{k,1}'\oplus \Delta
x_{k,2}'\oplus\ldots)$ using the chi-square 
\begin{equation}
\chi^2_{\text{comb}}(x,y_1,y_2,\ldots) \equiv \sum_k\,
\frac{1}{\delta x_k^2}\left[
x_k' - \left(x + \sum_i\,(y_i-y_i')\frac{\Delta x_{i,k}'}{\Delta y_i'}\right)
\right]^2 + \sum_i\,
\left(\frac{y_i - y_i'}{\Delta y_i'}\right)^2 \; ,
\end{equation}
and then minimize this $\chi^2$ to obtain the best values of $x$ and
$y_i$ and their uncertainties, as illustrated in
Fig.~\ref{fig:fit12}. Although this method determines new values for
the $y_i$, we do not report them since the $\Delta x_{i,k}$ reported
by each experiment are generally not intended for this purpose (for
example, they may represent a conservative upper limit rather than a
true reflection of a 68\% confidence level).

For comparison, the exact method we would
perform if we had the unconstrained likelihoods ${\cal L}_k(x,y_1,y_2,\ldots)$
available for each
measurement is to minimize the simultaneous constrained likelihood
\begin{equation}
{\cal L}_{\text{comb}}(x,y_1,y_2,\ldots) \equiv \prod_k\,{\cal
  L}_k(x,y_1,y_2,\ldots)\,\prod_{i}\,{\cal 
  L}_i(y_i) \; ,
\end{equation}
with an independent Gaussian external constraint on each $y_i$
\begin{equation}
{\cal L}_i(y_i) \equiv \exp\left[-\frac{1}{2}\,\left(\frac{y_i-y_i'}{\Delta
 y_i'}\right)^2\right] \; .
\end{equation}
The results of this exact method are illustrated by the filled ellipses
in Figs.~\ref{fig:fit12}(a,b), and agree with our method in the limit that
each ${\cal L}_k$ is parabolic and that each $\Delta
y_i' \ll \sigma(y_i)$. In the case of a non-parabolic unconstrained
likelihood, experiments would have to provide a description of ${\cal
  L}_k$ itself to allow an improved combination. In the case of some
$\sigma(y_i)\simeq \Delta y_i'$, experiments are advised to perform a
simultaneous measurement of both $x$ and $y$ so that their data will
improve the world knowledge about $y$. 

 The algorithm described above is used as a default in the averages
reported in the following sections.  For some cases, somewhat simplified
or more complex algorithms are used and noted in the corresponding 
sections. 

Following the prescription described above,
the central values and errors are rescaled
to a common set of input parameters in the averaging procedures, 
according to the dependency on
any of these input parameters.
We try to use the most up-to-date values for these common inputs and 
the same values among the HFAG subgroups.
For the parameters whose averages are produced by the HFAG, we use 
the updated values in the current update cycle.  For other external
parameters, we use the most recent PDG values. 

  The parameters and values used in this update cycle are listed in
each subgroup section.


%
%
%
%

%

%
%
%
%

%

\renewcommand{\topfraction}{0.9}

\newcommand{\auth}[1]{#1,}
\newcommand{\coll}[1]{#1 Collaboration,}
\newcommand{\authcoll}[2]{#1 \etal\ (#2 Collaboration),}
\newcommand{\titl}[1]{``#1'',} 
\newcommand{\J}[4]{{#1} {\bf #2}, #3 (#4)}
\newcommand{\subJ}[1]{submitted to #1}
\newcommand{\PRL}[3]{\J{Phys.\ Rev.\ Lett.}{#1}{#2}{#3}}
\newcommand{\subPRL}{\subJ{Phys.\ Rev.\ Lett.}}
\newcommand{\PRD}[3]{\J{Phys.\ Rev.\ D}{#1}{#2}{#3}}
\newcommand{\subPRD}{\subJ{Phys.\ Rev.\ D}}
\newcommand{\ZPC}[3]{\J{Z.\ Phys.\ C}{#1}{#2}{#3}}
\newcommand{\PLB}[3]{\J{Phys.\ Lett.\ B}{#1}{#2}{#3}}
\newcommand{\subPLB}{\subJ{Phys.\ Lett.\ B}}
\newcommand{\EPJC}[3]{\J{Eur.\ Phys.\ J.\ C}{#1}{#2}{#3}}
\newcommand{\NPB}[3]{\J{Nucl.\ Phys.\ B}{#1}{#2}{#3}}
\newcommand{\subNPB}{\subJ{Nucl.\ Phys.\ B}}
\newcommand{\NIMA}[3]{\J{Nucl.\ Instrum.\ Methods A}{#1}{#2}{#3}}
\newcommand{\subNIMA}{\subJ{Nucl.\ Instrum.\ Methods A}}
\newcommand{\JHEP}[3]{\J{J.\ of High Energy Physics }{#1}{#2}{#3}}
\newcommand{\ARNS}[3]{\J{Ann.\ Rev.\ Nucl.\ Sci.}{#1}{#2}{#3}}
\newcommand{\newref}{\\}

\newcommand{\particle}[1]{\ensuremath{#1}\xspace}
\renewcommand{\ee}{\particle{e^+e^-}}
\newcommand{\Ups}{\particle{\Upsilon(4S)}}
\renewcommand{\b}{\particle{b}}
\renewcommand{\B}{\particle{B}}
\newcommand{\Bd}{\particle{B^0}}
\renewcommand{\Bs}{\particle{B^0_s}}
\renewcommand{\Bu}{\particle{B^+}}
\newcommand{\Bc}{\particle{B^+_c}}
\newcommand{\Bdbar}{\particle{\bar{B}^0}}
\newcommand{\Bsbar}{\particle{\bar{B}^0_s}}
\newcommand{\Lb}{\particle{\Lambda_b^0}}
\newcommand{\Xib}{\particle{\Xi_b}}
\newcommand{\Lc}{\particle{\Lambda_c^+}}

\newcommand{\fBs}{\ensuremath{f_{\particle{s}}}\xspace}
\newcommand{\fBd}{\ensuremath{f_{\particle{d}}}\xspace}
\newcommand{\fBu}{\ensuremath{f_{\particle{u}}}\xspace}
\newcommand{\fbb}{\ensuremath{f_{\rm baryon}}\xspace}

\newcommand{\dmd}{\ensuremath{\Delta m_{\particle{d}}}\xspace}
\newcommand{\dms}{\ensuremath{\Delta m_{\particle{s}}}\xspace}
\newcommand{\xd}{\ensuremath{x_{\particle{d}}}\xspace}
\newcommand{\xs}{\ensuremath{x_{\particle{s}}}\xspace}
\newcommand{\yd}{\ensuremath{y_{\particle{d}}}\xspace}
\newcommand{\ys}{\ensuremath{y_{\particle{s}}}\xspace}
\newcommand{\chibar}{\ensuremath{\overline{\chi}}\xspace}
\newcommand{\chid}{\ensuremath{\chi_{\particle{d}}}\xspace}
\newcommand{\chis}{\ensuremath{\chi_{\particle{s}}}\xspace}
\newcommand{\Gd}{\ensuremath{\Gamma_{\particle{d}}}\xspace}
\newcommand{\DGd}{\ensuremath{\Delta\Gd}\xspace}
\newcommand{\DGGd}{\ensuremath{\DGd/\Gd}\xspace}
\newcommand{\Gs}{\ensuremath{\Gamma_{\particle{s}}}\xspace}
\newcommand{\DGs}{\ensuremath{\Delta\Gs}\xspace}
\newcommand{\DGGs}{\ensuremath{\Delta\Gs/\Gs}\xspace}

\renewcommand{\BR}[1]{\particle{{\cal B}(#1)}}
\newcommand{\CL}[1]{#1\%~\mbox{CL}}
\newcommand{\Qjet}{\ensuremath{Q_{\rm jet}}\xspace}

\newcommand{\labe}[1]{\label{equ:#1}}
\newcommand{\labs}[1]{\label{sec:#1}}
\newcommand{\labf}[1]{\label{fig:#1}}
\newcommand{\labt}[1]{\label{tab:#1}}
\newcommand{\refe}[1]{\ref{equ:#1}}
\newcommand{\refs}[1]{\ref{sec:#1}}
\newcommand{\reff}[1]{\ref{fig:#1}}
\newcommand{\reft}[1]{\ref{tab:#1}}
\newcommand{\Ref}[1]{Ref.~\cite{#1}}
\newcommand{\Refs}[1]{Refs.~\cite{#1}}
\newcommand{\Refss}[2]{Refs.~\cite{#1} and \cite{#2}}
\newcommand{\Refsss}[3]{Refs.~\cite{#1}, \cite{#2} and \cite{#3}}
\newcommand{\eq}[1]{(\refe{#1})}
\newcommand{\Eq}[1]{Eq.~(\refe{#1})}
\newcommand{\Eqs}[1]{Eqs.~(\refe{#1})}
\newcommand{\Eqss}[2]{Eqs.~(\refe{#1}) and (\refe{#2})}
\newcommand{\Eqssor}[2]{Eqs.~(\refe{#1}) or (\refe{#2})}
\newcommand{\Eqsss}[3]{Eqs.~(\refe{#1}), (\refe{#2}), and (\refe{#3})}
\newcommand{\Figure}[1]{Figure~\reff{#1}}
\newcommand{\Figuress}[2]{Figures~\reff{#1} and \reff{#2}}
\newcommand{\Fig}[1]{Fig.~\reff{#1}}
\newcommand{\Figs}[1]{Figs.~\reff{#1}}
\newcommand{\Figss}[2]{Figs.~\reff{#1} and \reff{#2}}
\newcommand{\Figsss}[3]{Figs.~\reff{#1}, \reff{#2}, and \reff{#3}}
\newcommand{\Section}[1]{Section~\refs{#1}}
\newcommand{\Sec}[1]{Sec.~\refs{#1}}
\newcommand{\Secs}[1]{Secs.~\refs{#1}}
\newcommand{\Secss}[2]{Secs.~\refs{#1} and \refs{#2}}
\newcommand{\Secsss}[3]{Secs.~\refs{#1}, \refs{#2}, and \refs{#3}}
\newcommand{\Table}[1]{Table~\reft{#1}}
\newcommand{\Tables}[1]{Tables~\reft{#1}}
\newcommand{\Tabless}[2]{Tables~\reft{#1} and \reft{#2}}
\newcommand{\Tablesss}[3]{Tables~\reft{#1}, \reft{#2}, and \reft{#3}}

\newcommand{\subsubsubsection}[1]{\vspace{2ex}\par\noindent {\bf\boldmath\em #1} \vspace{2ex}\par}


\newcommand{\definemath}[2]{\newcommand{#1}{\ensuremath{#2}\xspace}}

\definemath{\hfagCHIBARLEPval}{0.1257}
\definemath{\hfagCHIBARLEPerr}{\pm0.0042}
\definemath{\hfagTAUBDval}{1.528}
\definemath{\hfagTAUBDerr}{\pm0.009}
\definemath{\hfagTAUBUval}{1.643}
\definemath{\hfagTAUBUerr}{\pm0.010}
\definemath{\hfagRTAUBUval}{1.076}
\definemath{\hfagRTAUBUerr}{\pm0.008}
\definemath{\hfagTAUBSval}{1.479}
\definemath{\hfagTAUBSerr}{\pm0.044}
\definemath{\hfagRTAUBSval}{0.968}
\definemath{\hfagRTAUBSerr}{\pm0.029}
\definemath{\hfagTAULBval}{1.232}
\definemath{\hfagTAULBerr}{\pm0.072}
\definemath{\hfagTAUXBval}{1.39}
\definemath{\hfagTAUXBerp}{^{+0.34}}
\definemath{\hfagTAUXBern}{_{-0.28}}
\definemath{\hfagTAUBBval}{1.210}
\definemath{\hfagTAUBBerr}{\pm0.048}
\definemath{\hfagRTAUBBval}{0.792}
\definemath{\hfagRTAUBBerr}{\pm0.032}
\definemath{\hfagTAUBval}{1.568}
\definemath{\hfagTAUBerr}{\pm0.009}
\definemath{\hfagTAUBCval}{0.45}
\definemath{\hfagTAUBCerr}{\pm0.12}
\definemath{\hfagTAUBSSLval}{1.472}
\definemath{\hfagTAUBSSLerr}{\pm0.045}
\definemath{\hfagTAUBSSLXval}{1.478}
\definemath{\hfagTAUBSSLXerr}{\pm0.047}
\definemath{\hfagTAUBSMEANCONval}{1.405}
\definemath{\hfagTAUBSMEANCONerp}{^{+0.043}}
\definemath{\hfagTAUBSMEANCONern}{_{-0.047}}
\definemath{\hfagTAUBSJFval}{1.404}
\definemath{\hfagTAUBSJFerr}{\pm0.066}
\definemath{\hfagRTAUBSSLval}{0.963}
\definemath{\hfagRTAUBSSLerr}{\pm0.030}
\definemath{\hfagRTAUBSMEANCONval}{0.920}
\definemath{\hfagRTAUBSMEANCONerr}{\pm0.030}
\definemath{\hfagRTAULBval}{0.806}
\definemath{\hfagRTAULBerr}{\pm0.047}
\definemath{\hfagTAUBVTXval}{1.572}
\definemath{\hfagTAUBVTXerr}{\pm0.009}
\definemath{\hfagTAUBLEPval}{1.537}
\definemath{\hfagTAUBLEPerr}{\pm0.020}
\definemath{\hfagTAUBJPval}{1.533}
\definemath{\hfagTAUBJPerp}{^{+0.038}}
\definemath{\hfagTAUBJPern}{_{-0.034}}
\definemath{\hfagSDGDGDval}{-0.009}
\definemath{\hfagSDGDGDerr}{\pm0.037}
\definemath{\hfagDGSGSval}{+0.35}
\definemath{\hfagDGSGSerp}{^{+0.12}}
\definemath{\hfagDGSGSern}{_{-0.16}}
\definemath{\hfagDGSGSlow}{+0.01}
\definemath{\hfagDGSGSupp}{+0.59}
\definemath{\hfagTAUBSMEANval}{1.42}
\definemath{\hfagTAUBSMEANerp}{^{+0.06}}
\definemath{\hfagTAUBSMEANern}{_{-0.07}}
\definemath{\hfagRHODGSGSTAUBSMEAN}{N/A}
\definemath{\hfagDGSval}{+0.25}
\definemath{\hfagDGSerp}{^{+0.09}}
\definemath{\hfagDGSern}{_{-0.11}}
\definemath{\hfagDGSlow}{+0.01}
\definemath{\hfagDGSupp}{+0.43}
\definemath{\hfagTAUBSLval}{1.21}
\definemath{\hfagTAUBSLerp}{^{+0.08}}
\definemath{\hfagTAUBSLern}{_{-0.09}}
\definemath{\hfagTAUBSHval}{1.72}
\definemath{\hfagTAUBSHerr}{\pm0.19}
\definemath{\hfagDGSGSCONBDval}{+0.39}
\definemath{\hfagDGSGSCONBDerp}{^{+0.11}}
\definemath{\hfagDGSGSCONBDern}{_{-0.12}}
\definemath{\hfagDGSGSCONval}{+0.33}
\definemath{\hfagDGSGSCONerp}{^{+0.09}}
\definemath{\hfagDGSGSCONern}{_{-0.11}}
\definemath{\hfagDGSGSCONlow}{+0.01}
\definemath{\hfagDGSGSCONupp}{+0.59}
\definemath{\hfagTAUBSMEANCONXval}{1.405}
\definemath{\hfagTAUBSMEANCONXerp}{^{+0.043}}
\definemath{\hfagTAUBSMEANCONXern}{_{-0.047}}
\definemath{\hfagRHODGSGSTAUBSMEANCON}{-0.76}
\definemath{\hfagDGSCONval}{+0.23}
\definemath{\hfagDGSCONerr}{\pm0.08}
\definemath{\hfagDGSCONlow}{+0.07}
\definemath{\hfagDGSCONupp}{+0.37}
\definemath{\hfagTAUBSLCONval}{1.21}
\definemath{\hfagTAUBSLCONerr}{\pm0.08}
\definemath{\hfagTAUBSHCONval}{1.68}
\definemath{\hfagTAUBSHCONerp}{^{+0.08}}
\definemath{\hfagTAUBSHCONern}{_{-0.09}}
\definemath{\hfagDGSGSCONBDCONval}{+0.22}
\definemath{\hfagDGSGSCONBDCONerp}{^{+0.07}}
\definemath{\hfagDGSGSCONBDCONern}{_{-0.09}}
\definemath{\hfagFCWval}{0.508}
\definemath{\hfagFCWerr}{\pm0.007}
\definemath{\hfagFNWval}{0.492}
\definemath{\hfagFNWerr}{\pm0.007}
\definemath{\hfagFFWval}{1.030}
\definemath{\hfagFFWerr}{\pm0.029}
\definemath{\hfagFCNval}{0.513}
\definemath{\hfagFCNerr}{\pm0.013}
\definemath{\hfagFNNval}{0.487}
\definemath{\hfagFNNerr}{\pm0.013}
\definemath{\hfagFFNval}{1.053}
\definemath{\hfagFFNerr}{\pm0.054}
\definemath{\hfagFCval}{0.505}
\definemath{\hfagFCerr}{\pm0.008}
\definemath{\hfagFNval}{0.495}
\definemath{\hfagFNerr}{\pm0.008}
\definemath{\hfagFFval}{1.021}
\definemath{\hfagFFerr}{\pm0.034}
\definemath{\hfagFPRODval}{0.497}
\definemath{\hfagFPRODerr}{\pm0.021}
\definemath{\hfagFSUMval}{0.984}
\definemath{\hfagFSUMerr}{\pm0.031}
\definemath{\hfagFBSNOMIXval}{0.087}
\definemath{\hfagFBSNOMIXerr}{\pm0.021}
\definemath{\hfagFBBNOMIXval}{0.107}
\definemath{\hfagFBBNOMIXerr}{\pm0.018}
\definemath{\hfagFBDNOMIXval}{0.403}
\definemath{\hfagFBDNOMIXerr}{\pm0.011}
\definemath{\hfagCHIBARTEVval}{0.152}
\definemath{\hfagCHIBARTEVerr}{\pm0.013}
\definemath{\hfagCHIBARSFACTOR}{1.9}
\definemath{\hfagCHIBARval}{0.1281}
\definemath{\hfagCHIBARerr}{\pm0.0076}
\definemath{\hfagCHIDUval}{0.182}
\definemath{\hfagCHIDUerr}{\pm0.015}
\definemath{\hfagCHIDWUval}{0.189}
\definemath{\hfagCHIDWUerr}{\pm0.002}
\definemath{\hfagXDWUval}{0.778}
\definemath{\hfagXDWUerr}{\pm0.008}
\definemath{\hfagDMDWval}{0.510}
\definemath{\hfagDMDWsta}{\pm0.003}
\definemath{\hfagDMDWsys}{\pm0.004}
\definemath{\hfagDMDWerr}{\pm0.005}
\definemath{\hfagDMDWUval}{0.509}
\definemath{\hfagDMDWUerr}{\pm0.004}
\definemath{\hfagFBSMIXval}{0.116}
\definemath{\hfagFBSMIXerr}{\pm0.021}
\definemath{\hfagFBSval}{0.102}
\definemath{\hfagFBSerr}{\pm0.014}
\definemath{\hfagFBBval}{0.100}
\definemath{\hfagFBBerr}{\pm0.017}
\definemath{\hfagFBDval}{0.399}
\definemath{\hfagFBDerr}{\pm0.010}
\definemath{\hfagRHOFBBFBS}{-0.162}
\definemath{\hfagRHOFBDFBS}{-0.564}
\definemath{\hfagRHOFBDFBB}{-0.724}
\definemath{\hfagDMDHval}{0.495}
\definemath{\hfagDMDHsta}{\pm0.010}
\definemath{\hfagDMDHsys}{\pm0.009}
\definemath{\hfagDMDHerr}{\pm0.014}
\definemath{\hfagDMDBval}{0.511}
\definemath{\hfagDMDBsta}{\pm0.003}
\definemath{\hfagDMDBsys}{\pm0.005}
\definemath{\hfagDMDBerr}{\pm0.005}
\definemath{\hfagDMDTWODval}{0.514}
\definemath{\hfagDMDTWODsta}{\pm0.003}
\definemath{\hfagDMDTWODsys}{\pm0.004}
\definemath{\hfagDMDTWODerr}{\pm0.005}
\definemath{\hfagTAUBDTWODval}{1.532}
\definemath{\hfagTAUBDTWODsta}{\pm0.006}
\definemath{\hfagTAUBDTWODsys}{\pm0.010}
\definemath{\hfagTAUBDTWODerr}{\pm0.011}
\definemath{\hfagRHOstaDMDTAUBD}{-0.13}
\definemath{\hfagRHOsysDMDTAUBD}{-0.41}
\definemath{\hfagRHODMDTAUBD}{-0.31}
\definemath{\hfagQPval}{1.0013}
\definemath{\hfagQPerr}{\pm0.0034}
\definemath{\hfagASLval}{-0.0026}
\definemath{\hfagASLerr}{\pm0.0067}
\definemath{\hfagREBval}{-0.0007}
\definemath{\hfagREBerr}{\pm0.0017}
\definemath{\hfagDMSWLIMval}{14.5}
\definemath{\hfagDMSWSENSval}{18.5}
\definemath{\hfagDMSXLIMval}{14.5}
\definemath{\hfagDMSXSENSval}{19.5}
\definemath{\hfagDMSDLIMval}{0.0}
\definemath{\hfagDMSDSENSval}{1.0}
\definemath{\hfagDMSWUPPval}{21.7}
\definemath{\hfagXSWLIMval}{21.1}
\definemath{\hfagCHISWLIMval}{0.49888}

\newcommand{\unit}[1]{~\ensuremath{\rm #1}\xspace}
\renewcommand{\ps}{\unit{ps}}
\newcommand{\invps}{\unit{ps^{-1}}}
\newcommand{\TeV}{\unit{TeV}}
\newcommand{\MeVcc}{\unit{MeV/\mbox{$c$}^2}}
\newcommand{\MeV}{\unit{MeV}}

\definemath{\hfagCHIBARLEP}{\hfagCHIBARLEPval\hfagCHIBARLEPerr}
\definemath{\hfagTAUBD}{\hfagTAUBDval\hfagTAUBDerr\ps}
\definemath{\hfagTAUBDnounit}{\hfagTAUBDval\hfagTAUBDerr}
\definemath{\hfagTAUBU}{\hfagTAUBUval\hfagTAUBUerr\ps}
\definemath{\hfagTAUBUnounit}{\hfagTAUBUval\hfagTAUBUerr}
\definemath{\hfagRTAUBU}{\hfagRTAUBUval\hfagRTAUBUerr}
\definemath{\hfagTAUBS}{\hfagTAUBSval\hfagTAUBSerr\ps}
\definemath{\hfagTAUBSnounit}{\hfagTAUBSval\hfagTAUBSerr}
\definemath{\hfagRTAUBS}{\hfagRTAUBSval\hfagRTAUBSerr}
\definemath{\hfagTAULB}{\hfagTAULBval\hfagTAULBerr\ps}
\definemath{\hfagTAULBnounit}{\hfagTAULBval\hfagTAULBerr}
\definemath{\hfagTAUXBerr}{\hfagTAUXBerp\hfagTAUXBern}
\definemath{\hfagTAUXB}{\hfagTAUXBval\hfagTAUXBerr\ps}
\definemath{\hfagTAUXBnounit}{\hfagTAUXBval\hfagTAUXBerr}
\definemath{\hfagTAUBB}{\hfagTAUBBval\hfagTAUBBerr\ps}
\definemath{\hfagTAUBBnounit}{\hfagTAUBBval\hfagTAUBBerr}
\definemath{\hfagRTAUBB}{\hfagRTAUBBval\hfagRTAUBBerr}
\definemath{\hfagTAUB}{\hfagTAUBval\hfagTAUBerr\ps}
\definemath{\hfagTAUBnounit}{\hfagTAUBval\hfagTAUBerr}
\definemath{\hfagTAUBC}{\hfagTAUBCval\hfagTAUBCerr\ps}
\definemath{\hfagTAUBCnounit}{\hfagTAUBCval\hfagTAUBCerr}
\definemath{\hfagTAUBSSL}{\hfagTAUBSSLval\hfagTAUBSSLerr\ps}
\definemath{\hfagTAUBSSLnounit}{\hfagTAUBSSLval\hfagTAUBSSLerr}
\definemath{\hfagTAUBSSLX}{\hfagTAUBSSLXval\hfagTAUBSSLXerr\ps}
\definemath{\hfagTAUBSSLXnounit}{\hfagTAUBSSLXval\hfagTAUBSSLXerr}
\definemath{\hfagTAUBSMEANCONerr}{\hfagTAUBSMEANCONerp\hfagTAUBSMEANCONern}
\definemath{\hfagTAUBSMEANCON}{\hfagTAUBSMEANCONval\hfagTAUBSMEANCONerr\ps}
\definemath{\hfagTAUBSMEANCONnounit}{\hfagTAUBSMEANCONval\hfagTAUBSMEANCONerr}
\definemath{\hfagTAUBSJF}{\hfagTAUBSJFval\hfagTAUBSJFerr\ps}
\definemath{\hfagTAUBSJFnounit}{\hfagTAUBSJFval\hfagTAUBSJFerr}
\definemath{\hfagRTAUBSSL}{\hfagRTAUBSSLval\hfagRTAUBSSLerr}
\definemath{\hfagRTAUBSMEANCON}{\hfagRTAUBSMEANCONval\hfagRTAUBSMEANCONerr}
\definemath{\hfagRTAULB}{\hfagRTAULBval\hfagRTAULBerr}
\definemath{\hfagTAUBVTX}{\hfagTAUBVTXval\hfagTAUBVTXerr\ps}
\definemath{\hfagTAUBVTXnounit}{\hfagTAUBVTXval\hfagTAUBVTXerr}
\definemath{\hfagTAUBLEP}{\hfagTAUBLEPval\hfagTAUBLEPerr\ps}
\definemath{\hfagTAUBLEPnounit}{\hfagTAUBLEPval\hfagTAUBLEPerr}
\definemath{\hfagTAUBJPerr}{\hfagTAUBJPerp\hfagTAUBJPern}
\definemath{\hfagTAUBJP}{\hfagTAUBJPval\hfagTAUBJPerr\ps}
\definemath{\hfagTAUBJPnounit}{\hfagTAUBJPval\hfagTAUBJPerr}
\definemath{\hfagSDGDGD}{\hfagSDGDGDval\hfagSDGDGDerr}
\definemath{\hfagDGSGSerr}{\hfagDGSGSerp\hfagDGSGSern}
\definemath{\hfagDGSGS}{\hfagDGSGSval\hfagDGSGSerr}
\definemath{\hfagTAUBSMEANerr}{\hfagTAUBSMEANerp\hfagTAUBSMEANern}
\definemath{\hfagTAUBSMEAN}{\hfagTAUBSMEANval\hfagTAUBSMEANerr\ps}
\definemath{\hfagTAUBSMEANnounit}{\hfagTAUBSMEANval\hfagTAUBSMEANerr}
\definemath{\hfagDGSerr}{\hfagDGSerp\hfagDGSern}
\definemath{\hfagDGS}{\hfagDGSval\hfagDGSerr\invps}
\definemath{\hfagDGSnounit}{\hfagDGSval\hfagDGSerr}
\definemath{\hfagTAUBSLerr}{\hfagTAUBSLerp\hfagTAUBSLern}
\definemath{\hfagTAUBSL}{\hfagTAUBSLval\hfagTAUBSLerr\ps}
\definemath{\hfagTAUBSLnounit}{\hfagTAUBSLval\hfagTAUBSLerr}
\definemath{\hfagTAUBSH}{\hfagTAUBSHval\hfagTAUBSHerr\ps}
\definemath{\hfagTAUBSHnounit}{\hfagTAUBSHval\hfagTAUBSHerr}
\definemath{\hfagDGSGSCONBDerr}{\hfagDGSGSCONBDerp\hfagDGSGSCONBDern}
\definemath{\hfagDGSGSCONBD}{\hfagDGSGSCONBDval\hfagDGSGSCONBDerr}
\definemath{\hfagDGSGSCONerr}{\hfagDGSGSCONerp\hfagDGSGSCONern}
\definemath{\hfagDGSGSCON}{\hfagDGSGSCONval\hfagDGSGSCONerr}
\definemath{\hfagTAUBSMEANCONXerr}{\hfagTAUBSMEANCONXerp\hfagTAUBSMEANCONXern}
\definemath{\hfagTAUBSMEANCONX}{\hfagTAUBSMEANCONXval\hfagTAUBSMEANCONXerr\ps}
\definemath{\hfagTAUBSMEANCONXnounit}{\hfagTAUBSMEANCONXval\hfagTAUBSMEANCONXerr}
\definemath{\hfagDGSCON}{\hfagDGSCONval\hfagDGSCONerr\invps}
\definemath{\hfagDGSCONnounit}{\hfagDGSCONval\hfagDGSCONerr}
\definemath{\hfagTAUBSLCON}{\hfagTAUBSLCONval\hfagTAUBSLCONerr\ps}
\definemath{\hfagTAUBSLCONnounit}{\hfagTAUBSLCONval\hfagTAUBSLCONerr}
\definemath{\hfagTAUBSHCONerr}{\hfagTAUBSHCONerp\hfagTAUBSHCONern}
\definemath{\hfagTAUBSHCON}{\hfagTAUBSHCONval\hfagTAUBSHCONerr\ps}
\definemath{\hfagTAUBSHCONnounit}{\hfagTAUBSHCONval\hfagTAUBSHCONerr}
\definemath{\hfagDGSGSCONBDCONerr}{\hfagDGSGSCONBDCONerp\hfagDGSGSCONBDCONern}
\definemath{\hfagDGSGSCONBDCON}{\hfagDGSGSCONBDCONval\hfagDGSGSCONBDCONerr}
\definemath{\hfagFCW}{\hfagFCWval\hfagFCWerr}
\definemath{\hfagFNW}{\hfagFNWval\hfagFNWerr}
\definemath{\hfagFFW}{\hfagFFWval\hfagFFWerr}
\definemath{\hfagFCN}{\hfagFCNval\hfagFCNerr}
\definemath{\hfagFNN}{\hfagFNNval\hfagFNNerr}
\definemath{\hfagFFN}{\hfagFFNval\hfagFFNerr}
\definemath{\hfagFC}{\hfagFCval\hfagFCerr}
\definemath{\hfagFN}{\hfagFNval\hfagFNerr}
\definemath{\hfagFF}{\hfagFFval\hfagFFerr}
\definemath{\hfagFPROD}{\hfagFPRODval\hfagFPRODerr}
\definemath{\hfagFSUM}{\hfagFSUMval\hfagFSUMerr}
\definemath{\hfagFBSNOMIX}{\hfagFBSNOMIXval\hfagFBSNOMIXerr}
\definemath{\hfagFBBNOMIX}{\hfagFBBNOMIXval\hfagFBBNOMIXerr}
\definemath{\hfagFBDNOMIX}{\hfagFBDNOMIXval\hfagFBDNOMIXerr}
\definemath{\hfagCHIBARTEV}{\hfagCHIBARTEVval\hfagCHIBARTEVerr}
\definemath{\hfagCHIBAR}{\hfagCHIBARval\hfagCHIBARerr}
\definemath{\hfagCHIDU}{\hfagCHIDUval\hfagCHIDUerr}
\definemath{\hfagCHIDWU}{\hfagCHIDWUval\hfagCHIDWUerr}
\definemath{\hfagXDWU}{\hfagXDWUval\hfagXDWUerr}
\definemath{\hfagDMDW}{\hfagDMDWval\hfagDMDWerr\invps}
\definemath{\hfagDMDWnounit}{\hfagDMDWval\hfagDMDWerr}
\definemath{\hfagDMDWfull}{\hfagDMDWval\hfagDMDWsta\hfagDMDWsys\invps}
\definemath{\hfagDMDWnounitfull}{\hfagDMDWval\hfagDMDWsta\hfagDMDWsys}
\definemath{\hfagDMDWU}{\hfagDMDWUval\hfagDMDWUerr\invps}
\definemath{\hfagDMDWUnounit}{\hfagDMDWUval\hfagDMDWUerr}
\definemath{\hfagFBSMIX}{\hfagFBSMIXval\hfagFBSMIXerr}
\definemath{\hfagFBS}{\hfagFBSval\hfagFBSerr}
\definemath{\hfagFBB}{\hfagFBBval\hfagFBBerr}
\definemath{\hfagFBD}{\hfagFBDval\hfagFBDerr}
\definemath{\hfagDMDH}{\hfagDMDHval\hfagDMDHerr\invps}
\definemath{\hfagDMDHnounit}{\hfagDMDHval\hfagDMDHerr}
\definemath{\hfagDMDHfull}{\hfagDMDHval\hfagDMDHsta\hfagDMDHsys\invps}
\definemath{\hfagDMDHnounitfull}{\hfagDMDHval\hfagDMDHsta\hfagDMDHsys}
\definemath{\hfagDMDB}{\hfagDMDBval\hfagDMDBerr\invps}
\definemath{\hfagDMDBnounit}{\hfagDMDBval\hfagDMDBerr}
\definemath{\hfagDMDBfull}{\hfagDMDBval\hfagDMDBsta\hfagDMDBsys\invps}
\definemath{\hfagDMDBnounitfull}{\hfagDMDBval\hfagDMDBsta\hfagDMDBsys}
\definemath{\hfagDMDTWOD}{\hfagDMDTWODval\hfagDMDTWODerr\invps}
\definemath{\hfagDMDTWODnounit}{\hfagDMDTWODval\hfagDMDTWODerr}
\definemath{\hfagDMDTWODfull}{\hfagDMDTWODval\hfagDMDTWODsta\hfagDMDTWODsys\invps}
\definemath{\hfagDMDTWODnounitfull}{\hfagDMDTWODval\hfagDMDTWODsta\hfagDMDTWODsys}
\definemath{\hfagTAUBDTWOD}{\hfagTAUBDTWODval\hfagTAUBDTWODerr\ps}
\definemath{\hfagTAUBDTWODnounit}{\hfagTAUBDTWODval\hfagTAUBDTWODerr}
\definemath{\hfagTAUBDTWODfull}{\hfagTAUBDTWODval\hfagTAUBDTWODsta\hfagTAUBDTWODsys\ps}
\definemath{\hfagTAUBDTWODnounitfull}{\hfagTAUBDTWODval\hfagTAUBDTWODsta\hfagTAUBDTWODsys}
\definemath{\hfagQP}{\hfagQPval\hfagQPerr}
\definemath{\hfagASL}{\hfagASLval\hfagASLerr}
\definemath{\hfagREB}{\hfagREBval\hfagREBerr}
\definemath{\hfagDMSWLIM}{\hfagDMSWLIMval\invps}
\definemath{\hfagDMSWSENS}{\hfagDMSWSENSval\invps}
\definemath{\hfagDMSXLIM}{\hfagDMSXLIMval\invps}
\definemath{\hfagDMSXSENS}{\hfagDMSXSENSval\invps}
\definemath{\hfagDMSDLIM}{\hfagDMSDLIMval\invps}
\definemath{\hfagDMSDSENS}{\hfagDMSDSENSval\invps}
\definemath{\hfagDMSWUPP}{\hfagDMSWUPPval\invps}
\definemath{\hfagXSWLIM}{\hfagXSWLIMval}
\definemath{\hfagCHISWLIM}{\hfagCHISWLIMval}


\mysection{\b-hadron production fractions, lifetimes and mixing parameters}
\labs{life_mix}


Quantities such as \b-hadron production fractions, \b-hadron lifetimes, 
and neutral \B-meson oscillation frequencies have been measured
for many years at high-energy colliders, namely at LEP and SLC 
(\ee colliders at $\sqrt{s}=m_{\particle{Z}}$) as well as at the 
first version of the Tevatron
(\particle{p\bar{p}} collider at $\sqrt{s}=1.8\TeV$). More recently, 
precise measurements of the \Bd and \Bu lifetimes, as well as of the 
\Bd oscillation frequency, have also been performed at the 
asymmetric \B factories, KEKB and PEPII
(\ee colliders at $\sqrt{s}=m_{\Ups}$).
In most cases, these basic quantities, although interesting by themselves,
can now be seen as necessary ingredients for the more complicated and 
refined analyses being currently performed at the asymmetric \B factories
and at the upgraded Tevatron ($\sqrt{s}=1.96\TeV$),
in particular the time-dependent \CP asymmetry measurements.
It is therefore important that the best experimental
values of these quantities continue to be kept up-to-date and improved. 

In several cases, the averages presented in this chapter are indeed
needed and used as input for the results given in the subsequent chapters. 
However, within this chapter, some averages need the knowledge of other 
averages in a circular way. This ``coupling'', which appears through the 
\b-hadron fractions whenever inclusive or semi-exclusive measurements 
have to be considered, has been reduced significantly in the last years 
with increasingly precise exclusive measurements becoming available. 
To cope with this circularity,
a rather involved averaging procedure had been developed, in the framework 
of the former LEP Heavy Flavour Steering Group. This is still in use now
(details can be found in~\cite{LEPHFS}), 
although simplifications can be envisaged in the future when even more 
precise exclusive measurements become available. 

\mysubsection{\b-hadron production fractions}
\labs{fractions}
 
We consider here the relative fractions of the different \b-hadron 
species found in an unbiased sample of weakly-decaying \b hadrons 
produced under some specific conditions. The knowledge of these fractions
is useful to characterize the signal composition in inclusive \b-hadron 
analyses, or to predict the background composition in exclusive analyses.
Many analyses in \B physics need these fractions as input. We distinguish 
here the following two conditions: \Ups decays and 
high-energy collisions. 

\mysubsubsection{\b-hadron production fractions in \Ups decays}
\labs{bfraction}

Only pairs of the two lightest (charged and neutral) \B mesons 
can be produced in \Ups decays, 
and it is enough to determine the following branching 
fractions:
\begin{eqnarray}
f^{+-} & = & \Gamma(\Ups \to \particle{B^+B^-})/
             \Gamma_{\rm tot}(\Ups)  \,, \\
f^{00} & = & \Gamma(\Ups \to \particle{B^0\bar{B}^0})/
             \Gamma_{\rm tot}(\Ups) \,.
\end{eqnarray}
In practice, most analyses measure their ratio
\begin{equation}
R^{+-/00} = f^{+-}/f^{00} = \Gamma(\Ups \to \particle{B^+B^-})/
             \Gamma(\Ups \to \particle{B^0\bar{B}^0}) \,,
\end{equation}
which is easier to access experimentally.
Since an inclusive (but separate) reconstruction of 
\Bu and \Bd is difficult, specific exclusive decay modes, 
${\Bu} \to x^+$ and ${\Bd} \to x^0$, are usually considered to perform 
a measurement of $R^{+-/00}$, whenever they can be related by 
isospin symmetry (for example \particle{\Bu \to J/\psi K^+} and 
\particle{\Bd \to J/\psi K^0}).
Under the assumption that $\Gamma(\Bu \to x^+) = \Gamma(\Bd \to x^0)$, 
\ie\ that isospin invariance holds in these \B decays,
the ratio of the number of reconstructed
$\Bu \to x^+$ and $\Bd \to x^0$ mesons is proportional to
\begin{equation}
\frac{f^{+-}\, \BR{\Bu\to x^+}}{f^{00}\, \BR{\Bd\to x^0}}
= \frac{f^{+-}\, \Gamma({\Bu}\to x^+)\, \tau(\Bu)}%
{f^{00}\, \Gamma({\Bd}\to x^0)\,\tau(\Bd)}
= \frac{f^{+-}}{f^{00}} \, \frac{\tau(\Bu)}{\tau(\Bd)}  \,, 
\end{equation} 
where $\tau(\Bu)$ and $\tau(\Bd)$ are the \Bu and \Bd 
lifetimes respectively.
Hence the primary quantity measured in these analyses 
is $R^{+-/00} \, \tau(\Bu)/\tau(\Bd)$, 
and the extraction of $R^{+-/00}$ with this method therefore 
requires the knowledge of the $\tau(\Bu)/\tau(\Bd)$ lifetime ratio. 

\begin{table}
\caption{Published measurements of the $\Bu/\Bd$ production ratio
in \Ups decays, together with their average (see text).
Systematic uncertainties due to the imperfect knowledge of 
$\tau(\Bu)/\tau(\Bd)$ are included.}
\labt{R_data}
\begin{center}
\begin{tabular}{@{}l@{}c@{\,}cll@{}}
\hline
Experiment & Ref. & Decay modes & Published value of & Assumed value \\
and year & & or method & $R^{+-/00}=f^{+-}/f^{00}$ & of $\tau(\Bu)/\tau(\Bd)$ \\
\hline
CLEO,   2001 & \cite{CLEO_R2001}  & \particle{J/\psi K^{(*)}} 
             & $1.04 \pm0.07 \pm0.04$ & $1.066 \pm0.024$ \\
\babar, 2002 & \cite{BABAR_R2002} & \particle{(c\bar{c})K^{(*)}}
             & $1.10 \pm0.06 \pm0.05$ & $1.062 \pm0.029$\\ 
CLEO,   2002 & \cite{CLEO_R2002}  & \particle{D^*\ell\nu}
             & $1.058 \pm0.084 \pm0.136$ & $1.074 \pm0.028$\\
\belle, 2003 & \cite{BELLE_dmd_dilepton} & dilepton events 
             & $1.01 \pm0.03 \pm0.09$ & $1.083 \pm0.017$\\
\babar, 2004 & \cite{BABAR_R2004} & \particle{J/\psi K}
             & $1.006 \pm0.036 \pm0.031$ & $1.083 \pm0.017$ \\
\hline
Average      & & & \hfagFF~(tot) & \hfagRTAUBU \\
\hline
\end{tabular}
\end{center}
\end{table}

The published measurements of $R^{+-/00}$ are listed 
in \Table{R_data} together with the corresponding assumed values of 
$\tau(\Bu)/\tau(\Bd)$.
All measurements are based on the above-mentioned method, 
except the one from \belle, which is a by-product of the 
\Bd mixing frequency analysis using dilepton events
(but note that it also assumes isospin invariance, 
namely $\Gamma(\Bu \to \ell^+{\rm X}) = \Gamma(\Bd \to \ell^+{\rm X})$).
The latter is therefore treated in a slightly different 
manner in the following procedure used to combine 
these measurements:
\begin{itemize} 
\item each published value of $R^{+-/00}$ from CLEO and \babar
      is first converted back to the original measurement of 
      $R^{+-/00} \, \tau(\Bu)/\tau(\Bd)$, using the value of the 
      lifetime ratio assumed in the corresponding analysis;
\item a simple weighted average of these original
      measurements of $R^{+-/00} \, \tau(\Bu)/\tau(\Bd)$ from 
      CLEO and \babar (which do not depend on the assumed value 
      of the lifetime ratio) is then computed, assuming no 
      statistical or systematic correlations between them;


\item the weighted average of $R^{+-/00} \, \tau(\Bu)/\tau(\Bd)$ 
      is converted into a value of $R^{+-/00}$, using the latest 
      average of the lifetime ratios, $\tau(\Bu)/\tau(\Bd)=\hfagRTAUBU$ 
      (see \Sec{lifetime_ratio});
\item the \belle measurement of $R^{+-/00}$ is adjusted to the 
      current values of $\tau(\Bd)=\hfagTAUBD$ and 
      $\tau(\Bu)/\tau(\Bd)=\hfagRTAUBU$ (see \Sec{lifetime_ratio}),
      using the quoted systematic uncertainties due to these parameters;
\item the combined value of $R^{+-/00}$ from CLEO and \babar is averaged 
      with the adjusted value of $R^{+-/00}$ from \belle, assuming a 100\% 
      correlation of the systematic uncertainty due to the limited 
      knowledge on $\tau(\Bu)/\tau(\Bd)$; no other correlation is considered. 
\end{itemize} 
The resulting global average, 
\begin{equation}
R^{+-/00} = \frac{f^{+-}}{f^{00}} =  \hfagFF \,,
\labe{Rplusminus}
\end{equation}
is consistent with an equal production of charged and neutral \B mesons.

On the other hand, the \babar collaboration has 
recently performed a direct measurement of the $f^{00}$ fraction 
using a novel method, which does not rely on isospin symmetry nor requires 
the knowledge of $\tau(\Bu)/\tau(\Bd)$. Its analysis, 
based on a comparison between the number of events where a single 
$B^0 \to D^{*-} \ell^+ \nu$ decay could be reconstructed and the number 
of events where two such decays could be reconstructed, yields~\cite{BABAR_f00}
\begin{equation}
f^{00}= 0.487 \pm 0.010\,\mbox{(stat)} \pm 0.008\,\mbox{(syst)} \,.
\labe{fzerozero}
\end{equation}

The two results of \Eqss{Rplusminus}{fzerozero} are of very different natures 
and completely independent of each other. 
Their product is equal to $f^{+-} = \hfagFPROD$, 
while another combination of them gives $f^{+-} + f^{00}= \hfagFSUM$, 
compatible with unity.
Assuming $f^{+-}+f^{00}= 1$, also consistent with 
CLEO's observation that the fraction of \Ups decays 
to \BB pairs is larger than 0.96 at \CL{95}~\cite{CLEO_frac_limit},
the results of \Eqss{Rplusminus}{fzerozero}
can be averaged (first converting \Eq{Rplusminus} 
into a value of $f^{00}=1/(R^{+-/00}+1)$) 
to yield the following more precise estimates:
\begin{equation}
f^{00} = \hfagFNW  \,,~~~ f^{+-} = 1 -f^{00} =  \hfagFCW \,,~~~
\frac{f^{+-}}{f^{00}} =  \hfagFFW \,.
\end{equation}

\mysubsubsection{\b-hadron production fractions at high energy}
\labs{fractions_high_energy}

At high energy, all species of weakly-decaying \b hadrons 
can be produced, either directly or in strong and electromagnetic 
decays of excited \b hadrons.
We assume here that the fractions of these different species 
are the same in unbiased samples of high-$p_{\rm T}$ \b jets 
originating from \particle{Z^0} decays or from \particle{p\bar{p}} 
collisions at the Tevatron.
This hypothesis is plausible considering that, in both cases, 
the last step of the jet hadronization is a non-perturbative
QCD process occurring at the scale of $\Lambda_{\rm QCD}$.
On the other hand, there is no strong argument to claim that these 
fractions should be strictly equal, so this assumption 
should be checked experimentally.
Although the available data is not sufficient at 
this time to perform a significant check, 
it is expected that the new data from 
Tevatron Run II may improve this situation and 
allow one to confirm or disprove this assumption with reasonable 
confidence. Meanwhile, the attitude adopted here is that these 
fractions are assumed to be equal at all high-energy colliders
until demonstrated otherwise by experiment.\footnote{It is not unlikely
that the \b-hadron fractions in low-$p_{\rm T}$ jets 
at a hadronic machine be different; in particular, beam-remnant effects may
enhance the \b-baryon production.}

Contrary to what happens in the charm sector where the fractions of \particle{D^+} 
and \particle{D^0} are different, the relative amount of \Bu and \Bd is not affected by the 
electromagnetic decays of excited ${\Bu}^*$ and ${\Bd}^*$ states and strong decays of excited
${\Bu}^{**}$ and ${\Bd}^{**}$ states. Decays of the type \particle{{\Bs}^{**} \to B^{(*)}K}
also contribute to the \Bu and \Bd rates, but with the same magnitude if mass effects
can be neglected.
We therefore assume equal production of \Bu and \Bd. We also  
neglect the production of weakly-decaying states
made of several heavy quarks (like \Bc and other heavy baryons) 
which is known to be very small. Hence, for the purpose of determining 
the \b-hadron fractions, we use the constraints
\begin{equation}
\fBu = \fBd ~~~~\mbox{and}~~~ \fBu + \fBd + \fBs + \fbb = 1 \,,
\labe{constraints}
\end{equation}
where \fBu, \fBd, \fBs and \fbb
are the unbiased fractions of \Bu, \Bd, \Bs and \b-baryons, respectively.

The LEP experiments have measured
$\fBs \times \BR{\Bs\to\particle{D_s^-} \ell^+ \nu_\ell \mbox{$X$}}$~\cite{LEP_fs}, 
$\BR{\b\to\Lb} \times \BR{\Lb\to\Lc\ell^-\bar{\nu}_\ell \mbox{$X$}}$~\cite{DELPHI_fla,ALEPH_fla}
and $\BR{\b\to\Xib^-} \times \BR{\Xi_b^- \to \Xi^-\ell^-\overline\nu_\ell 
\mbox{$X$}}$~\cite{DELPHI_fxi,ALEPH_fxi}
from partially reconstructed final states 
including a lepton, \fbb
from protons identified in \b events~\cite{ALEPH-fbar}, and the 
production rate of charged \b hadrons~\cite{DELPHI-fch}. 
The various \b-hadron fractions 
have also been measured at CDF using electron-charm final states~\cite{CDF_f_ec}
and double semileptonic decays with \particle{\phi\ell} and 
\particle{K^*\ell} final states~\cite{CDF_f_phil_Kstl}.
All these published results have been combined 
following the procedure and assumptions described in~\cite{LEPHFS}
to yield $\fBu=\fBd=\hfagFBDNOMIX$, 
$\fBs=\hfagFBSNOMIX$ and $\fbb=\hfagFBBNOMIX$
under the constraints of \Eq{constraints}.
For this combination, other external inputs are used, \eg\ the branching 
ratios of \B mesons to final states with a \particle{D}, \particle{D^*} or 
\particle{D^{**}} in semileptonic decays, which are needed to evaluate the 
fraction of semileptonic \Bs decays with a \particle{D_s^-} in the final state.


Time-integrated mixing analyses performed with lepton pairs 
from \particle{b\bar{b}} 
events produced at high-energy colliders measure the quantity 
\begin{equation}
\chibar = f'_{\particle{d}} \,\chid + f'_{\particle{s}} \,\chis \,,
\end{equation}
where $f'_{\particle{d}}$ and $f'_{\particle{s}}$ are 
the fractions of \Bd and \Bs hadrons 
in a sample of semileptonic \b-hadron decays, and where \chid and \chis 
are the \Bd and \Bs time-integrated mixing probabilities.
Assuming that all \b hadrons have the same semileptonic decay width implies 
$f'_i = f_i R_i$, where $R_i = \tau_i/\tau_{\particle{b}}$ is the ratio of the lifetime 
$\tau_i$ of species $i$ to the average \b-hadron lifetime 
$\tau_{\particle{b}} = \sum_i f_i \tau_i$.
Hence measurements of the mixing probabilities
\chibar, \chid and \chis can be used to improve our 
knowledge of \fBu, \fBd, \fBs and \fbb.
In practice, the above relations yield another determination of 
\fBs obtained from \fbb and mixing information, 
\begin{equation}
\fBs = \frac{1}{R_{\particle{s}}}
\frac{(1+r)\overline{\chi}-(1-\fbb R_{\rm baryon}) \chid}{(1+r)\chis - \chid} \,,
\labe{fBs-mixing}
\end{equation}
where $r=R_{\particle{u}}/R_{\particle{d}} = \tau(\Bu)/\tau(\Bd)$.

\labs{chibar}
The published measurements of \chibar performed by the LEP
experiments have been combined by the LEP Electroweak Working Group to yield 
$\chibar = \hfagCHIBARLEP$~\cite{LEPEWWG}. This can be compared with a 
recent measurement from CDF, $\chibar = \hfagCHIBARTEV$~\cite{CDF-chibar}, 
obtained from an analysis of the Run I data. The two estimates deviate
from each other by $\hfagCHIBARSFACTOR\,\sigma$,
and could be an indication that the production fractions of \b hadrons 
at the \particle{Z} peak or at the Tevatron are not the same. 
Although this discrepancy 
is not very significant it should be carefully monitored in the future. 
We choose to combine these two results in a simple weighted average,
assuming no correlations, and, following the PDG prescription, we 
multiply the combined uncertainty by \hfagCHIBARSFACTOR to account 
for the discrepancy. Our world average is then
\begin{equation}
\chibar = \hfagCHIBAR \,.
\end{equation}

\begin{table}
\caption{Fractions of the different \b-hadron species in an unbiased sample of 
weakly-decaying \b hadrons produced at high energy, obtained from both direct
and mixing measurements.}
\labt{fractions}
\begin{center}
\begin{tabular}{crcc}
\hline
\b-hadron & \multicolumn{1}{c}{Fraction} & \multicolumn{2}{l}{Correlation coefficients} \\
species   &          & with $\fBd=\fBu$ & and \fBs\\
\hline
\Bd, \Bu   & $\fBd=\fBu = \hfagFBD$  & & \\
\Bs        & $\fBs = \hfagFBS$       & \hfagRHOFBDFBS & \\
\b baryons & $\fbb = \hfagFBB$       & \hfagRHOFBDFBB & \hfagRHOFBBFBS \\
\hline
\end{tabular}
\end{center}
\end{table}

Introducing the latter result in \Eq{fBs-mixing}, together with our world average 
$\chid = \hfagCHIDWU$ (see \Eq{chid} of \Sec{dmd}), the assumption $\chis= 1/2$ 
(justified by the large value of \dms, see \Eq{chis} in \Sec{dms}), the 
best knowledge of the lifetimes (see \Sec{lifetimes}) and the estimate of \fbb given above, 
yields $\fBs = \hfagFBSMIX$, an estimate dominated by the mixing information. 
Taking into account all known correlations (including the one introduced by \fbb), 
this result is then combined with the set of fractions obtained from direct measurements 
(given above), to yield the 
improved estimates of \Table{fractions}, 
still under the constraints of \Eq{constraints}. 
As can be seen, our knowledge on the mixing parameters 
substantially reduces the uncertainty on \fBs, despite the rather strong 
deweighting introduced in the computation of the world average of \chibar.
It should be noted that the results 
are correlated, as indicated in \Table{fractions}.


%
%

\mysubsection{\b-hadron lifetimes}
\labs{lifetimes}

In the spectator model the decay of \b-flavored hadrons $H_b$ is
governed entirely by the flavor changing \particle{b\to Wq} transition
($\particle{q}=\particle{c,u}$).  For this very reason, lifetimes of all
\b-flavored hadrons are the same in the spectator approximation
regardless of the (spectator) quark content of the $H_b$.  In the early
1990's experiments became sophisticated enough to start seeing the
differences of the lifetimes among various $H_b$ species.  The first
theoretical calculations of the spectator quark effects on $H_b$
lifetime emerged only few years earlier.

Currently, most of such calculations are performed in the framework of
the Heavy Quark Expansion, HQE.  In the HQE, under certain assumptions
(most important of which is that of quark-hadron duality), the decay
rate of an $H_b$ to an inclusive final state $f$ is expressed as the sum
of a series of expectation values of operators of increasing dimension,
multiplied by the correspondingly higher powers of $\Lambda_{\rm
QCD}/m_b$:
\begin{equation}
\Gamma_{H_b\to f} = |CKM|^2\sum_n c_n^{(f)}
\Bigl(\frac{\Lambda_{\rm QCD}}{m_b}\Bigr)^n\langle H_b|O_n|H_b\rangle,
\labe{hqe}
\end{equation}
where $|CKM|^2$ is the relevant combination of the CKM matrix elements.
Coefficients $c_n^{(f)}$ of this expansion, known as Operator Product
Expansion~\cite{OPE}, can be calculated perturbatively.  Hence, the HQE
predicts $\Gamma_{H_b\to f}$ in the form of an expansion in both
$\Lambda_{\rm QCD}/m_{\b}$ and $\alpha_s(m_{\b})$.  The precision of
current experiments makes it mandatory to go to the next-to-leading
order in QCD, {\em i.e.}\ to include correction of the order of
$\alpha_s(m_{\b})$ to the $c_n^{(f)}$'s.  All non-perturbative physics
is shifted into the expectation values $\langle H_b|O_n|H_b\rangle$ of
operators $O_n$.  These can be calculated using lattice QCD or QCD sum
rules, or can be related to other observables via the
HQE~\cite{Bigi_1995}.  One may reasonably expect that powers of
$\Lambda_{\rm QCD}/m_{\b}$ provide enough suppression that only the
first few terms of the sum in \Eq{hqe} matter.

Theoretical predictions are usually made for the ratios of the lifetimes
(with $\tau(\Bd)$ chosen as the common denominator) rather than for the
individual lifetimes, for this allows several uncertainties to cancel.
The precision of the current HQE calculations (see
\Refs{nlo_lifetimes,tarantino,Gabbiani_et_al} for the latest updates)
is in some instances already surpassed by the measurements,
\eg\ in the case of $\tau(\Bu)/\tau(\Bd)$.  Also, HQE calculations are
not assumption-free.  More accurate predictions are a matter of progress
in the evaluation of the non-perturbative hadronic matrix elements and
verifying the assumptions that the calculations are based upon.
However, the HQE, even in its present shape, draws a number of important
conclusions, which are in agreement with experimental observations:
\begin{itemize}
\item The heavier the mass of the heavy quark the smaller is the
  variation in the lifetimes among different hadrons containing this
  quark, which is to say that as $m_{\b}\to\infty$ we retrieve the
  spectator picture in which the lifetimes of all $H_b$'s are the same.
  This is well illustrated by the fact that lifetimes in the $b$ sector
  are all very similar, while in the $c$ sector
  ($m_{\particle{c}}<m_{\b}$) lifetimes differ by as much as a factor of
  2.
\item The non-perturbative corrections arise only at the order of
  $\Lambda_{\rm QCD}^2/m_{\b}^2$, which translates into 
  differences among $H_b$ lifetimes of only a few percent.
\item It is only the difference between meson and baryon lifetimes that
  appears at the $\Lambda_{\rm QCD}^2/m_{\b}^2$ level.  The splitting of the
  meson lifetimes occurs at the $\Lambda_{\rm QCD}^3/m_{\b}^3$ level, yet it is
  enhanced by a phase space factor $16\pi^2$ with respect to the leading
  free \b decay.
\end{itemize}

To ensure that certain sources of systematic uncertainty cancel, 
lifetime analyses are sometimes designed to measure a 
ratio of lifetimes.  However, because of the differences in decay
topologies, abundance (or lack thereof) of decays of a certain kind,
{\em etc.}, measurements of the individual lifetimes are more 
common.  In the following section we review the most common
types of the lifetime measurements.  This discussion is followed by the
presentation of the averaging of the various lifetime measurements, each
with a brief description of its particularities.



\mysubsubsection{Lifetime measurements, uncertainties and correlations}

In most cases lifetime of an $H_b$ is estimated from a flight distance
and a $\beta\gamma$ factor which is used to convert the geometrical
distance into the proper decay time.  Methods of accessing lifetime
information can roughly be divided in the following five categories:
\begin{enumerate}
\item {\bf\em Inclusive (flavor blind) measurements}.  These
  measurements are aimed at extracting the lifetime from a mixture of
  \b-hadron decays, without distinguishing the decaying species.  Often
  the knowledge of the mixture composition is limited, which makes these
  measurements experiment-specific.  Also, these
  measurements have to rely on Monte Carlo for estimating the
  $\beta\gamma$ factor, because the decaying hadrons are not fully
  reconstructed.  On the bright side, these usually are the largest
  statistics \b-hadron lifetime measurements that are accessible to a
  given experiment, and can, therefore, serve as an important
  performance benchmark.
\item {\it\bf Measurements in semileptonic decays of a specific
  {\boldmath $H_b$\unboldmath}}.  \particle{W}from \particle{\b\to Wc}
  produces $\ell\nu_l$ pair (\particle{\ell=e,\mu}) in about 21\% of the
  cases.  Electron or muon from such decays is usually a well-detected
  signature, which provides for clean and efficient trigger.
  \particle{c} quark from \particle{b\to Wc} transition and the other
  quark(s) making up the decaying $H_b$ combine into a charm hadron,
  which is reconstructed in one or more exclusive decay channels.
  Knowing what this charmed hadron is allows one to separate, at least
  statistically, different $H_b$ species.  The advantage of these
  measurements is in statistics, which usually is superior to that of the
  exclusively reconstructed $H_b$ decays.  Some of the main
  disadvantages are related to the difficulty of estimating lepton+charm
  sample composition and Monte Carlo reliance for the $\beta\gamma$
  factor estimate.
\item {\bf\em Measurements in exclusively reconstructed decays}.  These
  have the advantage of complete reconstruction of decaying $H_b$, which
  allows one to infer the decaying species as well as to perform precise
  measurement of the $\beta\gamma$ factor.  Both lead to generally
  smaller systematic uncertainties than in the above two categories.
  The downsides are smaller branching ratios, larger combinatoric
  backgrounds, especially in $H_b\rightarrow H_c\pi(\pi\pi)$ and
  multi-body $H_c$ decays, or in a hadron collider environment with
  non-trivial underlying event.  $H_b\to J/\psi H_s$ are relatively
  clean and easy to trigger on $J/\psi\to \ell^+\ell^-$, but their
  branching fraction is only about 1\%.
\item {\bf\em Measurements at asymmetric B factories}. In 
  the $\Ups\rightarrow B \bar{B}$ decay, the \B mesons (\Bu or \Bd) are
  essentially at rest in the \Ups rest frame.  This makes lifetime
  measurements impossible with experiments, such as CLEO, in which \Ups
  produced at rest.  At asymmetric \B factories \Ups is boosted
  resulting in \B and \particle{\bar{B}} moving nearly parallel to each
  other.  The lifetime is inferred from the distance $\Delta z$
  separating \B and \particle{\bar{B}} decay vertices and \Ups boost
  known from colliding beam energies.  In order to maximize the
  precision of the measurement, one \B meson is reconstructed in the
  \particle{D^{(*)}\ell\nu_{\ell}} decay.  The other \B is typically not
  fully reconstructed, only position of its decay vertex is determined.
  These measurements benefit from very large statistics, but suffer from
  poor $\Delta z$ resolution.
\item {\bf\em Direct measurement of lifetime ratios}.  This method has
  so far been only applied in the measurement of $\tau(\Bu)/\tau(\Bd)$.
  The ratio of the lifetimes is extracted from the dependence of the
  observed relative number of \Bu and \Bd candidates (both reconstructed
  in semileptonic decays) on the proper decay time.
\end{enumerate}

In some of the latest analyses, measurements of two (\eg\ $\tau(\Bu)$ and
$\tau(\Bu)/\tau(\Bd)$) or three (\eg\ $\tau(\Bu)$,
$\tau(\Bu)/\tau(\Bd)$, and \dmd) quantities are combined.  This
introduces correlations among measurements.  Another source of
correlations among the measurements are the systematic effects, which
could be common to an experiment or to an analysis technique across the
experiments.  When calculating the averages, such correlations are taken
into account per general procedure, described in
\Ref{lifetime_details}.

\mysubsubsection{Inclusive \b-hadron lifetimes}

The inclusive \b hadron lifetime is defined as $\tau_{\b} = \sum_i f_i
\tau_i$ where $\tau_i$ are the individual species lifetimes and $f_i$ are
the fractions of the various species present in an unbiased sample of
weakly-decaying \b hadrons produced at a high-energy
collider.\footnote{In principle such a quantity could be slightly
different in \particle{Z} decays and a the Tevatron, in case the
fractions of \b-hadron species are not exactly the same; see the
discussion in \Sec{fractions_high_energy}.}  This quantity is certainly
less fundamental than the lifetimes of the individual species, the
latter being much more useful in comparisons of the measurements with
the theoretical predictions.  Nonetheless, we perform the averaging of
the inclusive lifetime measurements for completeness as well as for the
reason that they might be of interest as ``technical numbers.''

\begin{table}[tp]
\caption{Measurements of average \b-hadron lifetimes.}
\labt{lifeincl}
\begin{center}
\begin{tabular}{lcccl} \hline
Experiment &Method           &Data set & $\tau_{\b}$ (ps)       &Ref.\\
\hline
ALEPH  &Dipole               &91     &$1.511\pm 0.022\pm 0.078$ &\cite{ALEIN2}\\
DELPHI &All track i.p.\ (2D) &91--92 &$1.542\pm 0.021\pm 0.045$ &\cite{DELIN0}$^a$\\
DELPHI &Sec.\ vtx            &91--93 &$1.582\pm 0.011\pm 0.027$ &\cite{DELIN}$^a$\\
DELPHI &Sec.\ vtx            &94--95 &$1.570\pm 0.005\pm 0.008$ &\cite{DELB04}\\
L3     &Sec.\ vtx + i.p.     &91--94 &$1.556\pm 0.010\pm 0.017$ &\cite{L3IN1}$^b$\\
OPAL   &Sec.\ vtx            &91--94 &$1.611\pm 0.010\pm 0.027$ &\cite{OPAIN2}\\
SLD    &Sec.\ vtx            &93     &$1.564\pm 0.030\pm 0.036$ &\cite{SLDIN}\\ 
\hline
\multicolumn{2}{l}{Average set 1 (\b vertex)} && \hfagTAUBVTXnounit &\\
\hline\hline
ALEPH  &Lepton i.p.\ (3D)    &91--93 &$1.533\pm 0.013\pm 0.022$ &\cite{ALEIN1}\\
L3     &Lepton i.p.\ (2D)    &91--94 &$1.544\pm 0.016\pm 0.021$ &\cite{L3IN1}$^b$\\
OPAL   &Lepton i.p.\ (2D)    &90--91 &$1.523\pm 0.034\pm 0.038$ &\cite{OPAIN1}\\ 
\hline
\multicolumn{2}{l}{Average set 2 ($\b\to\ell$)} && \hfagTAUBLEPnounit &\\
\hline\hline
CDF    &\particle{J/\psi} vtx&92--95 &$1.533\pm 0.015^{+0.035}_{-0.031}$ &\cite{CDFIN_BS1} \\ 
\hline\hline
\multicolumn{2}{l}{Average of all above} && \hfagTAUBnounit & \\
\hline
\multicolumn{5}{l}{$^a$ \footnotesize The combined DELPHI result quoted in
\cite{DELIN} is 1.575 $\pm$ 0.010 $\pm$ 0.026 ps.} \\[-0.5ex]
\multicolumn{5}{l}{$^b$ \footnotesize The combined L3 result quoted in \cite{L3IN1} 
is 1.549 $\pm$ 0.009 $\pm$ 0.015 ps.}
\end{tabular}
\end{center}
\end{table}

In practice, an unbiased measurement of the inclusive lifetime is
difficult to achieve, because it would imply an efficiency which is
guaranteed to be the same across species.  So most of the measurements
are biased.  In an attempt to group analyses which are expected to
select the same mixture of \b hadrons, the available results (given in
\Table{lifeincl}) are divided into the following three sets:
\begin{enumerate}
\item measurements at LEP and SLD that accept any \b-hadron decay, based 
      on topological reconstruction (secondary vertex or track impact
      parameters);
\item measurements at LEP based on the identification
      of a lepton from a \b decay; and
\item measurements at the Tevatron based on inclusive 
      \particle{H_b\to J/\psi X} reconstruction, where the
      \particle{J/\psi} is fully reconstructed.
\end{enumerate}

The measurements of the first set are generally considered as estimates
of $\tau_{\b}$, although the efficiency to reconstruct a secondary
vertex most probably depends, in an analysis-specific way, on the number
of tracks coming from the vertex, thereby depending on the type of the
$H_b$.  Even though these efficiency variations can in principle be
accounted for using Monte Carlo simulations (which inevitably contain
assumptions on branching fractions), the $H_b$ mixture in that case can
remain somewhat ill-defined and could be slightly different among
analyses in this set.

On the contrary, the mixtures corresponding to the other two sets of
measurements are better defined in the limit where the reconstruction
and selection efficiency of a lepton or a \particle{J/\psi} from an
$H_b$ does not depend on the decaying hadron type.  These mixtures are
given by the production fractions and the inclusive branching fractions
for each $H_b$ species to give a lepton or a \particle{J/\psi}.  In
particular, under the assumption that all \b hadrons have the same
semileptonic decay width, the analyses of the second set should measure
$\tau(\b\to\ell) = (\sum_i f_i \tau_i^2) /(\sum_i f_i \tau_i)$ which is
necessarily larger than $\tau_{\b}$ if lifetime differences exist.
Given the present knowledge on $\tau_i$ and $f_i$,
$\tau(\b\to\ell)-\tau_{\b}$ is expected to be of the order of 0.01\ps.

Measurements by SLC and LEP experiments are subject to a number of
common systematic uncertainties, such as those due to (lack of knowledge
of) \b and \particle{c} fragmentation, \b and \particle{c} decay models,
\BR{B\to\ell}, \BR{B\to c\to\ell}, \BR{c\to\ell}, $\tau_{\particle{c}}$,
and $H_b$ decay multiplicity.  In the averaging, these systematic
uncertainties are assumed to be 100\% correlated.  The averages for the
sets defined above (also given in \Table{lifeincl}) are
\begin{eqnarray}
\tau(\b~\mbox{vertex}) &=& \hfagTAUBVTX \,,\\
\tau(\b\to\ell) &=& \hfagTAUBLEP  \,, \\
\tau(\b\to\particle{J/\psi}) &=& \hfagTAUBJP\,,
\end{eqnarray}
whereas an average of all measurements, ignoring mixture differences, 
yields \hfagTAUB.

\mysubsubsection{\Bd and \Bu lifetimes and their ratio}
\labs{taubd}
\labs{taubu}
\labs{lifetime_ratio}

\begin{table}[tp]
\caption{Measurements of the \Bd lifetime.}
\labt{lifebd}
\begin{center}
\begin{tabular}{lcccl} \hline
Experiment &Method                    &Data set &$\tau(\Bd)$ (ps)                  &Ref.\\
\hline
ALEPH  &\particle{D^{(*)} \ell}       &91--95 &$1.518\pm 0.053\pm 0.034$          &\cite{ALEB01}\\
ALEPH  &Exclusive                     &91--94 &$1.25^{+0.15}_{-0.13}\pm 0.05$     &\cite{ALEB0}\\
ALEPH  &Partial rec.\ $\pi^+\pi^-$    &91--94 &$1.49^{+0.17+0.08}_{-0.15-0.06}$   &\cite{ALEB0}\\
DELPHI &\particle{D^{(*)} \ell}       &91--93 &$1.61^{+0.14}_{-0.13}\pm 0.08$     &\cite{DELB01}\\
DELPHI &Charge sec.\ vtx              &91--93 &$1.63 \pm 0.14 \pm 0.13$           &\cite{DELB02}\\
DELPHI &Inclusive \particle{D^* \ell} &91--93 &$1.532\pm 0.041\pm 0.040$          &\cite{DELB03}\\
DELPHI &Charge sec.\ vtx              &94--95 &$1.531 \pm 0.021\pm0.031$          &\cite{DELB04}\\
L3     &Charge sec.\ vtx              &94--95 &$1.52 \pm 0.06 \pm 0.04$           &\cite{L3B01}\\
OPAL   &\particle{D^{(*)} \ell}       &91--93 &$1.53 \pm 0.12 \pm 0.08$           &\cite{OPAB0}\\
OPAL   &Charge sec.\ vtx              &93--95 &$1.523\pm 0.057\pm 0.053$          &\cite{OPAB1}\\
OPAL   &Inclusive \particle{D^* \ell} &91--00 &$1.541\pm 0.028\pm 0.023$          &\cite{OPAB2}\\
SLD    &Charge sec.\ vtx $\ell$       &93--95 &$1.56^{+0.14}_{-0.13} \pm 0.10$    &\cite{SLDB01}$^a$\\
SLD    &Charge sec.\ vtx              &93--95 &$1.66 \pm 0.08 \pm 0.08$           &\cite{SLDB01}$^a$\\
CDF    &\particle{D^{(*)} \ell}       &92--95 &$1.474\pm 0.039^{+0.052}_{-0.051}$ &\cite{CDFB1}\\
CDF    &Excl. \particle{J/\psi K^{*0}}&92--95 &$1.497\pm 0.073\pm 0.032$          &\cite{CDFB2}\\
CDF    &Excl. \particle{J/\psi K^{*0}}&02--04 &$1.539\pm 0.051\pm0.008$           &\cite{CDFB3}$^p$\\
CDF    &Incl.\ \particle{D^{(*)} \ell}&02--04 &$1.473\pm 0.036\pm0.054$           &\cite{CDFB4}$^p$\\
CDF    &Excl.\ \particle{D^-(3)\pi}   &02--04 &$1.511\pm 0.023\pm0.013$           &\cite{CDFB5}$^p$\\
\dzero &Excl. \particle{J/\psi K^{*0}}&02--04 &$1.473^{+0.052}_{-0.050}\pm0.023$  &\cite{D0BS1}\\
\dzero &Excl. \particle {J/\psi K_S}  &02--04 &$1.400^{+0.110}_{-0.100}\pm0.030$  &\cite{D0LAMB}\\
\babar &Exclusive                     &99--00 &$1.546\pm 0.032\pm 0.022$          &\cite{BABAR1}\\
\babar &Inclusive \particle{D^* \ell} &99--01 &$1.529\pm 0.012\pm 0.029$          &\cite{BABAR2}\\
\babar &Exclusive \particle{D^* \ell} &99--02 &$1.523^{+0.024}_{-0.023}\pm 0.022$ &\cite{BABAR3}\\
\babar &Incl.\ \particle{D^*\pi}, \particle{D^*\rho} 
                                      &99--01 &$1.533\pm 0.034 \pm 0.038$         &\cite{BABAR4}\\
\babar &Inclusive \particle{D^* \ell} &99--04 &$1.501\pm0.008\pm0.030$            &\cite{BABAR5}$^p$\\
\belle & Exclusive                     & 00--03 & $1.534\pm 0.008\pm0.010$        & \cite{BELLE2}\\
\hline
Average&                               &        & \hfagTAUBDnounit & \\
\hline\hline           
\multicolumn{5}{l}{$^a$ \footnotesize The combined SLD result 
quoted in \cite{SLDB01} is 1.64 $\pm$ 0.08 $\pm$ 0.08 ps.}\\[-0.5ex]
\multicolumn{5}{l}{$^p$ {\footnotesize Preliminary.}}
\end{tabular}
\end{center}
\end{table}

After a number of years of dominating these averages LEP experiments
culminated with the recent publication~\cite{DELB04} by DELPHI
collaboration and yielded the scene to the asymmetric \B~factories and
the Tevatron experiments.  The \B~factories have been very successful in
utilizing their potential -- in only a few years of running, \babar and,
to a greater extent, \belle, have struck a balance between the
statistical and the systematic uncertainties, with both being close to
(or even better than) the impressive 1\%.  In the meanwhile, CDF and
\dzero have emerged as significant contributors to the field as the
Tevatron Run~II data flowed in.  Both appear to enjoy relatively small
systematic effects, and while current statistical uncertainties of their
measurements are factors of 2 to 4 larger than those of their \B-factory
counterparts, both Tevatron experiments stand to increase their samples
by an order of magnitude.

\begin{table}[tbp]
\caption{Measurements of the \Bu lifetime.}
\labt{lifebu}
\begin{center}
\begin{tabular}{lcccl} \hline
Experiment &Method                 &Data set &$\tau(\Bu)$ (ps)                 &Ref.\\
\hline
ALEPH  &\particle{D^{(*)} \ell}    &91--95 &$1.648\pm 0.049\pm 0.035$          &\cite{ALEB01}\\
ALEPH  &Exclusive                  &91--94 &$1.58^{+0.21+0.04}_{-0.18-0.03}$   &\cite{ALEB0}\\
DELPHI &\particle{D^{(*)} \ell}    &91--93 &$1.61\pm 0.16\pm 0.12$             &\cite{DELB01}$^a$\\
DELPHI &Charge sec.\ vtx           &91--93 &$1.72\pm 0.08\pm 0.06$             &\cite{DELB02}$^a$\\
DELPHI &Charge sec.\ vtx           &94--95 &$1.624\pm 0.014\pm 0.018$          &\cite{DELB04}\\
L3     &Charge sec.\ vtx           &94--95 &$1.66\pm  0.06\pm 0.03$            &\cite{L3B01}\\
OPAL   &\particle{D^{(*)} \ell}    &91--93 &$1.52 \pm 0.14\pm 0.09$            &\cite{OPAB0}\\
OPAL   &Charge sec.\ vtx           &93--95 &$1.643\pm 0.037\pm 0.025$          &\cite{OPAB1}\\
SLD    &Charge sec.\ vtx $\ell$    &93--95 &$1.61^{+0.13}_{-0.12}\pm 0.07$     &\cite{SLDB01}$^b$\\
SLD    &Charge sec.\ vtx           &93--95 &$1.67\pm 0.07\pm 0.06$             &\cite{SLDB01}$^b$\\
CDF    &\particle{D^{(*)} \ell}    &92--95 &$1.637\pm 0.058^{+0.045}_{-0.043}$ &\cite{CDFB1}\\
CDF    &Excl.\ \particle{J/\psi K} &92--95 &$1.636\pm 0.058\pm 0.025$          &\cite{CDFB2}\\
CDF    &Excl.\ \particle{J/\psi K} &02--04 &$1.662\pm 0.033\pm 0.008$          &\cite{CDFB3}$^p$\\
CDF    &Incl.\ \particle{D^0 \ell} &02--04 &$1.653\pm 0.029^{+0.033}_{-0.031}$ &\cite{CDFB4}$^p$\\
CDF    &Excl.\ \particle{D^0 \pi}  &02--04 &$1.661\pm 0.027\pm0.013$           &\cite{CDFB5}$^p$\\
\babar &Exclusive                  &99--00 &$1.673\pm 0.032\pm 0.023$          &\cite{BABAR1}\\
\belle &Exclusive                  &00--03 &$1.635\pm 0.011\pm 0.011$          &\cite{BELLE2}\\
\hline
Average&                           &       &\hfagTAUBUnounit &\\
\hline\hline
\multicolumn{5}{l}{$^a$ \footnotesize The combined DELPHI result quoted 
in~\cite{DELB02} is $1.70 \pm 0.09$ ps.} \\[-0.5ex]
\multicolumn{5}{l}{$^b$ \footnotesize The combined SLD result 
quoted in~\cite{SLDB01} is $1.66 \pm 0.06 \pm 0.05$ ps.}\\[-0.5ex]
\multicolumn{5}{l}{$^p$ {\footnotesize Preliminary.}}
\end{tabular}
\end{center}
\end{table}

At present time we are in an interesting position of having three sets
of measurements (from LEP/SLC, \B factories and the Tevatron) that
originate from different environments, obtained using substantially
different techniques and are precise enough for incisive comparison.


\begin{table}[tb]
\caption{Measurements of the ratio $\tau(\Bu)/\tau(\Bd)$.}
\labt{liferatioBuBd}
\begin{center}
\begin{tabular}{lcccl} 
\hline
Experiment &Method                 &Data set &Ratio $\tau(\Bu)/\tau(\Bd)$      &Ref.\\
\hline
ALEPH  &\particle{D^{(*)} \ell}    &91--95 &$1.085\pm 0.059\pm 0.018$          &\cite{ALEB01}\\
ALEPH  &Exclusive                  &91--94 &$1.27^{+0.23+0.03}_{-0.19-0.02}$   &\cite{ALEB0}\\
DELPHI &\particle{D^{(*)} \ell}    &91--93 &$1.00^{+0.17}_{-0.15}\pm 0.10$     &\cite{DELB01}\\
DELPHI &Charge sec.\ vtx           &91--93 &$1.06^{+0.13}_{-0.11}\pm 0.10$     &\cite{DELB02}\\
DELPHI &Charge sec.\ vtx           &94--95 &$1.060\pm 0.021 \pm 0.024$         &\cite{DELB04}\\
L3     &Charge sec.\ vtx           &94--95 &$1.09\pm 0.07  \pm 0.03$           &\cite{L3B01}\\
OPAL   &\particle{D^{(*)} \ell}    &91--93 &$0.99\pm 0.14^{+0.05}_{-0.04}$     &\cite{OPAB0}\\
OPAL   &Charge sec.\ vtx           &93--95 &$1.079\pm 0.064 \pm 0.041$         &\cite{OPAB1}\\
SLD    &Charge sec.\ vtx $\ell$    &93--95 &$1.03^{+0.16}_{-0.14} \pm 0.09$    &\cite{SLDB01}$^a$\\
SLD    &Charge sec.\ vtx           &93--95 &$1.01^{+0.09}_{-0.08} \pm0.05$     &\cite{SLDB01}$^a$\\
CDF    &\particle{D^{(*)} \ell}    &92--95 &$1.110\pm 0.056^{+0.033}_{-0.030}$ &\cite{CDFB1}\\
CDF    &Excl.\ \particle{J/\psi K} &92--95 &$1.093\pm 0.066 \pm 0.028$         &\cite{CDFB2}\\
CDF    &Excl.\ \particle{J/\psi K} &02--04 &$1.080\pm 0.042$                   &\cite{CDFB3}$^p$\\
CDF    &Incl.\ \particle{D \ell}   &02--04 &$1.123\pm0.040^{+0.041}_{-0.039}$  &\cite{CDFB4}$^p$\\
CDF    &Excl.\ \particle{D \pi}    &02--04 &$1.10\pm 0.02\pm 0.01$             &\cite{CDFB5}$^p$\\
\dzero &\particle{D^{*+} \mu} \particle{D^0 \mu} ratio
	                           &02--04 &$1.080\pm 0.016\pm 0.014$          &\cite{D0B01}$^p$\\
\babar &Exclusive                  &99--00 &$1.082\pm 0.026\pm 0.012$          &\cite{BABAR1}\\
\belle &Exclusive                  &00--03 &$1.066\pm 0.008\pm 0.008$          &\cite{BELLE2}\\
\hline
Average&                           &       & \hfagRTAUBU & \\   
\hline\hline
\multicolumn{5}{l}{$^a$ \footnotesize The combined SLD result quoted
	   in~\cite{SLDB01} is $1.01 \pm 0.07 \pm 0.06$.} \\[-0.5ex] 
\multicolumn{5}{l}{$^p$ {\footnotesize Preliminary.}}
\end{tabular}
\end{center}
\end{table}

The averaging of $\tau(\Bu)$, $\tau(\Bd)$ and $\tau(\Bu)/\tau(\Bd)$
measurements is summarized in \Tablesss{lifebd}{lifebu}{liferatioBuBd}.
For $\tau(\Bu)/\tau(\Bd)$ we averaged only the measurements of this
quantity provided by experiments rather than using all available
knowledge, which would have included, for example, $\tau(\Bu)$ and
$\tau(\Bd)$ measurements which did not contribute to any of the ratio
measurements.

The following sources of correlated (within experiment/machine)
systematic uncertainties have been considered:
\begin{itemize}
\item for SLC/LEP measurements -- \particle{D^{**}} branching ratio uncertainties~\cite{LEPHFS},
momentum estimation of \b mesons from \particle{Z^0} decays
(\b-quark fragmentation parameter $\langle X_E \rangle = 0.702 \pm 0.008$~\cite{LEPHFS}),
\Bs and \b baryon lifetimes (see \Secss{taubs}{taulb}),
and \b hadron fractions at high energy (see \Table{fractions}).  
\item for \babar measurements -- alignment, $z$ scale, PEP-II boost,
sample composition (where applicable) 
\item for \dzero and CDF Run~II measurements -- alignment (separately
within each experiment)
\end{itemize}
The resultant averages are:
\begin{eqnarray}
\tau(\Bd) & = & \hfagTAUBD \,, \\
\tau(\Bu) & = & \hfagTAUBU \,, \\
\tau(\Bu)/\tau(\Bd) & = & \hfagRTAUBU \,.
\end{eqnarray}
%
%
%

\mysubsubsection{\Bs lifetime}
\labs{taubs}

Similar to the kaon system, neutral \B mesons contain
short- and long-lived components, since the
light (L) and heavy  (H)
eigenstates, $\B_{\rm L}$ and $\B_{\rm H}$, differ not only
in their masses, but also in their widths 
with\footnote{The sign convention used here for
\DGs is the one adopted by the authors of the 
analyses measuring \DGs and is opposite to 
that used for \DGd in \Sec{DGd}.}
$\Delta\Gamma = \Gamma_{\rm L} - \Gamma_{\rm H}$. 
In the case of the \Bs system, $\DGs$ can
be particularly large. The current theoretical
prediction in the Standard Model for
the fractional width difference is
$\DGs/\Gs = 0.12 \pm 0.05$~\cite{delta_gams},
where $\Gs = (\Gamma_{\rm L} + \Gamma_{\rm H})/2$.
Specific measurements of \DGs and \Gs are explained
in \Sec{DGs}, but the result for
\Gs is quoted here.

Neglecting \CP violation, which is expected to be
small in the \Bs system~\cite{delta_gams}, the
\Bs mass eigenstates are also \CP eigenstates. In
the Standard Model assuming no \CP violation in
the \Bs system,
$\Gamma_{\rm L}$ is the width of
the \CP-even state and
$\Gamma_{\rm H}$ the width of
the \CP-odd state.
Final states can be decomposed into
\CP-even and \CP-odd components, each with a different
lifetime.

In view of a possibly substantial width difference,
and the fact that various
decay channels will have different proportions of 
the $\B_{\rm L}$ and $\B_{\rm H}$ eigenstates,
the straight average of all available 
\Bs lifetime measurements
is rather ill-defined.  Therefore,
the \Bs lifetime measurements are broken down into
three categories and averaged separately.

\begin{itemize}
\item 
{\bf\em Flavor-specific decays}, such as semileptonic
$\particle{B_s} \to \particle{D_s \ell \nu}$
or $\particle{B_s} \to \particle {D_s \pi}$, will
have equal 
fractions of $\B_{\rm L}$ and $\B_{\rm H}$ at time
zero, where
$\tau_{\rm L} = 1/\Gamma_{\rm L}$ 
is expected to be the shorter-lived component and
$\tau_{\rm H} = 1/\Gamma_{\rm H}$ 
expected
to be the longer-lived component.  A superposition
of two exponentials thus results with decay
widths $\Gs \pm \DGs /2$.
Fitting to a single exponential results in a
measure of a flavor-specific 
lifetime, one obtains~\cite{Hartkorn_Moser}:
\begin{equation}
\tau(\Bs)_{\rm fs} = \frac{1}{\Gs}
\frac{{1+\left(\frac{\DGs}{2\Gs}\right)^2}}{{1-\left(\frac{\DGs}{2\Gs}\right)^2}
}.
\end{equation}
As given in \Table{lifebs}, the flavor-specific 
\Bs lifetime world average is:
\begin{equation}
\tau(\Bs)_{\rm fs} = \hfagTAUBSSL \,.
\end{equation}
This world average will be used later in \Sec{DGs} in combination
with other measurements to find
$\bar{\tau}(\Bs) = 1/\Gs$ and $\DGs$.

The following correlated systematic errors were considered:
average \B lifetime used in backgrounds,
\Bs decay multiplicity, and branching ratios used to determine 
backgrounds (\eg\ \BR{B\to D_s D}).
A knowledge of the multiplicity of \Bs decays is important for
measurements that partially reconstruct the final state such as 
\particle{\B\to D_s \mbox{$X$}} (where $X$ is not a lepton). 
The boost deduced from Monte Carlo simulation depends on the multiplicity used.
Since this is not well known, the multiplicity in the simulation is
varied and this range of values observed is taken to be a systematic.
Similarly not all the branching ratios for the potential background
processes are measured. Where they are available, the PDG values are
used for the error estimate. Where no measurements are available
estimates can usually be made by using measured branching ratios of
related processes and using some reasonable extrapolation.
\end{itemize}



\begin{table}[tb]
\caption{Measurements of the \Bs lifetime.}
\labt{lifebs}
\begin{center}
\begin{tabular}{lcccl} \hline
Experiment & Method           & Data set & $\tau(\Bs)$ (ps)               & Ref. \\
\hline
ALEPH  & \particle{D_s \ell}  & 91--95 & $1.54^{+0.14}_{-0.13}\pm 0.04$   & \cite{ALEBS1}          \\
CDF    & \particle{D_s \ell}  & 92--96 & $1.36\pm 0.09 ^{+0.06}_{-0.05}$  & \cite{CDFBS}           \\
DELPHI & \particle{D_s \ell}  & 91--95 & $1.42^{+0.14}_{-0.13}\pm 0.03$   & \cite{DELBS0}          \\
OPAL   & \particle{D_s \ell}  & 90--95 & $1.50^{+0.16}_{-0.15}\pm 0.04$   & \cite{OPABS1_OPALAM2}  \\
\dzero & \particle{D_s \mu}  & 02--04 & $1.420 \pm 0.043 \pm 0.057   $   & \cite{D0BS2}$^p$       \\ 
CDF    & \particle{D_s \pi, D_s \pi \pi \pi} 
                              & 02--04 & $1.60 \pm 0.10 \pm 0.02      $   & \cite{CDFBS2}$^p$      \\ \hline
\multicolumn{3}{l}{Average of flavor-specific measurements} &  \hfagTAUBSSLnounit & \\  
\hline
ALEPH  & \particle{D_s h}     & 91--95 & $1.47\pm 0.14\pm 0.08$           & \cite{ALEBS2}          \\
DELPHI & \particle{D_s h}     & 91--95 & $1.53^{+0.16}_{-0.15}\pm 0.07$   & \cite{DELBS1_dms_excl} \\
OPAL   & \particle{D_s} incl. & 90--95 & $1.72^{+0.20+0.18}_{-0.19-0.17}$ & \cite{OPABS2}          \\ 
\hline
\multicolumn{3}{l}{Average of all above \particle{D_s} measurements} &  \hfagTAUBSnounit & \\ 
\hline\hline
CDF      & \particle{J/\psi\phi} & 92--95  & $1.34^{+0.23}_{-0.19}    \pm 0.05$ & \cite{CDFIN_BS1} \\
CDF      & \particle{J/\psi\phi} & 02--04  & $1.369 \pm 0.100 ^{+0.008}_{-0.010}$ & \cite{CDFB3}$^p$ \\
\dzero   & \particle{J/\psi\phi} & 02--04  & $1.444^{+0.098}_{-0.090} \pm 0.02$ & \cite{D0BS1}  \\ \hline 
\multicolumn{3}{l}{Average of \particle{J/\psi \phi} measurements} &  \hfagTAUBSJFnounit & \\ 
\hline
\multicolumn{5}{l}{$^p$ \footnotesize Preliminary.}
\end{tabular}
\end{center}
\end{table}

\begin{itemize}
\item
{\bf\em \boldmath $\Bs\to\Ds X$ decays}.
Included in \Table{lifebs} are measurements
of lifetimes using samples of \particle{\Bs} decays to
\particle{D_s} plus
hadrons, and hence into a less known mixture
of \CP-states.  A lifetime
weighted this way can still be a useful input
for analyses examining such an inclusive sample.
These are separated in \Table{lifebs} and combined
with the semileptonic lifetime to obtain:
\begin{equation}
\tau(\Bs)_{\particle{D_s {\rm X}}} = \hfagTAUBS \,.
\end{equation}

\item
{\bf\em Fully exclusive 
{\boldmath \Bs $\to J/\psi \phi$ \unboldmath}decays}
are expected to be
dominated by the \CP-even state and its lifetime.
First measurements of the \CP mix for this decay mode
are outlined in \Sec{DGs}.
CDF and \dzero measurements from this particular mode
\particle{\Bs\to J/\psi\phi} are combined into an
average
given in \Table{lifebs}.  There are no correlations
between the measurements for this fully exclusive
channel, and the world average for this 
specific decay is:
\begin{equation}
\tau(\Bs)_{\particle{J/\psi \phi}} = \hfagTAUBSJF \,.
\end{equation}
A caveat is that different experimental acceptances
will likely lead to different admixtures of the 
\CP-even and \CP-odd states, and fits to a single
exponential may result in inherently different 
measurements of these quantities.
\end{itemize}

Finally, as will be shown in \Sec{DGs}, measurements
of $\DGs$, including separation into
\CP-even and \CP-odd components, give
\begin{equation}
\bar{\tau}(\Bs) = 1/\Gs = \hfagTAUBSMEAN \,,
\end{equation}
and when combined with the flavor-specific lifetime
measurements:
\begin{equation}
\bar{\tau}(\Bs) = 1/\Gs = \hfagTAUBSMEANCON \,.
\end{equation}

\mysubsubsection{\Bc lifetime}
\labs{taubc}

There are currently two measurements of the lifetime of the \Bc meson
from CDF~\cite{CDFBC1} and \dzero~\cite{D0BC1} using the semileptonic decay
mode \particle{\Bc \to J/\psi \ell} and fitting
simultaneously to the mass and lifetime using the vertex formed
with the leptons from the decay of the \particle{J/\psi} and
the third lepton. Correction factors
to estimate the boost due to the missing neutrino are used.
Mass values of
$6.40 \pm 0.39 \pm 0.13$~GeV/$c^2$ and 
$5.95^{+0.14}_{-0.13} \pm 0.34$~GeV/$c^2$, respectively, are
found by fitting
to the tri-lepton invariant mass spectrum. These mass measurements
are consistent to within uncertainties.
Correlated systematic errors include the impact
of the uncertainty of the \Bc $p_T$ spectrum on the correction
factors, the level of feed-down from $\psi(2S)$, 
MC modeling of the decay model varying from phase space
to the ISGW model, and mass variations.
Values of the \particle{\Bc} lifetime are given
in \Table{lifebc} and the world average is
determined to be:
\begin{equation}
\tau(\Bc) = \hfagTAUBC \,.
\end{equation}

\begin{table}[tb]
\caption{Measurements of the \Bc lifetime.}
\labt{lifebc}
\begin{center}
\begin{tabular}{lcccl} \hline
Experiment & Method                    & Data set  & $\tau(\Bc)$ (ps)
      & Ref.\\   \hline
CDF        & \particle{J/\psi \ell} & 92--95  & $0.46^{+0.18}_{-0.16} \pm
 0.03$   & \cite{CDFBC1}  \\ 
 \dzero & \particle{J/\psi \mu} & 02--04  & $0.448^{+0.123}_{-0.096} 
\pm  0.121$   & \cite{D0BC1}$^p$  \\ \hline
  \multicolumn{2}{l}{Average} &   &  \hfagTAUBCnounit
                 &    \\   \hline
\multicolumn{5}{l}{$^p$ \footnotesize Preliminary.}
\end{tabular}
\end{center}
\end{table}

\mysubsubsection{\Lb and \b-baryon lifetimes}
\labs{taulb}

The most precise measurements of the \b-baryon lifetime
originate from two classes of partially reconstructed decays.
In the first class, decays with an exclusively 
reconstructed \Lc baryon
and a lepton of opposite charge are used. These products are
more likely to occur in the decay of \Lb baryons.
In the second class, more inclusive final states with a baryon
(\particle{p}, \particle{\bar{p}}, $\Lambda$, or $\bar{\Lambda}$) 
and a lepton have been used, and these final states can generally
arise from any \b baryon.

The following sources of correlated systematic uncertainties have 
been considered:
experimental time resolution within a given experiment, \b-quark
fragmentation distribution into weakly decaying \b baryons,
\Lb polarization, decay model,
and evaluation of the \b-baryon purity in the selected event samples.
In computing the averages
the central values of the masses are scaled to 
$M(\Lb) = 5624 \pm 9\MeVcc$~\cite{PDGmass} and
$M(\mbox{\b-baryon}) = 5670 \pm 100\MeVcc$.

The meaning of decay model and the correlations are not always clear.
Uncertainties related to the decay model are dominated by
assumptions on the fraction of $n$-body decays.
To be conservative it is assumed
that it is correlated whenever given as an error.
DELPHI varies the fraction of 4-body decays from 0.0 to 0.3. 
In computing the average, the DELPHI
result is corrected for $0.2 \pm 0.2$.

Furthermore, in computing the average,
the semileptonic decay results are corrected for a polarization of 
$-0.45^{+0.19}_{-0.17}$~\cite{LEPHFS} and a 
\Lb fragmentation parameter
$\langle X_E \rangle =0.70\pm 0.03$~\cite{LBFRAG}.




Inputs to the averages are given in \Table{lifelb}.
The world average lifetime of \b baryons is then:
\begin{equation}
\langle\tau(\mbox{\b-baryon})\rangle = \hfagTAUBB \,.
\end{equation}
Keeping only \particle{\Lambda^{\pm}_c \ell^{\mp}}
and $\Lambda \ell^- \ell^+$ final states, as representative of
the \Lb baryon, the following lifetime is obtained:
\begin{equation}
\tau(\Lb) = \hfagTAULB \,. 
\end{equation}

Averaging the measurements based on the $\Xi^{\mp} \ell^{\mp}$
final states~\cite{ALEPH_fxi,DELPHI_fxi} gives
a lifetime value for a sample of events
containing $\Xib^0$ and $\Xib^-$ baryons:
\begin{equation}
\langle\tau(\Xib)\rangle = \hfagTAUXB \,.
\end{equation}

\begin{table}[t]
\caption{Measurements of the \b-baryon lifetimes.
}
\labt{lifelb}
\begin{center}
\begin{tabular}{lcccl} 
\hline
Experiment&Method                &Data set& Lifetime (ps) & Ref. \\\hline
ALEPH  &$\Lc\ell$             & 91--95 &$1.18^{+0.13}_{-0.12} \pm 0.03$ & \cite{ALEPH_fla}\\
ALEPH  &$\Lambda\ell^-\ell^+$ & 91--95 &$1.30^{+0.26}_{-0.21} \pm 0.04$ & \cite{ALEPH_fla}\\
CDF    &$\Lc\ell$             & 91--95 &$1.32 \pm 0.15        \pm 0.06$ & \cite{CDFLAM}\\
CDF    &$J/\psi \Lambda$      & 02--03 &$1.25 \pm 0.26 \pm 0.10$        & \cite{CDFLAM2}$^p$ \\
\dzero &$J/\psi \Lambda$      & 02--04 &$1.22^{+0.22}_{-0.18} \pm 0.04$ & \cite{D0LAMB} \\
DELPHI &$\Lc\ell$             & 91--94 &$1.11^{+0.19}_{-0.18} \pm 0.05$ & \cite{DELLAM0}$^a$\\
OPAL   &$\Lc\ell$, $\Lambda\ell^-\ell^+$ 
                                 & 90--95 & $1.29^{+0.24}_{-0.22} \pm 0.06$ & \cite{OPABS1_OPALAM2}\\ 
\hline
\multicolumn{3}{l}{Average of above 7 (\Lb lifetime)} & \hfagTAULBnounit & \\
\hline
ALEPH  &$\Lambda\ell$         & 91--95 &$1.20^{+0.08}_{-0.08} \pm 0.06$ & \cite{ALEPH_fla}\\
DELPHI &$\Lambda\ell\pi$ vtx  & 91--94 &$1.16 \pm 0.20 \pm 0.08$        & \cite{DELLAM0}$^a$\\
DELPHI &$\Lambda\mu$ i.p.     & 91--94 &$1.10^{+0.19}_{-0.17} \pm 0.09$ & \cite{DELLAM1}$^a$ \\
DELPHI &\particle{p\ell}      & 91--94 &$1.19 \pm 0.14 \pm 0.07$        & \cite{DELLAM0}$^a$\\
OPAL   &$\Lambda\ell$ i.p.    & 90--94 &$1.21^{+0.15}_{-0.13} \pm 0.10$ & \cite{OPALAM1}$^b$  \\
OPAL   &$\Lambda\ell$ vtx     & 90--94 &$1.15 \pm 0.12 \pm 0.06$        & \cite{OPALAM1}$^b$ \\ 
\hline
\multicolumn{3}{l}{Average of above 13 (\b-baryon lifetime)} & \hfagTAUBBnounit & \\  
\hline\hline
ALEPH  &$\Xi\ell$             & 90--95 &$1.35^{+0.37+0.15}_{-0.28-0.17}$ & \cite{ALEPH_fxi}\\
DELPHI &$\Xi\ell$             & 91--93 &$1.5 ^{+0.7}_{-0.4} \pm 0.3$     & \cite{DELPHI_fxi} \\
\hline
\multicolumn{3}{l}{Average of above 2 (\Xib lifetime)} & \hfagTAUXBnounit & \\
\hline
\multicolumn{5}{l}{$^a$ \footnotesize The combined DELPHI result quoted 
in \cite{DELLAM0} is $1.14 \pm 0.08 \pm 0.04$ ps.} \\[-0.5ex]
\multicolumn{5}{l}{$^b$ \footnotesize The combined OPAL result quoted 
in \cite{OPALAM1} is $1.16 \pm 0.11 \pm 0.06$ ps.} \\[-0.5ex]
\multicolumn{5}{l}{$^p$ \footnotesize Preliminary.}
\end{tabular}
\end{center}
\end{table}

\mysubsubsection{Summary and comparison with theoretical predictions}
\labs{lifesummary}

Averages of lifetimes of specific \b hadron species are collected
in \Table{sumlife}.
\begin{table}[t]
\caption{Summary of lifetimes of different \b hadron species.}
\labt{sumlife}
\begin{center}
\begin{tabular}{lc} \hline
\b hadron species & Measured lifetime \\ \hline
\Bu                         & \hfagTAUBU   \\
\Bd                         & \hfagTAUBD   \\
\Bs ($\to$ flavor specific) & \hfagTAUBSSL \\
\Bs ($\to J/\psi\phi$)      & \hfagTAUBSJF \\
\Bs ($1/\Gs$)               & \hfagTAUBSMEANCON \\
\Bc                         & \hfagTAUBC   \\ 
\Lb                         & \hfagTAULB   \\
\Xib mixture                & \hfagTAUXB   \\
\b-baryon mixture           & \hfagTAUBB   \\
\b-hadron mixture           & \hfagTAUB    \\
\hline
\end{tabular}
\end{center}
\caption{Ratios of \b-hadron lifetimes relative to
the \Bd lifetime and theoretical ranges predicted
by theory~\cite{Gabbiani_et_al}.}
\labt{liferatio}
\begin{center}
\begin{tabular}{lcc} \hline
Lifetime ratio & Measured value & Predicted range \\ \hline
$\tau(\Bu)/\tau(\Bd)$ & \hfagRTAUBU & 1.04 -- 1.08 \\
$\bar{\tau}(\Bs)/\tau(\Bd)^a$ & \hfagRTAUBSMEANCON & 0.99 -- 1.01 \\
$\tau(\Lb)/\tau(\Bd)$ & \hfagRTAULB & 0.81 -- 0.91    \\
$\tau(\mbox{\b-baryon})/\tau(\Bd)$  & \hfagRTAUBB & 0.81 -- 0.91 \\
\hline
\multicolumn{3}{l}{$^a$ \footnotesize 
Using $\bar{\tau}(\Bs) = 1/\Gs = 2/(\Gamma_{\rm L} + \Gamma_{\rm H})$.
}
\end{tabular}
\end{center}
\end{table}
As described in \Sec{lifetimes},
Heavy Quark Effective Theory
can be employed to explain the hierarchy of
$\tau(\Bc) \ll \tau(\Lb) < \tau(\Bs) \approx \tau(\Bd) < \tau(\Bu)$,
and used to predict the ratios between lifetimes.
Typical predictions are compared to the measured 
lifetime ratios in \Table{liferatio}.

A recent prediction of the ratio between the \Bu and \Bd lifetimes,
is $1.06 \pm 0.02$~\cite{tarantino}, in good agreement with experiment. 


The total widths of the \Bs and \Bd mesons
are expected to be very close and differ by at most 
1\%~\cite{equal_lifetimes,Gabbiani_et_al}.
However, the experimental ratio $\bar{\tau}(\Bs)/\tau(\Bd)$,
where $\bar{\tau}(\Bs)=1/\Gs$ is obtained from \DGs and 
flavour-specific lifetime measurements, now appears to be 
smaller than 1, at deviation with respect to the prediction. 
At present this discrepancy is not very significant. 

The ratio $\tau(\Lb)/\tau(\Bd)$ has particularly
been the source of theoretical
scrutiny since earlier calculations~\cite{lblife_early}
predicted a value greater than 0.90, almost two sigma higher
than the world average at the time. Recent calculations
of this ratio that include higher order effects predict a
ratio between the 
\Lb and \Bd lifetimes of $0.86 \pm 0.05$~\cite{Gabbiani_et_al}
and reduces this difference.
\Ref{Gabbiani_et_al} presents probability density functions
of its predictions with variation of theoretical inputs, and the
indicated errors (and ranges in \Table{liferatio}) 
are the RMS of the distributions.


\mysubsection{Neutral \B-meson mixing}
\labs{mixing}

There are two neutral $\B-\bar{\B}$ systems, $\Bd-\Bdbar$ and $\Bs-\Bsbar$, which 
both exhibit the phenomenon of particle-antiparticle mixing. For each of these systems, 
there are two mass eigenstates which are linear combinations of the two flavour states,
\B or $\bar{\B}$. 
We consider the case where a neutral \B meson is produced and 
detected in a flavour state, through its decay to a flavour-specific final state. 
There are four different time-dependent probabilities; if \CPT is conserved (which  
will be assumed throughout), they can be written as 
\begin{equation}
\left\{
\begin{array}{rcl}
{\cal P}(\B\to\B) & = &  \frac{e^{-\Gamma t}}{2} 
\left[ \cosh\!\left(\frac{\Delta\Gamma}{2}t\right) + \cos\!\left(\Delta m t\right)\right]  \\
{\cal P}(\B\to\bar{\B}) & = &  \frac{e^{-\Gamma t}}{2} 
\left[ \cosh\!\left(\frac{\Delta\Gamma}{2}t\right) - \cos\!\left(\Delta m t\right)\right] 
\left|\frac{q}{p}\right|^2 \\
{\cal P}(\bar{\B}\to\B) & = &  \frac{e^{-\Gamma t}}{2} 
\left[ \cosh\!\left(\frac{\Delta\Gamma}{2}t\right) - \cos\!\left(\Delta m t\right)\right] 
\left|\frac{p}{q}\right|^2 \\
{\cal P}(\bar{\B}\to\bar{\B}) & = &  \frac{e^{-\Gamma t}}{2} 
\left[ \cosh\!\left(\frac{\Delta\Gamma}{2}t\right) + \cos\!\left(\Delta m t\right)\right] 
\end{array} \right. \,,
\labe{oscillations}
\end{equation}
where $t$ is the proper time of the system (\ie\ the time interval between the production 
and the decay in the rest frame of the \B meson) and $\Gamma = 1/\tau(\B)$ 
is the average decay width.
At the \B factories, only the proper-time difference $\Delta t$ between the decays
of the two neutral \B mesons from the \Ups can be determined, but, 
because the two \B mesons evolve coherently (keeping opposite flavours as long as
none of them has decayed), the 
above formulae remain valid 
if $t$ is replaced with $\Delta t$ and the production flavour is replaced by the flavour 
at the time of the decay of the accompanying \B meson in a flavour specific state.
As can be seen in the above expressions,
the mixing probabilities 
depend on the following three observables: the mass difference $\Delta m$ and the decay 
width difference $\Delta\Gamma$ between the two mass eigenstates, and the parameter 
$|q/p|^2$ which signals \CP violation in the mixing if $|q/p|^2 \ne 1$.

In the following sections we review in turn the experimental knowledge
on these three parameters, separately 
for the \Bd meson (\dmd, \DGd, $|q/p|_{\particle{d}}$) 
and the \Bs meson (\dms, \DGs, $|q/p|_{\particle{s}}$). 

\mysubsubsection{\Bd mixing parameters}

\subsubsubsection{\boldmath \CP violation parameter $|q/p|_{\particle{d}}$}
\labs{qpd}

Evidence for \CP violation in \Bd mixing
has been searched for,
both with flavor-specific and inclusive \Bd decays, 
in samples where the initial 
flavor state is tagged. In the case of semileptonic 
(or other flavor-specific) decays, 
where the final state tag is 
also available, the following asymmetry
\begin{equation} 
 {\cal A}_{\rm SL} = 
\frac{
N(\hbox{\Bdbar}(t) \to \ell^+      \nu_{\ell} X) -
N(\hbox{\Bd}(t)    \to \ell^- \bar{\nu}_{\ell} X) }{
N(\hbox{\Bdbar}(t) \to \ell^+      \nu_{\ell} X) +
N(\hbox{\Bd}(t)    \to \ell^- \bar{\nu}_{\ell} X) } 
= \frac{|p/q|_{\particle{d}}^2 - |q/p|_{\particle{d}}^2}%
{|p/q|_{\particle{d}}^2 + |q/p|_{\particle{d}}^2}
\labe{ASL}
\end{equation} 
has been measured, either in time-integrated analyses at 
CLEO~\cite{CLEO_chid_CP,CLEO_chid_CP_y,CLEO_CP_semi} 
and CDF~\cite{CDF_CP_semi}, or in time-dependent analyses at 
OPAL~\cite{OPAL_CP_semi}, ALEPH~\cite{ALEPH_CP}, 
\babar~\cite{BABAR_DGd_qp,BABAR_CP_semi} and 
\belle~\cite{BELLE_CP_preliminary}.
In the inclusive case, also investigated and published
at ALEPH~\cite{ALEPH_CP} and OPAL~\cite{OPAL_CP_incl},
no final state tag is used, and the asymmetry~\cite{incl_asym}
\begin{equation} 
\frac{
N(\hbox{\Bd}(t) \to {\rm all}) -
N(\hbox{\Bdbar}(t) \to {\rm all}) }{
N(\hbox{\Bd}(t) \to {\rm all}) +
N(\hbox{\Bdbar}(t) \to {\rm all}) } 
\simeq
{\cal A}_{\rm SL} \left[ \frac{\dmd}{2\Gd} \sin(\dmd \,t) - 
\sin^2\left(\frac{\dmd \,t}{2}\right)\right] 
\labe{ASLincl}
\end{equation} 
must be measured as a function of the proper time to extract information 
on \CP violation.
In all cases asymmetries compatible with zero have been found,  
with a precision limited by the available statistics. A simple 
average of all published 
results for the \Bd 
meson~\cite{CLEO_chid_CP_y,CLEO_CP_semi,OPAL_CP_semi,ALEPH_CP,BABAR_DGd_qp,BABAR_CP_semi,OPAL_CP_incl}
and of the preliminary \belle result~\cite{BELLE_CP_preliminary}
yields 
\begin{equation}
{\cal A}_{\rm SL} = \hfagASL 
\end{equation}
or, equivalently through \Eq{ASL},
\begin{equation}
|q/p|_{\particle{d}} = \hfagQP \,.
\end{equation}
This result\footnote{Early analyses and (perhaps hence) the PDG use the complex
parameter $\epsilon_{\B} = (p-q)/(p+q)$; if \CP violation in the mixing in small, 
${\cal A}_{\rm SL} \cong 4 {\rm Re}(\epsilon_{\B})/(1+|\epsilon_{\B}|^2)$ and our 
current world average  
is ${\rm Re}(\epsilon_{\B})/(1+|\epsilon_{\B}|^2)=\hfagREB$.}, 
summarized in \Table{qoverp},
is compatible 
with no \CP violation in the mixing, an assumption we make for the rest 
of this section.

\begin{table}
\caption{Measurements of \CP violation in \Bd mixing and their average in terms of 
both ${\cal A}_{\rm SL}$ and $|q/p|_{\particle{d}}$.
The individual results are listed as quoted in the original publications, 
or converted\addtocounter{footnote}{-1}\protect\footnotemark\
to an ${\cal A}_{\rm SL}$ value.
When two errors are quoted, the first one is statistical and the second one systematic.
}
\labt{qoverp}
\begin{center}
\begin{tabular}{@{}rcl@{$\,\pm$}l@{$\pm$}ll@{$\,\pm$}l@{$\pm$}l@{}}
\hline
Exp.\ \& Ref. & Method & \multicolumn{3}{c}{Measured ${\cal A}_{\rm SL}$} 
                       & \multicolumn{3}{c}{Measured $|q/p|_{\particle{d}}$} \\
\hline
CLEO   \cite{CLEO_chid_CP_y} & partial hadronic rec. 
                             & $+0.017$ & 0.070 & 0.014 
                             & \multicolumn{3}{c}{} \\
CLEO   \cite{CLEO_CP_semi}   & dileptons 
                             & $+0.013$ & 0.050 & 0.005 
                             & \multicolumn{3}{c}{} \\
CLEO   \cite{CLEO_CP_semi}   & average of above two 
                             & $+0.014$ & 0.041 & 0.006 
                             & \multicolumn{3}{c}{} \\
OPAL   \cite{OPAL_CP_semi}   & leptons     
                             & $+0.008$ & 0.028 & 0.012 
                             & \multicolumn{3}{c}{} \\
OPAL   \cite{OPAL_CP_incl}   & inclusive (\Eq{ASLincl}) 
                             & $+0.005$ & 0.055 & 0.013 
                             & \multicolumn{3}{c}{} \\
ALEPH  \cite{ALEPH_CP}       & leptons 
                             & $-0.037$ & 0.032 & 0.007 
                             & \multicolumn{3}{c}{} \\
ALEPH  \cite{ALEPH_CP}       & inclusive (\Eq{ASLincl}) 
                             & $+0.016$ & 0.034 & 0.009 
                             & \multicolumn{3}{c}{} \\
ALEPH  \cite{ALEPH_CP}       & average of above two 
                             & $-0.013$ & \multicolumn{2}{l}{0.026 (tot)} 
                             & \multicolumn{3}{c}{} \\
\babar \cite{BABAR_CP_semi}  & dileptons
                             & $+0.005$ & 0.012 & 0.014 
                             & 0.998 & 0.006 & 0.007 \\ 
\babar \cite{BABAR_DGd_qp}   & full hadronic rec. 
                             & \multicolumn{3}{c}{}  
                             & $1.029$ & 0.013 & 0.011  \\
\belle \cite{BELLE_CP_preliminary} & dileptons (prel.) 
                             & $-0.0013$ & 0.0060 & 0.0056 
                             & 1.0006 & 0.0030 & 0.0028 \\
\hline
& Average of all above       & \multicolumn{3}{l}{\hfagASL\ (tot)} 
                             & \multicolumn{3}{l}{\hfagQP\  (tot)} \\ 
\hline
\end{tabular}
\end{center}
\end{table}

\subsubsubsection{\boldmath Mass and decay width differences \dmd and \DGd}
\labs{dmd}
\labs{DGd}

\begin{table}
\caption{Time-dependent measurements included in the \dmd average.
The results obtained from multi-dimensional fits involving also 
the \Bd (and \Bu) lifetimes
as free parameter(s)~\cite{BABAR3,BABAR5,BELLE2} 
have been converted into one-dimensional measurements of \dmd.
All the measurements have then been adjusted to a common set of physics
parameters before being combined. 
The latest \babar result~\cite{BABAR5} 
as well as the CDF2 and D0 results are preliminary.}
\labt{dmd}
\begin{center}
\begin{tabular}{@{}rc@{}cc@{}c@{}cc@{}c@{}c@{}}
\hline
Experiment & \multicolumn{2}{c}{Method} & \multicolumn{3}{l}{\dmd in\invps}   
                                        & \multicolumn{3}{l}{\dmd in\invps}     \\
and Ref.   &  rec. & tag                & \multicolumn{3}{l}{before adjustment} 
                                        & \multicolumn{3}{l}{after adjustment} \\
\hline
 ALEPH~\cite{ALEPH_dmd}  & \particle{ \ell  } & \particle{ \Qjet  } & $  0.404 $ & $ \pm  0.045 $ & $ \pm  0.027 $ & & & \\
 ALEPH~\cite{ALEPH_dmd}  & \particle{ \ell  } & \particle{ \ell  } & $  0.452 $ & $ \pm  0.039 $ & $ \pm  0.044 $ & & & \\
 ALEPH~\cite{ALEPH_dmd}  & \multicolumn{2}{c}{above two combined} & $  0.422 $ & $ \pm  0.032 $ & $ \pm  0.026 $ & $  0.440 $ & $ \pm  0.032 $ & $ ^{+  0.021 }_{-  0.020 } $ \\
 ALEPH~\cite{ALEPH_dmd}  & \particle{ D^*  } & \particle{ \ell,\Qjet  } & $  0.482 $ & $ \pm  0.044 $ & $ \pm  0.024 $ & $  0.482 $ & $ \pm  0.044 $ & $ \pm  0.024 $ \\
 DELPHI~\cite{DELPHI_dmd}  & \particle{ \ell  } & \particle{ \Qjet  } & $  0.493 $ & $ \pm  0.042 $ & $ \pm  0.027 $ & $  0.505 $ & $ \pm  0.042 $ & $ \pm  0.024 $ \\
 DELPHI~\cite{DELPHI_dmd}  & \particle{ \pi^*\ell  } & \particle{ \Qjet  } & $  0.499 $ & $ \pm  0.053 $ & $ \pm  0.015 $ & $  0.501 $ & $ \pm  0.053 $ & $ \pm  0.015 $ \\
 DELPHI~\cite{DELPHI_dmd}  & \particle{ \ell  } & \particle{ \ell  } & $  0.480 $ & $ \pm  0.040 $ & $ \pm  0.051 $ & $  0.491 $ & $ \pm  0.040 $ & $ ^{+  0.049 }_{-  0.048 } $ \\
 DELPHI~\cite{DELPHI_dmd}  & \particle{ D^*  } & \particle{ \Qjet  } & $  0.523 $ & $ \pm  0.072 $ & $ \pm  0.043 $ & $  0.518 $ & $ \pm  0.072 $ & $ \pm  0.043 $ \\
 DELPHI~\cite{DELPHI_dmd_dms_vtx}  & \particle{ \mbox{vtx}  } & \particle{ \mbox{comb}  } & $  0.531 $ & $ \pm  0.025 $ & $ \pm  0.007 $ & $  0.530 $ & $ \pm  0.025 $ & $ \pm  0.006 $ \\
 L3~\cite{L3_dmd}  & \particle{ \ell  } & \particle{ \ell  } & $  0.458 $ & $ \pm  0.046 $ & $ \pm  0.032 $ & $  0.472 $ & $ \pm  0.046 $ & $ \pm  0.029 $ \\
 L3~\cite{L3_dmd}  & \particle{ \ell  } & \particle{ \Qjet  } & $  0.427 $ & $ \pm  0.044 $ & $ \pm  0.044 $ & $  0.435 $ & $ \pm  0.044 $ & $ \pm  0.042 $ \\
 L3~\cite{L3_dmd}  & \particle{ \ell  } & \particle{ \ell\mbox{(IP)}  } & $  0.462 $ & $ \pm  0.063 $ & $ \pm  0.053 $ & $  0.485 $ & $ \pm  0.063 $ & $ \pm  0.047 $ \\
 OPAL~\cite{OPAL_dmd_dilepton}  & \particle{ \ell  } & \particle{ \ell  } & $  0.430 $ & $ \pm  0.043 $ & $ ^{+  0.028 }_{-  0.030 } $ & $  0.461 $ & $ \pm  0.043 $ & $ ^{+  0.018 }_{-  0.017 } $ \\
 OPAL~\cite{OPAL_dmd_lepton}  & \particle{ \ell  } & \particle{ \Qjet  } & $  0.444 $ & $ \pm  0.029 $ & $ ^{+  0.020 }_{-  0.017 } $ & $  0.466 $ & $ \pm  0.029 $ & $ ^{+  0.015 }_{-  0.014 } $ \\
 OPAL~\cite{OPAL_dmd_dstar}  & \particle{ D^*\ell  } & \particle{ \Qjet  } & $  0.539 $ & $ \pm  0.060 $ & $ \pm  0.024 $ & $  0.544 $ & $ \pm  0.060 $ & $ \pm  0.023 $ \\
 OPAL~\cite{OPAL_dmd_dstar}  & \particle{ D^*  } & \particle{ \ell  } & $  0.567 $ & $ \pm  0.089 $ & $ ^{+  0.029 }_{-  0.023 } $ & $  0.571 $ & $ \pm  0.089 $ & $ ^{+  0.028 }_{-  0.022 } $ \\
 OPAL~\cite{OPAL_dmd_slowpion}  & \particle{ \pi^*\ell  } & \particle{ \Qjet  } & $  0.497 $ & $ \pm  0.024 $ & $ \pm  0.025 $ & $  0.496 $ & $ \pm  0.024 $ & $ \pm  0.025 $ \\
 CDF1~\cite{CDF1_dmd_dlepton_SST}  & \particle{ D\ell  } & \particle{ \mbox{SST}  } & $  0.471 $ & $ ^{+  0.078 }_{-  0.068 } $ & $ ^{+  0.033 }_{-  0.034 } $ & $  0.470 $ & $ ^{+  0.078 }_{-  0.068 } $ & $ ^{+  0.033 }_{-  0.034 } $ \\
 CDF1~\cite{CDF1_dmd_dimuon}  & \particle{ \mu  } & \particle{ \mu  } & $  0.503 $ & $ \pm  0.064 $ & $ \pm  0.071 $ & $  0.512 $ & $ \pm  0.064 $ & $ \pm  0.070 $ \\
 CDF1~\cite{CDF1_dmd_lepton}  & \particle{ \ell  } & \particle{ \ell,\Qjet  } & $  0.500 $ & $ \pm  0.052 $ & $ \pm  0.043 $ & $  0.534 $ & $ \pm  0.052 $ & $ \pm  0.036 $ \\
 CDF1~\cite{CDF1_dmd_dstarlepton}  & \particle{ D^*\ell  } & \particle{ \ell  } & $  0.516 $ & $ \pm  0.099 $ & $ ^{+  0.029 }_{-  0.035 } $ & $  0.523 $ & $ \pm  0.099 $ & $ ^{+  0.028 }_{-  0.035 } $ \\
 CDF2~\cite{CDF2_dmd_dlepton_preliminary}  & \particle{ D^{(*)}\ell  } & \particle{ \mbox{OST}  } & $  0.497 $ & $ \pm  0.028 $ & $ \pm  0.015 $ & $  0.497 $ & $ \pm  0.028 $ & $ \pm  0.015 $ \\
 CDF2~\cite{CDF2_dmd_exclusive_preliminary}  & \particle{ \Bd  } & \particle{ \mbox{SST}  } & $  0.526 $ & $ \pm  0.056 $ & $ \pm  0.005 $ & $  0.526 $ & $ \pm  0.056 $ & $ \pm  0.005 $ \\
 \dzero~\cite{D0_dmd_preliminary}  & \particle{ D^*\mu  } & \particle{ \mbox{comb}  } & $  0.456 $ & $ \pm  0.034 $ & $ \pm  0.025 $ & $  0.456 $ & $ \pm  0.034 $ & $ \pm  0.025 $ \\
 \babar~\cite{BABAR_dmd_full}  & \particle{ \Bd  } & \particle{ \ell,K,\mbox{NN}  } & $  0.516 $ & $ \pm  0.016 $ & $ \pm  0.010 $ & $  0.520 $ & $ \pm  0.016 $ & $ \pm  0.008 $ \\
 \babar~\cite{BABAR_dmd_dilepton}  & \particle{ \ell  } & \particle{ \ell  } & $  0.493 $ & $ \pm  0.012 $ & $ \pm  0.009 $ & $  0.489 $ & $ \pm  0.012 $ & $ \pm  0.006 $ \\
 \babar~\cite{BABAR5}  & \particle{ D^*\ell\nu\mbox{(part)}  } & \particle{ \ell  } & $  0.523 $ & $ \pm  0.004 $ & $ \pm  0.007 $ & $  0.521 $ & $ \pm  0.004 $ & $ \pm  0.007 $ \\
 \babar~\cite{BABAR3}  & \particle{ D^*\ell\nu  } & \particle{ \ell,K,\mbox{NN}  } & $  0.492 $ & $ \pm  0.018 $ & $ \pm  0.014 $ & $  0.491 $ & $ \pm  0.018 $ & $ \pm  0.013 $ \\
 \belle~\cite{BELLE_dmd_dstarpi_partial}  & \particle{ D^*\pi\mbox{(part)}  } & \particle{ \ell  } & $  0.509 $ & $ \pm  0.017 $ & $ \pm  0.020 $ & $  0.512 $ & $ \pm  0.017 $ & $ \pm  0.019 $ \\
 \belle~\cite{BELLE_dmd_dilepton}  & \particle{ \ell  } & \particle{ \ell  } & $  0.503 $ & $ \pm  0.008 $ & $ \pm  0.010 $ & $  0.506 $ & $ \pm  0.008 $ & $ \pm  0.009 $ \\
 \belle~\cite{BELLE2}  & \particle{ \Bd,D^*\ell\nu  } & \particle{ \mbox{comb}  } & $  0.511 $ & $ \pm  0.005 $ & $ \pm  0.006 $ & $  0.512 $ & $ \pm  0.005 $ & $ \pm  0.007 $ \\
 \hline \\[-2.0ex]
 \multicolumn{6}{l}{World average (all above measurements included):} & $  0.510 $ & $ \pm  0.003 $ & $ \pm  0.004 $ \\

\\[-2.0ex]
\multicolumn{6}{l}{~~~ -- ALEPH, DELPHI, L3, OPAL and CDF1 only:}
     & \hfagDMDHval & \hfagDMDHsta & \hfagDMDHsys \\
\multicolumn{6}{l}{~~~ -- Above measurements of \babar and \belle only:}
     & \hfagDMDBval & \hfagDMDBsta & \hfagDMDBsys \\
\hline
\end{tabular}
\end{center}
\end{table}

Many time-dependent \Bd--\Bdbar oscillation analyses have been performed by the 
ALEPH, \babar, \belle, CDF, \dzero, DELPHI, L3 and OPAL collaborations. 
The corresponding measurements of \dmd are summarized in 
\Table{dmd},
where only the most recent results
are listed (\ie\ measurements superseded by more recent ones have been omitted). 
Although a variety of different techniques have been used, the 
individual \dmd
results obtained at high-energy colliders have remarkably similar precision.
Their average is compatible with the recent and more precise measurements 
from the asymmetric \B factories.
The systematic uncertainties are not negligible; 
they are often dominated by sample composition, mistag probability,
or \b-hadron lifetime contributions.
Before being combined, the measurements are adjusted on the basis of a 
common set of input values, including the averages of the 
\b-hadron fractions and lifetimes given in this report 
(see \Secss{fractions}{lifetimes}).
Some measurements are statistically correlated. 
Systematic correlations arise both from common physics sources 
(fractions, lifetimes, branching ratios of \b hadrons), and from purely 
experimental or algorithmic effects (efficiency, resolution, flavour tagging, 
background description). Combining all published measurements
listed in \Table{dmd}
and accounting for all identified correlations
as described in~\cite{LEPHFS} yields $\dmd = \hfagDMDWfull$.

On the other hand, ARGUS and CLEO have published 
measurements of the time-integrated mixing probability 
\chid~\cite{ARGUS_chid,CLEO_chid_CP,CLEO_chid_CP_y}, 
which average to $\chid =\hfagCHIDU$.
Following \Ref{CLEO_chid_CP_y}, 
the width difference \DGd could 
in principle be extracted from the
measured value of $\Gd=1/\tau(\Bd)$ and the above averages for 
\dmd and \chid 
(provided that \DGd has a negligible impact on 
the \dmd analyses that have assumed $\DGd=0$), 
using the relation
\begin{equation}
\chid = \frac{\xd^2+\yd^2}{2(\xd^2+1)} ~~~ \mbox{with} ~~ \xd=\frac{\dmd}{\Gd} 
~~~ \mbox{and} ~~ \yd=\frac{\DGd}{2\Gd} \,.
\labe{chid_definition}
\end{equation}
However, direct time-dependent studies provide much stronger constraints: 
DELPHI published the result
$|\DGd|/\Gd < 18\%$ at \CL{95}~\cite{DELPHI_dmd_dms_vtx}, 
while \babar recently obtained 
$-8.4\% < {\rm sign}({\rm Re} \lambda_{\CP}) \DGGd < 6.8\%$
at \CL{90}~\cite{BABAR_DGd_qp}, 
where 
$\lambda_{\CP} = (q/p)_{\particle{d}} (\bar{A}_{\CP}/A_{\CP})$
is defined for a \CP-even final state and where 
\DGd is defined as\footnote{This sign convention for 
\DGd, taken from \Ref{BABAR_DGd_qp},
is opposite to that used for \DGs in \Secss{taubs}{DGs}.}
$\DGd = \Gamma(\Bd_{\rm H})-\Gamma(\Bd_{\rm L})$
(the sensitivity to the overall sign of 
${\rm sign}({\rm Re} \lambda_{\CP}) \DGGd$ comes
from the use of \Bd decays to \CP final states).
Combining these two results after adjustment to 
$1/\Gd=\tau(\Bd)=\hfagTAUBD$ yields
\begin{equation}
{\rm sign}({\rm Re} \lambda_{\CP}) \DGGd  = \hfagSDGDGD \,.
\end{equation}
The sign of ${\rm Re} \lambda_{\CP}$ is not measured,
but expected to be positive from the global fits
of the Unitarity Triangle within the Standard Model.

Assuming $\DGd=0$ 
and using $1/\Gd=\tau(\Bd)=\hfagTAUBD$,
the \dmd and \chid results are combined through \Eq{chid_definition} 
to yield the 
world average
\begin{equation} 
\dmd = \hfagDMDWU \,,
\labe{dmd}
\end{equation} 
or, equivalently,
\begin{equation} 
\xd= \hfagXDWU ~~~ \mbox{and} ~~~ \chid=\hfagCHIDWU \,.  
\labe{chid}
\end{equation}
\Figure{dmd} compares the \dmd values obtained by the different experiments.

\begin{figure}
\begin{center}
\epsfig{figure=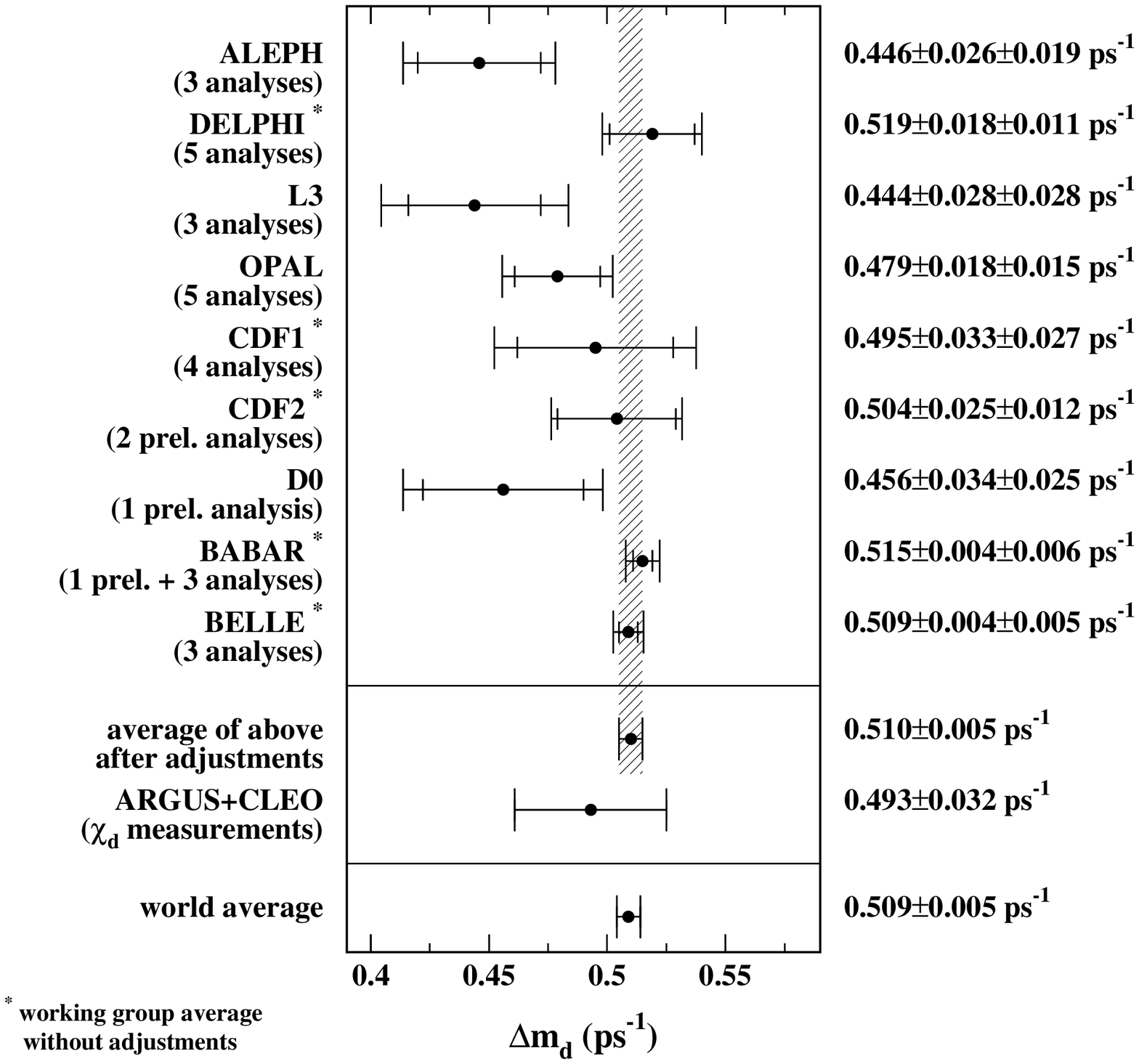,width=\textwidth}
\caption{The \Bd--\Bdbar oscillation frequency \dmd as measured by the different experiments. 
The averages quoted for ALEPH, L3 and OPAL are taken from the original publications, while the 
ones for DELPHI, CDF, \babar, and \belle have been computed from the individual results 
listed in \Table{dmd} without performing any adjustments. The time-integrated measurements 
of \chid from the symmetric \B factory experiments ARGUS and CLEO have been converted 
to a \dmd value using $\tau(\Bd)=\hfagTAUBD$. The two global averages have been obtained 
after adjustments of all the individual \dmd results of \Table{dmd} (see text).}
\labf{dmd}
\end{center}
\end{figure}

The \Bd mixing averages given in \Eqss{dmd}{chid}
and the \b-hadron fractions of \Table{fractions} have been obtained in a fully 
consistent way, taking into account the fact that the fractions are computed using 
the \chid value of \Eq{chid} and that many individual measurements of \dmd
at high energy depend on the assumed values for the \b-hadron fractions.
Furthermore, this set of averages is consistent with the lifetime averages 
of \Sec{lifetimes}.

It should be noted that the most recent (and precise) analyses at the 
asymmetric \B factories measure \dmd
as a result of a multi-dimensional fit. 
The preliminary \babar
analysis~\cite{BABAR5},  
based on partially reconstructed $\Bd \to D^*\ell\nu$ decays, 
extracts simultaneously \dmd and $\tau(\Bd)$
in a way similar to the published \babar 
analysis based on fully reconstructed $\Bd\to D^* \ell\nu$ 
decays~\cite{BABAR3}, 
while the latest \belle published analysis~\cite{BELLE2},  
based on fully reconstructed hadronic \Bd decays and $\Bd \to D^*\ell\nu$ decays, 
extracts simultaneously \dmd, $\tau(\Bd)$ and $\tau(\Bu)$.
The measurements of \dmd and $\tau(\Bd)$ of these three analyses 
are displayed in \Table{dmd2D} and in \Fig{dmd2D}. Their two-dimensional average, 
taking into account all statistical and systematic correlations, and expressed
at $\tau(\Bu)=\hfagTAUBU$, is
\begin{equation}
\left.
\begin{array}{r@{}l}
\dmd = \hfagDMDTWODnounit & \invps \\
\tau(\Bd) = \hfagTAUBDTWODnounit & \ps
\end{array}
\right\}
~\mbox{with a total correlation of \hfagRHODMDTAUBD.}
\end{equation}

\begin{table}
\caption{Simultaneous measurements of \dmd and $\tau(\Bd)$, and their average.
The \belle analysis also 
measures $\tau(\Bu)$ at the same time, but it is converted here into a two-dimensional measurement 
of \dmd and $\tau(\Bd)$, for an assumed value of $\tau(\Bu)$. 
The first quoted error on the measurements is statistical
and the second one systematic; in the case of adjusted measurements, the 
latter includes a contribution obtained from the variation of $\tau(\Bu)$ or 
$\tau(\Bu)/\tau(\Bd)$ in the indicated range. Units are\invps\ for \dmd
and\unit{ps} for the lifetimes. 
The three different values of $\rho(\dmd,\tau(\Bd))$ correspond 
to the statistical, systematic and total correlation coefficients
between the adjusted measurements of \dmd and $\tau(\Bd)$.
The second \babar result~\cite{BABAR5} is still preliminary.}
\labt{dmd2D}
\begin{center}
\begin{tabular}{@{}r@{~}c@{}c@{}c@{~}c@{}c@{}c@{~}c@{}c@{}c@{\hspace{0ex}}c@{}}
\hline
Exp.\ \& Ref.
& \multicolumn{3}{c}{Measured \dmd}   
& \multicolumn{3}{c}{Measured $\tau(\Bd)$}   
& \multicolumn{3}{c}{Measured $\tau(\Bu)$}   
&  Assumed $\tau(\Bu)$ \\
\hline
\babar \cite{BABAR3}  
      & $0.492$ & $\pm 0.018$ & $\pm 0.013$ 
      & $1.523$ & $\pm 0.024$ & $\pm 0.022$ 
      & \multicolumn{3}{c}{---}
      & $(1.083$$\pm 0.017)\tau(\Bd)$ \\  
\babar \cite{BABAR5}  
      & $0.523$ & $\pm 0.004$ & $\pm 0.007$ 
      & $1.501$ & $\pm 0.008$ & $\pm 0.030$
      & \multicolumn{3}{c}{---}
      & $1.671$$\pm 0.018$ \\  
\belle \cite{BELLE2}  
      & $0.511$ & $\pm 0.005$ & $\pm 0.006$
      & $1.534$ & $\pm 0.008$ & $\pm 0.010$
      & $1.635$ & $\pm 0.011$ & $\pm 0.011$
      & --- \\  
\cline{2-10}
& \multicolumn{3}{c}{Adjusted \dmd}   
& \multicolumn{3}{c}{Adjusted $\tau(\Bd)$}   
& \multicolumn{3}{c}{$\rho(\dmd,\Bd)$} 
\\
\cline{2-10}
\babar \cite{BABAR3}  
      & $0.492$ & $\pm 0.018$ & $\pm 0.013$ 
      & $1.524$ & $\pm 0.025$ & $\pm 0.022$ 
      & $-0.22$ & $+0.74$ & $+0.16$ 
      & $(\hfagRTAUBUval$$\hfagRTAUBUerr)\tau(\Bd)$ \\  
\babar \cite{BABAR5}  
      & $0.521$ & $\pm 0.004$ & $\pm 0.007$ 
      & $1.505$ & $\pm 0.009$ & $\pm 0.030$
      & $-0.01$ & $-0.71$ & $-0.58$ 
      & $\hfagTAUBUval$$\hfagTAUBUerr$ \\  
\belle \cite{BELLE2}  
      & $0.511$ & $\pm 0.007$ & $\pm 0.005$
      & $1.534$ & $\pm 0.009$ & $\pm 0.009$
      & $-0.27$ & $-0.09$ & $-0.19$ 
      & $\hfagTAUBUval$$\hfagTAUBUerr$ \\  
\hline
\multicolumn{1}{l}{Average} 
      & \hfagDMDTWODval   & \hfagDMDTWODsta   & \hfagDMDTWODsys
      & \hfagTAUBDTWODval & \hfagTAUBDTWODsta & \hfagTAUBDTWODsys
      & \hfagRHOstaDMDTAUBD & \hfagRHOsysDMDTAUBD & \hfagRHODMDTAUBD 
      & $\hfagTAUBUval$$\hfagTAUBUerr$ \\  
\hline 
\end{tabular}
\end{center}
\end{table}

\begin{figure}
\begin{center}
\epsfig{figure=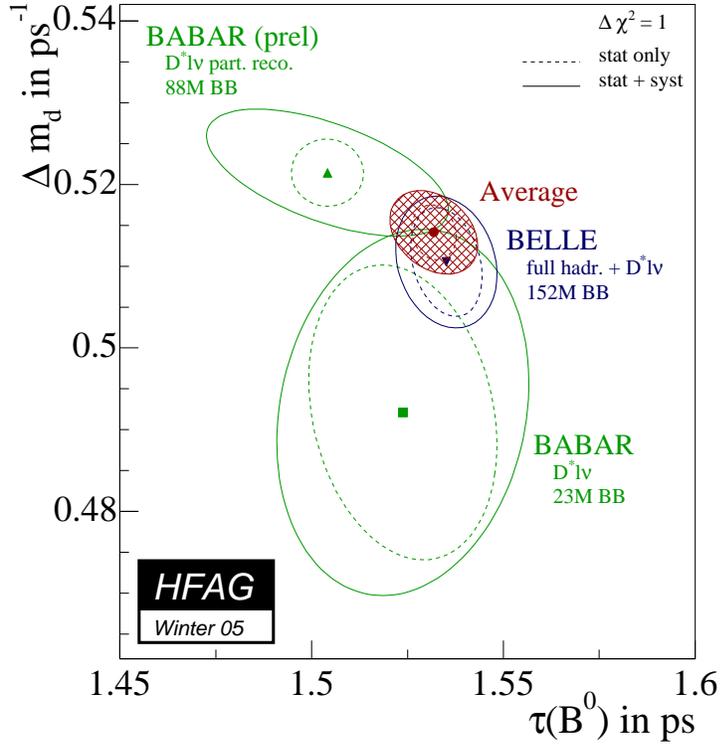,width=0.6\textwidth}
\caption{Simultaneous measurements of
\dmd and $\tau(\Bd)$~\cite{BABAR3,BABAR5,BELLE2}, 
after adjustment to a common set of parameters (see text). 
Statistical and total uncertainties are represented as dashed and
solid contours respectively.
The average of the three measurements
is indicated by a hatched ellipse.}
\labf{dmd2D}
\end{center}
\end{figure}

\mysubsubsection{\Bs mixing parameters}

\subsubsubsection{\boldmath \CP violation parameter $|q/p|_{\particle{s}}$}
\labs{qps}

No measurement or experimental limit exists on $|q/p|_{\particle{s}}$,
except in the form of a relatively weak constraint from CDF 
on a combination of $|q/p|_{\particle{d}}$ and $|q/p|_{\particle{s}}$,
$f'_{\particle{d}} \,\chid(1-|q/p|^2_{\particle{d}})+
 f'_{\particle{s}} \,\chis(1-|q/p|^2_{\particle{s}})=
0.006\pm 0.017$~\cite{CDF_CP_semi},
using inclusive semileptonic decays of \b hadrons. 
The result is compatible with no \CP violation in the 
mixing, an assumption made in all results described below. 

\subsubsubsection{\boldmath Mass difference \dms}
\labs{dms}

The time-integrated measurements of \chibar (see \Sec{chibar}), when compared to our knowledge
of \chid and the \b-hadron fractions, indicate that \Bs mixing is large, with a value of 
\chis close to its maximal possible value of $1/2$.
However, the time dependence of this mixing (called \Bs oscillations) has not been 
observed yet, mainly because the period of these oscillations turns out to be so small 
that it can't be resolved with the proper-time resolutions achieved so far. 

The statistical significance ${\cal S}$ of a \Bs oscillation signal can be
approximated as~\cite{amplitude}
\begin{equation}
{\cal S} \approx \sqrt{\frac{N}{2}} \,f_{\rm sig}\, (1-2w)\,
\exp{\left(-\left(\dms\sigma_t\right)^2/2\right)}\,,
\labe{significance}
\end{equation}
where $N$ is 
the number of selected and tagged \Bs candidates, 
$f_{\rm sig}$ is the fraction of \Bs signal
in the selected and tagged sample, $w$ is the total mistag probability, 
and $\sigma_t$ is the resolution on proper time.
As can be seen, the quantity ${\cal S}$ decreases very quickly as 
\dms increases: this dependence is controlled by $\sigma_t$, 
which is therefore the most critical parameter for \dms analyses. 
The method widely used for \Bs oscillation searches
consists of measuring a \Bs oscillation amplitude ${\cal A}$
at several different test values of \dms, 
using a maximum likelihood fit based on the functions 
of \Eq{oscillations} where the cosine terms have been multiplied 
by ${\cal A}$.
One expects ${\cal A}=1$ at the true 
value of \dms and to ${\cal A}=0$ at a test value of \dms 
(far) below the true value.
To a good approximation, the statistical uncertainty on ${\cal A}$
is Gaussian and equal to $1/{\cal S}$~\cite{amplitude}.

\begin{figure}
\begin{center}
\epsfig{figure=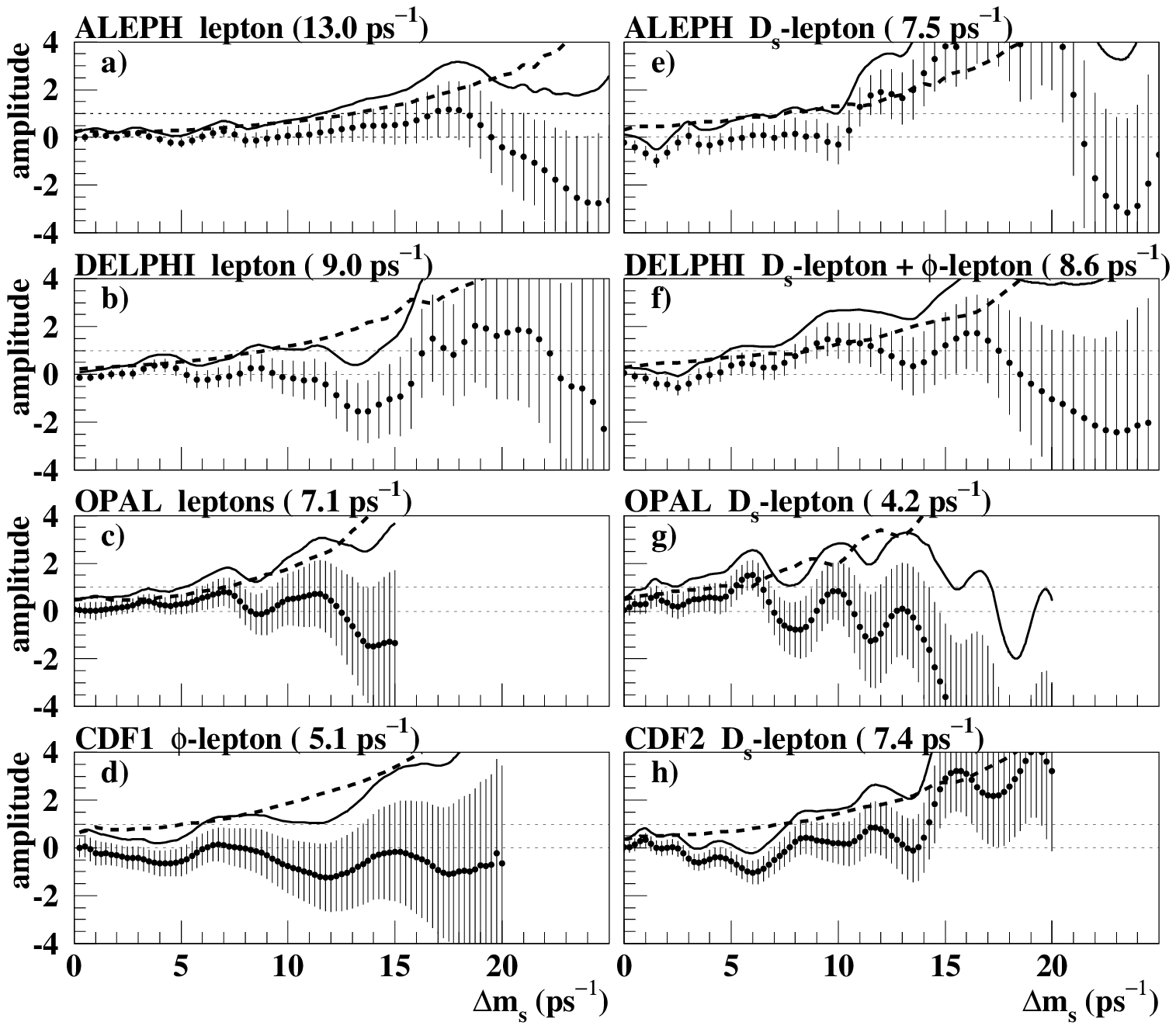,width=\textwidth,%
bbllx=55,bblly=55,bburx=490,bbury=490}
\caption{\Bs-oscillation amplitude spectra, displayed separately for each 
\Bs oscillation analysis. 
The points and error bars represent the measurements of the amplitude ${\cal A}$ and 
their total uncertainties $\sigma_{\cal A}$, adjusted to a set of physics parameters
common to all analyses (including $\fBs=\hfagFBS$).
Values of \dms where the solid curve 
(${\cal A}+1.645\,\sigma_{\cal A}$) is below 1 are excluded at \CL{95}. 
The dashed curve shows $1.645\,\sigma_{\cal A}$; the number in parenthesis indicates where 
this curve is equal to 1, and is a measure of the sensitivity of the analysis. 
a) ALEPH inclusive lepton~\cite{ALEPH_dms},
b) DELPHI inclusive lepton~\cite{DELPHI_dms_last}, 
c) OPAL inclusive lepton and dilepton~\cite{OPAL_dms_l},
d) CDF1 $\phi$-$\ell$~\cite{CDF1_dms}, 
e) ALEPH \particle{D_s}-$\ell$~\cite{ALEPH_dms}, 
f) DELPHI \particle{D_s}-$\ell$~\cite{DELPHI_dms_last} and $\phi$-$\ell$~\cite{DELPHI_dms_dgs},
g) OPAL \particle{D_s}-$\ell$~\cite{OPAL_dms_dsl},
h) CDF2 \particle{D_s}-$\ell$ (preliminary)~\cite{CDF2_dms_dslnu_prel}.
Continuation on \Fig{individual_amplitudes_2}).}
\labf{individual_amplitudes_1}
\end{center}
\end{figure}

\begin{figure}
\begin{center}
\epsfig{figure=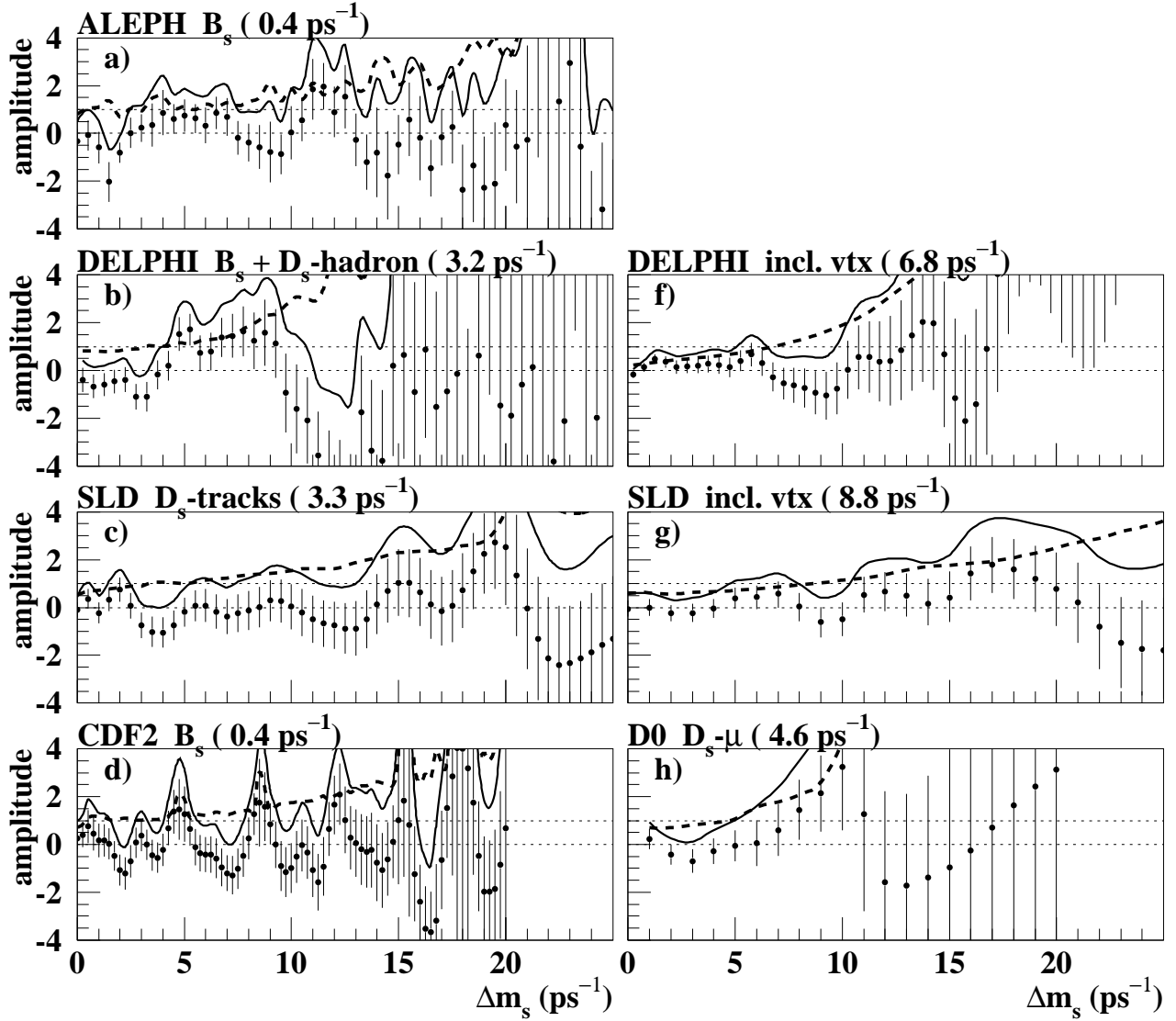,width=\textwidth,%
bbllx=55,bblly=55,bburx=490,bbury=490}
\caption{(continuation of \Fig{individual_amplitudes_1})
\Bs-oscillation amplitude spectra, displayed separately for each 
\Bs oscillation analysis, in the same manner as in \Fig{individual_amplitudes_1}. 
a) ALEPH fully reconstructed \Bs~\cite{ALEPH_dms}, 
b) DELPHI fully reconstructed \Bs and \particle{D_s}-hadron~\cite{DELBS1_dms_excl},
c) SLD \particle{D_s}+tracks~\cite{SLD_dms_ds},
d) CDF2 fully reconstructed \Bs (preliminary)~\cite{CDF2_dms_dspi_prel},
f) DELPHI inclusive vertex~\cite{DELPHI_dmd_dms_vtx}, 
g) SLD inclusive vertex dipole~\cite{SLD_dms_dipole},
h) \dzero \particle{D_s}-$\mu$ (preliminary)~\cite{D0_dms_dsmux_prel}.}
\labf{individual_amplitudes_2}
\end{center}
\end{figure}

\begin{figure}
\begin{center}
\epsfig{figure=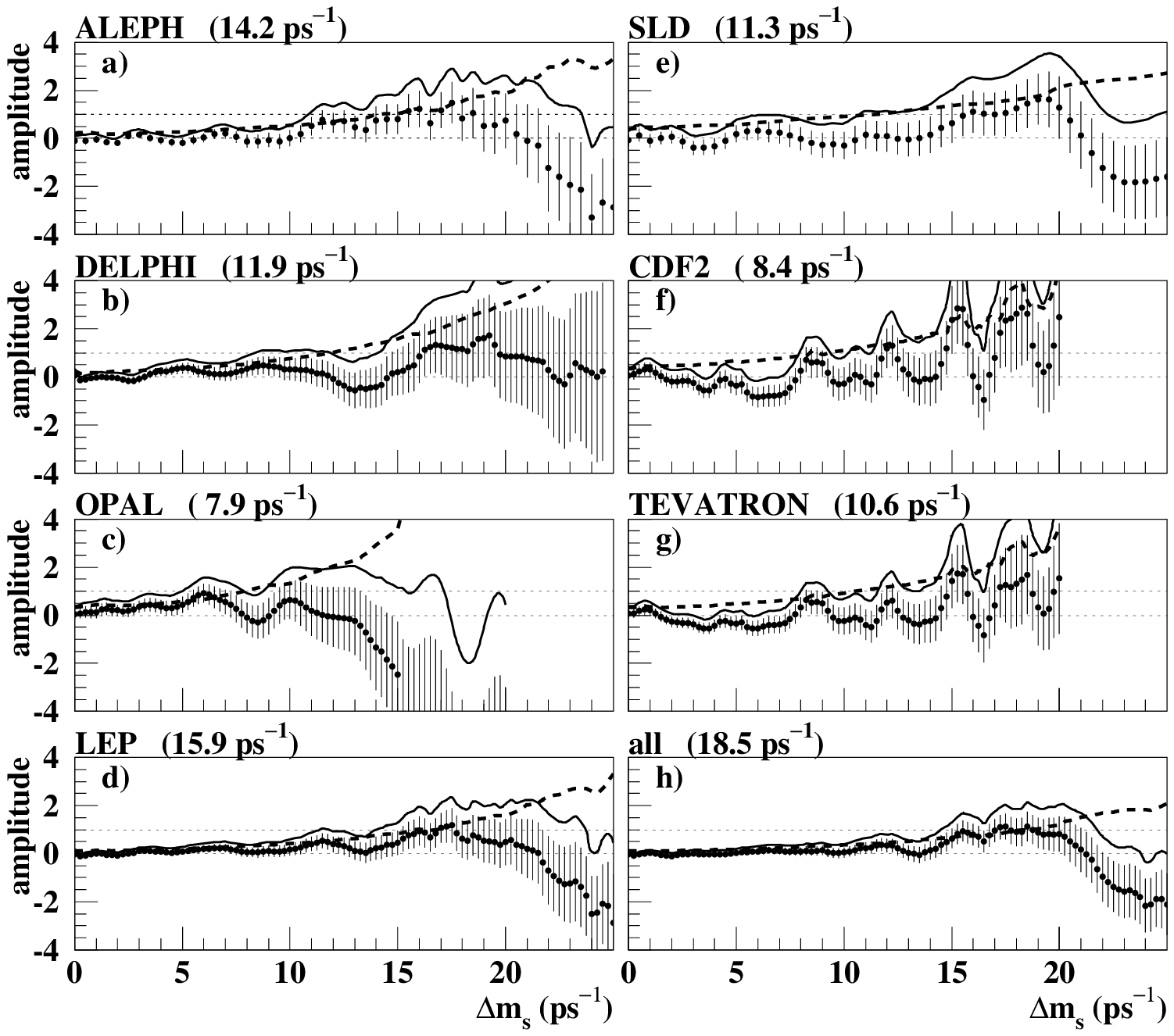,width=\textwidth,%
bbllx=55,bblly=55,bburx=490,bbury=490}
\caption{Combined \Bs-oscillation amplitude spectra, displayed separately for each 
experiment and collider, in the same manner as in \Fig{individual_amplitudes_1}. 
a) ALEPH~\cite{ALEPH_dms}, 
b) DELPHI~\cite{DELBS1_dms_excl,DELPHI_dms_dgs,DELPHI_dmd_dms_vtx,DELPHI_dms_last},
c) OPAL~\cite{OPAL_dms_l,OPAL_dms_dsl}, 
d) LEP~\cite{ALEPH_dms,DELBS1_dms_excl,DELPHI_dms_dgs,DELPHI_dmd_dms_vtx,DELPHI_dms_last,OPAL_dms_l,OPAL_dms_dsl},
e) SLD~\cite{SLD_dms_ds,SLD_dms_dipole},
f) CDF2~\cite{CDF2_dms_dslnu_prel,CDF2_dms_dspi_prel}, 
g) TEVATRON~\cite{CDF1_dms,CDF2_dms_dslnu_prel,CDF2_dms_dspi_prel,D0_dms_dsmux_prel}, 
f) all experiments together.
}
\labf{individual_amplitudes_3}
\end{center}
\end{figure}

\Figuress{individual_amplitudes_1}{individual_amplitudes_2} show the
amplitude spectra obtained by ALEPH~\cite{ALEPH_dms},
CDF~\cite{CDF1_dms,CDF2_dms_dslnu_prel,CDF2_dms_dspi_prel},
D0~\cite{D0_dms_dsmux_prel}, 
DELPHI~\cite{DELPHI_dmd_dms_vtx,DELPHI_dms_dgs,DELBS1_dms_excl,DELPHI_dms_last},
OPAL~\cite{OPAL_dms_l,OPAL_dms_dsl} and 
SLD~\cite{SLD_dms_dipole,SLD_dms_ds}.\footnote{An unpublished analysis 
from SLD~\cite{SLD_dms_leptDvtx_unpublished}, 
based on an inclusive reconstruction from a 
lepton and a topologically reconstructed \particle{D} meson, 
is not included in the plots or
combined results quoted in this section. However, nothing is known 
to be wrong about this analysis, and including it would increase the 
combined \dms limit of \Eq{dmslimit} by \hfagDMSDLIM and the combined 
sensitivity by \hfagDMSDSENS.}
In each analysis, a particular value of \dms
can be excluded at \CL{95} if ${\cal A}+ 1.645\,\sigma_{\cal A} < 1$, 
where $\sigma_{\cal A}$ is the total uncertainty on ${\cal A}$.
Because of the proper time resolution, the quantity $\sigma_{\cal A}(\dms)$
is an increasing function of \dms (see \Eq{significance} which merely models  
$1/\sigma_{\cal A}(\dms)$ since all results are limited 
by the available statistics). Therefore, 
if the true value of \dms were infinitely large, one 
expects to be able to exclude all values of \dms up to $\dms^{\rm sens}$, 
where $\dms^{\rm sens}$, called here the
sensitivity of the analysis, is defined by
$1.645\,\sigma_{\cal A}(\dms^{\rm sens}) = 1$. 
The most sensitive analyses appear to be the ones based 
on inclusive lepton samples at LEP, where reasonable statistics is available. 
Because of their better proper time resolution, the small data samples 
analyzed inclusively at SLD, as well as the few fully reconstructed \Bs decays 
at LEP, turn out to be also very useful to explore the high \dms region.
Recent preliminary analyses are available from CDF and \dzero, which presently
are the only experiments active in this area. 

\begin{figure}
\begin{center}
\epsfig{figure=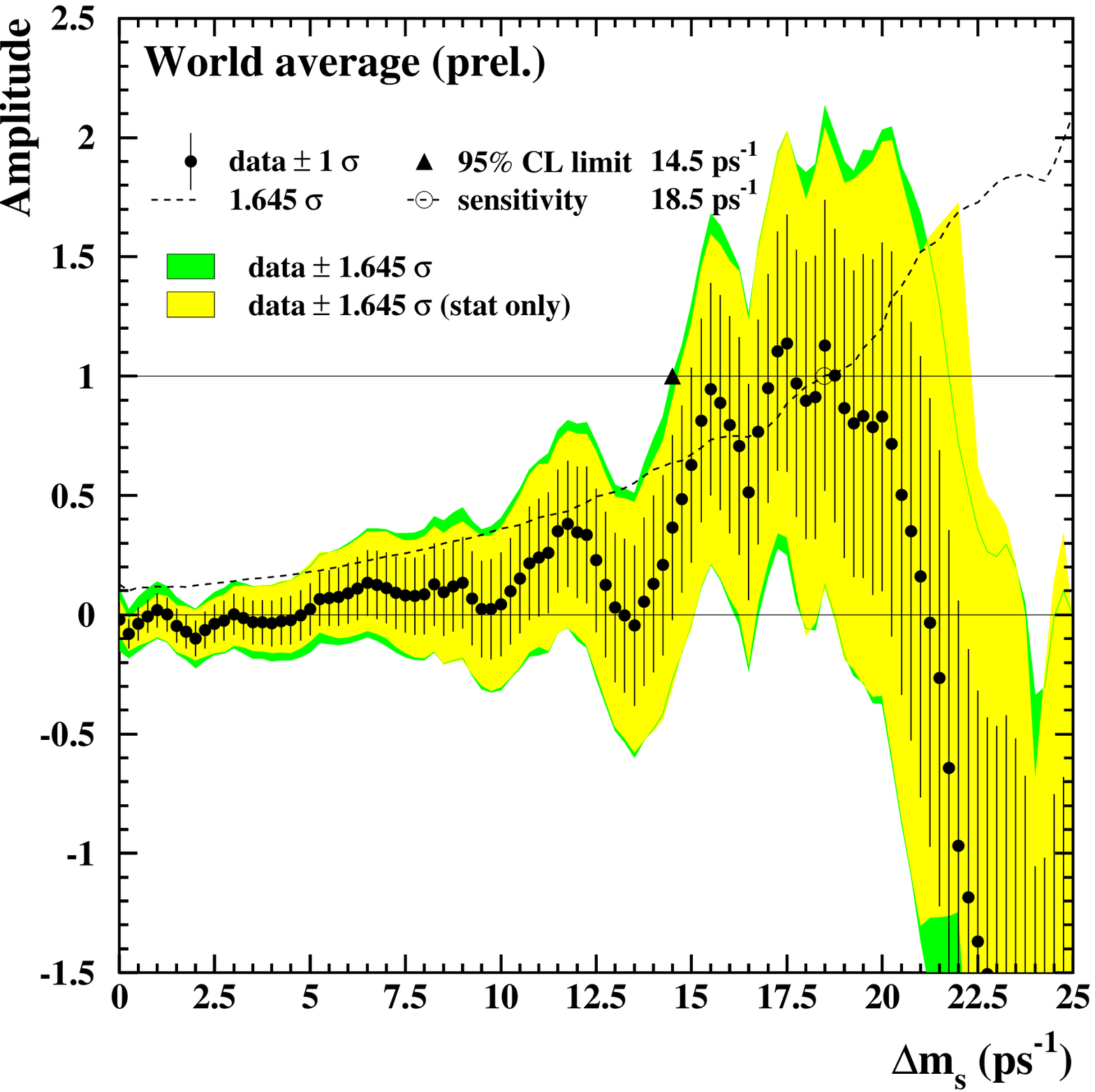,width=\textwidth}
\caption{Combined measurements of the \Bs oscillation amplitude as a 
function of \dms, including all published results and preliminary 
results presented at the Winter 2005 conferences~%
\cite{ALEPH_dms,CDF1_dms,CDF2_dms_dslnu_prel,CDF2_dms_dspi_prel,D0_dms_dsmux_prel,DELPHI_dmd_dms_vtx,DELPHI_dms_dgs,DELBS1_dms_excl,DELPHI_dms_last,OPAL_dms_l,OPAL_dms_dsl,SLD_dms_dipole,SLD_dms_ds}.
The measurements are dominated by statistical uncertainties. 
Neighboring points are statistically correlated.}
\labf{amplitude}
\end{center}
\end{figure}

These oscillation searches can easily be combined 
by averaging the measured amplitudes ${\cal A}$ at each test value 
of \dms. The combined amplitude spectra for the individual experiments are 
displayed in \Fig{individual_amplitudes_3}, and the world average spectrum is 
displayed in \Fig{amplitude}.
The individual results have been adjusted to common physics inputs, 
and all known correlations have been accounted for; 
in the case of the inclusive analyses, the sensitivities (\ie\ 
the statistical uncertainties on ${\cal A}$), which depend directly 
through \Eq{significance} on the assumed fraction $f_{\rm sig}\sim\fBs$
of \Bs mesons in an unbiased sample of weakly-decaying \b hadrons, 
have also been rescaled to a common average of $\fBs = \hfagFBS$.
The combined sensitivity for \CL{95} exclusion of \dms values is found 
to be \hfagDMSWSENS.
All values of \dms below \hfagDMSWLIM\ are excluded at \CL{95},
which we express as
\begin{equation}
\dms > \rm \hfagDMSWLIM~at~\CL{95} \,.
\labe{dmslimit}
\end{equation}
The values between \hfagDMSWLIM\ and \hfagDMSWUPP\ cannot be excluded, because 
the data is compatible with a signal in this region. However,
no deviation from ${\cal A}=0$ is seen in \Fig{amplitude} that would
indicate the observation of a signal.

It should be noted that most \dms analyses assume no decay-width difference in the \Bs system.
Due to the presence of the $\cosh$ terms in \Eq{oscillations}, a non-zero value of 
\DGs would reduce the oscillation amplitude with a small time-dependent factor that would be 
very difficult to distinguish from time resolution effects.

Convoluting the average \Bs lifetime, \hfagTAUBS, with the limit of \Eq{dmslimit} yields
\begin{equation}
\xs = \dms\,\tau(\Bs) > \rm \hfagXSWLIM~at~\CL{95} \,. 
\labe{xs}
\end{equation}
Under the assumption $\DGs=0$, \ie\ $\ys=\DGs/(2\Gs)=0$ 
(and no \CP violation in the mixing), this is 
equivalent to
\begin{equation}
\chis = \frac{\xs^2+\ys^2}{2(\xs^2+1)} > \rm \hfagCHISWLIM~at~\CL{95} \,.
\labe{chis}
\end{equation}


%
%
%

\subsubsubsection{Decay width difference \DGs}
\labs{DGs}



Definitions and an introduction to \DGs can also 
be found in \Sec{taubs}.
Neglecting \CP violation, the mass eigenstates are
also \CP eigenstates, with the long-lived state being
\CP-even and the short-lived one being \CP-odd.
Information on \DGs can be obtained by studying the proper time 
distribution of untagged data samples enriched in 
\Bs mesons~\cite{Hartkorn_Moser}.
In the case of an inclusive \Bs selection~\cite{L3B01} or a semileptonic 
\Bs decay selection~\cite{DELPHI_dms_dgs,CDFBS,D0BS2}, 
both the short- and long-lived
components are present, and the proper time distribution is a superposition 
of two exponentials with decay constants 
$\Gs\pm\DGs/2$.
In principle, this provides sensitivity to both \Gs and 
$(\DGGs)^2$. Ignoring \DGs and fitting for 
a single exponential leads to an estimate of \Gs with a 
relative bias proportional to $(\DGGs)^2$. 
An alternative approach, which is directly sensitive to first order in \DGGs, 
is to determine the lifetime of \Bs candidates decaying to \CP
eigenstates; measurements exist for 
\particle{\Bs\to J/\psi\phi}~\cite{CDFIN_BS1,CDFB3,D0BS1} and 
\particle{\Bs\to D_s^{(*)+} D_s^{(*)-}}~\cite{ALEPH_DGs}, which are 
mostly \CP-even states~\cite{Aleksan}. 
However, more recent
time-dependent angular analyses of \particle{\Bs\to J/\psi\phi} 
allow the simultaneous extraction of \DGGs and the \CP-even and \CP-odd 
amplitudes~\cite{CDF2_DGs,D01_DGs}. 
An estimate of \DGGs
has also been obtained directly from a measurement of the 
\particle{\Bs\to D_s^{(*)+} D_s^{(*)-}} branching ratio~\cite{ALEPH_DGs}, 
under the assumption that 
these decays account for all the \CP-even final states 
(however, no systematic uncertainty due to this assumption is given, so 
the average quoted below will not include this estimate).

\begin{table}
\caption{Experimental constraints on \DGGs. The upper limits,
which have been obtained by the working group, are quoted at the \CL{95}.}
\labt{dgammat}
\begin{center}
\begin{tabular}{l|c|c|c}
\hline
Experiment & Method            & $\Delta \Gs/\Gs$ & Ref.  \\
\hline
L3         & lifetime of inclusive \b-sample              
           & $<0.67$   & \cite{L3B01}      \\
DELPHI     & $\Bsb\to D_s^+\ell^- \overline{\nu_{\ell}} X$, lifetime
	   & $<0.46$   & \cite{DELBS0} \\
ALEPH      & $\Bs\to\phi\phi X$ , 
	     \BR{\Bs \to D_s^{(*)+} D_s^{(*)-}}
	   & $0.26^{+0.30}_{-0.15}$ & \cite{ALEPH_DGs} \\
ALEPH      & $\Bs\to\phi\phi X$, lifetime 
           & $0.45^{+0.80}_{-0.49}$ & \cite{ALEPH_DGs}\\
DELPHI     & $\Bsb \to D_s^+$ hadron, lifetime
           & $<0.69$ & \cite{DELBS0}   \\
CDF        & $\Bs \to J/\psi\phi$, lifetime
	   & $0.33^{+0.45}_{-0.42}$ & \cite{CDFIN_BS1} \\ \hline
CDF        & $\Bs \to J/\psi\phi$, time-dependent angular analysis
           & $0.65^{+0.25}_{-0.33} \pm 0.01$ & \cite{CDF2_DGs} \\
\dzero     & $\Bs \to J/\psi\phi$, time-dependent angular analysis
           & $0.21^{+0.33}_{-0.45}$ & \cite{D01_DGs}$^p$ \\
	 \hline
	 \multicolumn{4}{l}{$^p$ \footnotesize
	 Preliminary}
	 \end{tabular}
	 \end{center}
	 \end{table}

Measurements quoting \DGs results are listed in \Table{dgammat}.
There is significant correlation
between \DGGs and $1/\Gamma_s$. In order to combine these measurements,
the two-dimensional log-likelihood for each measurement
in the $(1/\Gs,\,\DGGs)$ plane is summed and the total
normalized with respect to its minimum.  The one-sigma contour (corresponding
to 0.5 units of log-likelihood greater than the minimum) and
95\% contour are found. 
Inputs as indicated in \Table{dgammat} were used in the combination, 
with the exception of the L3~\cite{L3B01} result since the likelihood
for the results was not available, and the 
ALEPH~\cite{ALEPH_DGs} branching ratio result for the reason
given above.

Results of the combination are shown as the one-sigma contour
labelled ``Direct" in both plots of \Fig{DGs}.  Transformation
of variables from $(1/\Gs,\,\DGGs)$ space to other pairs
of variables such as $(1/\Gs,\,\DGs)$ and 
$(\tau_{\rm L} = 1/\Gamma_{\rm L},\,\tau_{\rm H} = 1/\Gamma_{\rm H})$
are also made.
The resulting one-sigma contour for the latter is shown in
\Fig{DGs}(b).

\begin{figure}
\begin{center}
\epsfig{figure=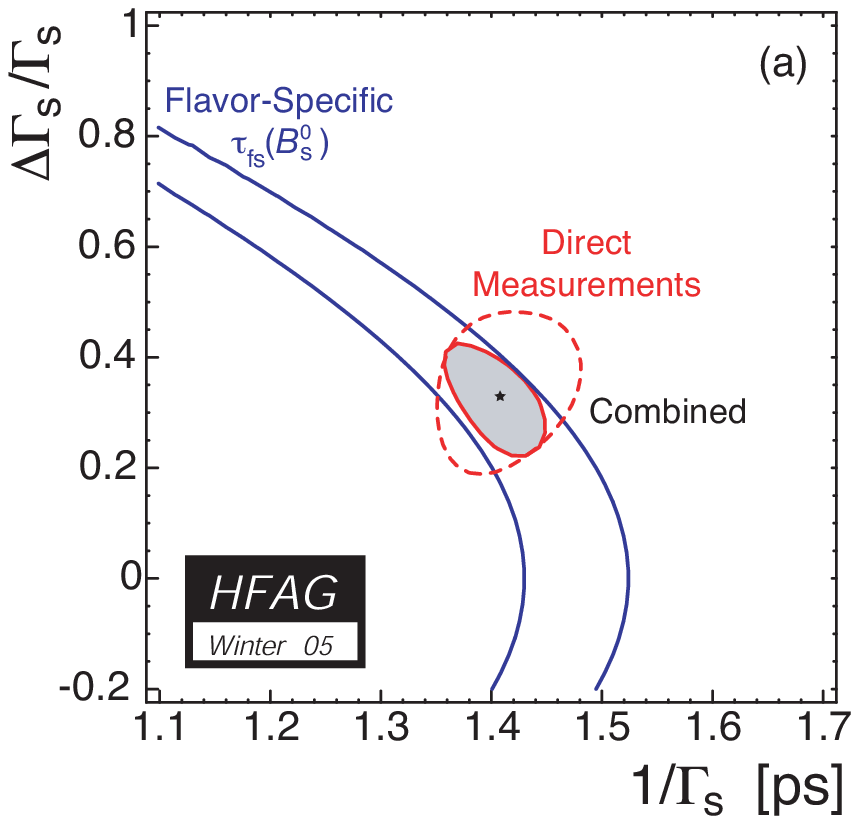,width=0.45\textwidth}
\hfill
\epsfig{figure=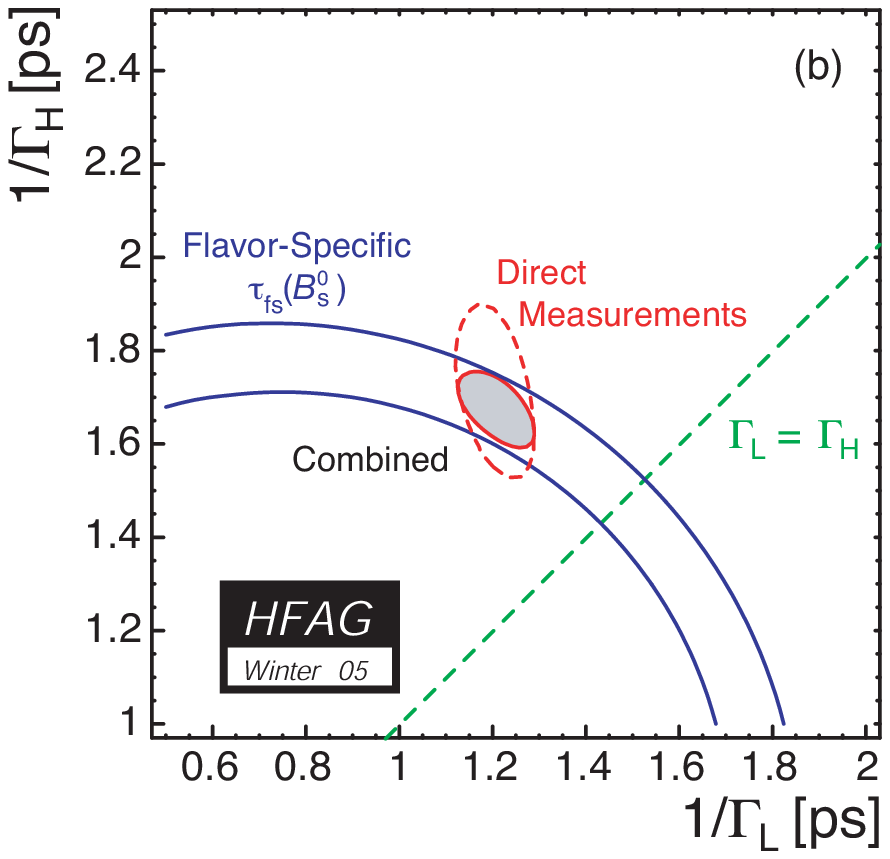,width=0.45\textwidth}
\caption{\DGs combination results with one-sigma contours
($\Delta\log\mathcal{L} = 0.5$) shown for (a) \DGGs versus
$\bar{\tau}(\Bs) = 1/\Gs$  and (b)
$\tau_{\rm H} = 1/\Gamma_{\rm H}$ versus $\tau_{\rm L} = 1/\Gamma_{\rm L}$.
Contours labelled ``Direct" are the result of the combination of
most measurements of \Table{dgammat}, the blue bands are the one-sigma
contours due to the world average of flavor-specific measurements,
and the shaded region the combination of both.
In (b), the diagonal dashed line indicates 
$\Gamma_{\rm L} = \Gamma_{\rm H}$, i.e., where $\DGs = 0$.}
\labf{DGs}
\end{center}
\end{figure}




\newcommand{\comment}[1]{}\comment{

{\em This text below taken by Donatella from previous \DGs notes \ldots}

\newcommand{\Bh}{B^{\rm heavy}_{d,s}}
\newcommand{\Bl}{B^{\rm light}_{d,s}}
\newcommand{\Mh}{m^{\rm heavy}_{d,s}}
\newcommand{\Ml}{m^{\rm light}_{d,s}}
\newcommand{\Gh}{\Gamma^{\rm heavy}_{d,s}}
\newcommand{\Gl}{\Gamma^{\rm light}_{d,s}}
\newcommand{\Gsho}{\Gamma^{\rm short}_{d,s}}
\newcommand{\Glon}{\Gamma^{\rm long}_{d,s}}
\newcommand{\G}{\Gamma_{d,s}}
\newcommand{\tb}{\tau(B^0_{d,s})}
\newcommand{\dg}{\Delta\Gamma_{d,s}}
\newcommand{\tbssemi}{\tau(B_s)_{\rm semi}}
\newcommand{\Bssh}{B^{\rm short}_s}
\newcommand{\tbsshort}{\tau(\Bssh)}

The Standard Model predicts that the \Bs and \Bd can mix before decay. 
The phenomenology of this interaction can be 
described in terms of a $2\times 2$ effective Hamiltonian matrix, 
$M - i\Gamma /2$.
This results in new states called heavy and light, 
$\Bh$  and $\Bl$, for \Bs and \Bd with masses $\Mh$ ,
$\Ml$. Also the  widths  $\Gh$ and $\Gl$ could be different.

Neglecting \CP violation, the mass eigenstates are also \CP eigenstates, the ``long''  state being 
\CP even and the short one being \CP odd.  For convenience of notation, in the following
we therefore substitute 
$\Gl \equiv \Gsho$ and $\Gh \equiv \Glon$, and 
define $\G=1/\tb=(\Glon+\Gsho)/2$ and 
$\Delta \G = \Gsho-\Glon$ which is positive.

$\dg$ is related to the off-diagonal matrix elements, which have been recently 
calculated at the NLO including NLO QCD correction~\cite{Ciuchini2}. 
The theoretical values are:
\begin{equation}
\DGGs = (7.4\pm 2.4 )\times 10^{-2} \,, \hspace{1truecm} \DGGd = (2.42 \pm 0.59 ) \times 10^{-3} \,.
\end{equation}

In the same work the ratio $\DGd/\DGs$ is evaluated since the uncertainties 
coming from higher orders of QCD and $\Lambda_{\rm QCD}/m_{\b}$ corrections cancel out:
\begin{equation}
\DGd/\DGs = (3.2\pm 0.8)\times10^{-2}
\end{equation}

Experimentally \DGs can be measured fitting the lifetime of the 
light and heavy component of the \Bs.
An alternative method is based on the measurement of the  
branching fraction \particle{\Bs\to D_s^{(*)+}D_s^{(*)-}}.
Methods based on lifetime measurements have two different approaches.
Double exponential lifetime fits to samples containing a mixture of \CP eigenstates like
inclusive or semileptonic \Bs decays or $\Bs\to D_s$-hadron have a quadratic sensitivity 
to \DGs.
Whereas the isolation of a single \CP eigenstate as $\Bs\to\phi\phi$ or 
\particle{\Bs\to J/\psi\phi} to extract the lifetime of the \CP-even or odd state have 
a linear dependence on \DGs and it is more sensitive to \DGs but tend 
to suffer from reduced statistics.
The branching fraction method, exploited by ALEPH~\cite{ALEPH-phiphi}, 
is based on several theoretical assumptions~\cite{theoBR}, and allows to have 
information on \DGs only through the branching fraction measurement:
\begin{equation}
\BR{B_s\to D_s^{(*)+}
D_s^{(*)-}} = \frac{\DGs}{\Gs\left(1+\frac{\DGs}{2\Gs}\right)} \,.
\label{eq:dg_ratio}
\end{equation}

The available results are summarized in \Table{dgammat}. 
The values of the limit on \DGGs quoted in the last column of this 
table have been obtained by the working group.

Details on how these measurements are included in the average can be found  
in the previous summaries~\cite{Pcomb}.

\begin{table}
\caption{Experimental constraints on \DGGs. The upper limits,
which have been obtained by the working group, are quoted at the \CL{95}.}
\labt{dgammat}
\begin{center}
\begin{tabular}{|l|c|c|c|} 
\hline
Experiment & Selection        & Measurement            & $\Delta \Gs/\Gs$ \\ 
\hline
L3~\cite{L3B01}         & inclusive \b-sample              &                               & $<0.67$         \\
DELPHI~\cite{DELBS0}     & $\Bsb\to D_s^+\ell^- \overline{\nu_{\ell}} X$ & $\tbssemi=(1.42^{+0.14}_{-0.13}\pm0.03)$~ps  & $<0.46$ \\
others~\cite{ref:others}& $\Bsb\to D_s^+\ell^-  \overline{\nu_{\ell}} X$  & $\tbssemi=(1.46\pm{0.07})$~ps & $<0.30$ \\
ALEPH~\cite{ALEPH-phiphi}      & $\Bs\to\phi\phi X$      & 
$\BR{\Bssh \to D_s^{(*)+} D_s^{(*)-}} =(23\pm10^{+19}_{-~9})\%$       & $0.26^{+0.30}_{-0.15}$ \\
ALEPH~\cite{ALEPH-phiphi}      & $\Bs\to\phi\phi X$      & $\tbsshort=(1.27\pm0.33\pm0.07)$~ps           & 
$0.45^{+0.80}_{-0.49}$ \\ 
DELPHI~\cite{DELBS0}$^a$    & $\Bsb \to D_s^+$ hadron    
&  $\tau_{\rm B^{D_s-had.}_s}=(1.53^{+0.16}_{-0.15}\pm0.07)$~ps                          & $<0.69$         \\
CDF~\cite{CDFB01} & $\Bs \to {\rm J}/\psi\phi$        
& $\tau_{\rm B^{{\rm J}/\psi \phi}_s}=(1.34^{+0.23}_{-0.19}\pm0.05)$~ps & $0.33^{+0.45}_{-0.42}$ \\ 
\hline
\multicolumn{4}{l}{$^a$ \footnotesize 
The value quoted for the measured lifetime differs
slightly from the one quoted in \Table{bs} because it} \\[-1ex]
\multicolumn{4}{l}{~~ \footnotesize 
corresponds to the present status of the analysis in which the information
on \DGs has been obtained.}
\end{tabular}
\end{center}
\end{table}

Here only a short description will be given.

L3 and DELPHI use inclusively reconstructed
\Bs and $\Bs\to \particle{D_s} \ell\nu X$ events respectively.
If those sample are fitted assuming a single exponential lifetime then,
assuming \DGGs is small, the measured lifetime is given by:
\begin{equation}
\tau(\Bs)_{\rm incl.} = \frac{1}{\Gs} \frac{1}{1-\left(\frac{\DGs}{2\Gs}\right)^2}
\quad \quad ; \quad \quad
\tau(\Bs)_{\rm semi.} = \frac{1}{\Gs} 
\frac{{1+\left(\frac{\DGs}{2\Gs}\right)^2}}{{1-\left(\frac{\DGs}{2\Gs}\right)^2}}.     
\end{equation}

The single lifetime fit is thus more sensitive to the effects of 
\DGs in the semileptonic case than in the fully inclusive case.

The same method is used for the \Bs world average lifetime
(recomputed without the DELPHI measurement) obtained
by using only the semileptonic decays and referenced in \Table{dgammat} as {\it others}.

The technique of reconstructing only decays at defined \CP
has been exploited by ALEPH, DELPHI and CDF.

ALEPH reconstructs the decay
$\Bs\to \particle{D_s^{(*)+}D_s^{(*)-}} \to \phi\phi X$
which is predominantly \CP even.
The proper time dependence of the \Bs component is a simple
exponential and the lifetime is related to \DGs via
\begin{equation}
\frac{\Delta \Gs}{\Gs}=2(\frac{1}{\Gs~\tbsshort}-1).  
\end{equation} 
 The same data have been used by ALEPH to exploit the branching fraction method.

DELPHI uses a sample of $\Bs\to D_s$-hadron,
which is expected to have an increased \CP-even component as the contribution
due to \particle{D_s^{(*)+}_s^{(*)-}} events is enhanced by
selection criteria.

CDF reconstructs \particle{\Bs\to J/\psi\phi} with
\particle{J/\psi\to\mu^+\mu^-} and \particle{\phi\to K^+K^-}
where the \CP-even component is equal to $0.84\pm 0.16$ obtained by
combining CLEO~\cite{cleo} measurement of \CP-even fraction in
\particle{\Bd\to J/\psi K^{*0}} and possible SU(3) symmetry
correction.

In order to combine all the measurements~\footnote{L3 is not 
included since the likelihood for the results
was not available} the two-dimensional log-likelihood in the ($1/\Gs$, \DGGs) 
plane is summed and normalized with respect to its minimum.
The 68\%, 95\% and \CL{99} contours of the combined negative 
log-likelihood are shown in \Fig{dgplot} (left)
The corresponding limit on \DGGs is:
\begin{eqnarray}
\DGGs & = & 0.16^{+0.15}_{-0.16}  \,, \\
\DGGs & < & 0.54~\mbox{at \CL{95}} \,. 
\end{eqnarray}

\begin{figure}
\begin{center}
\epsfig{figure=figures/osc/dg_w_notaubd_bw_2d.eps,width=\textwidth}
\epsfig{figure=figures/osc/dg_w_taubd_bw_2d.eps,width=\textwidth}
\end{center}
\caption{Top: 68\%, 95\% and \CL{99} contours of the negative log-likelihood 
distribution in the plane ($1/\Gs$, \DGGs).
Bottom: Same, but with the constraint $1/\Gs \equiv\tau_{\Bd}$} 
\labf{dgplot}
\end{figure}

\begin{figure}
\begin{center}
\epsfig{figure=figures/osc/dg_w_taubd_col_1d,width=\textwidth}
\end{center}
\caption{Probability density distribution for \DGGs after applying the constraint; 
the three shaded regions show the limits at the 68\%, 95\% and \CL{99} respectively.} 
\labf{dgprobplot}
\end{figure}

An improved limit on \DGGs can be obtained by applying the $\tau_{\Bd}=\HFAGtauBd$ constraint.
The world average \Bs lifetime is not used, as its meaning 
is not clear if $\Delta \Gs$ is non-zero.
This is well motivated theoretically, as 
the total widths of the \Bs and \Bd mesons
are expected to be 
equal within less than one percent~\cite{bigilife}, \cite{Beneke}
and \DGd is expected to be small. 
 
The two-dimensional log-likelihood obtained, after including the constraint is shown in 
\Fig{dgplot} (right). The resulting probability density distribution for \DGGs is 
shown in \Fig{dgprobplot}. The corresponding limit on \DGGs is:
\begin{eqnarray}
\DGGs & = & 0.07^{+0.09}_{-0.07} \,, \\
\DGGs & < & 0.29~~\mbox{at \CL{95}} \,.
\end{eqnarray}

} 




Numerical results of the combination of the described inputs
of \Table{dgammat} are:
\begin{eqnarray}
\DGGs &\in& [\hfagDGSGSlow,\hfagDGSGSupp] ~ \mbox{at \CL{95}} \,, \\
\DGGs &=& \hfagDGSGS \,, \\
\DGs &=& \hfagDGS \,, \\
\bar{\tau}(\Bs) = 1/\Gs &=& \hfagTAUBSMEAN \,, \\
1/\Gamma_{\rm L} &=& \tau_{\rm short} = \hfagTAUBSL \,, \\
1/\Gamma_{\rm H} &=& \tau_{\rm long} = \hfagTAUBSH \,. 
\end{eqnarray}

Flavor-specific lifetime measurements are of an equal mix
of \CP-even and \CP-odd states at time zero, and  
if a single exponential function is used in the likelihood
lifetime fit of such a sample~\cite{Hartkorn_Moser}, 
\begin{equation}
\tau(\Bs)_{\rm fs} = \frac{1}{\Gs}
\frac{{1+\left(\frac{\DGs}{2\Gs}\right)^2}}{{1-\left(\frac{\DGs}{2\Gs}\right)^2}
}.
\end{equation}
Using the world average flavor-specific 
lifetime\footnote{The world average of all \Bs lifetime 
measurements using flavour-specific final states is \hfagTAUBSSL; however,
for the purpose of the \DGs extraction, we remove from this average one
DELPHI analysis that is already included in the set of ``direct 
measurements'' and obtain \hfagTAUBSSLX, 
shown as the blue bands on the two plots of \Fig{DGs}.} of \Sec{taubs}
the one-sigma blue bands shown in \Fig{DGs} are obtained. 
Higher-order corrections were checked to be negligible in the
combination.
When these flavor-specific measurements
are combined with the measurements of \Table{dgammat}, the shaded
regions of \Fig{DGs} are obtained, with numerical results:
\begin{eqnarray}
\DGGs &\in& [\hfagDGSGSCONlow,\hfagDGSGSCONupp] ~ \mbox{at \CL{95}} \,, \\
\DGGs &=& \hfagDGSGSCON \,, \\
\DGs &=& \hfagDGSCON \,, \\
\bar{\tau}(\Bs) = 1/\Gs &=& \hfagTAUBSMEANCON \,, \\
\rho(\DGGs, 1/\DGs) &=& \hfagRHODGSGSTAUBSMEANCON \,, \\
1/\Gamma_{\rm L} &=& \tau_{\rm short} = \hfagTAUBSLCON \,, \\
1/\Gamma_{\rm H} &=& \tau_{\rm long} = \hfagTAUBSHCON \,. 
\end{eqnarray}

\begin{figure}
\begin{center}
\epsfig{figure=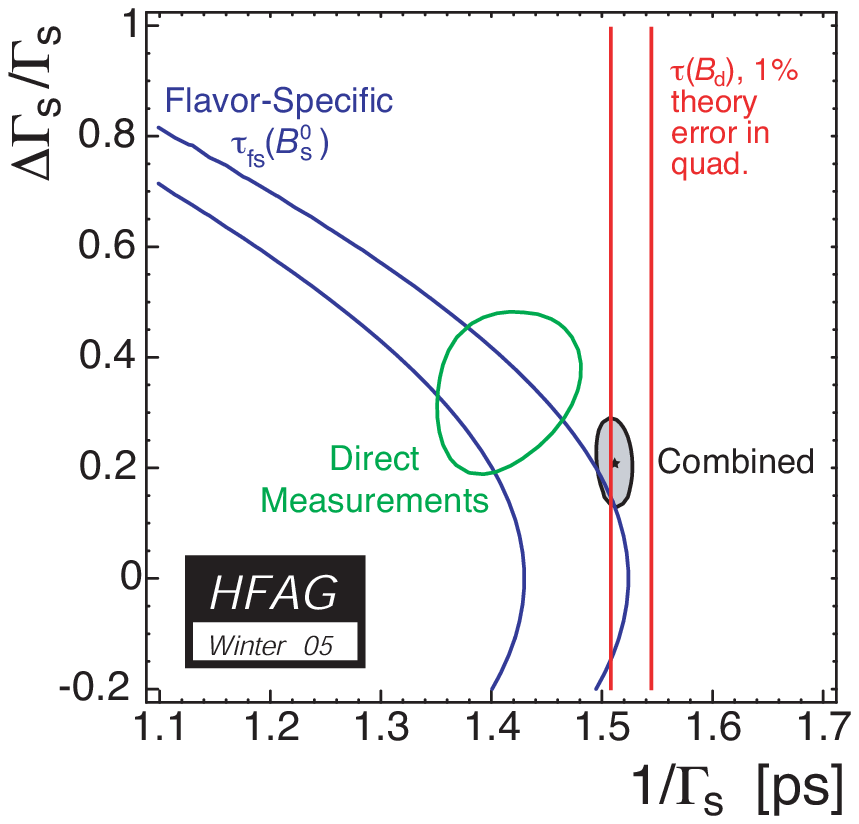,width=0.45\textwidth}
\caption{\DGs combination results with one-sigma contours
($\Delta\log\mathcal{L} = 0.5$) shown for \DGGs versus
$\bar{\tau}(\Bs) = 1/\Gs$.
The contour labelled ``Direct" is the result of the combination of
most measurements of \Table{dgammat}, the blue band is the one-sigma
contour due to the world average of flavor-specific measurements,
the red vertical band the one-sigma constraint of
$\bar{\tau}(\Bs) = 1/\Gs = \tau(\Bd) = 1/\Gd$ (with an additional
1\% error added in quadrature),
and the shaded region the combination.}
\labf{DGscon}
\end{center}
\end{figure}

The {\em average} \Bs and \Bd
lifetimes are predicted to be equal within 
1\%~\cite{equal_lifetimes,Gabbiani_et_al}
and an informative constraint to apply is to 
set $\Gs = \Gd$.
The constraint $1/\Gs = \tau(\Bd)$ is used,  
where $\tau(\Bd) = \hfagTAUBD$ is the world average of experimental results,
including an additional relative 1\% theoretical uncertainty added
in quadrature with the indicated error.
With this constraint, one obtains the one-sigma contour shown in
\Fig{DGscon}, and a numerical value of
$\DGGs = \hfagDGSGSCONBDCON$. This and the results above can
be compared with the theoretical prediction of 
$\DGs/\Gs = 0.12 \pm 0.05$~\cite{delta_gams}.



\mysection{Semileptonic $B$ decays}
\label{slbdecays}

In  this  edition  of the  HFAG  updates,  we  present updated values for
the total semileptonic branching fraction, the branching fractions
$\cbf(\BzbDstarlnu)$ and $\cbf(\BzbDplnu)$, and the measurements of
$F(1)\vcb$ versus $\rho^2$ and $G(1)\vcb$ versus $\rho^2$.  For other
quantities the 
reader is referred to the previous HFAG document.~\cite{hfag_hepex}

The determination of \vub\ from inclusive decays is the subject of
ongoing intense activity in both experiment and theory.  HFAG
subgroups are currently determining updated values for the heavy quark
parameters $m_b$ and $\mu_{\pi}^2$ based on moments measured in
inclusive $\Bxclnu$ and $B\to X_s\gamma$ decays.  In addition, the
theoretical tools have improved~\cite{ref:blnp} and these improvements are
being incorporated by the experiments.  A comprehensive determination
of \vub\ from inclusive decays will be made based on the results
presented at the 2005 summer conferences.

Some new measurements of exclusive $\Bxulnu$ decays were presented at
the winter 2005 conferences.  A subgroup is working on the averaging
of these results and the determination of \vub\ from them.  In this
article the results are summarized, but no average is quoted.

In the following a detailed description of all parameters and
published analyses (including preliminary results) relevant for the
determination of the combined results is provided.  The description is
based on the information available on the web page at

\medskip
 \centerline{\tt http://www.slac.stanford.edu/xorg/hfag/semi/winter05/winter05.shtml}
\medskip

In the combination of the published results, the central values and
errors are rescaled to a common set of input parameters, summarized in
Table~\ref{tab:common.param} and provided in the file {\tt
common.param} (accessible from the web page).  All measurements with a
dependency on any of these parameters are rescaled to the central
values given in Table~\ref{tab:common.param}, and their error is
recalculated based on the error provided in the column ``Excursion''.
The detailed dependency for each measurement is contained in files
(provided by the experiments) accessible from the web page.

\begin{table}[!htb]
\caption{Common input parameters for the combination of semileptonic
$B$ decays.  Most of the parameters are taken from
Ref.~\cite{Eidelman:2004wy}.  This table is encoded in the file {\tt
common.param}. The units are picoseconds for lifetimes and percentage
for branching fractions.}
\begin{center}
\begin{tabular}{|l|c|c|l|}\hline
Parameter    &Assumed Value  &Excursion             &Description\\
\hline\hline 
rb           &$21.646$       &$\pm0.065$            &$R_b$\\
bdst         &$1.27$         &$\pm0.021$            &$\cbf(\Bb\to \Dstar\tau\nub)$\\
bdsd         &$1.62$         &$\pm0.040$            &$\cbf(\Bb\to \Dstar D)$\\
bdst2        &$0.65$         &$\pm0.013$            &$\cbf(b\to \Dstar \tau)$ (OPAL incl) \\
bdsd2        &$4.2$          &$\pm1.5$              &$\cbf(b\to \Dstar D)$ (OPAL incl)\\
bdsd3        &$0.87$         &${}^{+0.23}_{-0.19}$  &$\cbf(b\to \Dstar D)$ (DELPHI incl)\\
xe           &$0.702$        &$\pm0.008$            &$B$ fragmentation:
$\langle E_B\rangle/E_{\rm beam}$\\
bdsi         &$17.3$         &$\pm2.0$              &$\cbf(b\to \Dstarp\ \mathrm{incl})$\\
cdsi         &$22.6$         &$\pm1.4$              &$\cbf(c\to \Dstarp\ \mathrm{incl})$\\
\hline
tb0          &$1.532$        &$\pm0.009$            &$\tau(\Bz)$\\
tbplus       &$1.638$        &$\pm0.011$            &$\tau(\Bp)$\\
tbps         &$1.469$        &$\pm0.059$            &$\tau(\Bs)$\\
\hline
fbd          &$39.9$         &$\pm1.0$              &$\Bz$ fraction at $\sqrt{s} = m_{Z^0}$\\
fbs          &$10.3$         &$\pm1.5$              &$\Bs$ fraction at $\sqrt{s} = m_{Z^0}$\\
fbar         &$10.0$          &$\pm1.7$              &Baryon fraction at $\sqrt{s} = m_{Z^0}$\\
\hline
dst          &$67.7$         &$\pm0.5$              &$\cbf(\Dstarp\to \Dz\pi^+)$\\
dkpp         &$9.20$         &$\pm0.6$              &$\cbf(\Dp \to K^-\pi^+\pi^+)$\\
dkp          &$3.80$         &$\pm0.09$             &$\cbf(\Dz \to K^-\pi^+)$\\
dkpzp        &$13.0$         &$\pm0.8$              &$\cbf(\Dz \to K^-\pi^+\piz)$\\
dkppp        &$7.46$         &$\pm0.31$             &$\cbf(\Dz \to K^-\pi^+\pi^+\pi^-)$\\
dkzpp        &$2.99$         &$\pm0.18$             &$\cbf(\Dz \to K^0\pi^+\pi^-)$\\
dkln         &$7.0$          &$\pm0.4$              &$\cbf(\Dz \to K^-\ell^+\nu)$\\
dkk          &$4.3$          &$\pm0.2$              &$\cbf(\Dz \to K^-K^+)$\\
dkx          &$1.100$        &$\pm0.025$            &$K^-\pi^+X$ rates\\
dkox         &$0.42$         &$\pm0.05$             &$\cbf(\Dz \to K^0X)$\\
dnlx         &$6.87$         &$\pm0.28$             &$\cbf(\Dz \to X\ell\nub)$\\
dkpcl        &$61.2$         &$\pm2.9$              &$\cbf(\Dstarz \to \Dz\piz)$\\
dssR         &$0.64$         &$\pm0.11$             &$\cbf(b\to D^{**}\ell\nub)\times\cbf(D^{**}\to D^{*+}X)$ \\
\hline
fb0          &$49.8$         &$\pm0.9$              &$f^{00} = \cbf(\FourS\to \Bz\Bzb)$\\
chid         &$0.187$        &$\pm0.003$            &$\chi_d$, time-integrated probability for \Bz\ mixing\\
chi          &$0.0930$        &$\pm0.0023$            &$\chi = \chi_d\times (f^{00}/100)$\\
\hline
\end{tabular}
\end{center}
\label{tab:common.param}
\end{table}


\mysubsection{Exclusive Cabibbo-favored decays}
\label{slbdecays_b2cexcl}

Aspects of the phenomenology of exclusive Cabibbo-favored $B$ decays
and their use in the determination of \vcb\ in the context of Heavy
Quark Effective Theory (HQET) are described in many places, \eg, in
Ref.~\cite{maeb} and will not be repeated here.

Averages are provided for the branching fractions
$\cbf(\BzbDstarlnu)$ plus $\cbf(\BzbDplnu)$.  In addition,
averages are provided for
$\vcb F(1)$.vs.$\rho^2$, where $F(1)$ and $\rho^2$
are the normalization and slope of the form factor at zero
recoil in $\BzbDstarlnu$ decays, and for the corresponding quantities
$\vcb G(1)$.vs.$\rho^2$ in $\BzbDplnu$ decays.

\mysubsubsection{\BzbDstarlnu}
\label{slbdecays_dstarlnu}

The measurements included in the average, shown in
Table~\ref{tab:dstarlnu} are scaled to a consistent set of input
parameters and their errors.  Therefore some of the (older)
measurements are subject to considerable adjustments.

\begin{itemize}
 
  \item In order to reduce the dependence on theoretical error
 estimates, the central values and errors for the form factors $R_1$
 and $R_2$ are taken from the measurement by
 CLEO~\cite{Duboscq:1995mv}. However, all experiments (except for the
 CLEO~\cite{Adam:2002uw}, the recent DELPHI~\cite{Abdallah:2004rz} and
 the \babar~\cite{Aubert:2004bw} measurements) quote in their
 abstracts $F(1)\vcb$ based on form factors (and their respective
 errors) from theory.  \belle provides a second result evaluated with
 the CLEO form factors.  All other experiments have recalculated
 $F(1)\vcb$ to rely on the CLEO form factors.  In the future, a
 substantial improvement in the error associated with form factors is
 expected by including form factor measurements at the $B$ factories.
 
  \item Updates in the branching fractions of $D$ mesons and the
 production of $D^{**}$ mesons in $B$ decays have generally lead to
 increased rates for $\cbf(\BzbDstarlnu)$.

  \item The average \Bz\ lifetime has changed considerably since the
 ALEPH measurement, especially with the much higher precision
 available at the $B$ factories. This effect is less visible in the
 other measurements as they are more recent.
 
  \item The production of \Bz\ mesons at $\sqrt{s} = m_{Z^0}$ has a
 direct impact on all LEP measurements.  Adjusting results on the
 \FourS\ for the branching fraction of $\cbf(\FourS\to \BzBzb) \equiv
 f^{00}$ yields an increase compared to the assumption of $f^{00} =
 0.5$.
 
  \item Many input parameters are now known with a much increased
 precision---this decreases some of the systematic errors of the
 rescaled results with respect to the original publication.
 
\end{itemize}

The largest correlated errors are the fraction of \Bz\ mesons (fbd and
fb0, respectively), the form factors $R_1(1)$ and $R_2(1)$ at zero
recoil, \Bz\ meson lifetime, branching fractions of $D$ mesons, and
the details of $D^{**}$ modeling.

\begin{figure}[!ht]
 \begin{center}
  \unitlength1.0cm 
  \begin{picture}(14.,8.0)  
   \put( -0.5,  0.0){\includegraphics[width=7.5cm]{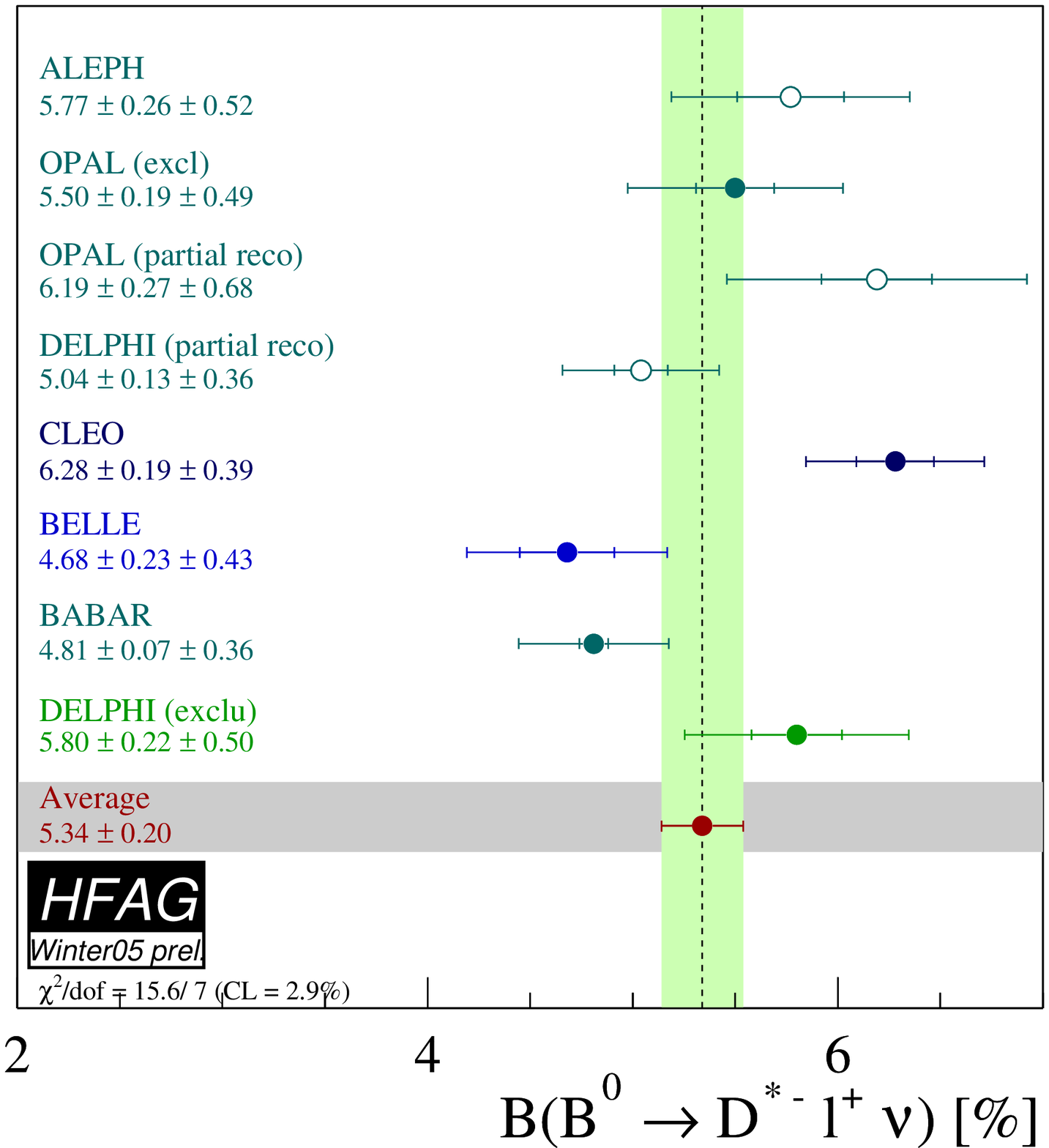}}

   \put(  8.0,  0.0){\includegraphics[width=7.5cm]{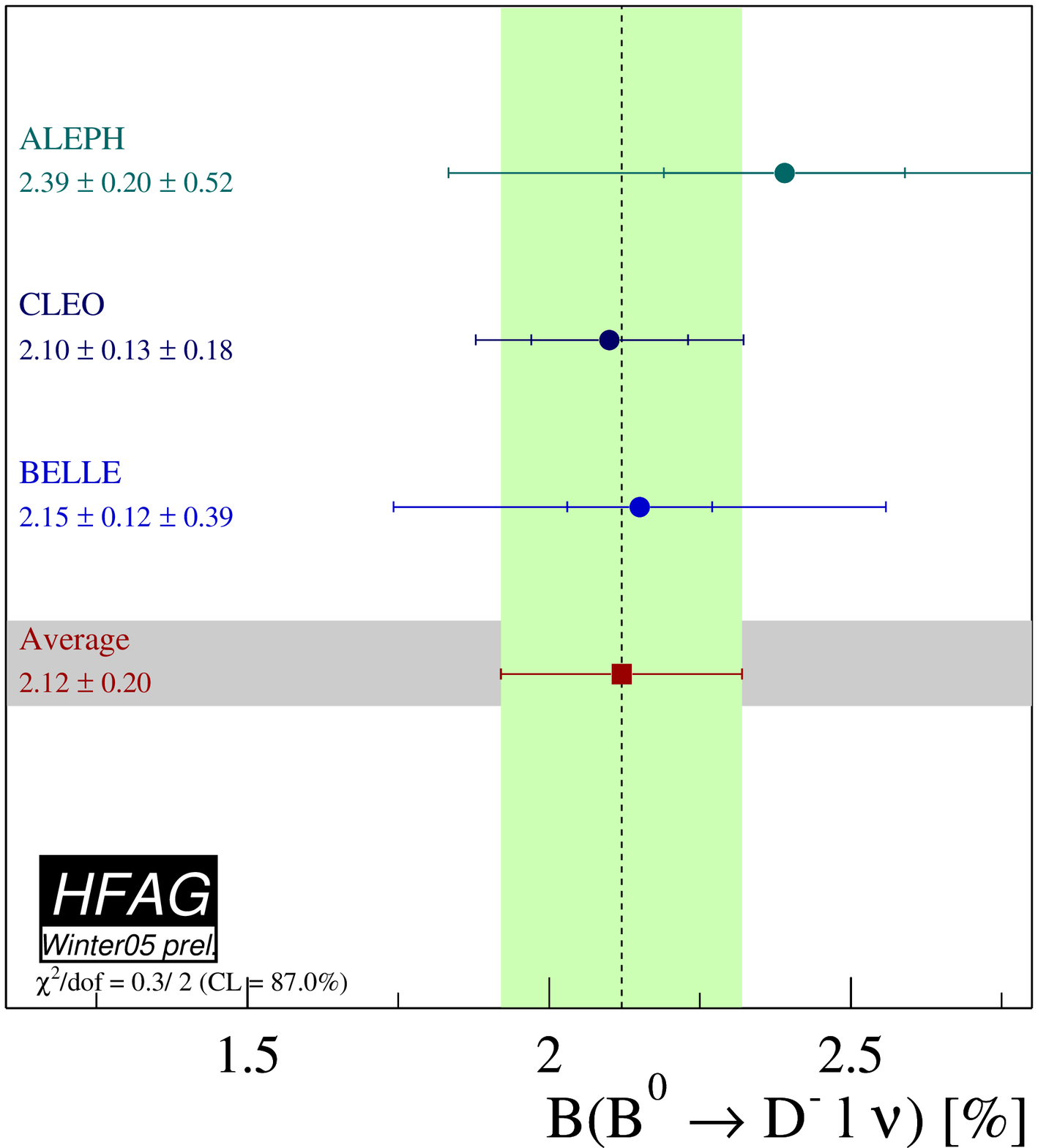}}
    
   \put(  5.5,  6.8){{\large\bf a)}}
   \put( 14.0,  6.8){{\large\bf b)}}
  \end{picture}
  \caption{Average branching fraction of  exclusive semileptonic $B$ decays. (a) \BzbDstarlnu\ and (b) \BzbDplnu.}
  \label{fig:brdsl}
 \end{center}
\end{figure}

At LEP, the measurements of \BzbDstarlnu\ decays have been done both
with ``inclusive'' analyses based on a partial reconstruction of the
\BzbDstarlnu\ decay and a full reconstruction of the exclusive decay.
The average branching fraction $\cbf(\BzbDstarlnu)$ is determined in a
one-dimensional fit from the measurements provided in
Table~\ref{tab:dstarlnu}.  The statistical correlation between two
analyses from the same experiment (DELPHI and OPAL, respectively) is
taken into account.  Figure~\ref{fig:brdsl}(a) illustrates the
measurements and the resulting average.  The $\chi^2/\dof = 15.6/7$
corresponds to a 2.9\%\ confidence level, suggesting some caution
in interpreting the errors on the average.

\begin{table}[!htb]
\caption{Average branching fraction $\cbf(\BzbDstarlnu)$ and individual
  results. See the description in the text for explanations why the
  published results are lower than the rescaled results. }
\begin{center}
\begin{tabular}{|l|c|c|}\hline
Experiment    &$\cbf(\BzbDstarlnu) [\%]$    (rescaled)     &$\cbf(\BzbDstarlnu) [\%]$  (published) \\
\hline\hline 
ALEPH (excl)~\hfill\cite{Buskulic:1996yq}  &$5.77\pm 0.26_{\rm stat} \pm 0.36_{\rm syst}$ &$5.53\pm 0.26_{\rm stat} \pm 0.52_{\rm syst}$ \\
OPAL (excl)~\hfill\cite{Abbiendi:2000hk}   &$5.50\pm 0.19_{\rm stat} \pm 0.39_{\rm syst}$ &$5.11\pm 0.19_{\rm stat} \pm 0.49_{\rm syst}$ \\
OPAL (incl)~\hfill\cite{Abbiendi:2000hk}   &$6.19\pm 0.27_{\rm stat} \pm 0.58_{\rm syst}$ &$5.92\pm 0.27_{\rm stat} \pm 0.68_{\rm syst}$ \\
DELPHI (incl)~\hfill\cite{Abreu:2001ic}    &$5.04\pm 0.13_{\rm stat} \pm 0.36_{\rm syst}$ &$4.70\pm 0.13_{\rm stat} \ {}^{+0.36}_{-0.31}\ {}_{\rm syst}$ \\
\belle  (excl)~\hfill\cite{Abe:2001cs}      &$4.68\pm 0.23_{\rm stat} \pm 0.42_{\rm syst}$ &$4.60\pm 0.23_{\rm stat} \pm 0.40_{\rm syst}$ \\
CLEO  (excl)~\hfill\cite{Adam:2002uw}      &$6.28\pm 0.19_{\rm stat} \pm 0.39_{\rm syst}$ &$6.09\pm 0.19_{\rm stat} \pm 0.40_{\rm syst}$ \\
DELPHI (excl)~\hfill\cite{Abdallah:2004rz} &$5.80\pm 0.22_{\rm stat} \pm 0.46_{\rm syst}$ &$5.90\pm 0.22_{\rm stat} \pm 0.50_{\rm syst}$ \\
\babar\ (excl)~\hfill\cite{Aubert:2004bw}  &$4.81\pm 0.07_{\rm stat} \pm 0.34_{\rm syst}$ &$4.90\pm 0.07_{\rm stat} \pm 0.36_{\rm syst}$ \\
\hline 
{\bf Average}                              &\mathversion{bold}$5.34\pm0.20$           &\mathversion{bold}$\chi^2/\dof = 15.6/7$ \\
\hline 
\end{tabular}
\end{center}
\label{tab:dstarlnu}
\end{table}

\begin{figure}[!ht]
 \begin{center}
  \unitlength1.0cm 
  \begin{picture}(14.,8.0)  
   \put( -0.5,  0.0){\includegraphics[width=7.5cm]{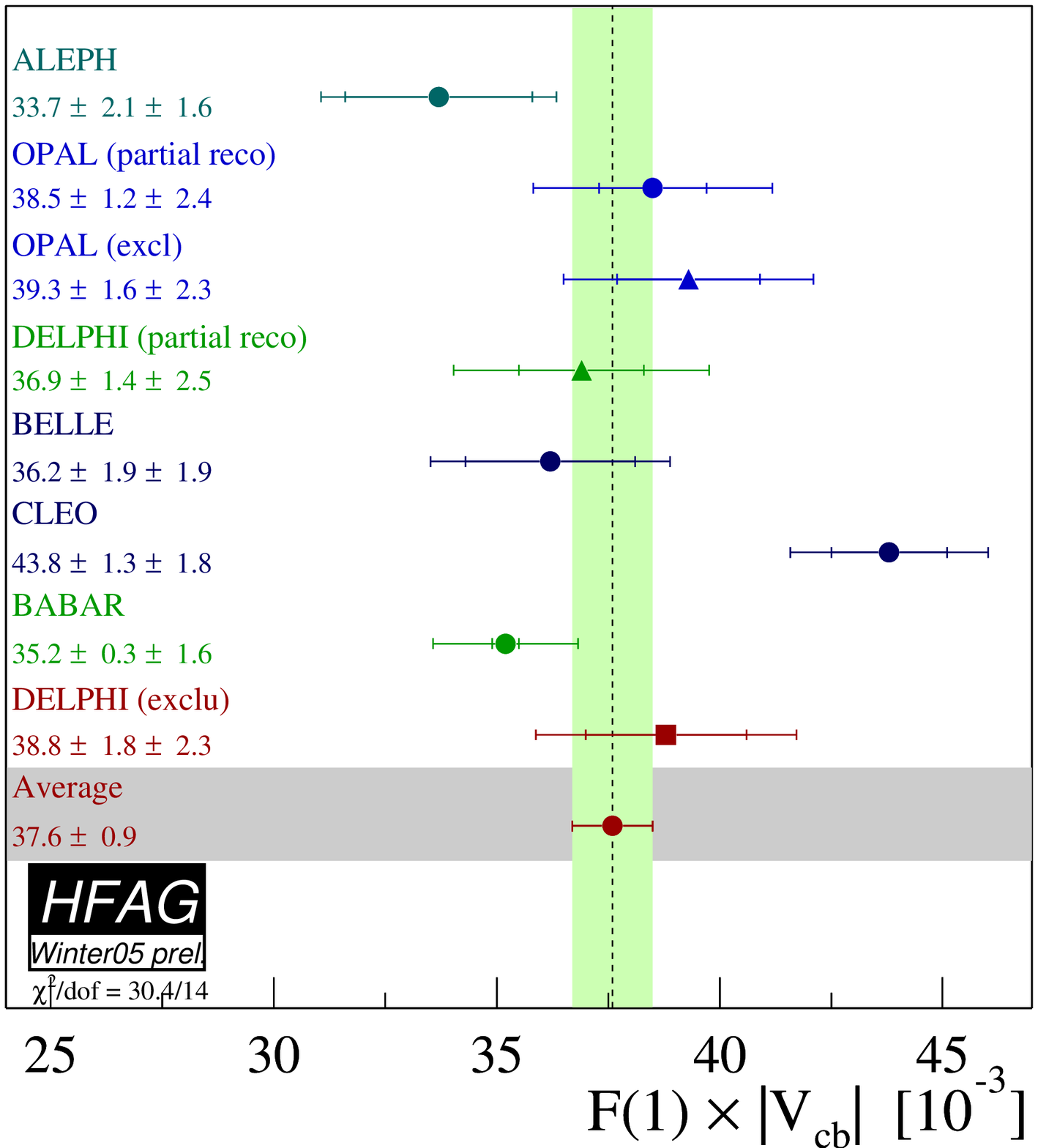}}
   \put(  8.0, -0.2){\includegraphics[width=8.0cm]{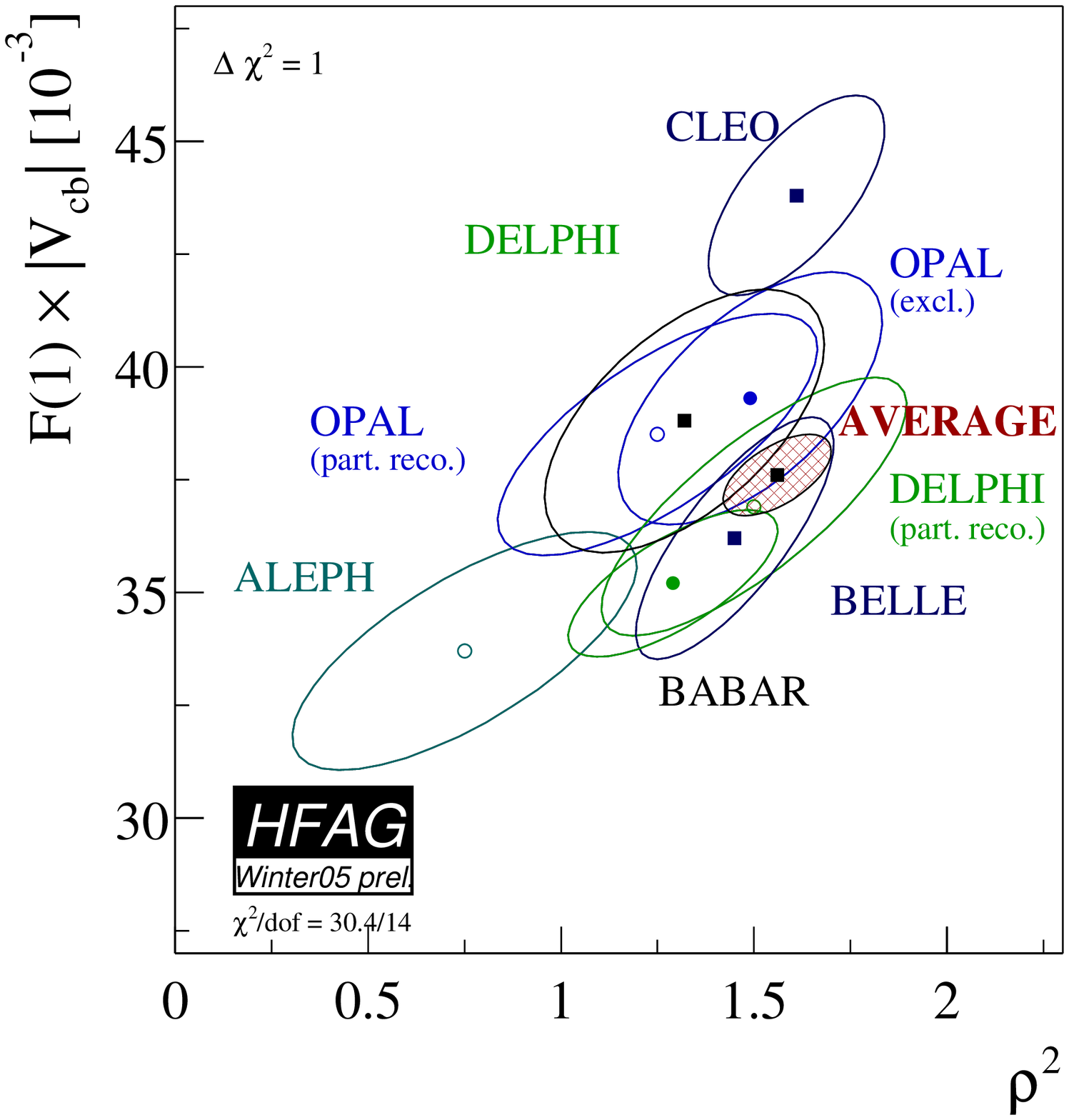}}
   \put(  5.5,  6.8){{\large\bf a)}}  
   \put( 14.4,  6.8){{\large\bf b)}}
   \end{picture} \caption{(a)  Illustration of the  average $F(1)\vcb$
   and   rescaled  measurements   of   exclusive  \BzbDstarlnu\   decays
   determined   in  a  two-dimensional   fit.   (b)   Illustration  of
   $F(1)\vcb$   vs.  $\rho^2$.  The   error  ellipses   correspond  to
   $\Delta\chi^2 = 1$.  }  \label{fig:vcbf1} \end{center}
\end{figure}

The average for $F(1)\vcb$ is determined by the two-dimensional
combination of the results provided in Table~\ref{tab:vcbf1}.  This
allows the correlation between $F(1)\vcb$ and $\rho^2$ to be
maintained. Figure~\ref{fig:vcbf1}(b) illustrates the average
$F(1)\vcb$ and the measurements included in the
average. Figure~\ref{fig:vcbf1}(a) provides a one-dimensional
projection for illustrative purposes.  The $\chi^2/\dof = 30.4/14$ 
corresponds to a 0.7\%\ confidence level; 
the errors on the average should be taken with caution
(no scale factor is applied).

\begin{table}[!htb]

\caption{Average of $F(1)\vcb$ determined in the decay \BzbDstarlnu\ and
individual  results.   The  fit  for  the average  has  $\chi^2/\dof  =
30.4/14$.  The total  correlation between  the average  $F(1)\vcb$ and
$\rho^2$ is 0.59.}

\begin{center}
\begin{tabular}{|l|c|c|}\hline
Experiment                                 &$F(1)\vcb [10^{-3}]$ (rescaled) &$\rho^2$ (rescaled) \\
                                           &$F(1)\vcb [10^{-3}]$ (published) &$\rho^2$ (published) \\
\hline\hline 
ALEPH (excl)~\hfill\cite{Buskulic:1996yq}  &$33.7\pm 2.1_{\rm stat} \pm 1.6_{\rm syst}$   &$0.75\pm 0.25_{\rm stat} \pm 0.37_{\rm syst}$ \\
                                           &$31.9\pm 1.8_{\rm stat} \pm 1.9_{\rm syst}$   &$0.37\pm 0.26_{\rm stat} \pm 0.14_{\rm syst}$ \\
\hline
OPAL (incl)~\hfill\cite{Abbiendi:2000hk}   &$38.5\pm 1.2_{\rm stat} \pm 2.4_{\rm syst}$   &$1.25\pm 0.14_{\rm stat} \pm 0.39_{\rm syst}$ \\
                                           &$37.5\pm 1.2_{\rm stat} \pm 2.5_{\rm syst}$   &$1.12\pm 0.14_{\rm stat} \pm 0.29_{\rm syst}$ \\
\hline
OPAL (excl)~\hfill\cite{Abbiendi:2000hk}   &$39.3\pm 1.6_{\rm stat} \pm 1.8_{\rm syst}$   &$1.49\pm 0.21_{\rm stat} \pm 0.26_{\rm syst}$ \\
                                           &$36.8\pm 1.6_{\rm stat} \pm 2.0_{\rm syst}$   &$1.31\pm 0.21_{\rm stat} \pm 0.16_{\rm syst}$ \\
\hline
DELPHI (incl)~\hfill\cite{Abreu:2001ic}    &$36.9\pm 1.4_{\rm stat} \pm 2.5_{\rm syst}$   &$1.50\pm 0.14_{\rm stat} \pm 0.37_{\rm syst}$ \\
                                  &$35.5\pm1.4_{\rm stat}\ {}^{+2.3}_{-2.4}{}_{\rm syst}$ &$1.34\pm0.14_{\rm stat}\ {}^{+0.24}_{-0.22}{}_{\rm syst}$\\
\hline
\belle  (excl)~\hfill\cite{Abe:2001cs}     &$36.2\pm 1.9_{\rm stat} \pm 1.9_{\rm syst}$   &$1.45\pm 0.16_{\rm stat} \pm 0.20_{\rm syst}$ \\
                                           &$35.8\pm 1.9_{\rm stat} \pm 1.9_{\rm syst}$   &$1.45\pm 0.16_{\rm stat} \pm 0.20_{\rm syst}$ \\
\hline
CLEO  (excl)~\hfill\cite{Adam:2002uw}      &$43.8\pm 1.3_{\rm stat} \pm 1.8_{\rm syst}$   &$1.61\pm 0.09_{\rm stat} \pm 0.21_{\rm syst}$ \\
                                           &$43.1\pm 1.3_{\rm stat} \pm 1.8_{\rm syst}$   &$1.61\pm 0.09_{\rm stat} \pm 0.21_{\rm syst}$ \\
\hline
DELPHI (excl)~\hfill\cite{Abdallah:2004rz} &$38.8\pm 1.8_{\rm stat} \pm 2.1_{\rm syst}$   &$1.32\pm 0.15_{\rm stat} \pm 0.33_{\rm syst}$ \\
                                           &$39.2\pm 1.8_{\rm stat} \pm 2.3_{\rm syst}$   &$1.32\pm 0.15_{\rm stat} \pm 0.33_{\rm syst}$ \\
\hline
\babar\ (excl)~\hfill\cite{Aubert:2004bw}  &$35.2\pm 0.3_{\rm stat} \pm 1.6_{\rm syst}$   &$1.29\pm 0.03_{\rm stat} \pm 0.27_{\rm syst}$ \\
                                           &$35.5\pm 0.3_{\rm stat} \pm 1.6_{\rm syst}$   &$1.29\pm 0.03_{\rm stat} \pm 0.27_{\rm syst}$ \\
\hline 
{\bf Average }                             &\mathversion{bold}$37.6\pm0.9$        &\mathversion{bold}$1.56\pm0.14$ \\
\hline 
\end{tabular}
\end{center}
\label{tab:vcbf1}
\end{table}

For a determination of \vcb, the form factor at zero recoil $F(1)$
needs to be computed.  A possible choice is $F(1) = 0.91\pm0.04_{\rm
theo}$~\cite{Battaglia:2003in}, resulting in

\begin{displaymath}
\vcb = (41.3 \pm 1.0_{\rm exp} \pm 1.8_{\rm theo}) \times 10^{-3}.
\end{displaymath}

\noindent
The value for $F(1)$ and its error  is based on a comparison
of estimates 
using OPE sum rules  
and with an HQET based lattice gauge calculation 
(see Ref.~\cite{Battaglia:2003in} for more details).

\mysubsubsection{\BzbDplnu}
\label{slbdecays_dlnu}

The average branching fraction $\cbf(\BzbDplnu)$ is determined by the
combination of the results provided in Table~\ref{tab:dlnu}.  The
error sources here are the same as discussed in
Sec.~\ref{slbdecays_dstarlnu}, but generally at a higher level due to
larger background levels, less stringent kinematic constraints, and
larger kinematic suppression at the endpoint.
Figure~\ref{fig:brdsl}(b) illustrates the measurements and the
resulting average.

\begin{table}[!htb]
\caption{Average of the branching fraction $\cbf(\BzbDplnu)$ and individual
results. }
\begin{center}
\begin{tabular}{|l|c|c|}\hline
Experiment                                 &$\cbf(\BzbDplnu) [\%]$ (rescaled) &$\cbf(\BzbDplnu) [\%]$ (published) \\
\hline\hline 
ALEPH ~\hfill\cite{Buskulic:1996yq}        &$2.39 \pm0.20_{\rm stat} \pm0.52_{\rm syst}$  &$2.35 \pm0.20_{\rm stat} \pm0.44_{\rm syst}$ \\
CLEO  ~\hfill\cite{Bartelt:1998dq}         &$2.10 \pm0.13_{\rm stat} \pm0.18_{\rm syst}$  &$2.20 \pm0.16_{\rm stat} \pm0.19_{\rm syst}$ \\
\belle  ~\hfill\cite{Abe:2001yf}           &$2.18 \pm0.12_{\rm stat} \pm0.39_{\rm syst}$  &$2.13 \pm0.12_{\rm stat} \pm0.39_{\rm syst}$ \\
\hline 
{\bf Average}                              &\mathversion{bold}$2.12 \pm0.20$      &\mathversion{bold}$\chi^2/\dof = 0.3/2$ \\
\hline 
\end{tabular}
\end{center}
\label{tab:dlnu}
\end{table}

\begin{figure}[!ht]
 \begin{center}
  \unitlength1.0cm 
  \begin{picture}(14.,8.) 
   \put( -0.5,  0.0){\includegraphics[width=7.5cm]{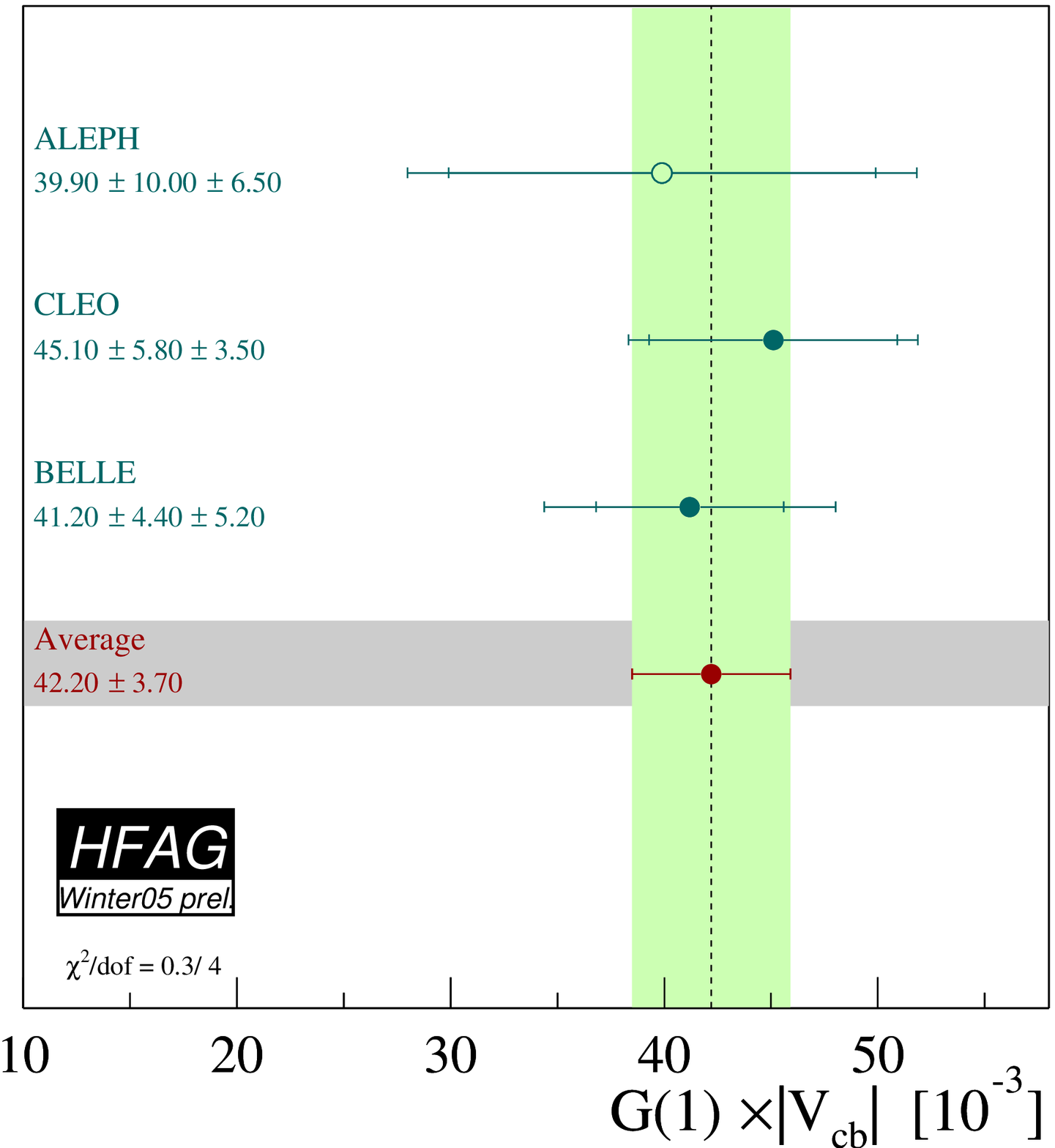}}
   \put(  8.0, -0.2){\includegraphics[width=8.0cm]{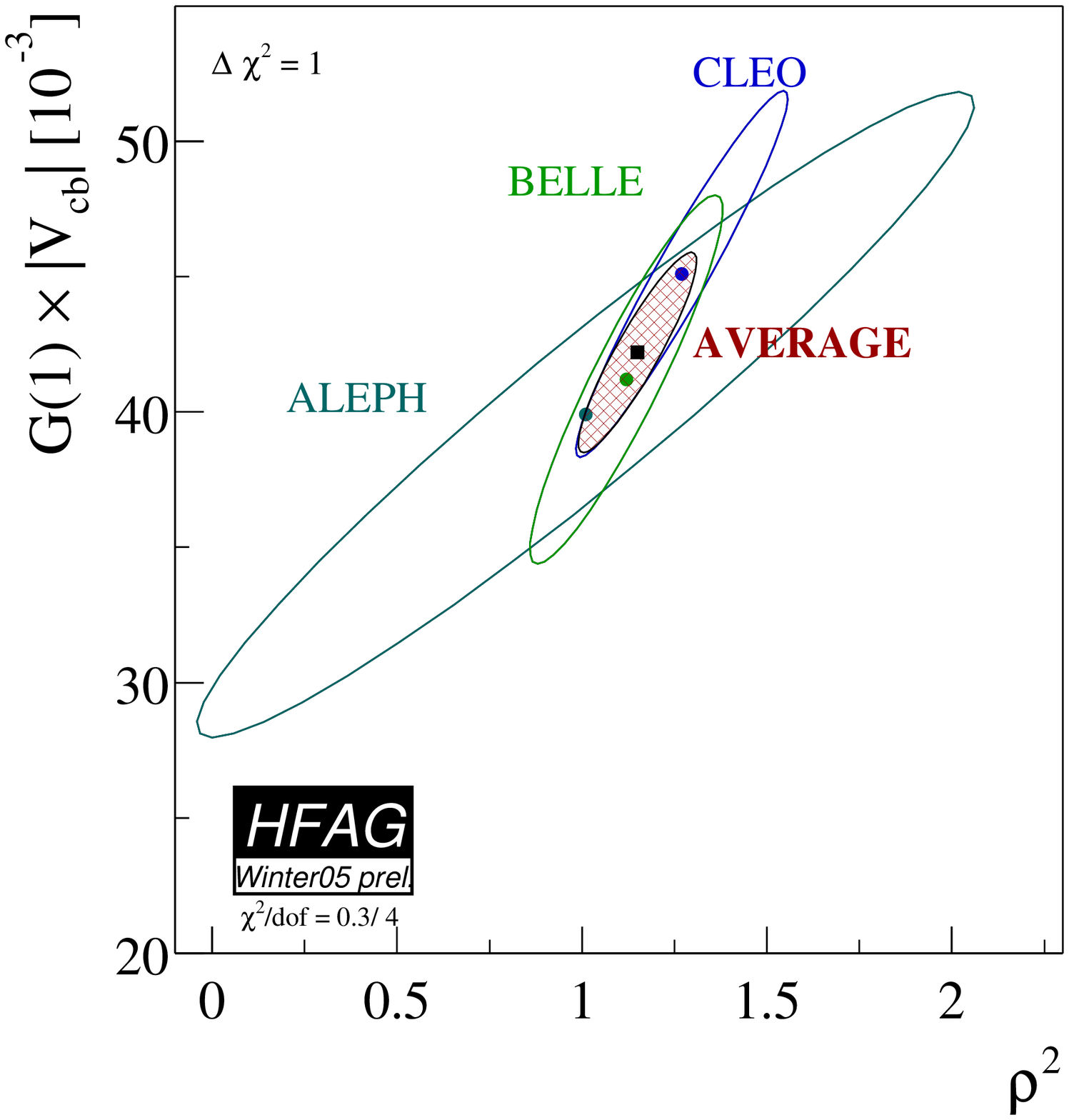}}
   \put(  5.5,  6.8){{\large\bf a)}}
   \put( 14.4,  6.8){{\large\bf b)}}
  \end{picture}
  \caption{(a)  Illustration of the  average $G(1)\vcb$
   and   rescaled  measurements   of   exclusive  \BzbDplnu\   decays
   determined   in  a  two-dimensional   fit.   (b)   Illustration  of
   $F(1)\vcb$   vs.  $\rho^2$.  The   error  ellipses   correspond  to
   $\Delta\chi^2 = 1$. }
  \label{fig:vcbg1}
 \end{center}
\end{figure}

The average for $G(1)\vcb$ is determined by the two-dimensional
combination of the results provided in Table~\ref{tab:vcbg1}.
Figure~\ref{fig:vcbg1}(b) illustrates the average $F(1)\vcb$ and the
measurements included in the average.  Figure~\ref{fig:vcbg1}(a)
provides a one-dimensional projection for illustrative purposes.

\begin{table}[!htb]
\caption{Average  of $G(1)\vcb$  determined  in the  decay \BzbDplnu\  and
individual  results. The  fit  for the  average  has $\chi^2/\dof  =
0.3/4$.}
\begin{center}
\begin{tabular}{|l|c|c|}\hline
Experiment &$G(1)\vcb [10^{-3}]$ (rescaled)  &$\rho^2$ (rescaled) \\ 
           &$G(1)\vcb [10^{-3}]$ (published) &$\rho^2$ (published) \\
\hline\hline 
ALEPH~\hfill\cite{Buskulic:1996yq}  &$39.9 \pm10.0_{\rm stat} \pm6.5_{\rm syst}$   &$1.01 \pm0.98_{\rm stat} \pm0.38_{\rm syst}$ \\
                                    &$31.1 \pm9.9_{\rm stat}  \pm8.6_{\rm syst}$   &$0.20 \pm0.98_{\rm stat} \pm0.50_{\rm syst}$ \\
\hline
CLEO ~\hfill\cite{Bartelt:1998dq}   &$45.1 \pm5.8_{\rm stat} \pm3.5_{\rm syst}$    &$1.27 \pm0.25_{\rm stat} \pm0.14_{\rm syst}$ \\
                                    &$44.8 \pm6.1_{\rm stat} \pm3.7_{\rm syst}$  &$1.30 \pm0.27_{\rm stat} \pm0.14_{\rm syst}$ \\
\hline
\belle~\hfill\cite{Abe:2001yf}       &$41.2 \pm4.4_{\rm stat} \pm5.2_{\rm syst}$    &$1.12 \pm0.22_{\rm stat} \pm0.14_{\rm syst}$ \\
                                    &$41.1 \pm4.4_{\rm stat} \pm5.1_{\rm syst}$    &$1.12 \pm0.22_{\rm stat} \pm0.14_{\rm syst}$ \\
\hline 
{\bf Average }                      &\mathversion{bold}$42.2 \pm3.7$       &\mathversion{bold}$1.15 \pm0.16$      \\
\hline 
\end{tabular}
\end{center}
\label{tab:vcbg1}
\end{table}

For a determination of \vcb, the form factor at zero recoil $G(1)$
needs to be computed.  A possible choice is $G(1) = 1.04\pm0.06_{\rm
theo}$~\cite{Battaglia:2003in}, resulting in

\begin{displaymath}
\vcb = (40.6 \pm 3.6_{\rm exp} \pm 2.3_{\rm theo}) \times 10^{-3}.
\end{displaymath}

\mysubsection{Inclusive Cabibbo-favored decays}
\label{slbdecays_b2cincl}

Aspects of the theory and phenomenology of inclusive Cabibbo-favored
$B$ decays and their use in the determination of \vcb\ in the context
of the Heavy Quark Expansion (HQE), an Operator Product Expansion
based on HQET, are described in many places (see, \eg,
Ref.~\cite{b2xintros} and references therein).

An updated average for the total semileptonic branching fraction
$\cbf(\Bb\to X\ell\nub)$ is provided here.  In future compilations
HFAG will provide averages of moments of the electron and hadron
spectra in inclusive semileptonic decays, and perform global fits of these
moments to OPE calculations to extract \vcb\ and HQE parameters.
At present several groups have published these global 
fits~\cite{Mahmood:2004kq,Battaglia:2002tm,Aubert:2004aw,Bauer:2004ve}.


\mysubsubsection{Total semileptonic branching fraction}
\label{slbdecays_b2cinclrate}

The measurements of the total semileptonic branching fraction
$\cbf(b\to X\ell\nub)$ at LEP (see, \eg, Ref~\cite{Eidelman:2004wy} or
Ref.~\cite{lepewwg}) represent a different analysis class with a more
explicit model dependence than the (lepton-)tagged analyses used at
the \FourS.  Therefore the LEP measurements are not used in the
averages computed here.

The average for the total branching fraction $\cbf(\Bb\to X \ell\nub)$
is determined by the combination of the results provided in
Table~\ref{tab:brisltot}.  In this average, the extrapolation of the
measured rate to the total decay rate is performed by each experiment,
usually with a fit of several components to the experimental spectrum.
In the two most precise measurements~\cite{Aubert:2004aw,Mahmood:2004kq}, 
the extrapolation is part of the
HQE fit to electron momentum and hadronic invariant mass moments in
$\Bxclnu$ decays.

\begin{table}[!htb]
\caption{Average of the total semileptonic branching fractions $\cbf(\Bb\to X \ell\nub)$
determined in tagged measurements on the \FourS.}
\begin{center}
\begin{tabular}{|l|c|c|}\hline
Experiment                                   &$\cbf_{tot}(\Bb\to X \ell\nub) [\%]$ (rescaled)    &$\cbf_{tot}(\Bb\to X \ell\nub) [\%]$ (published) \\
\hline\hline 
ARGUS ($\ell$-tag)~\hfill\cite{Albrecht:1993pu}      &$9.77  \pm0.50 \pm0.39$            &$9.70  \pm0.50 \pm0.60$ \\
\babar (\breco-tag)~\hfill\cite{Langenegger:2001qp}  &$10.40 \pm0.50 \pm0.46$            &$10.40 \pm0.50 \pm0.46$ \\
\belle ($\ell$-tag)~\hfill\cite{Abe:2002du}           &$10.98 \pm0.12 \pm0.50$            &$10.90 \pm0.12 \pm0.49$ \\
\babar ($e$-tag HQE)~\hfill\cite{Aubert:2004aw}          &$10.87 \pm0.06 \pm0.16$            &$10.83 \pm0.06 \pm0.16$ \\
\belle (\breco-tag)~\hfill\cite{belle:breco}          &$11.19 \pm0.20 \pm0.31$            &$11.19 \pm0.20 \pm0.31$ \\
CLEO  ($\ell$-tag HQE)~\hfill\cite{Mahmood:2004kq}       &$10.95 \pm0.09 \pm0.24$            &$10.91 \pm0.09 \pm0.24$ \\
\hline 
{\bf Average}                                        &\mathversion{bold}$10.95\pm0.15$   &\mathversion{bold}$\chi^2/\dof = 5.0/5$ \\
\hline 
\end{tabular}
\end{center}
\label{tab:brisltot}
\end{table}

\mysubsection{Exclusive Cabibbo-suppressed decays}
\label{slbdecays_b2uexcl}

Here we list results on exclusive determinations of \vub. An average
of (these) exclusive $b\to u\ell\nub$ results is envisioned for the
future. The measurements are separated into two classes: a first one
averaging over the entire $\qq$ range (shown in
Table~\ref{tab:xslvuballqq}), and a second class, where the decay rate
is measured differentially in (few) bins of $\qq$ (shown in
Table~\ref{tab:xslvublimqq}).

\begin{table}[!htb]
\caption{Summary of exclusive determinations of $\cbf(\Bb\to X
\ell\nub)$ and \vub\ using the entire $\qq$ range. The errors quoted
on \vub\ correspond to statistical, experimental systematic and
theoretical systematic, respectively.}
\begin{center}
\begin{small}
\begin{tabular}{|l|l|c|c|}\hline
Experiment             &Mode                       &$\cbf [10^{-4}]$           &\vub\ $[10^{-3}]$ (rescaled) \\
\hline\hline 
CLEO  ~\hfill\cite{Behrens:1999vv}  &$\Bz\to \rho^-\ell^+\nu$   &$2.69 \pm0.41 \ {}^{+0.35}_{-0.40} \pm0.50$   
                                                                               &$3.24  \pm0.25 \ {}^{+0.21}_{-0.24}\pm0.58$ \\
\babar ~\hfill\cite{Aubert:2003zd}  &$\Bz\to \rho^-\ell^+\nu$   &$3.29 \pm0.42 \pm0.47            \pm0.60$     
                                                                               &$3.59  \pm0.23 \pm0.26 \pm0.66$ \\
\belle ~\hfill\cite{Schwanda:2004fa} &$\Bp\to \omega\ell^+\nu$   &$1.3  \pm0.4  \pm0.2             \pm0.3$      
                                                                               &$3.1   \pm0.2  \pm0.2  \pm0.6$ \\
\babar ~\hfill\cite{Dingfelder:ckm05a} &$\Bp\to \pi^0\ell^+\nu$   &$1.80 \pm0.37  \pm0.23$
& \\   
\hline 
\end{tabular}
\end{small}
\end{center}
\label{tab:xslvuballqq}
\end{table}

\begin{table}[!htb]
\caption{Summary of exclusive determinations of $\cbf(\Bb\to  X
\ell\nub)$ and \vub\ binned in $\qq$. The  errors quoted on \vub\ correspond to  statistical,
experimental systematic, theoretical systematic, and signal form-factor shape
and normalization, respectively.}
\begin{center}
\begin{small}
\begin{tabular}{|l|l|c|c|}\hline
Experiment             &Mode                     &$\cbf [10^{-4}]$  &\vub\ $[10^{-3}]$ (rescaled) \\
\hline\hline 
CLEO  ~\hfill\cite{Athar:2003yg}   &$\Bz\to \pi^-\ell^+\nu$  &$1.33 \pm0.18 \pm0.11            \pm0.01 \pm0.07$   
                                                                    &$2.88  \pm0.55 \pm0.30 \ {}^{+0.45}_{-0.35} \pm0.18$ \\
\belle ~\hfill\cite{Abe:2004zm}     &$\Bz\to \pi^-\ell^+\nu$ &$1.76 \pm0.28\ \pm0.20 \pm0.03$   
                                                                    &$3.90  \pm0.71 \pm0.23\ {}^{+0.62}_{-0.48}$ \\
\babar~\hfill\cite{danieledelre}   &$\Bz\to \pi^-\ell^+\nu$ &$1.46 \pm0.27\ \pm0.28$ & \\
\babar ~\hfill\cite{Dingfelder:ckm05} &$\Bz\to \pi-\ell^+\nu$   &$1.38\pm0.10  \pm0.18             \pm0.08$  
& $3.82\pm 0.14\pm 0.24\pm 0.66$ \\   
\babar ~\hfill\cite{Dingfelder:ckm05a} &$\Bz\to \pi-\ell^+\nu$   &$1.03 \pm0.25  \pm0.13$
& \\   
CLEO  ~\hfill\cite{Athar:2003yg}   &$\Bz\to \rho^-\ell^+\nu$ &$2.17 \pm0.34\ {}^{+0.47}_{-0.54} \pm0.41 \pm0.01$   
                                                                    &$3.34  \pm0.32\ {}^{+0.27}_{-0.36}\ {}^{+0.50}_{-0.40}$ \\
\belle ~\hfill\cite{Abe:2004zm}     &$\Bz\to \rho^-\ell^+\nu$ &$2.54 \pm0.78\ \pm0.85 \pm0.30$ & \\
\babar ~\hfill\cite{Dingfelder:ckm05} &$\Bz\to \rho-\ell^+\nu$   &$2.14 \pm0.21  \pm0.51             \pm0.28$    
& \\
\hline 
\end{tabular}
\end{small}
\end{center}
\label{tab:xslvublimqq}
\end{table}


\mysubsection{Inclusive Cabibbo-suppressed decays}
\label{slbdecays_b2uincl}

Since the last HFAG update, there has been considerable change in
this area.  The theoretical tools have been updated,~\cite{ref:blnp} 
and the use
of HQE parameters determined from global fits to $\Bxclnu$ moments
as input to shape function parameterizations is now established.
A bit more time is needed, however, to integrate these changes into
an updated average value for \vub.  Preliminary results shown at
the CKM2005 workshop~\cite{inclvub:ckm05} suggest that a substantial
downward shift in \vub\ can be expected.

 

\mysection{Measurements related to Unitarity Triangle angles
}
\label{sec:cp_uta}

The charge of the ``$\CP(t)$ and Unitarity Triangle angles'' group
is to provide averages of measurements related (mostly) to the 
angles of the Unitarity Triangle (UT).
To date, most of the measurements that can be used to 
obtain model-independent information on the UT angles
come from time-dependent $\CP$ asymmetry analyses.
In cases where considerable theoretical input is required to 
extract the fundamental quantities, no attempt is made to do so at 
this stage. However, straightforward interpretations of the averages 
are given, where possible.

In Sec.~\ref{sec:cp_uta:introduction} 
a brief introduction to the relevant phenomenology is given.
In Sec.~\ref{sec:cp_uta:notations}
an attempt is made to clarify the various different notations in use.
In Sec.~\ref{sec:cp_uta:common_inputs}
the common inputs to which experimental results are rescaled in the
averaging procedure are listed. We also briefly introduce the treatment
of experimental errors. 
In the remainder of this section,
the experimental results and their averages are given,
divided into subsections based on the underlying quark-level decays.

\mysubsection{Introduction
}
\label{sec:cp_uta:introduction}

The Standard Model Cabibbo-Kobayashi-Maskawa (CKM) quark mixing matrix $\VCKM$ 
must be unitary. A $3 \times 3$ unitary matrix has four free parameters,\footnote{
  In the general case there are nine free parameters,
  but five of these are absorbed into unobservable quark phases.}
and these are conventionally written by the product
of three (complex) rotation matrices~\cite{ref:cp_uta:chau}, where the rotations are 
characterized by the Euler angles $\theta_{12}$, $\theta_{13}$ 
and $\theta_{23}$, which are the mixing angles
between the generations, and one overall phase $\delta$,
\begin{equation}
\label{eq:ckmPdg}
\VCKM =
        \left(
          \begin{array}{ccc}
            V_{ud} & V_{us} & V_{ub} \\
            V_{cd} & V_{cs} & V_{cb} \\
            V_{td} & V_{ts} & V_{tb} \\
          \end{array}
        \right)
        =
        \left(
        \begin{array}{ccc}
        c_{12}c_{13}    
                &    s_{12}c_{13}   
                        &   s_{13}e^{-i\delta}  \\
        -s_{12}c_{23}-c_{12}s_{23}s_{13}e^{i\delta} 
                &  c_{12}c_{23}-s_{12}s_{23}s_{13}e^{i\delta} 
                        & s_{23}c_{13} \\
        s_{12}s_{23}-c_{12}c_{23}s_{13}e^{i\delta}  
                &  -c_{12}s_{23}-s_{12}c_{23}s_{13}e^{i\delta} 
                        & c_{23}c_{13} 
        \end{array}
        \right)
\end{equation}
where $c_{ij}=\cos\theta_{ij}$, $s_{ij}=\sin\theta_{ij}$ for 
$i<j=1,2,3$. 

Following the observation of a hierarchy between the different
matrix elements, the Wolfenstein parameterization~\cite{ref:cp_uta:wolfenstein}
is an expansion of $\VCKM$ in terms of the four real parameters $\lambda$
(the expansion parameter), $A$, $\rho$ and $\eta$. Defining to 
all orders in $\lambda$~\cite{ref:cp_uta:buras}
\begin{eqnarray}
  \label{eq:burasdef}
  s_{12}             &\equiv& \lambda,\nonumber \\ 
  s_{23}             &\equiv& A\lambda^2, \\
  s_{13}e^{-i\delta} &\equiv& A\lambda^3(\rho -i\eta),\nonumber
\end{eqnarray}
and inserting these into the representation of Eq.~(\ref{eq:ckmPdg}), 
unitarity of the CKM matrix is achieved to all orders.
A Taylor expansion of $\VCKM$ leads to the familiar approximation
\begin{equation}
  \label{eq:cp_uta:ckm}
  \VCKM
  = 
  \left(
    \begin{array}{ccc}
      1 - \lambda^2/2 & \lambda & A \lambda^3 ( \rho - i \eta ) \\
      - \lambda & 1 - \lambda^2/2 & A \lambda^2 \\
      A \lambda^3 ( 1 - \rho - i \eta ) & - A \lambda^2 & 1 \\
    \end{array}
  \right) + {\cal O}\left( \lambda^4 \right).
\end{equation}
The non-zero imaginary part of the CKM matrix,
which is the origin of $\CP$ violation in the Standard Model,
is encapsulated in a non-zero value of $\eta$.

The unitarity relation $\VCKM^\dagger\VCKM = {\mathit 1}$
results in nine expressions which can be written
$\sum_{i=u,c,t} V^*_{ij}V_{ik} = \delta_{jk}$,
where $\delta_{jk}$ is the Kronecker symbol.
Of the off-diagonal expressions ($j \neq k$),
three can be trivially transformed into the other three 
(under $j \leftrightarrow k$),
leaving six relations, in which three complex numbers sum to zero,
which therefore can be expressed as triangles in the complex plane.

\begin{figure}[t]
  \begin{center}
    \resizebox{0.55\textwidth}{!}{\includegraphics{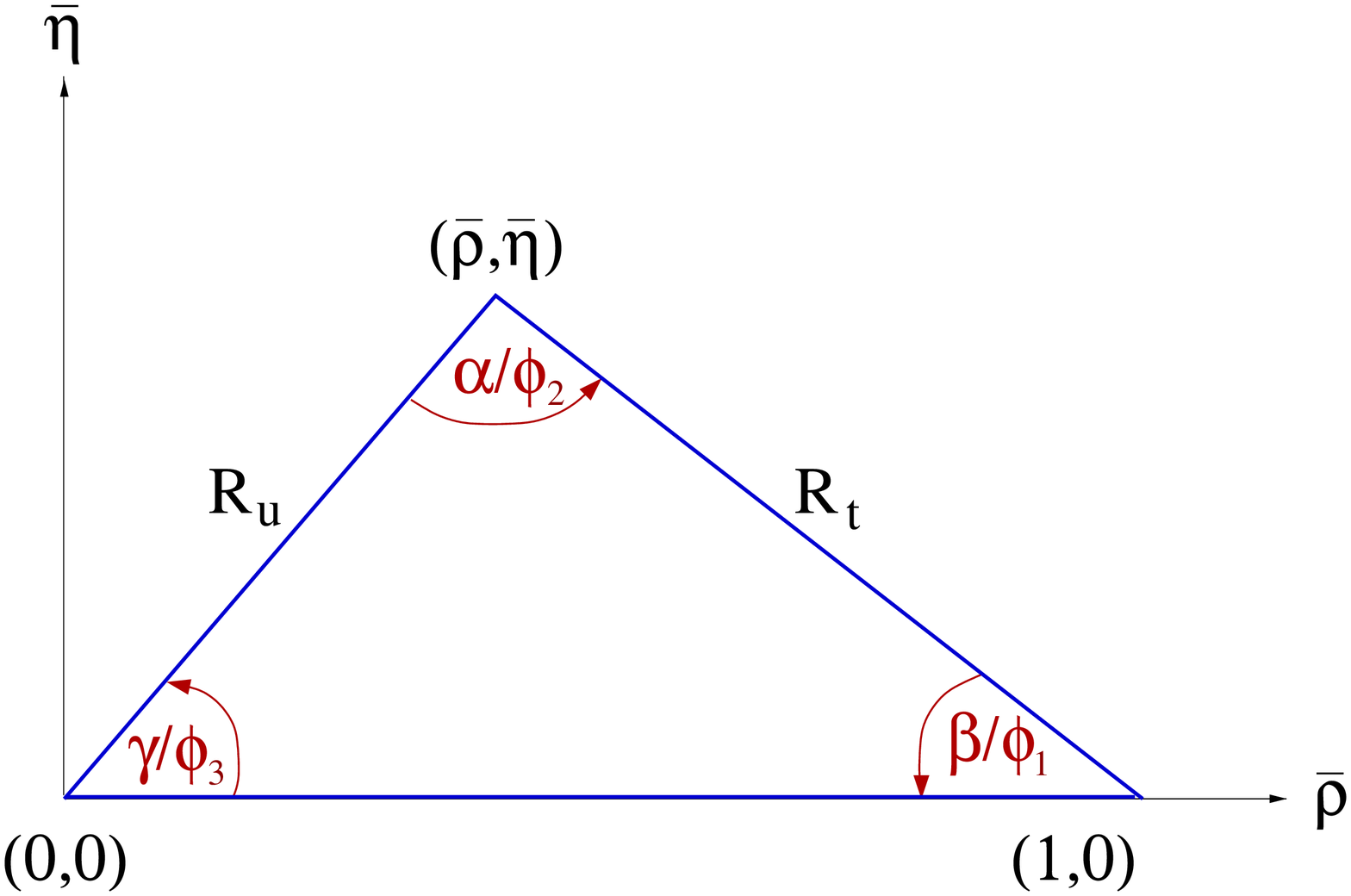}}
    \caption{The Unitarity Triangle.}
    \label{fig:cp_uta:ut}
  \end{center}
\end{figure}

One of these,
\begin{equation}
  \label{eq:cp_uta:ut}
  V_{ud}V^*_{ub} + V_{cd}V^*_{cb} + V_{td}V^*_{tb} = 0,
\end{equation}
is specifically related to $\B$ decays.
The three terms in Eq.~(\ref{eq:cp_uta:ut}) are of the same order 
(${\cal O}\left( \lambda^3 \right)$),
and this relation is commonly known as the Unitarity Triangle.
For presentational purposes,
it is convenient to rescale the triangle by $(V_{cd}V^*_{cb})^{-1}$,
as shown in Fig.~\ref{fig:cp_uta:ut}.

Two popular naming conventions for the UT angles exist in the literature:
\begin{equation}
  \label{eq:cp_uta:abc}
  \alpha  \equiv  \phi_2  = 
  \arg\left[ - \frac{V_{td}V_{tb}^*}{V_{ud}V_{ub}^*} \right],
  \hspace{0.5cm}
  \beta   \equiv   \phi_1 =  
  \arg\left[ - \frac{V_{cd}V_{cb}^*}{V_{td}V_{tb}^*} \right],
  \hspace{0.5cm}
  \gamma  \equiv   \phi_3  =  
  \arg\left[ - \frac{V_{ud}V_{ub}^*}{V_{cd}V_{cb}^*} \right].
  \nonumber
\end{equation}
In this document the $\left( \alpha, \beta, \gamma \right)$ set is used.

The apex of the Unitarity Triangle is given by the following 
definition~\cite{ref:cp_uta:buras}  
\begin{eqnarray}
  \label{eq:rhoetabar}
  \rhobar + i\etabar
  \;\equiv\;-\frac{V_{ud}V_{ub}^*}{V_{cd}V_{cb}^*}
  = (\rho + i\eta)
       (1 - \frac{1}{2}\lambda^{2}) + {\cal O}(\lambda^4).
\end{eqnarray}
The sides $R_u$ and $R_t$ of the Unitarity Triangle 
(the third side being normalized to unity) 
are given by
\begin{eqnarray}
  \label{eq:ru}
  R_u &=& 
  \left|\frac{V_{ud}V_{ub}^*}{V_{cd}V_{cb}^*} \right|
  \;=\; \sqrt{\rhobar^2+\etabar^2}, \\
  \label{eq:rt}
  R_t &=& 
  \left|\frac{V_{td}V_{tb}^*}{V_{cd}V_{cb}^*}\right| 
  \;=\; \sqrt{(1-\rhobar)^2+\etabar^2}.
\end{eqnarray} 

\mysubsection{Notations
}
\label{sec:cp_uta:notations}

Several different notations for $\CP$ violation parameters
are commonly used.
This section reviews those found in the experimental literature,
in the hope of reducing the potential for confusion, 
and to define the frame that is used for the averages.

In some cases, when $\B$ mesons decay into 
multibody final states via broad resonances ($\rho$, $\Kstar$, \etc),
the experiments ignore interference effects in the analyses.
This is referred to as the quasi-two-body (Q2B) approximation
in the following.

\mysubsubsection{$\CP$ asymmetries
}
\label{sec:cp_uta:notations:pra}

The $\CP$ asymmetry is defined as the difference between the rate 
involving a $b$ quark and that involving a $\bar b$ quark, divided 
by the sum. For example, the partial rate (or charge) asymmetry for 
a charged $\B$ decay would be given as 
\begin{equation}
  \label{eq:cp_uta:pra}
  \Acp_{f} \;\equiv\; 
  \frac{\Gamma(\Bm \to f)-\Gamma(\Bp \to \bar{f})}{\Gamma(\Bm \to f)+\Gamma(\Bp \to \bar{f})}.
\end{equation}

\mysubsubsection{Time-dependent \CP asymmetries in decays to $\CP$ eigenstates
}
\label{sec:cp_uta:notations:cp_eigenstate}

If the amplitudes for $\Bz$ and $\Bzb$ to decay to a final state $f$, 
which is a $\CP$ eigenstate with eigenvalue $\etacpf$,
are given by $\Af$ and $\Abarf$, respectively, 
then the decay distributions for neutral $\B$ mesons, 
with known flavour at time $\Delta t =0$,
are given by
\begin{eqnarray}
  \Gamma_{\Bzb \to f} (\Delta t) & = &
  \frac{e^{-| \Delta t | / \tau(\Bz)}}{4\tau(\Bz)}
  \left[ 
    1 +
      \frac{2\, \Im(\lambda_f)}{1 + |\lambda_f|^2} \sin(\Delta m \Delta t) -
      \frac{1 - |\lambda_f|^2}{1 + |\lambda_f|^2} \cos(\Delta m \Delta t)
  \right], \\
  \Gamma_{\Bz \to f} (\Delta t) & = &
  \frac{e^{-| \Delta t | / \tau(\Bz)}}{4\tau(\Bz)}
  \left[ 
    1 -
      \frac{2\, \Im(\lambda_f)}{1 + |\lambda_f|^2} \sin(\Delta m \Delta t) +
      \frac{1 - |\lambda_f|^2}{1 + |\lambda_f|^2} \cos(\Delta m \Delta t)
  \right].
\end{eqnarray}
Here $\lambda_f = \frac{q}{p} \frac{\Abarf}{\Af}$ 
contains terms related to $\Bz$-$\Bzb$ mixing and to the decay amplitude
(the eigenstates of the effective Hamiltonian in the $\BzBzb$ system 
are $\left| B_\pm \right> = p \left| \Bz \right> \pm q \left| \Bzb \right>$).
This formulation assumes $\CPT$ invariance, 
and neglects possible lifetime differences in the neutral $\B$ meson system.
The time-dependent $\CP$ asymmetry,
again defined as the difference between the rate 
involving a $b$ quark and that involving a $\bar b$ quark,
is then given by
\begin{equation}
  \label{eq:cp_uta:td_cp_asp}
  \Acp_{f} \left(\Delta t\right) \; \equiv \;
  \frac{
    \Gamma_{\Bzb \to f} (\Delta t) - \Gamma_{\Bz \to f} (\Delta t)
  }{
    \Gamma_{\Bzb \to f} (\Delta t) + \Gamma_{\Bz \to f} (\Delta t)
  } \; = \;
  \frac{2\, \Im(\lambda_f)}{1 + |\lambda_f|^2} \sin(\Delta m \Delta t) -
  \frac{1 - |\lambda_f|^2}{1 + |\lambda_f|^2} \cos(\Delta m \Delta t).
\end{equation}

While the coefficient of the $\sin(\Delta m \Delta t)$ term in 
Eq.~(\ref{eq:cp_uta:td_cp_asp}) is everywhere\footnote
{
  Occasionally one also finds Eq.~(\ref{eq:cp_uta:td_cp_asp}) written as
  $\Acp_{f} \left(\Delta t\right) = 
  {\cal A}^{\rm mix}_f \sin(\Delta m \Delta t) + {\cal A}^{\rm dir}_f \cos(\Delta m \Delta t)$,
  or similar.
} denoted $S_f$:
\begin{equation}
  \label{eq:cp_uta:s_def}
  S_f \;\equiv\; \frac{2\, \Im(\lambda_f)}{1 + \left|\lambda_f\right|^2},
\end{equation}
different notations are in use for the
coefficient of the $\cos(\Delta m \Delta t)$ term:
\begin{equation}
  \label{eq:cp_uta:c_def}
  C_f \;\equiv\; - A_f \;\equiv\; \frac{1 - \left|\lambda_f\right|^2}{1 + \left|\lambda_f\right|^2}.
\end{equation}
The $C$ notation is used by the \babar\  collaboration 
(see \eg~\cite{ref:cp_uta:ccs:babar}), 
and also in this document.
The $A$ notation is used by the \belle\ collaboration
(see \eg~\cite{BELLE2}).

Neglecting effects due to $\CP$ violation in mixing 
(\ie, taking $|q/p| = 1$),
if the decay amplitude contains terms with a single weak phase
then $\left|\lambda_f\right| = 1$ and one finds
$S_f = -\etacpf \sin(\phi_{\rm mix} + \phi_{\rm dec})$, $C_f = 0$,
where $\phi_{\rm mix}=\arg(q/p)$ and $\phi_{\rm dec}=\arg(\Abarf/\Af)$.
Note that $\phi_{\rm mix}\approx2\beta $ 
in the Standard Model (in the usual phase convention). 
If amplitudes with different weak phases contribute to the decay, 
no clean interpretation of $S_f$ is possible. If the decay amplitudes
have in addition different $\CP$ conserving strong phases,
then $\left| \lambda_f \right| \neq 1$ and no clean interpretation is possible.
The coefficient of the cosine term becomes non-zero,
indicating direct $\CP$ violation.
The sign of $A_f$ as defined above is consistent with that of $\Acp_{f}$ in 
Eq.~(\ref{eq:cp_uta:pra}).

\mysubsubsection{Time-dependent \CP asymmetries in decays to vector-vector final states
}
\label{sec:cp_uta:notations:vv}

Consider \B decays to states consisting of two vector particles,
such as $\jpsi K^{*0}(\to\KS\piz)$, $D^{*+}D^{*-}$ and $\rho^+\rho^-$,
which are eigenstates of charge conjugation but not of parity.\footnote{
  \noindent
  This is not true of all vector-vector final states,
  \eg, $D^{*\pm}\rho^{\mp}$ is clearly not an eigenstate of 
  charge conjugation.
}
In fact, for such a system, there are three possible final states;
in the helicity basis these can be written $h_{-1}, h_0, h_{+1}$.
The $h_0$ state is an eigenstate of parity, and hence of $\CP$;
however, $\CP$ transforms $h_{+1} \leftrightarrow h_{-1}$ (up to 
an unobservable phase). In the transversity basis, these states 
are transformed into  $h_\parallel =  (h_{+1} + h_{-1})/2$ and 
$h_\perp = (h_{+1} - h_{-1})/2$.
In this basis all three states are $\CP$ eigenstates, 
and $h_\perp$ has the opposite $\CP$ to the others.

The amplitudes to these states are usually given by $A_{0,\perp,\parallel}$
(here we use a normalization such that 
$| A_0 |^2 + | A_\perp |^2 + | A_\parallel |^2 = 1$).
Then the effective $\CP$ of the vector-vector state is known if 
$| A_\perp |^2$ is measured.
An alternative strategy is to measure just the longitudinally polarized 
component,  $| A_0 |^2$
(sometimes denoted by $f_{\rm long}$), 
which allows a limit to be set on the effective $\CP$ since
$| A_\perp |^2 \leq | A_\perp |^2 + | A_\parallel |^2 = 1 - | A_0 |^2$.
The most complete treatment for 
neutral $\B$ decays to vector-vector final states
is time-dependent angular analysis 
(also known as time-dependent transversity analysis).
In such an analysis, 
the interference between the $\CP$ even and $\CP$ odd states 
provides additional sensitivity to the weak and strong phases involved.

\mysubsubsection{Time-dependent \CP asymmetries in decays to non-$\CP$ eigenstates
}
\label{sec:cp_uta:notations:non_cp}

Consider a non-$\CP$ eigenstate $f$, and its conjugate $\bar{f}$. 
For neutral $\B$ decays to these final states,
there are four amplitudes to consider:
those for $\Bz$ to decay to $f$ and $\bar{f}$
($\Af$ and $\Afbar$, respectively),
and the equivalents for $\Bzb$
($\Abarf$ and $\Abarfbar$).
If $\CP$ is conserved in the decay, then
$\Af = \Abarfbar$ and $\Afbar = \Abarf$.


The time-dependent decay distributions can be written in many different ways.
Here, we follow Sec.~\ref{sec:cp_uta:notations:cp_eigenstate}
and define $\lambda_f = \frac{q}{p}\frac{\Abarf}{\Af}$ and
$\lambda_{\bar f} = \frac{q}{p}\frac{\Abarfbar}{\Afbar}$.
The time-dependent \CP asymmetries then follow Eq.~(\ref{eq:cp_uta:td_cp_asp}):
\begin{eqnarray}
\label{eq:cp_uta:non-cp-obs}
  {\cal A}_f (\Delta t) \; \equiv \;
  \frac{
    \Gamma_{\Bzb \to f} (\Delta t) - \Gamma_{\Bz \to f} (\Delta t)
  }{
    \Gamma_{\Bzb \to f} (\Delta t) + \Gamma_{\Bz \to f} (\Delta t)
  } & = & S_f \sin(\Delta m \Delta t) - C_f \cos(\Delta m \Delta t), \\
  {\cal A}_{\bar{f}} (\Delta t) \; \equiv \;
  \frac{
    \Gamma_{\Bzb \to \bar{f}} (\Delta t) - \Gamma_{\Bz \to \bar{f}} (\Delta t)
  }{
    \Gamma_{\Bzb \to \bar{f}} (\Delta t) + \Gamma_{\Bz \to \bar{f}} (\Delta t)
  } & = & S_{\bar{f}} \sin(\Delta m \Delta t) - C_{\bar{f}} \cos(\Delta m \Delta t),
\end{eqnarray}
with the definitions of the parameters 
$C_f$, $S_f$, $C_{\bar{f}}$ and $S_{\bar{f}}$,
following Eqs.~(\ref{eq:cp_uta:s_def}) and~(\ref{eq:cp_uta:c_def}).

The time-dependent decay rates are given by
\begin{eqnarray}
  \Gamma_{\Bzb \to f} (\Delta t) & = &
  \frac{e^{-\left| \Delta t \right| / \tau(\Bz)}}{8\tau(\Bz)} 
  ( 1 + \Adirnoncp ) 
  \left\{ 
    1 + S_f \sin(\Delta m \Delta t) - C_f \cos(\Delta m \Delta t) 
  \right\},
  \\
  \Gamma_{\Bz \to f} (\Delta t) & = &
  \frac{e^{-\left| \Delta t \right| / \tau(\Bz)}}{8\tau(\Bz)} 
  ( 1 + \Adirnoncp ) 
  \left\{ 
    1 - S_f \sin(\Delta m \Delta t) + C_f \cos(\Delta m \Delta t) 
  \right\},
  \\
  \Gamma_{\Bzb \to \bar{f}} (\Delta t) & = &
  \frac{e^{-\left| \Delta t \right| / \tau(\Bz)}}{8\tau(\Bz)} 
  ( 1 - \Adirnoncp ) 
  \left\{ 
    1 + S_{\bar{f}} \sin(\Delta m \Delta t) - C_{\bar{f}} \cos(\Delta m \Delta t) 
  \right\},
  \\
  \Gamma_{\Bz \to \bar{f}} (\Delta t) & = &
    \frac{e^{-\left| \Delta t \right| / \tau(\Bz)}}{8\tau(\Bz)} 
  ( 1 - \Adirnoncp ) 
  \left\{ 
    1 - S_{\bar{f}} \sin(\Delta m \Delta t) + C_{\bar{f}} \cos(\Delta m \Delta t) 
  \right\},
\end{eqnarray}
where the time-independent parameter \Adirnoncp
represents an overall asymmetry in the production of the 
$f$ and $\bar{f}$ final states,\footnote{
  This parameter is often denoted ${\cal A}_f$ (or ${\cal A}_{\CP}$),
  but here we avoid this notation to prevent confusion with the
  time-dependent $\CP$ asymmetry.
}
\begin{equation}
  \Adirnoncp = 
  \frac{
    \left( 
      \left| \Af \right|^2 + \left| \Abarf \right|^2
    \right) - 
    \left( 
      \left| \Afbar \right|^2 + \left| \Abarfbar \right|^2
    \right)
  }{
    \left( 
      \left| \Af \right|^2 + \left| \Abarf \right|^2
    \right) +
    \left( 
      \left| \Afbar \right|^2 + \left| \Abarfbar \right|^2
    \right)
  }.
\end{equation}
Assuming $|q/p| = 1$,
the parameters $C_f$ and $C_{\bar{f}}$
can also be written in terms of the decay amplitudes as follows:
\begin{equation}
  C_f = 
  \frac{
    \left| \Af \right|^2 - \left| \Abarf \right|^2 
  }{
    \left| \Af \right|^2 + \left| \Abarf \right|^2
  }
  \hspace{5mm}
  {\rm and}
  \hspace{5mm}
  C_{\bar{f}} = 
  \frac{
    \left| \Afbar \right|^2 - \left| \Abarfbar \right|^2
  }{
    \left| \Afbar \right|^2 + \left| \Abarfbar \right|^2
  },
\end{equation}
giving asymmetries in the decay amplitudes of $\Bz$ and $\Bzb$
to the final states $f$ and $\bar{f}$ respectively.
In this notation, the direct $\CP$ invariance conditions are
$\Adirnoncp = 0$ and $C_f = - C_{\bar{f}}$.
Note that $C_f$ and $C_{\bar{f}}$ are typically non-zero;
\eg, for a flavour-specific final state, 
$\Abarf = \Afbar = 0$ ($\Af = \Abarfbar = 0$), they take the values
$C_f = - C_{\bar{f}} = 1$ ($C_f = - C_{\bar{f}} = -1$).

The coefficients of the sine terms
contain information about the weak phase. 
In the case that each decay amplitude contains only a single weak phase
(\ie, no direct $\CP$ violation),
these terms can be written
\begin{equation}
  S_f = 
  \frac{ 
    - 2 \left| \Af \right| \left| \Abarf \right| 
    \sin( \phi_{\rm mix} + \phi_{\rm dec} - \delta_f )
  }{
    \left| \Af \right|^2 + \left| \Abarf \right|^2
  } 
  \hspace{5mm}
  {\rm and}
  \hspace{5mm}
  S_{\bar{f}} = 
  \frac{
    - 2 \left| \Afbar \right| \left| \Abarfbar \right| 
    \sin( \phi_{\rm mix} + \phi_{\rm dec} + \delta_f )
  }{
    \left| \Afbar \right|^2 + \left| \Abarfbar \right|^2
  },
\end{equation}
where $\delta_f$ is the strong phase difference between the decay amplitudes.
If there is no $\CP$ violation, the condition $S_f = - S_{\bar{f}}$ holds.
If amplitudes with different weak and strong phases contribute,
no clean interpretation of $S_f$ and $S_{\bar{f}}$ is possible.

Since two of the $\CP$ invariance conditions are 
$C_f = - C_{\bar{f}}$ and $S_f = - S_{\bar{f}}$,
there is motivation for a rotation of the parameters:
\begin{equation}
\label{eq:cp_uta:non-cp-s_and_deltas}
  S_{f\bar{f}} = \frac{S_{f} + S_{\bar{f}}}{2},
  \hspace{4mm}
  \Delta S_{f\bar{f}} = \frac{S_{f} - S_{\bar{f}}}{2},
  \hspace{4mm}
  C_{f\bar{f}} = \frac{C_{f} + C_{\bar{f}}}{2},
  \hspace{4mm}
  \Delta C_{f\bar{f}} = \frac{C_{f} - C_{\bar{f}}}{2}.
\end{equation}
With these parameters, the $\CP$ invariance conditions become
$S_{f\bar{f}} = 0$ and $C_{f\bar{f}} = 0$. 
The parameter $\Delta C_{f\bar{f}}$ gives a measure of the ``flavour-specificity''
of the decay:
$\Delta C_{f\bar{f}}=\pm1$ corresponds to a completely flavour-specific decay,
in which no interference between decays with and without mixing can occur,
while $\Delta C_{f\bar{f}} = 0$ results in 
maximum sensitivity to mixing-induced $\CP$ violation.
The parameter $\Delta S_{f\bar{f}}$ is related to the strong phase difference 
between the decay amplitudes of $\Bz$ to $f$ and to $\bar f$. 
We note that the observables of Eq.~(\ref{eq:cp_uta:non-cp-s_and_deltas})
exhibit experimental correlations 
(typically of $\sim 20\%$, depending on the tagging purity, and other effects)
between $S_{f\bar{f}}$ and  $\Delta S_{f\bar{f}}$, 
and between $C_{f\bar{f}}$ and $\Delta C_{f\bar{f}}$. 
On the other hand, 
the final state specific observables of Eq.~(\ref{eq:cp_uta:non-cp-obs})
tend to have low correlations.

Alternatively, if we recall that the $\CP$ invariance
conditions at the amplitude level are
$\Af = \Abarfbar$ and $\Afbar = \Abarf$,
we are led to consider the parameters~\cite{ref:cp_uta:uud:charles}
\begin{equation}
  \label{eq:cp_uta:non-cp-directcp}
  {\cal A}_{f\bar{f}} = 
  \frac{
    \left| \Abarfbar \right|^2 - \left| \Af \right|^2 
  }{
    \left| \Abarfbar \right|^2 + \left| \Af \right|^2
  }
  \hspace{5mm}
  {\rm and}
  \hspace{5mm}
  {\cal A}_{\bar{f}f} = 
  \frac{
    \left| \Abarf \right|^2 - \left| \Afbar \right|^2
  }{
    \left| \Abarf \right|^2 + \left| \Afbar \right|^2
  }.
\end{equation}
These are sometimes considered more physically intuitive parameters
since they characterize direct $\CP$ violation 
in decays with particular topologies.
For example, in the case of $\Bz \to \rho^\pm\pi^\mp$
(choosing $f =  \rho^+\pi^-$ and $\bar{f} = \rho^-\pi^+$),
${\cal A}_{f\bar{f}}$ (also denoted ${\cal A}^{+-}_{\rho\pi}$)
parameterizes direct $\CP$ violation
in decays in which the produced $\rho$ meson does not contain the 
spectator quark,
while ${\cal A}_{\bar{f}f}$ (also denoted ${\cal A}^{-+}_{\rho\pi}$)
parameterizes direct $\CP$ violation 
in decays in which it does.
Note that we have again followed the sign convention that the asymmetry 
is the difference between the rate involving a $b$ quark and that
involving a $\bar{b}$ quark, \cf\ Eq.~(\ref{eq:cp_uta:pra}). 
Of course, these parameters are not independent of the 
other sets of parameters given above, and can be written
\begin{equation}
  {\cal A}_{f\bar{f}} =
  - \frac{
    \Adirnoncp + C_{f\bar{f}} + \Adirnoncp \Delta C_{f\bar{f}} 
  }{
    1 + \Delta C_{f\bar{f}} + \Adirnoncp C_{f\bar{f}} 
  }
  \hspace{5mm}
  {\rm and}
  \hspace{5mm}
  {\cal A}_{\bar{f}f} =
  \frac{
    - \Adirnoncp + C_{f\bar{f}} + \Adirnoncp \Delta C_{f\bar{f}} 
  }{
    - 1 + \Delta C_{f\bar{f}} + \Adirnoncp C_{f\bar{f}}  
  }.
\end{equation}
They usually exhibit strong correlations.

We now consider the various notations which have been used 
in experimental studies of
time-dependent $\CP$ asymmetries in decays to non-$\CP$ eigenstates.

\mysubsubsubsection{$\Bz \to D^{*\pm}D^\mp$
}
\label{sec:cp_uta:notations:non_cp:dstard}

The above set of parameters 
($\Adirnoncp$, $C_f$, $S_f$, $C_{\bar{f}}$, $S_{\bar{f}}$),
has been used by both
\babar~\cite{ref:cp_uta:ccd:babar:dstard} and
\belle~\cite{ref:cp_uta:ccd:belle:dstard} in the $D^{*\pm}D^{\mp}$ system
($f = D^{*+}D^-$, $\bar{f} = D^{*-}D^+$).
However, slightly different names for the parameters are used:
\babar\ uses 
(${\cal A}$, $C_{+-}$, $S_{+-}$, $C_{-+}$, $S_{-+}$);
\belle\ uses
(${\cal A}$, $C_{+}$,  $S_{+}$,  $C_{-}$,  $S_{-}$).
In this document, we follow the notation used by \babar.

\mysubsubsubsection{$\Bz \to \rho^{\pm}\pi^\mp$
}
\label{sec:cp_uta:notations:non_cp:rhopi}

In the $\rho^\pm\pi^\mp$ system, the 
($\Adirnoncp$, $C_{f\bar{f}}$, $S_{f\bar{f}}$, $\Delta C_{f\bar{f}}$, 
$\Delta S_{f\bar{f}}$)
set of parameters has been used 
originally by 
\babar~\cite{ref:cp_uta:uud:babar:rhopi_old}, and more recently by
\belle~\cite{ref:cp_uta:uud:belle:rhopi}, in the Q2B approximation;
the exact names\footnote{
  \babar\ has used the notations
  $A_{\CP}^{\rho\pi}$~\cite{ref:cp_uta:uud:babar:rhopi_old} and 
  ${\cal A}_{\rho\pi}$~\cite{ref:cp_uta:uud:babar:rhopi}
  in place of ${\cal A}_{\CP}^{\rho\pi}$.
}
used in this case are
$\left( 
  {\cal A}_{\CP}^{\rho\pi}, C_{\rho\pi}, S_{\rho\pi}, \Delta C_{\rho\pi}, \Delta S_{\rho\pi}
\right)$,
and these names are also used in this document.

Since $\rho^\pm\pi^\mp$ is reconstructed in the final state $\pi^+\pi^-\pi^0$,
the interference between the $\rho$ resonances
can provide additional information about the phases.
\babar~\cite{ref:cp_uta:uud:babar:rhopi} has performed 
a time-dependent Dalitz plot analysis, 
from which the weak phase $\alpha$ is directly extracted.
In such an analysis, the measured Q2B parameters are 
also naturally corrected for interference effects.

\mysubsubsubsection{$\Bz \to D^{\pm}\pi^{\mp}, D^{*\pm}\pi^{\mp}, D^{\pm}\rho^{\mp}$
}
\label{sec:cp_uta:notations:non_cp:dstarpi}

Time-dependent $\CP$ analyses have also been performed for the
final states $D^{\pm}\pi^{\mp}$, $D^{*\pm}\pi^{\mp}$ and $D^{\pm}\rho^{\mp}$.
In these theoretically clean cases, no penguin contributions are possible,
so there is no direct $\CP$ violation.
Furthermore, due to the smallness of the ratio of the magnitudes of the 
suppressed ($b \to u$) and favoured ($b \to c$) amplitudes (denoted $R_f$),
to a very good approximation, $C_f = - C_{\bar{f}} = 1$
(using $f = D^{(*)-}h^+$, $\bar{f} = D^{(*)+}h^-$ $h = \pi,\rho$),
and the coefficients of the sine terms are given by
\begin{equation}
  S_f = - 2 R_f \sin( \phi_{\rm mix} + \phi_{\rm dec} - \delta_f )
  \hspace{5mm}
  {\rm and}
  \hspace{5mm}
  S_{\bar{f}} = - 2 R_f \sin( \phi_{\rm mix} + \phi_{\rm dec} + \delta_f ).
\end{equation}
Thus weak phase information can be cleanly obtained from measurements
of $S_f$ and $S_{\bar{f}}$, 
although external information on at least one of $R_f$ or $\delta_f$ is necessary.
(Note that $\phi_{\rm mix} + \phi_{\rm dec} = 2\beta + \gamma$ for all the decay modes 
in question, while $R_f$ and $\delta_f$ depend on the decay mode.)

Again, different notations have been used in the literature.
\babar\xspace\cite{ref:cp_uta:cud:babar:full,ref:cp_uta:cud:babar:partial}
defines the time-dependent probability function by
\begin{equation}
  f^\pm (\eta, \Delta t) = \frac{e^{-|\Delta t|/\tau}}{4\tau} 
  \left[  
    1 \mp S_\zeta \sin (\Delta m \Delta t) \mp \eta C_\zeta \cos(\Delta m \Delta t) 
  \right],
\end{equation} 
where the upper (lower) sign corresponds to 
the tagging meson being a $\Bz$ ($\Bzb$). 
[Note here that a tagging $\Bz$ ($\Bzb$) corresponds to $-S_\xi$ ($+S_\xi$).]
The parameters $\eta$ and $\zeta$ take the values $+1$ and $+$ ($-1$ and $-$) 
when the final state is, \eg, $D^-\pi^+$ ($D^+\pi^-$). 
However, in the fit, the substitutions $C_\zeta = 1$ and 
$S_\zeta = a \mp \eta b_i - \eta c_i$ are made.\footnote{
  The subscript $i$ denotes tagging category.
}
[Note that, neglecting $b$ terms, $S_+ = a - c$ and $S_- = a + c$, 
so that $a = (S_+ + S_-)/2$, $c = (S_- - S_+)/2$, in analogy to 
the parameters of Eq.~(\ref{eq:cp_uta:non-cp-s_and_deltas}).] 
The subscript $i$ denotes the tagging category. 
These are motivated by the possibility of 
$\CP$ violation on the tag side~\cite{ref:cp_uta:cud:tagside}, 
which is absent for semileptonic $\B$ decays (mostly lepton tags). 
The parameter $a$ is not affected by tag side $\CP$ violation. 

The parameters used by \belle\ in the analysis using 
partially reconstructed $\B$ decays~\cite{ref:cp_uta:cud:belle:partial}, 
are similar to the $S_\zeta$ parameters defined above. 
However, in the \belle\ convention, 
a tagging $\Bz$ corresponds to a $+$ sign in front of the sine coefficient; 
furthermore the correspondence between the super/subscript 
and the final state is opposite, so that $S_\pm$ (\babar) = $- S^\mp$ (\belle). 
In this analysis, only lepton tags are used, 
so there is no effect from tag side $\CP$ violation. 
In the \belle\ analysis using 
fully reconstructed $\B$ decays~\cite{ref:cp_uta:cud:belle:full}, 
this effect is measured and taken into account using $\Dstar l \nu$ decays; 
in neither \belle\ analysis are the $a$, $b$ and $c$ parameters used. 
In the latter case, the measured parameters are 
$2 R_{D^{(*)}\pi} \sin( 2\phi_1 + \phi_3 \pm \delta_{D^{(*)}\pi} )$; 
the definition is such that 
$S^\pm$ (\belle) = $- 2 R_{\Dstar \pi} \sin( 2\phi_1 + \phi_3 \pm \delta_{\Dstar \pi} )$. 
However, the definition includes an 
angular momentum factor $(-1)^L$~\cite{ref:cp_uta:cud:fleischer}, 
and so for the results in the $D\pi$ system, 
there is an additional factor of $-1$ in the conversion.

Explicitly, the conversion then reads as given in 
Table~\ref{tab:cp_uta:notations:non_cp:dstarpi}, 
where we have neglected the $b_i$ terms used by \babar.
For the averages in this document,
we use the $a$ and $c$ parameters,
and give the explicit translations used in 
Table~\ref{tab:cp_uta:notations:non_cp:dstarpi2}.
It is to be fervently hoped that the experiments will
converge on a common notation in future.

\begin{table}
  \begin{center} 
    \caption{
      Conversion between the various notations used for 
      $\CP$ violation parameters in the 
      $D^{\pm}\pi^{\mp}$, $D^{*\pm}\pi^{\mp}$ and $D^{\pm}\rho^{\mp}$ systems.
      The $b_i$ terms used by \babar\ have been neglected.
    }
    \vspace{0.2cm}
    \setlength{\tabcolsep}{0.0pc}
    \begin{tabular*}{\textwidth}{@{\extracolsep{\fill}}cccc} \hline 
      & \babar\ & \belle\ partial rec. & \belle\ full rec. \\
      \hline
      $S_{D^+\pi^-}$    & $- S_- = - (a + c_i)$ &  N/A  &
      $2 R_{D\pi} \sin( 2\phi_1 + \phi_3 + \delta_{D\pi} )$ \\
      $S_{D^-\pi^+}$    & $- S_+ = - (a - c_i)$ &  N/A  &
      $2 R_{D\pi} \sin( 2\phi_1 + \phi_3 - \delta_{D\pi} )$ \\
      $S_{D^{*+}\pi^-}$ & $- S_- = - (a + c_i)$ & $S^+$ &   
      $- 2 R_{\Dstar \pi} \sin( 2\phi_1 + \phi_3 + \delta_{\Dstar \pi} )$ \\
      $S_{D^{*-}\pi^+}$ & $- S_+ = - (a - c_i)$ & $S^-$ &
      $- 2 R_{\Dstar \pi} \sin( 2\phi_1 + \phi_3 - \delta_{\Dstar \pi} )$ \\
      $S_{D^+\rho^-}$    & $- S_- = - (a + c_i)$ &  N/A  &  N/A  \\
      $S_{D^-\rho^+}$    & $- S_+ = - (a - c_i)$ &  N/A  &  N/A  \\
      \hline 
    \end{tabular*}
    \label{tab:cp_uta:notations:non_cp:dstarpi}
  \end{center}
\end{table}
   
\begin{table}
  \begin{center} 
    \caption{
      Translations used to convert the parameters measured by \belle
      to the parameters used for averaging in this document.
      The angular momentum factor $L$ is $-1$ for $\Dstar\pi$ and $+1$ for $D\pi$.
    }
    \vspace{0.2cm}
    \setlength{\tabcolsep}{0.0pc}
    \begin{tabular*}{\textwidth}{@{\extracolsep{\fill}}ccc} \hline 
        & $\Dstar\pi$ partial rec. & $D^{(*)}\pi$ full rec. \\
        \hline
        $a$ & $- (S^+ + S^-)$ &
        $\frac{1}{2} (-1)^{L+1}
        \left(
          2 R_{D^{(*)}\pi} \sin( 2\phi_1 + \phi_3 + \delta_{D^{(*)}\pi} ) + 
          2 R_{D^{(*)}\pi} \sin( 2\phi_1 + \phi_3 - \delta_{D^{(*)}\pi} )
        \right)$ \\
        $c$ & $- (S^+ - S^-)$ & 
        $\frac{1}{2} (-1)^{L+1}
        \left(
          2 R_{D^{(*)}\pi} \sin( 2\phi_1 + \phi_3 + \delta_{D^{(*)}\pi} ) -
          2 R_{D^{(*)}\pi} \sin( 2\phi_1 + \phi_3 - \delta_{D^{(*)}\pi} )
        \right)$ \\
        \hline 
      \end{tabular*}
    \label{tab:cp_uta:notations:non_cp:dstarpi2}
  \end{center}
\end{table}

\mysubsubsubsection{Time-dependent asymmetries in radiative $\B$ decays
}
\label{sec:cp_uta:notations:non_cp:radiative}

As a special case of decays to non-$\CP$ eigenstates,
let us consider radiative $\B$ decays.
Here, the emitted photon has a distinct helicity,
which is in principle observable, but in practice is not usually measured.
Thus the measured time-dependent decay rates 
are given by~\cite{ref:cp_uta:bsg:ags,ref:cp_uta:bsg:aghs}
\begin{eqnarray}
  \Gamma_{\Bzb \to X \gamma} (\Delta t) & = &
  \Gamma_{\Bzb \to X \gamma_L} (\Delta t) + \Gamma_{\Bzb \to X \gamma_R} (\Delta t) \\ \nonumber
  & = &
  \frac{e^{-\left| \Delta t \right| / \tau(\Bz)}}{4\tau(\Bz)} 
  \left\{ 
    1 + 
    \left( S_L + S_R \right) \sin(\Delta m \Delta t) - 
    \left( C_L + C_R \right) \cos(\Delta m \Delta t) 
  \right\},
  \\
  \Gamma_{\Bz \to X \gamma} (\Delta t) & = & 
  \Gamma_{\Bz \to X \gamma_L} (\Delta t) + \Gamma_{\Bz \to X \gamma_R} (\Delta t) \\ \nonumber 
  & = &
  \frac{e^{-\left| \Delta t \right| / \tau(\Bz)}}{8\tau(\Bz)} 
  \left\{ 
    1 - 
    \left( S_L + S_R \right) \sin(\Delta m \Delta t) + 
    \left( C_L + C_R \right) \cos(\Delta m \Delta t) 
  \right\},
\end{eqnarray}
where in place of the subscripts $f$ and $\bar{f}$ we have used $L$ and $R$
to indicate the photon helicity.
In order for interference between decays with and without $\Bz$-$\Bzb$ mixing
to occur, the $X$ system must not be flavour-specific,
\eg, in case of $\Bz \to K^{*0}\gamma$, the final state must be $\KS \pi^0 \gamma$.
The sign of the sine term depends on the $C$ eigenvalue of the $X$ system.
The photons from $b \to q \gamma$ ($\bar{b} \to \bar{q} \gamma$) are predominantly
left (right) polarized, with corrections of order of $m_q/m_b$,
thus interference effects are suppressed.
The predicted smallness of the $S$ terms in the Standard Model
results in sensitivity to new physics contributions.

\mysubsubsection{Asymmetries in $\B \to \DorDstar K^{(*)}$ decays
}
\label{sec:cp_uta:notations:cus}

$\CP$ asymmetries in $\B \to \DorDstar K^{(*)}$ decays are sensitive to $\gamma$.
The neutral $D^{(*)}$ meson produced 
is an admixture of $\DorDstarz$ (produced by a $b \to c$ transition) and 
$\DorDstarzb$ (produced by a colour-suppressed $b \to u$ transition) states.
If the final state is chosen so that both $\DorDstarz$ and $\DorDstarzb$ 
can contribute, the two amplitudes interfere,
and the resulting observables are sensitive to $\gamma$, 
the relative weak phase between 
the two $\B$ decay amplitudes~\cite{ref:cp_uta:cus:bs}.
Various methods have been proposed to exploit this interference,
including those where the neutral $D$ meson is reconstructed 
as a $\CP$ eigenstate (GLW)~\cite{ref:cp_uta:cus:glw},
in a suppressed final state (ADS)~\cite{ref:cp_uta:cus:ads},
or in a self-conjugate three-body final state, 
such as $\KS \pi^+\pi^-$ (Dalitz)~\cite{ref:cp_uta:cus:dalitz}.
It should be emphasised that while each method 
differs in the choice of $D$ decay,
they are all sensitive to the same parameters of the $B$ decay,
and can be considered as variations of the same technique.

Consider the case of $\Bmp \to D \Kmp$,
with $D$ decaying to a final state $f$,
which is accessible to both $\Dz$ and $\Dzb$.
We can write the decay rates for $\Bm$ and $\Bp$ ($\Gamma_\mp$), 
the charge averaged rate ($\Gamma = (\Gamma_- + \Gamma_+)/2$)
and the charge asymmetry 
(${\cal A} = (\Gamma_- - \Gamma_+)/(\Gamma_- + \Gamma_+)$, see Eq.~(\ref{eq:cp_uta:pra})) as 
\begin{eqnarray}
  \label{eq:cp_uta:dk:rate_def}
  \Gamma_\mp  & \propto & 
  r_B^2 + r_D^2 + 2 r_B r_D \cos \left( \delta_B + \delta_D \mp \gamma \right), \\
  \label{eq:cp_uta:dk:av_rate_def}
  \Gamma & \propto &  
  r_B^2 + r_D^2 + 2 r_B r_D \cos \left( \delta_B + \delta_D \right) \cos \left( \gamma \right), \\
  \label{eq:cp_uta:dk:acp_def}
  {\cal A} & = & 
  \frac{
    2 r_B r_D \sin \left( \delta_B + \delta_D \right) \sin \left( \gamma \right)
  }{
    r_B^2 + r_D^2 + 2 r_B r_D \cos \left( \delta_B + \delta_D \right) \cos \left( \gamma \right),  
  }
\end{eqnarray}
where the ratio of $\B$ decay amplitudes\footnote{
  Note that here we use the notation $r_B$ to denote the ratio
  of $\B$ decay amplitudes, 
  whereas in Sec.~\ref{sec:cp_uta:notations:non_cp:dstarpi} 
  we used, \eg, $R_{D\pi}$, for a rather similar quantity.
  The reason is that here we need to be concerned also with 
  $D$ decay amplitudes,
  and so it is convenient to use the subscript to denote the decaying particle.
  Hopefully, using $r$ in place of $R$ will help reduce potential confusion.
} 
is usually defined to be less than one,
\begin{equation}
  \label{eq:cp_uta:dk:rb_def}
  r_B = 
  \frac{
    \left| A\left( \Bm \to \Dzb K^- \right) \right|
  }{
    \left| A\left( \Bm \to \Dz  K^- \right) \right|
  },
\end{equation}
and the ratio of $D$ decay amplitudes is correspondingly defined by
\begin{equation}
  \label{eq:cp_uta:dk:rd_def}
  r_D = 
  \frac{
    \left| A\left( \Dz  \to f \right) \right|
  }{
    \left| A\left( \Dzb \to f \right) \right|
  }.
\end{equation}
The strong phase differences between the $\B$ and $D$ decay amplitudes 
are given by $\delta_B$ and $\delta_D$, respectively.
The values of $r_D$ and $\delta_D$ depend on the final state $f$:
for the GLW analysis, $r_D = 1$ and $\delta_D$ is trivial (either zero or $\pi$),
in the Dalitz plot analysis $r_D$ and $\delta_D$ vary across the Dalitz plot,
and depend on the $D$ decay model used,
for the ADS analysis, the values of $r_D$ and $\delta_D$ are not trivial.

Note that, for given values of $r_B$ and $r_D$, 
the maximum size of ${\cal A}$ (at $\sin \left( \delta_B + \delta_D \right) = 1$)
is $2 r_B r_D \sin \left( \gamma \right) / \left( r_B^2 + r_D^2 \right)$.
Thus even for $D$ decay modes with small $r_D$, 
large asymmetries, and hence sensitivity to $\gamma$, 
may occur for $B$ decay modes with similar values of $r_B$.
For this reason, the ADS analysis of the decay $B^\mp \to D \pi^\mp$ 
is also of interest.

In the GLW analysis, the measured quantities are the 
partial rate asymmetry, and the charge averaged rate,
which are measured both for $\CP$ even and $\CP$ odd $D$ decays.
For the latter, it is experimentally convenient to measure a double ratio,
\begin{equation}
  \label{eq:cp_uta:dk:double_ratio}
  R_{\CP} = 
  \frac{
    \Gamma\left( \Bm \to D_{\CP} \Km  \right) \, / \, \Gamma\left( \Bm \to \Dz \Km \right)
  }{
    \Gamma\left( \Bm \to D_{\CP} \pim \right) \, / \, \Gamma\left( \Bm \to \Dz \pim \right)
  }
\end{equation}
that is normalized both to the rate for the favoured $\Dz \to \Km\pip$ decay, 
and to the equivalent quantities for $\Bm \to D\pim$ decays
(charge conjugate modes are implicitly 
included in Eq.~(\ref{eq:cp_uta:dk:double_ratio})).
In this way the constant of proportionality drops out of 
Eq.~(\ref{eq:cp_uta:dk:av_rate_def}).

For the ADS analysis, using a suppressed $D \to f$ decay,
the measured quantities are again the partial rate asymmetry, 
and the charge averaged rate.
In this case it is sufficient to measure the rate in a single ratio
(normalized to the favoured $D \to \bar{f}$ decay)
since detection systematics cancel naturally.
In the ADS analysis, there are an additional two unknowns ($r_D$ and $\delta_D$)
compared to the GLW case.  
However, the value of $r_D$ can be measured using 
decays of $D$ mesons of known flavour.

The relations between the measured quantities and the
underlying parameters are summarized in Table~\ref{tab:cp_uta:notations:dk}.
Note carefully that the hadronic factors $r_B$ and $\delta_B$ 
are different, in general, for each $\B$ decay mode.
In the Dalitz plot analysis,
once a model is assumed for the $D$ decay, 
which gives the values of $r_D$ and $\delta_D$ across the Dalitz plot,
the values of ($\gamma$, $r_B$, $\delta_B$) can be directly extracted from
a simultaneous fit to the $\Bm$ and $\Bp$ data.

\begin{table}
  \begin{center} 
    \caption{
      Summary of relations between measured and physical parameters 
      in GLW and ADS analyses of $\B \to \DorDstar K^{(*)}$.
    }
    \vspace{0.2cm}
    \setlength{\tabcolsep}{1.0pc}
    \begin{tabular}{cc} \hline 
      \mc{2}{c}{GLW analysis} \\
      $R_{\CP\pm}$ & $1 + r_B^2 \pm 2 r_B \cos \left( \delta_B \right) \cos \left( \gamma \right)$ \\
      $A_{\CP\pm}$ & $\pm 2 r_B \sin \left( \delta_B \right) \sin \left( \gamma \right) / R_{\CP\pm}$ \\
      \hline
      \mc{2}{c}{ADS analysis} \\
      $R_{ADS}$ & $r_B^2 + r_D^2 + 2 r_B r_D \cos \left( \delta_B + \delta_D \right) \cos \left( \gamma \right)$ \\
      $A_{ADS}$ & $2 r_B r_D \sin \left( \delta_B + \delta_D \right) \sin \left( \gamma \right) / R_{ADS}$ \\
      \hline
    \end{tabular}
    \label{tab:cp_uta:notations:dk}
  \end{center}
\end{table}

\mysubsection{Common inputs and error treatment
}
\label{sec:cp_uta:common_inputs}

The common inputs used for rescaling are listed in 
Table~\ref{tab:cp_uta:common_inputs}.
The $\Bz$ lifetime ($\tau(\Bz)$) and mixing parameter ($\Delta m_d$)
averages are provided by the HFAG Lifetimes and Oscillations 
subgroup (Sec.~\ref{sec:life_mix}).
The fraction of the perpendicularly polarized component 
($\left| A_{\perp} \right|^2$) in $\B \to \jpsi \Kstar(892)$ decays,
which determines the $\CP$ composition, 
is averaged from results by 
\babar~\cite{ref:cp_uta:ccs:babar:psi_kstar} and
\belle~\cite{ref:cp_uta:ccs:belle:psi_kstar}.

At present, we only rescale to a common set of input parameters
for modes with reasonably small statistical errors
($b \to c\bar{c}s$ and $b \to q\bar{q}s$ transitions).
Correlated systematic errors are taken into account
in these modes as well.
For all other modes, the effect of such a procedure is 
currently negligible.

\begin{table}
  \begin{center}
    \caption{
      Common inputs used in calculating the averages.
    }
    \vspace{0.2cm}
    \setlength{\tabcolsep}{1.0pc}
    \begin{tabular}{cc} \hline 
      $\tau(\Bz)$ $({\rm ps})$ 	& $1.536 \pm 0.014$  \\
      $\Delta m_d$ $({\rm ps}^{-1})$ & $0.502 \pm 0.007$ \\
      $\left| A_{\perp} \right|^2 (\jpsi \Kstar)$ & $0.211 \pm 0.011$ \\
      \hline
    \end{tabular}
    \label{tab:cp_uta:common_inputs}
  \end{center}
\end{table}

As explained in Sec.~\ref{sec:intro},
we do not apply a rescaling factor on the error of an average
that has $\chi^2/\dof > 1$ 
(unlike the procedure currently used by the PDG~\cite{Eidelman:2004wy}).
We provide a confidence level of the fit so that
one can know the consistency of the measurements included in the average,
and attach comments in case some care needs to be taken in the interpretation.
Note that, in general, results obtained from data samples with low statistics
will exhibit some non-Gaussian behaviour.
For measurements where one error is given, 
it represents the total error, 
where statistical and systematic uncertainties have been added in quadrature.
If two errors are given, the first is statistical and the second systematic.
If more than two errors are given,
the origin of the additional uncertainty will be explained in the text.

Averages are computed by maximizing a log-likelihood function 
${\cal L}$ assuming Gaussian statistical and systematic errors.
When observables exhibit significant correlatations 
(\eg, sine and cosine coefficients in some time-dependent \CP asymmetries), 
a combined minimization is performed, 
taking into account the correlations. 
Asymmetric errors are treated by defining an asymmetric log-likelihood
function: ${\cal L}_i = (x - x_i)^2/(2\sigma_{i}^2)$,
where $\sigma_i=\sigma_{i,+}$ ($\sigma_i=\sigma_{i,-}$) if $x>x_i$ ($x<x_i$), 
and where $x_i$ is the $i$th measurement of 
the observable $x$ that is averaged.
This example assumes no correlations between observables. 
The correlated case is a straightforward extension to this.

\mysubsection{Time-dependent $\CP$ asymmetries in $b \to c\bar{c}s$ transitions
}
\label{sec:cp_uta:ccs}

\begin{figure}
  \begin{center}
    \resizebox{0.65\textwidth}{!}{\includegraphics{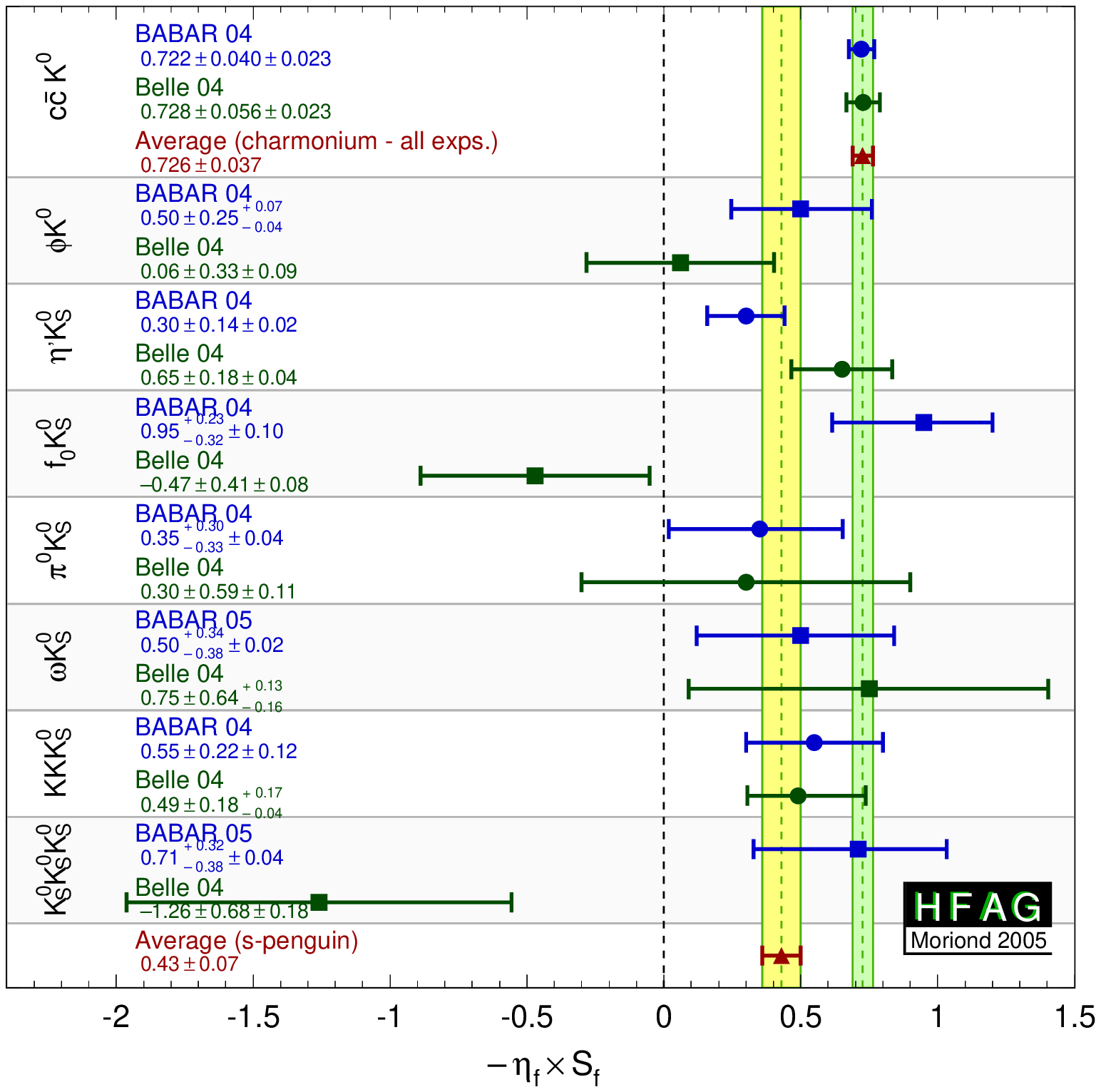}}
  \end{center}
  \vspace{-0.8cm}
  \caption{
    $S_{b \to c\bar c s}$ and $S_{b \to q\bar q s}$.
  }
  \label{fig:cp_uta:s_qqs}
\end{figure}

\begin{figure}
  \begin{center}
    \begin{tabular}{cc}
      \resizebox{0.46\textwidth}{!}{\includegraphics{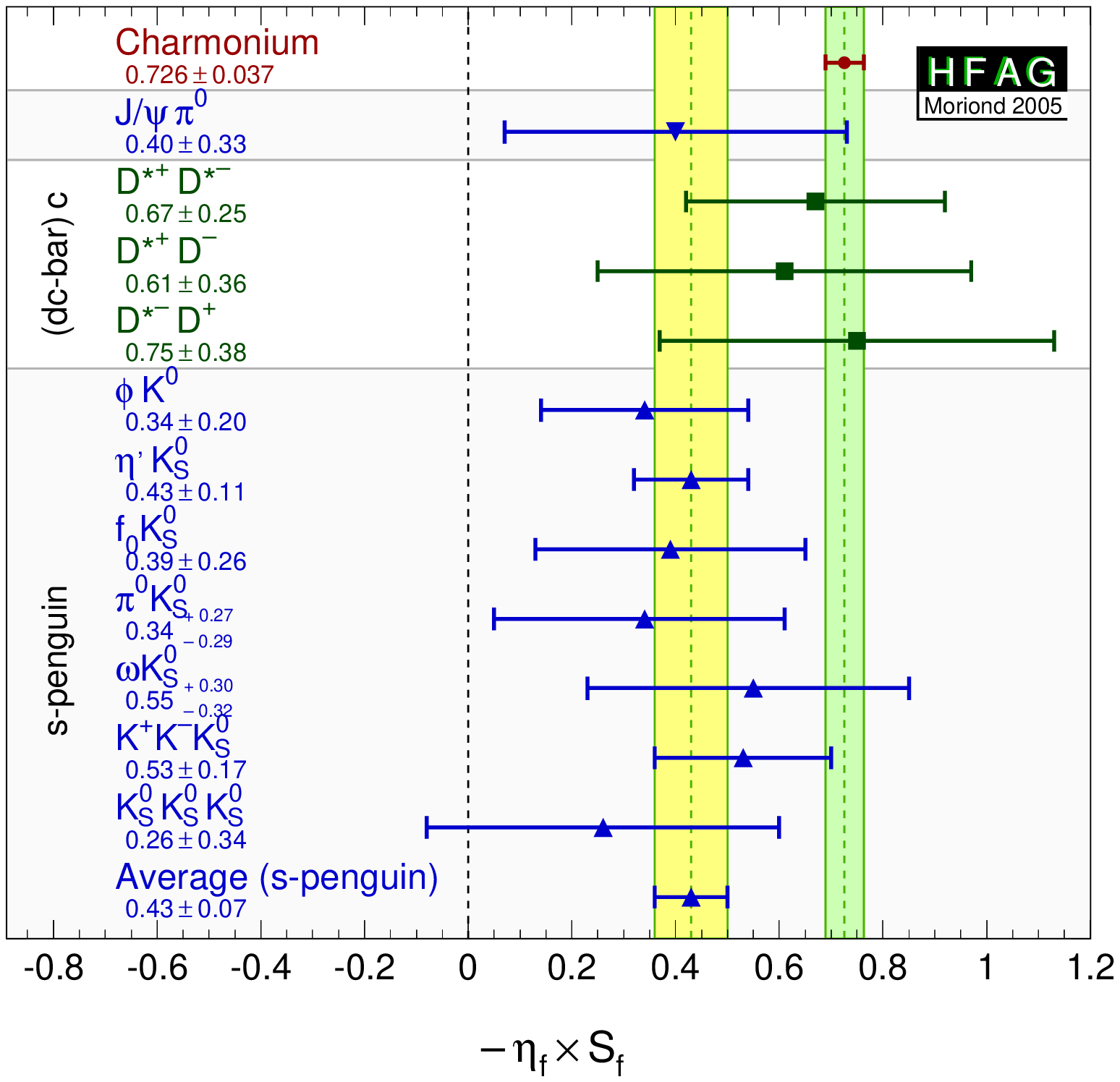}}
      &
      \resizebox{0.46\textwidth}{!}{\includegraphics{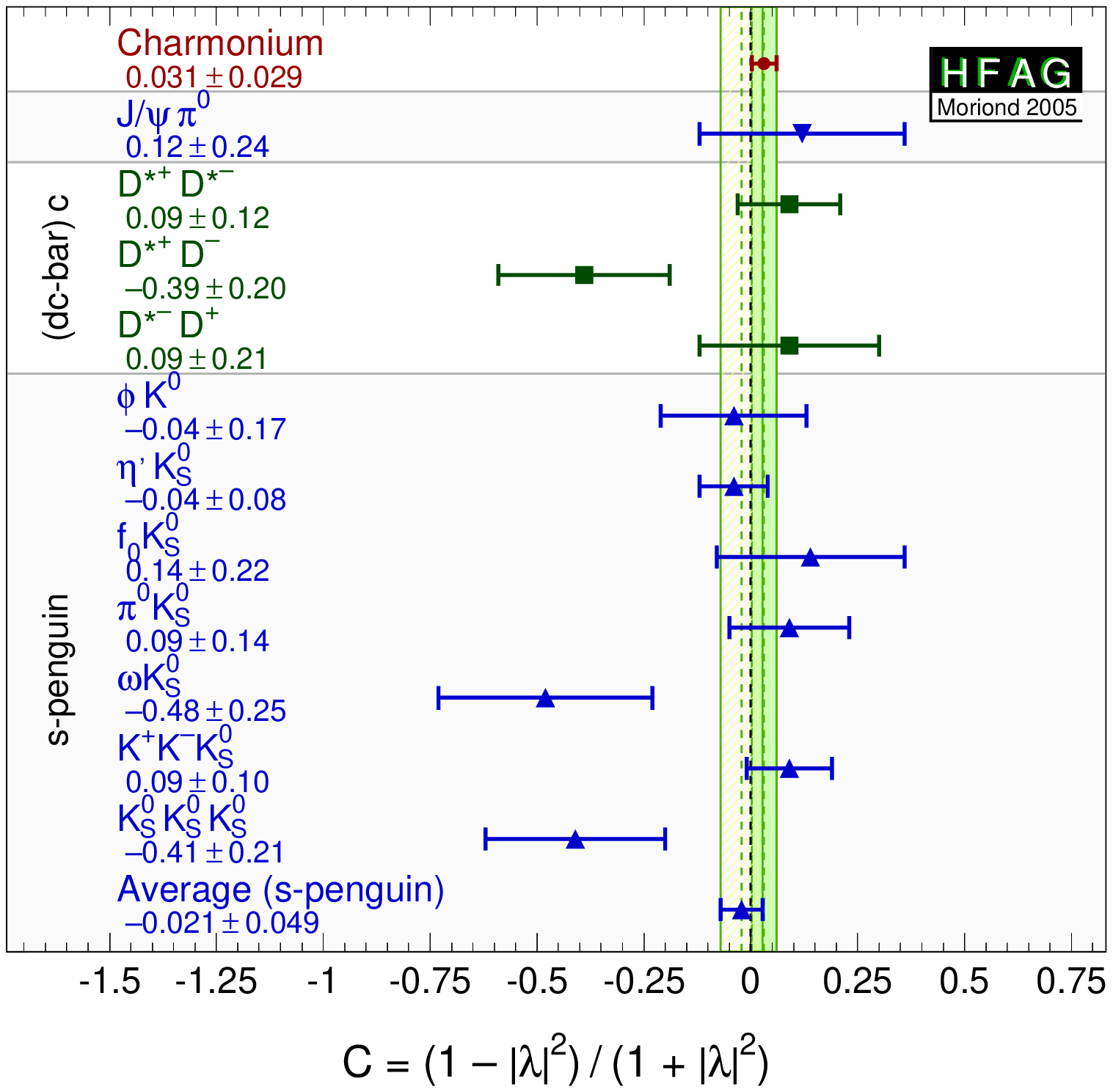}}
    \end{tabular}
  \end{center}
  \vspace{-0.8cm}
  \caption{
    Comparisons of the averages of (left)
    $S_{b \to c\bar c s}$, $S_{b \to c\bar c d}$ and $S_{b \to q\bar q s}$,
    and (right)
    $C_{b \to c\bar c s}$, $C_{b \to c\bar c d}$ and $C_{b \to q\bar q s}$.
  }
  \label{fig:cp_uta:qqs_ccd}
\end{figure}

In the Standard Model, the time-dependent parameters for
$b \to c\bar c s$ transitions are predicted to be: 
$S_{b \to c\bar c s} = - \etacp \sin(2\beta)$,
$C_{b \to c\bar c s} = 0$ to very good accuracy.
The averages for $S_{b \to c\bar c s}$ and $C_{b \to c\bar c s}$
are provided in Table~\ref{tab:cp_uta:ccs}.
The averages for $S_{b \to c\bar c s}$ are shown in Fig.~\ref{fig:cp_uta:s_qqs};
averages for $C_{b \to c\bar c s}$ are included in Fig.~\ref{fig:cp_uta:qqs_ccd}.

Both \babar\  and \belle\ use the $\etacp = -1$ modes
$\jpsi \KS$, $\psi(2S) \KS$, $\chi_{c1} \KS$ and $\eta_c \KS$, 
as well as $\jpsi \KL$, which has $\etacp = +1$
and $\jpsi K^{*0}(892)$, which is found to have $\etacp$ close to $+1$
based on the measurement of $\left| A_\perp \right|$ 
(see Sec.~\ref{sec:cp_uta:common_inputs}).
ALEPH, OPAL and CDF use only the $\jpsi \KS$ final state.

\begin{table}
  \begin{center}
    \caption{
      $S_{b \to c\bar c s}$ and $C_{b \to c\bar c s}$.
    }
    \vspace{0.2cm}
    \setlength{\tabcolsep}{0.0pc}
    \begin{tabular*}{\textwidth}{@{\extracolsep{\fill}}lrcc} \hline 
      \mc{2}{l}{Experiment} & 
      $- \etacp S_{b \to c\bar c s}$ & $C_{b \to c\bar c s}$ \\
      \hline
      \babar & \cite{ref:cp_uta:ccs:babar} & 
      $0.722 \pm 0.040 \pm 0.023$ & $\ph{-}0.051 \pm 0.033 \pm 0.014$ \\
      \belle & \cite{BELLE2} & 
      $0.728 \pm 0.056 \pm 0.023$ & $-0.007 \pm 0.041 \pm 0.033$ \\
      \hline
      \mc{2}{l}{\bf \boldmath $\B$ factory average} & 
      $0.725 \pm 0.037$ & $0.031 \pm 0.029$ \\
      \mc{2}{l}{\small Confidence level} & 
      \small $0.91$ & \small $0.30$ \\
      \hline
      ALEPH & \cite{ref:cp_uta:ccs:aleph} & 
      $0.84 \, ^{+0.82}_{-1.04} \pm 0.16$ \\
      OPAL  & \cite{ref:cp_uta:ccs:opal}  & 
      $3.2 \, ^{+1.8}_{-2.0} \pm 0.5$ \\
      CDF   & \cite{ref:cp_uta:ccs:cdf}   & 
      $0.79 \, ^{+0.41}_{-0.44}$ \\
      \hline
      \mc{2}{l}{\bf Average} & 
      $0.726 \pm 0.037$ & $0.031 \pm 0.029$ \\
      \hline
    \end{tabular*}
    \label{tab:cp_uta:ccs}
  \end{center}
\end{table}

\begin{figure}[p]
  \begin{center}
    \resizebox{0.80\textwidth}{!}{
      \includegraphics{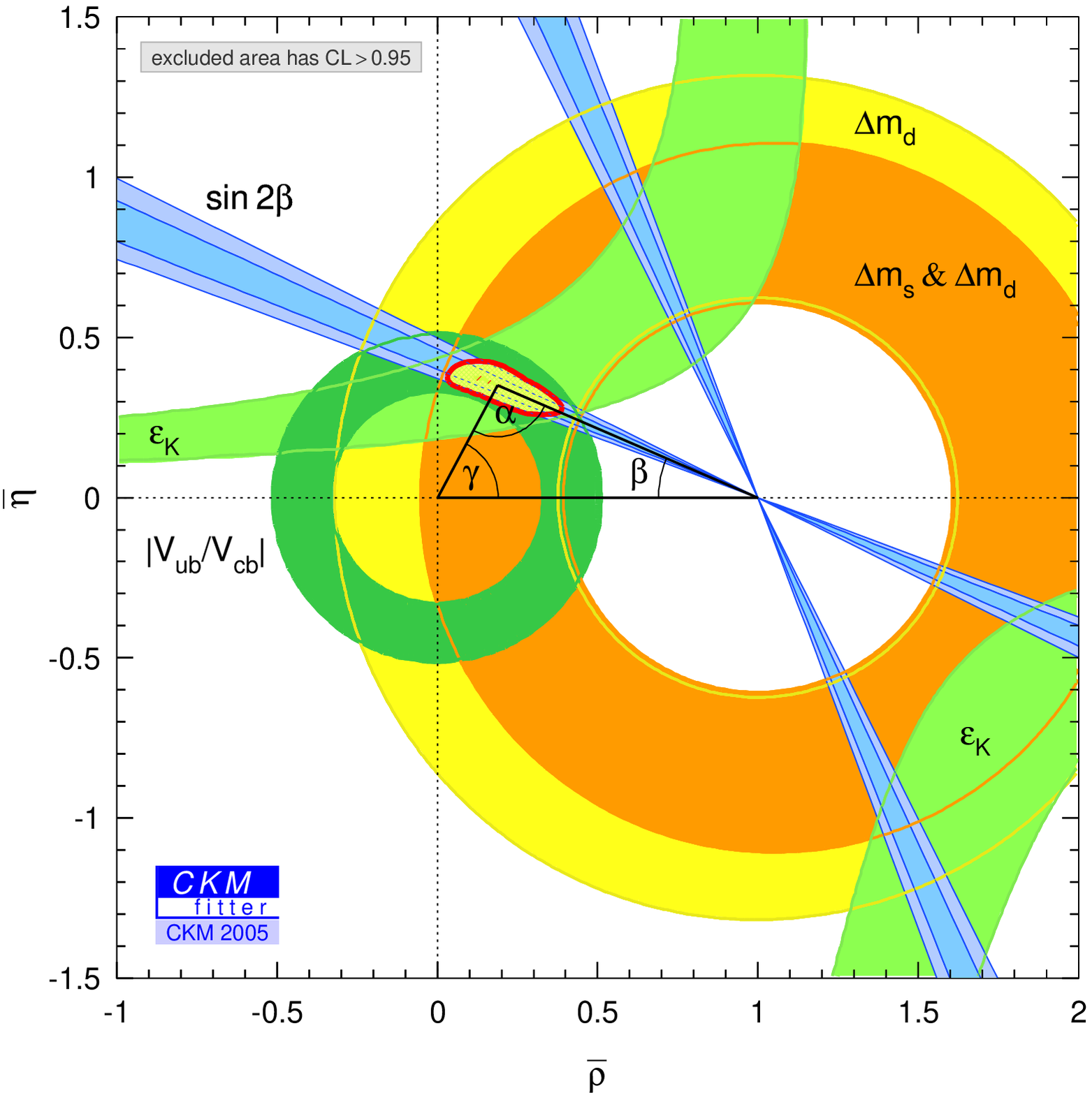}
    }
    \resizebox{0.95\textwidth}{!}{\hspace{1cm}
      \includegraphics{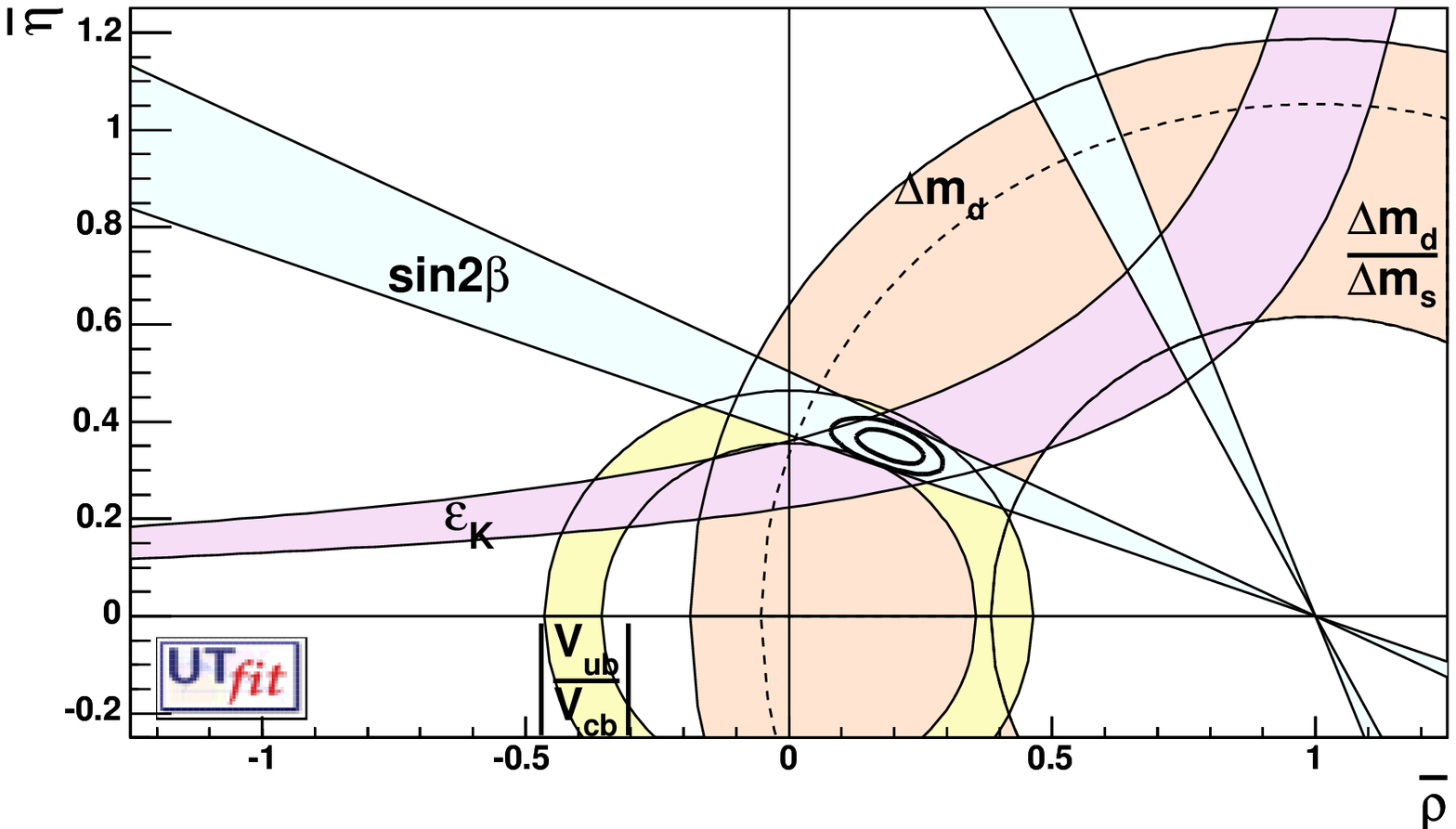}
    }
  \end{center}
  \vspace{-0.5cm}
  \caption{
    Standard Model constraints on the $(\rhobar,\etabar)$ plane,
    from (top)~\cite{ref:cp_uta:ckmfitter} and 
    (bottom)~\cite{ref:cp_uta:utfit}.
  }
  \label{fig:cp_uta:ckmfitter_sin2beta}
\end{figure}

These results give a precise constraint on the $(\rhobar,\etabar)$ plane,
in remarkable agreement with other constraints from 
$\CP$ conserving quantities, 
and with $\CP$ violation in the kaon system, in the form of the parameter $\epsilon_K$.
Such comparisons have been performed by various phenomenological groups,
such as CKMfitter~\cite{ref:cp_uta:ckmfitter} 
and UTFit~\cite{ref:cp_uta:utfit}.
Figure~\ref{fig:cp_uta:ckmfitter_sin2beta} displays the constraints 
obtained from these two groups.

\mysubsection{Time-dependent transversity analysis of $\Bz \to J/\psi K^{*0}$
}
\label{sec:cp_uta:ccs_vv}

$\B$ meson decays to the vector-vector final state $J/\psi K^{*0}$
are also mediated by the $b \to c \bar c s$ transition.
When a final state which is not flavour-specific ($K^{*0} \to \KS \pi^0$) is used,
a time-dependent transversity analysis can be performed 
allowing sensitivity to both $\sin(2\beta)$ and $\cos(2\beta)$.
Such analyses have been performed by both $\B$ factory experiments.
In principle, the strong phases between the transversity amplitudes
are not uniquely determined by such an analysis, 
leading to a discrete ambiguity in the sign of $\cos(2\beta)$.
The \babar~\cite{ref:cp_uta:ccs:babar:psi_kstar} collaboration resolves 
this ambiguity using the known variation~\cite{ref:cp_uta:ccs:lass}
of the P-wave phase (fast) relative to the S-wave phase (slow) 
with the invariant mass of the $K\pi$ system 
in the vicinity of the $K^*(892)$ resonance. 
The result is in agreement with the prediction from 
$s$ quark helicity conservation,
and corresponds to Solution II defined by Suzuki~\cite{ref:cp_uta:ccs:suzuki}.
We use this phase convention for the averages given in 
Table~\ref{tab:cp_uta:ccs:psi_kstar}.

\begin{table}
  \begin{center}
    \caption{
      Averages from $\Bz \to J/\psi K^{*0}$ transversity analyses.
    }
    \vspace{0.2cm}
    \setlength{\tabcolsep}{0.0pc}
    \begin{tabular*}{\textwidth}{@{\extracolsep{\fill}}lrcc} \hline 
      \mc{2}{l}{Experiment} & $\sin(2\beta)$ & $\cos(2\beta)$ \\
      \hline
      \babar & \cite{ref:cp_uta:ccs:babar:psi_kstar} & 
      $-0.10 \pm 0.57 \pm 0.14$ & $ 3.32 \, ^{+0.76}_{-0.96} \pm 0.27$ \\
      \belle & \cite{ref:cp_uta:ccs:belle:psi_kstar} &
      $\ph{-}0.30 \pm 0.32 \pm 0.02$ & $ 0.31 \pm 0.91 \pm 0.11$ \\
      \hline
      \mc{2}{l}{\bf Average} & 
      $0.21 \pm 0.28$ & $1.69 \pm 0.67 $ \\
      \mc{2}{l}{\small Confidence level} & 
      \small $0.55~(0.6\sigma)$ & \small $0.026~(2.2\sigma)$ \\
      \hline
    \end{tabular*}
    \label{tab:cp_uta:ccs:psi_kstar}
  \end{center}
\end{table}

While the statistical errors are large, 
and exhibit non-Gaussian behaviour, 
$\cos(2\beta)>0$ is preferred 
by the experimental data in $J/\psi \Kstar$.

\mysubsection{Time-dependent $\CP$ asymmetries in $b \to q\bar{q}s$ transitions
}
\label{sec:cp_uta:qqs}

The flavour changing neutral current $b \to s$ penguin
can be mediated by any up-type quark in the loop, 
and hence the amplitude can be written as
\begin{equation}
  \label{eq:cp_uta:b_to_s}
  \begin{array}{ccccc}
    A_{b \to s} & = & 
    \mc{3}{l}{F_u V_{ub}V^*_{us} + F_c V_{cb}V^*_{cs} + F_t V_{tb}V^*_{ts}} \\
    & = & (F_u - F_c) V_{ub}V^*_{us} & + & (F_t - F_c) V_{tb}V^*_{ts} \\
    & = & {\cal O}(\lambda^4) & + & {\cal O}(\lambda^2) \\
  \end{array}
\end{equation}
using the unitarity of the CKM matrix.
Therefore, in the Standard Model, 
this amplitude is dominated by $V_{tb}V^*_{ts}$, 
and to within a few degrees ($\delta\beta\lesssim2^\circ$ for 
$\beta\simeq23.3^\circ$) 
the time-dependent parameters 
can be written\footnote
{
  	The presence of a small (${\cal O}(\lambda^2)$) weak phase in 
	the dominant amplitude of the $s$ penguin decays introduces 
	a phase shift given by
	$S_{b \to q\bar q s} = -\eta\sin(2\beta)\cdot(1 + \Delta)$. 
	Using the CKMfitter results for the Wolfenstein 
	parameters~\cite{ref:cp_uta:uud:charles}, one finds: 
	$\Delta \simeq 0.033$, which corresponds to a shift of 
	$2\beta$ of $+2.1$ degrees. Nonperturbative contributions
      	can alter this result.
}
$S_{b \to q\bar q s} \approx - \etacp \sin(2\beta)$,
$C_{b \to q\bar q s} \approx 0$,
assuming $b \to s$ penguin contributions only ($q = u,d,s$).

Due to the large virtual mass scales occurring in the penguin loops,
additional diagrams from physics beyond the Standard Model,
with heavy particles in the loops, may contribute.
In general, these contributions will affect the values of 
$S_{b \to q\bar q s}$ and $C_{b \to q\bar q s}$.
A discrepancy between the values of 
$S_{b \to c\bar c s}$ and $S_{b \to q\bar q s}$
can therefore provide a clean indication of new physics.

However, there is an additional consideration to take into account.
The above argument assumes only the $b \to s$ penguin contributes
to the $b \to q\bar q s$ transition.
For $q = s$ this is a good assumption, 
which neglects only rescattering effects.
However, for $q = u$ there is a colour-suppressed $b \to u$ tree diagram
(of order ${\cal O}(\lambda^4)$), which has a different weak 
(and possibly strong) phase.
In the case $q = d$, any light neutral meson that is formed from
$d \bar{d}$ also has a $u \bar{u}$ component,
and so again there is ``tree pollution''. The \Bz decays to 
$\piz\KS$ and $\omega\KS$ belong to this category.
The mesons $f_0$ and $\etapr$ are expected to have predominant $s\bar s$ parts,
which reduces the possible tree pollution. If the inclusive 
decay $\Bz\to\Kp\Km\Kz$ (excluding $\phi\Kz$) is dominated by
a non-resonant three-body transition, an OZI-rule suppressed 
tree-level diagram can occur through insertion of an 
$s\sbar$ pair. The corresponding penguin-type transition 
proceeds via insertion of a $u\ubar$ pair, which is expected
to be favored over the $s\sbar$ insertion by fragmentation models.
Neglecting rescattering, the final state $\Kz\Kzb\Kz$ has no 
tree pollution.

The averages for $S_{b \to q\bar q s}$ and $C_{b \to q\bar q s}$
can be found in Table~\ref{tab:cp_uta:qqs}.
The averages for $S_{b \to q\bar q s}$ are shown in Fig.~\ref{fig:cp_uta:s_qqs};
averages for $C_{b \to q\bar q s}$ are included in Fig.~\ref{fig:cp_uta:qqs_ccd}.
Results from both \babar\  and \belle\ are averaged for the modes
$\phi K^0$ (both $\phi\KS$ and $\phi\KL$ are used), 
$\etapr \KS$, $K^+K^-\KS$, $f_0 \KS$, $\pi^0 \KS$, $\omega\KS$ and $\KS\KS\KS$. 
Of these modes,
$\phi\KS$, $\etapr \KS$, $\pi^0 \KS$ and $\omega\KS$ have $\CP$ eigenvalue $\etacp = -1$, 
while $\phi\KL$, $f_0 \KS$ and $\KS\KS\KS$ have $\etacp = +1$.

The final state $K^+K^-\KS$ 
(contributions from $\phi \KS$ are implicitly excluded) 
is not a $\CP$ eigenstate.
However, the $\CP$ composition can be determined using either an 
isospin argument (used by \belle\ to determine a $\CP$ even fraction of 
$1.03 \pm 0.15 \pm 0.05$~\cite{ref:cp_uta:qqs:belle:kkks_cp})
or a moments analysis (used by \babar\ to find a 
$\CP$ even fraction of $0.89 \pm 0.08 \pm 0.06$~\cite{ref:cp_uta:qqs:babar:kkk0}).
The uncertainty in the $\CP$ even fraction leads to an 
asymmetric error on $S_{b \to q\bar q s}$, which is taken to be 
correlated among the experiments.
To combine, we rescale the results to the 
average $\CP$ even fraction of $0.93 \pm 0.09$.

\begin{table}
  \begin{center}
    \caption{
      $S_{b \to q\bar q s}$ and $C_{b \to q\bar q s}$.
    }
    \vspace{0.2cm}
    \setlength{\tabcolsep}{0.0pc}
    \begin{tabular*}{\textwidth}{@{\extracolsep{\fill}}lrcc} \hline 
      \mc{2}{l}{Experiment} & 
      $- \etacp S_{b \to q\bar q s}$ & $C_{b \to q\bar q s}$ \\
      \hline
      \mc{4}{c}{$\phi K^0$} \\
      \babar & \cite{ref:cp_uta:qqs:babar:kkk0} & 
      $0.50 \pm 0.25 \, ^{+0.07}_{-0.04}$ & $\ph{-}0.00 \pm 0.23 \pm 0.05$ \\
      \belle & \cite{ref:cp_uta:qqs:belle} & 
      $0.06 \pm 0.33 \pm 0.09$ & $ -0.08 \pm 0.22 \pm 0.09$ \\
      \mc{2}{l}{\bf Average} & 
      $0.34 \pm 0.20$ & $-0.04 \pm 0.17$ \\
      \mc{2}{l}{\small Confidence level} & 
      \small $0.30$ & \small $0.81$ \\
      \hline
      \mc{4}{c}{$\etapr \KS$} \\
      \babar & \cite{ref:cp_uta:qqs:babar:etapks} & 
      $0.30 \pm 0.14 \pm 0.02$ & $-0.21 \pm 0.10 \pm 0.02$ \\
      \belle & \cite{ref:cp_uta:qqs:belle} & 
      $0.65 \pm 0.18 \pm 0.04$ & $\ph{-}0.19 \pm 0.11 \pm 0.05$ \\
      \mc{2}{l}{\bf Average} & 
      $0.43 \pm 0.11$ & $-0.04 \pm 0.08$ \\
      \mc{2}{l}{\small Confidence level} & 
      \small $0.13~(1.5\sigma)$ & \small $0.011~(2.5\sigma)$ \\
      \hline
      \mc{4}{c}{$f_0 \KS$} \\
      \babar & \cite{ref:cp_uta:qqs:babar:f0ks} &
      $0.95 \, ^{+0.23}_{-0.32} \pm 0.10$ & $-0.24 \pm 0.31 \pm 0.15$ \\
      \belle & \cite{ref:cp_uta:qqs:belle} & 
      $-0.47 \pm 0.41 \pm 0.08$ & $\ph{-}0.39 \pm 0.27 \pm 0.08$ \\
      \mc{2}{l}{\bf Average} & 
      $0.39 \pm 0.26$ & $ 0.14 \pm 0.22$ \\
      \mc{2}{l}{\small Confidence level} & 
      \small $0.008~(2.7\sigma)$ & \small $0.16~(1.4\sigma)$ \\
      \hline
      \mc{4}{c}{$\pi^0 \KS$} \\
      \babar & \cite{ref:cp_uta:qqs:babar:pi0ks} &
      $0.35 \, ^{+0.30}_{-0.33} \pm 0.04$ & $0.06 \pm 0.18 \pm 0.03$ \\
      \belle & \cite{ref:cp_uta:qqs:belle} & 
      $0.30 \pm 0.59 \pm 0.11$ & $0.12 \pm 0.20 \pm 0.07$ \\
      \mc{2}{l}{\bf Average} & 
      $0.34 \, ^{+0.27}_{-0.29}$ & $0.09 \pm 0.14$ \\ 
      \mc{2}{l}{\small Confidence level} & 
      \small $0.94$ & \small $0.83$ \\
      \hline
      \mc{4}{c}{$\omega \KS$} \\
      \babar & \cite{ref:cp_uta:qqs:babar:omegaks} & 
      $0.50 ^{+0.34}_{-0.38} \pm 0.02$ & $-0.56 \, ^{+0.29}_{-0.27} \pm 0.03$ \\
      \belle & \cite{ref:cp_uta:qqs:belle} & 
      $0.75 \pm 0.64 \, ^{+0.13}_{-0.16}$ & $-0.26 \pm 0.48 \pm 0.15$ \\
      \mc{2}{l}{\bf Average} & 
      $0.55 \, ^{+0.30}_{-0.32}$ & $-0.48 \pm 0.25$ \\
      \mc{2}{l}{\small Confidence level} & 
      \small $0.74$ & \small $0.61$ \\
      \hline
      \mc{4}{c}{$K^+K^-\KS$} \\
      \babar & \cite{ref:cp_uta:qqs:babar:kkk0} &
      $0.55 \pm 0.22 \pm 0.04 \pm 0.11$ & $0.10 \pm 0.14 \pm 0.06$ \\
      \belle & \cite{ref:cp_uta:qqs:belle} & 
      $0.49 \pm 0.18 \pm 0.04 \, ^{+0.17}_{-0.00}$ & $0.08 \pm 0.12 \pm 0.07$ \\
      \mc{2}{l}{\bf Average} & 
      $0.53 \pm 0.17$ & $0.09 \pm 0.10$ \\ 
      \mc{2}{l}{\small Confidence level} & 
      \small $0.72 $ & \small $0.92$ \\
      \hline
      \mc{4}{c}{$\KS\KS\KS$} \\
      \babar & \cite{ref:cp_uta:qqs:babar:ksksks} &
      $0.71 \, ^{+0.32}_{-0.38} \pm 0.04$ & $-0.34 \, ^{+0.28}_{-0.25} \pm 0.05$ \\
      \belle & \cite{ref:cp_uta:qqs:belle:ksksks} & 
      $-1.26 \pm 0.68 \pm 0.20$ & $-0.54 \pm 0.34 \pm 0.09$ \\
      \mc{2}{l}{\bf Average} & 
      $0.26 \pm 0.34$ & $-0.41 \pm 0.21$ \\
      \mc{2}{l}{\small Confidence level} & 
      \small $0.014~(2.5\sigma)$ & \\
      \hline 
      \mc{2}{l}{\bf \boldmath Average of all $b \to q\bar q s$} & 
      $0.43 \pm 0.07$ & $-0.021 \pm 0.049$ \\
      \mc{2}{l}{\small Confidence level} & 
      \small $0.17~(1.4\sigma)$ & \small $0.15~(1.4\sigma)$ \\
      \hline 
      \mc{2}{l}{\bf \boldmath Average including $b \to c\bar c s$} &
      $0.665 \pm 0.0033$ & $ 0.018 \pm 0.025$ \\
      \mc{2}{l}{\small Confidence level} & 
      \small $0.006 (2.7\sigma)$ & \small $0.17~(1.4\sigma)$ \\
      \hline
    \end{tabular*}
    \label{tab:cp_uta:qqs}
  \end{center}
\end{table}

If we treat the combined error as a Gaussian quantity,
we note that the average of $- \etacp S_{b \to q\bar q s}$ 
of all $b \to q\bar q s$ dominated modes ($0.43 \pm 0.07$)
is more than $5\sigma$ from zero, 
and hence $\CP$ violation in $b \to q\bar q s$ transitions is established.
Furthermore, the averages of $- \etacp S_{b \to q\bar q s}$ 
for the modes $\etapr\KS$ and $K^+K^-\KS$ are more than $3\sigma$ from zero.

\begin{figure}
  \begin{center}
    \resizebox{0.60\textwidth}{!}{\includegraphics{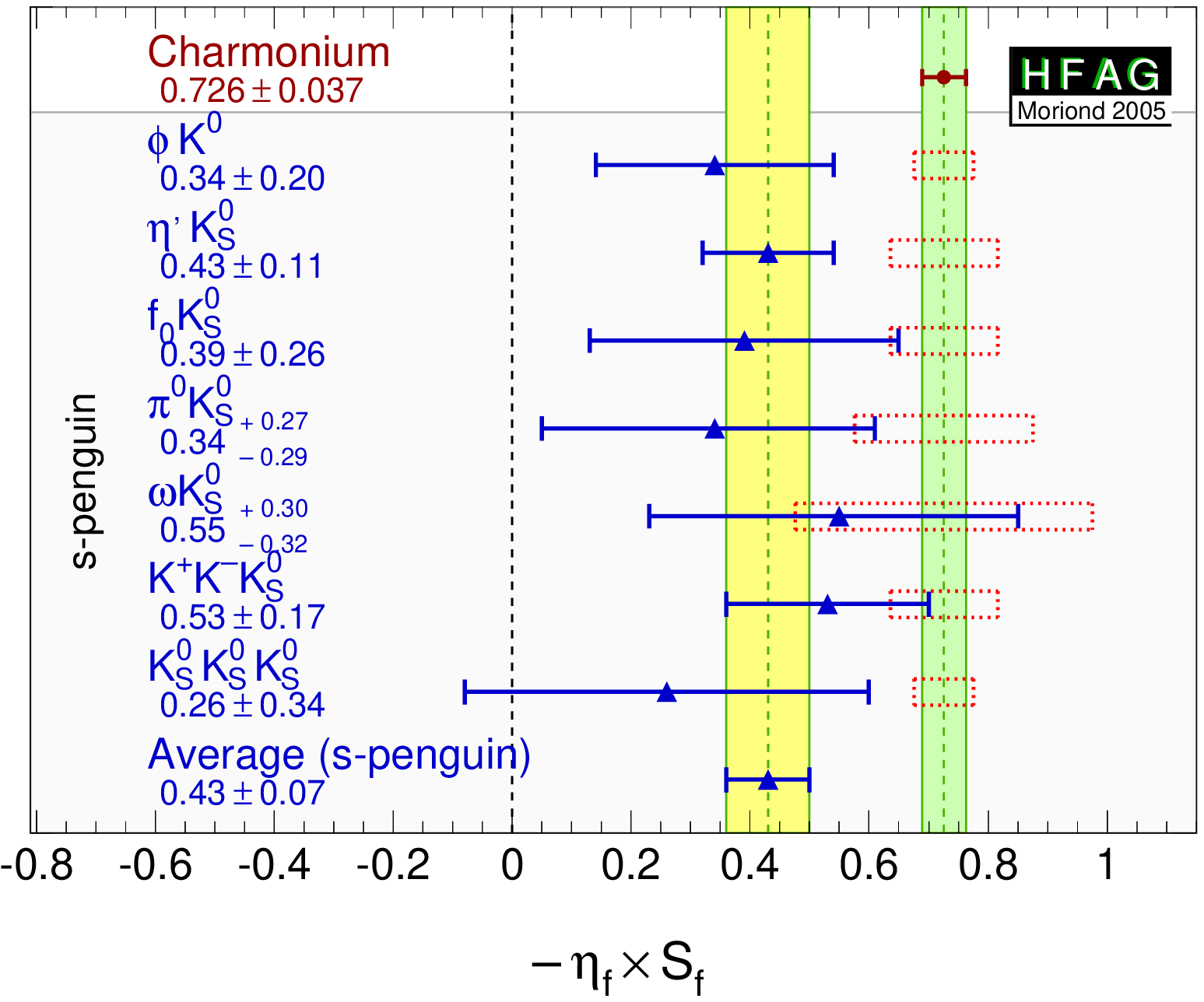}}
  \end{center}
  \vspace{-0.8cm}
  \caption{
    Averages for $S_{b \to q\bar q s}$ compared to $S_{b \to c\bar c s}$,
    with coarse dimensional estimates of associated theoretical uncertainties 
    indicated by the open boxes.
  }
  \label{fig:cp_uta:s_qqs_theo}
\end{figure}

Neglecting theory errors due to suppressed contributions with 
different weak phases,
the difference between $S_{b \to c\bar c s}$ and $S_{b \to q\bar q s}$
can be calculated.
We find the confidence level (CL) of 
a direct comparison of the $S_{b \to c\bar c s}$ and $S_{b \to q\bar q s}$ 
averages to be $0.00021$, which corresponds to a $3.7\sigma$ discrepancy.
To give an idea of the theoretical uncertainties involved,
Fig.~\ref{fig:cp_uta:s_qqs_theo} shows 
coarse estimates of the theoretical errors
associated with the non-charmonium modes.
These crude estimates are obtained from dimensional arguments only, 
based on the CKM suppression of the $V_{ub}$ penguin, 
and on the naive contribution from tree diagrams. Including these 
estimates according to the procedure defined in Ref.~\cite{ref:cp_uta:uud:charles} 
improves the CL of the joint average to about $3\sigma$.

\mysubsection{Time-dependent $\CP$ asymmetries in $b \to c\bar{c}d$ transitions
}
\label{sec:cp_uta:ccd}

The transition $b \to c\bar c d$ can occur via either a $b \to c$ tree
or a $b \to d$ penguin amplitude.  
Similarly to Eq.~(\ref{eq:cp_uta:b_to_s}), the amplitude for 
the $b \to d$ penguin can be written
\begin{equation}
  \label{eq:cp_uta:b_to_d}
  \begin{array}{ccccc}
    A_{b \to d} & = & 
    \mc{3}{l}{F_u V_{ub}V^*_{ud} + F_c V_{cb}V^*_{cd} + F_t V_{tb}V^*_{td}} \\
    & = & (F_u - F_c) V_{ub}V^*_{ud} & + & (F_t - F_c) V_{tb}V^*_{td} \\
    & = & {\cal O}(\lambda^3) & + & {\cal O}(\lambda^3). \\
  \end{array}
\end{equation}
From this it can be seen that the $b \to d$ penguin amplitude 
does not have a dominant weak phase.

In the above, we have followed Eq.~(\ref{eq:cp_uta:b_to_s}) 
by eliminating the $F_c$ term using unitarity.
However, we could equally well write
\begin{equation}
  \label{eq:cp_uta:b_to_d_alt}
  \begin{array}{ccccc}
    A_{b \to d} 
    & = & (F_u - F_t) V_{ub}V^*_{ud} & + & (F_c - F_t) V_{cb}V^*_{cd}, \\
    & = & (F_c - F_u) V_{cb}V^*_{cd} & + & (F_t - F_u) V_{tb}V^*_{td}. \\
  \end{array}
\end{equation}
Since the $b \to c\bar{c}d$ tree amplitude 
has the weak phase of $V_{cb}V^*_{cd}$,
either of the above expressions allow the penguin to be decomposed into 
parts with weak phases the same and different to the tree amplitude
(the relative weak phase can be chosen to be either $\beta$ or $\gamma$).
However, if the tree amplitude dominates,
there is little sensitivity to any phase 
other than that from $\Bz$-$\Bzb$ mixing.

\begin{figure}
  \begin{center}
    \begin{tabular}{cc}
      \resizebox{0.46\textwidth}{!}{\includegraphics{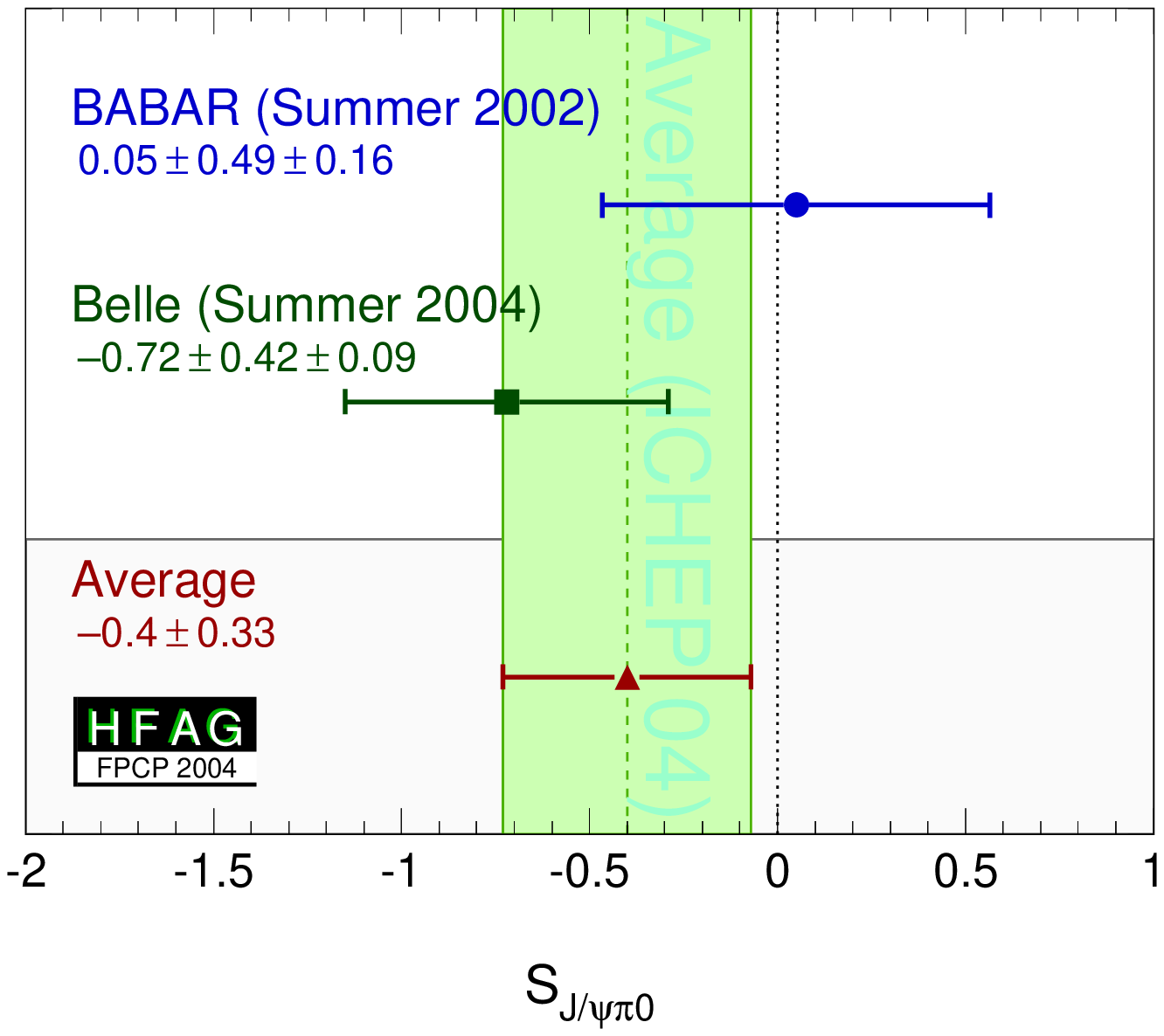}}
      &
      \resizebox{0.46\textwidth}{!}{\includegraphics{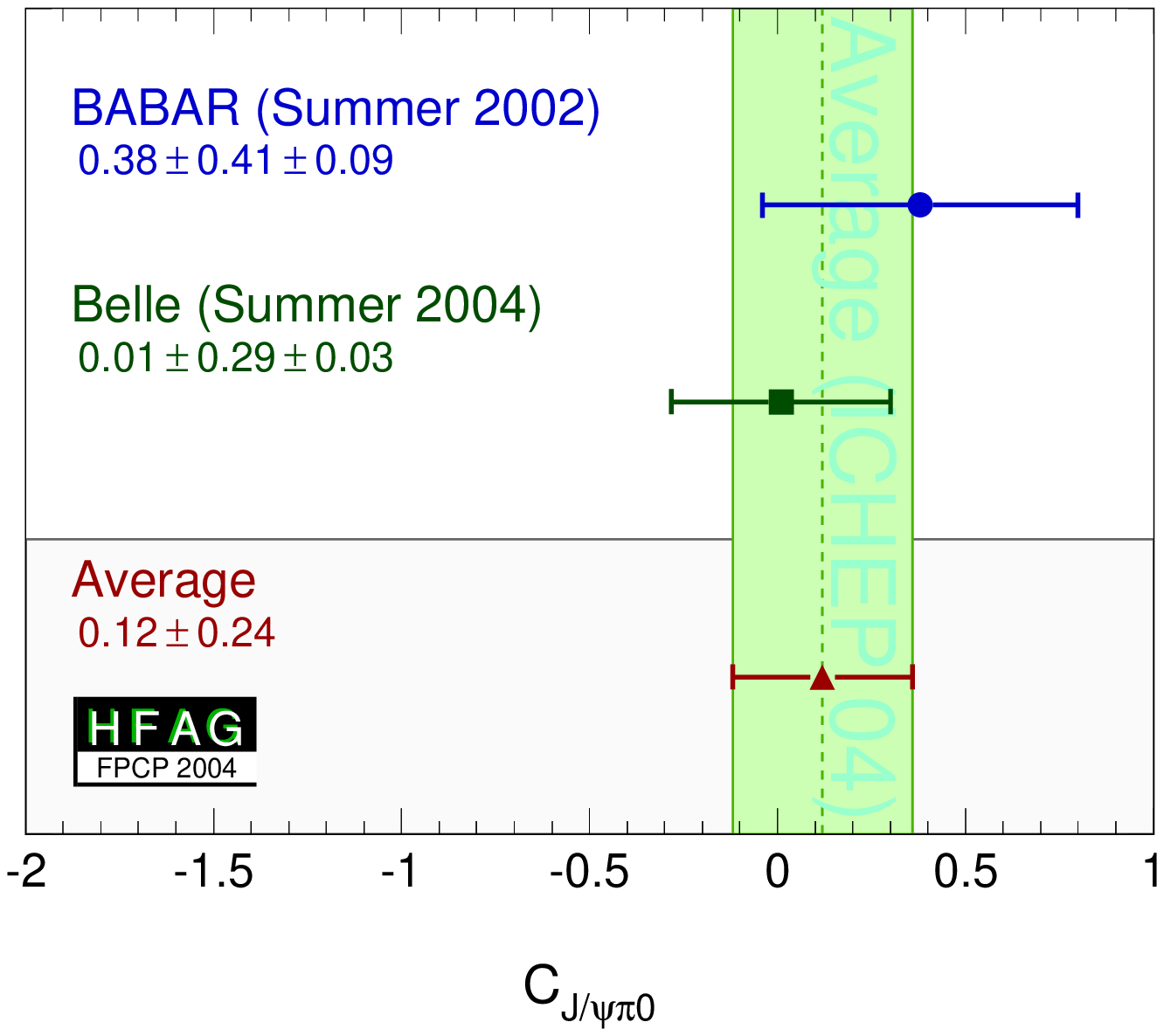}}
    \end{tabular}
  \end{center}
  \vspace{-0.8cm}
  \caption{
    Averages of 
    (left) $S_{b \to c\bar c d}$ and (right) $C_{b \to c\bar c d}$ 
    for the mode $\Bz \to J/ \psi \pi^0$.
  }
  \label{fig:cp_uta:ccd:psipi0}
\end{figure}

The $b \to c\bar{c}d$ transitions can be investigated with studies 
of various different final states. 
Results are available from both \babar\  and \belle\ 
using the final states $\jpsi \pi^0$, $D^{*+}D^{*-}$ and $D^{*\pm}D^{\mp}$;
the averages of these results are given in Table~\ref{tab:cp_uta:ccd}.
The results using the $\CP$ eigenstate ($\etacp = +1$) $\jpsi \pi^0$
are shown in Fig.~\ref{fig:cp_uta:ccd:psipi0}.
The vector-vector mode $D^{*+}D^{*-}$ 
is found to be dominated by the $\CP$ even longitudinally polarized component;
\babar\ measures a $\CP$ odd fraction of 
$0.124 \pm 0.044 \pm 0.007$~\cite{ref:cp_uta:ccd:babar:dstardstar} while
\belle\ measures a $\CP$ odd fraction of 
$0.19  \pm 0.08  \pm 0.01 $~\cite{ref:cp_uta:ccd:belle:dstardstar}
(here we do not average these fractions and rescale the inputs,
however the average is almost independent of the treatment).
We treat the uncertainty due to the error in the $\CP$-odd fractions
(quoted as a third uncertainty) as a correlated systematic error.
For the non-$\CP$ eigenstate mode $D^{*\pm}D^{\mp}$
\babar\ uses fully reconstructed events while 
\belle\ combines both fully and partially reconstructed samples.

\begin{table}
  \begin{center}
    \caption{
      Averages for $b \to c \bar c d$ modes.
    }
    \vspace{0.2cm}
    \setlength{\tabcolsep}{0.0pc}
    \begin{tabular*}{\textwidth}{@{\extracolsep{\fill}}lrcc} \hline 
      \mc{2}{l}{Experiment} & 
      $S_{b \to c\bar c d}$ & $C_{b \to c\bar c d}$ \\
      \hline
      \mc{4}{c}{$\jpsi \pi^0$} \\
      \babar & \cite{ref:cp_uta:ccd:babar:psipi0} & 
      $\ph{-}0.05 \pm 0.49 \pm 0.16$ & $ 0.38 \pm 0.41 \pm 0.09$ \\
      \belle & \cite{ref:cp_uta:ccd:belle:psipi0} & 
      $-0.72 \pm 0.42 \pm 0.09$ & $ 0.01 \pm 0.29 \pm 0.03$ \\
      \mc{2}{l}{\bf Average} & 
      $-0.40 \pm 0.33$ & $ 0.12 \pm 0.24$ \\
      \mc{2}{l}{\small Confidence level} & 
      \mc{2}{c}{\small combined average: $0.36$} \\
      \hline
      \mc{4}{c}{$D^{*+}D^{*-}$} \\
      \babar & \cite{ref:cp_uta:ccd:babar:dstardstar} &
      $-0.65 \pm 0.26 \pm 0.04 ^{+0.09}_{-0.07}$ & $0.04 \pm 0.14 \pm 0.02$ \\
      \belle & \cite{ref:cp_uta:ccd:belle:dstardstar} &
      $-0.75 \pm 0.56 \pm 0.10 \pm 0.06$ & $ 0.26 \pm 0.26 \pm 0.05 \pm 0.01$ \\
      \mc{2}{l}{\bf Average} &
      $-0.67 \pm 0.25$ & $ 0.09 \pm 0.12$ \\
      \mc{2}{l}{\small Confidence level} & 
      \small $0.89$ & \small $0.47$ \\
      \hline
    \end{tabular*}

    \vspace{2ex}

    \resizebox{\textwidth}{!}{
      \setlength{\tabcolsep}{0.0pc}
      \begin{tabular}{@{\extracolsep{2mm}}lrccccc} \hline 
        \mc{2}{l}{Experiment} & 
        $S_{+-}$ & $C_{+-}$ & $S_{-+}$ & $C_{-+}$ & $A$ \\
        \hline
        \mc{7}{c}{$D^{*\pm}D^{\mp}$} \\        
        \babar & \cite{ref:cp_uta:ccd:babar:dstard} &
        $-0.82 \pm 0.75 \pm 0.14$ & $-0.47 \pm 0.40 \pm 0.12$ & 
        $-0.24 \pm 0.69 \pm 0.12$ & $-0.22 \pm 0.37 \pm 0.10$ & $-0.03 \pm 0.11 \pm 0.05$ \\
        \belle & \cite{ref:cp_uta:ccd:belle:dstard} &
        $-0.55 \pm 0.39 \pm 0.12$ & $-0.37 \pm 0.22 \pm 0.06$ & 
        $-0.96 \pm 0.43 \pm 0.12$ & $\ph{-}0.23 \pm 0.25 \pm 0.06$ & $\ph{-}0.07 \pm 0.08 \pm 0.04$ \\
        \mc{2}{l}{\bf Average} & 
        $-0.61 \pm 0.36$ & $-0.39 \pm 0.20$ & 
        $-0.75 \pm 0.38$ & $ 0.09 \pm 0.21$ & $ 0.03 \pm 0.07$ \\
        \hline 
      \end{tabular}
    }

    \label{tab:cp_uta:ccd}
  \end{center}
\end{table}

In the absence of the penguin contribution,
the time-dependent parameters would be given by
$S_{b \to c\bar c d} = - \etacp \sin(2\beta)$,
$C_{b \to c\bar c d} = 0$,
$S_{+-} = \sin(2\beta + \delta)$,
$S_{-+} = \sin(2\beta - \delta)$,
$C_{+-} = - C_{-+}$ and 
$A_{+-} = 0$,
where $\delta$ is the strong phase difference between the 
$D^{*+}D^-$ and $D^{*-}D^+$ decay amplitudes.
In the presence of the penguin contribution,
there is no clean interpretation in terms of CKM parameters,
however
direct $\CP$ violation may be observed as any of
$C_{b \to c\bar c d} \neq 0$, $C_{+-} \neq - C_{-+}$ or $A_{+-} \neq 0$.

\begin{figure}
  \begin{center}
    \begin{tabular}{cc}
      \resizebox{0.46\textwidth}{!}{\includegraphics{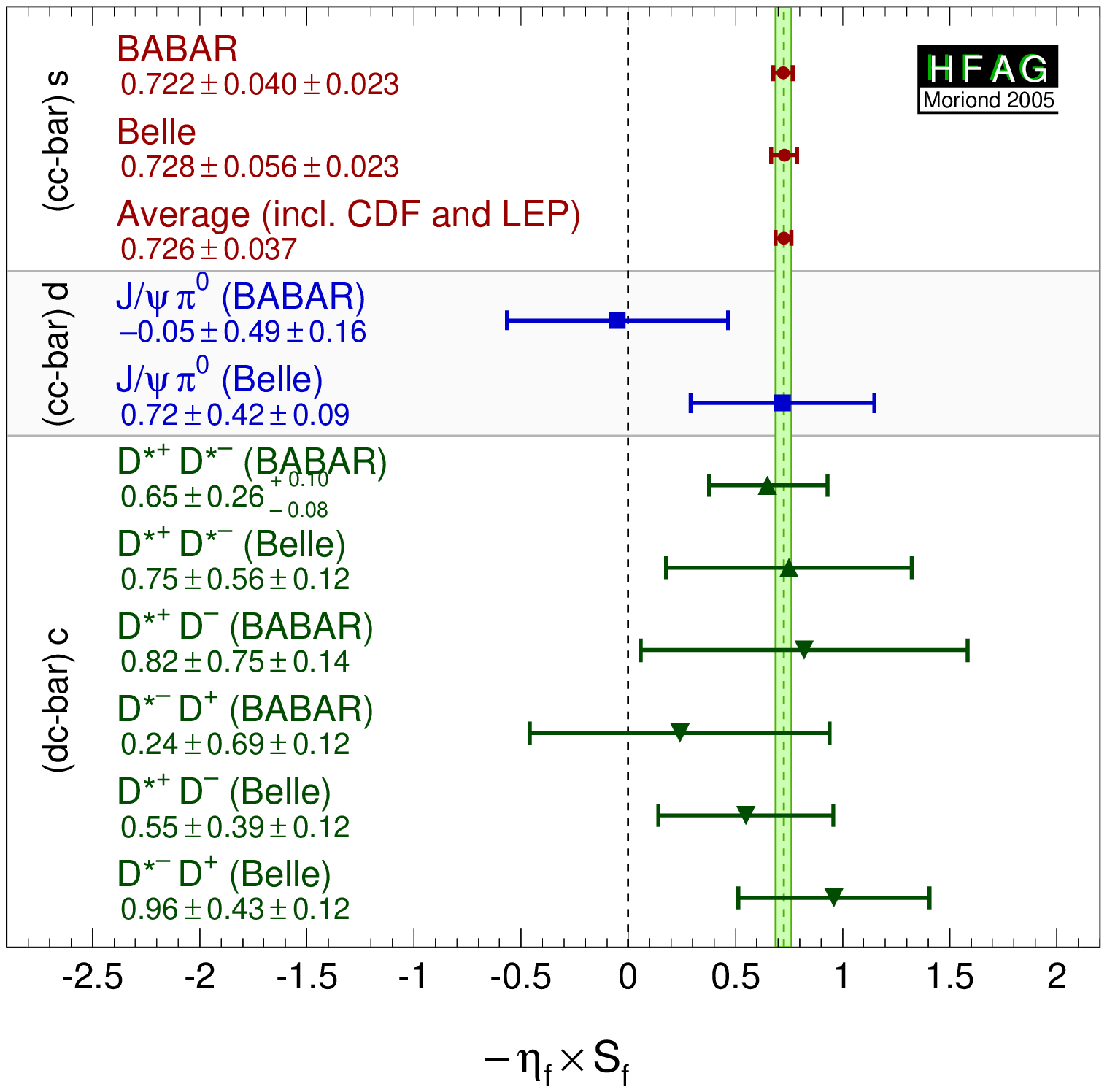}}
      &
      \resizebox{0.46\textwidth}{!}{\includegraphics{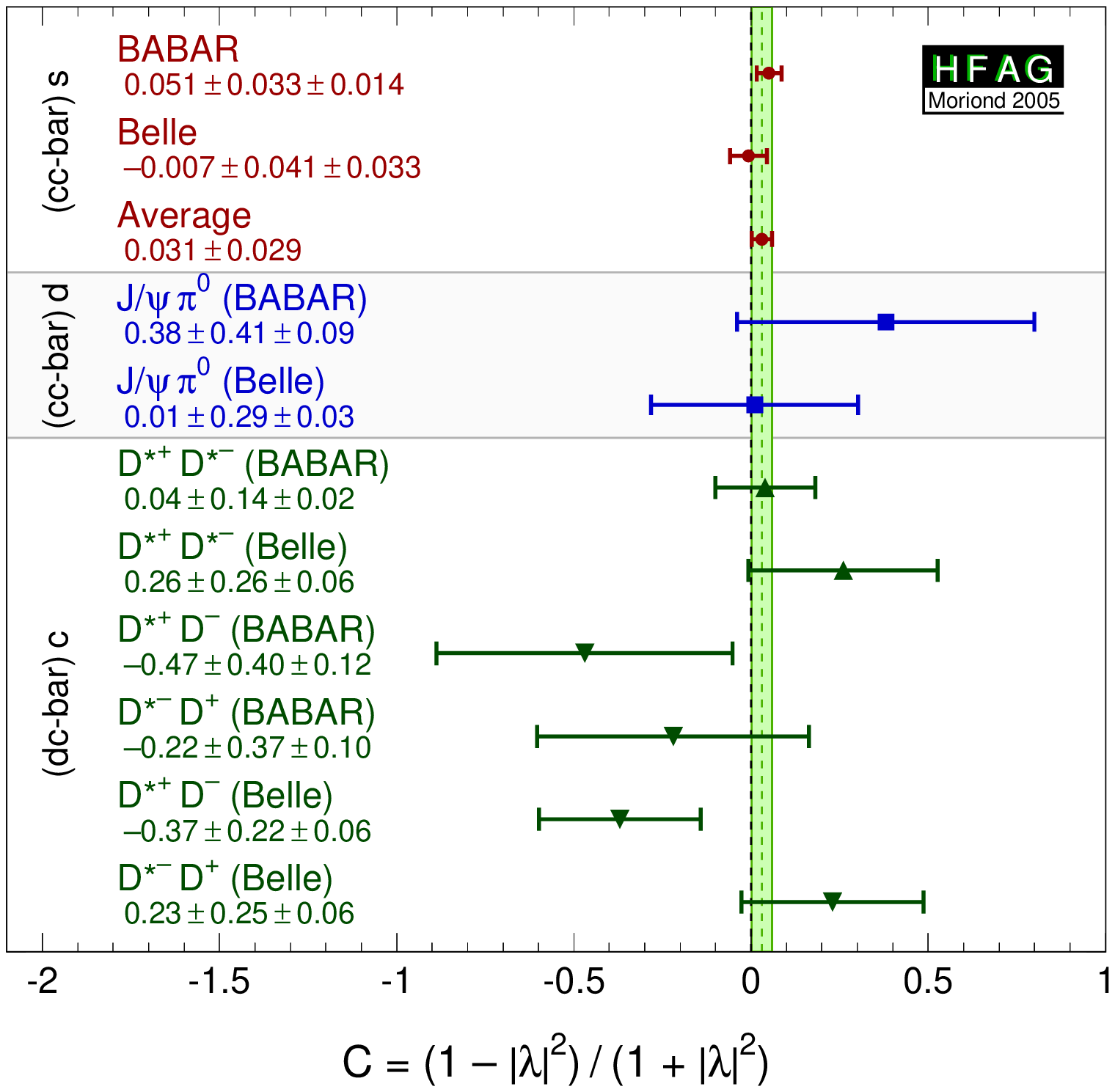}}
    \end{tabular}
  \end{center}
  \vspace{-0.8cm}
  \caption{
    Averages of 
    (left) $S_{b \to c\bar c d}$ and (right) $C_{b \to c\bar c d}$,
    compared to $S_{b \to c\bar c s}$ and $C_{b \to c\bar c s}$, respectively.
  }
  \label{fig:cp_uta:ccd}
\end{figure}

The averages for the $b \to c\bar c d$ modes 
are shown in Fig.~\ref{fig:cp_uta:ccd}.
Comparisons of the results for the $b \to c\bar c d$ modes 
to the $b \to c\bar c s$ and $b \to q\bar q s$ modes,
can be seen in Fig.~\ref{fig:cp_uta:qqs_ccd}.

\mysubsection{Time-dependent asymmetries in $b \to s\gamma$ transitions
}
\label{sec:cp_uta:bsg}

The radiative decays $b \to s\gamma$ produce photons 
which are highly polarized in the Standard Model.
The decays $\Bz \to F \gamma$ and $\Bzb \to F \gamma$ 
produce photons with opposite helicities, 
and since the polarization is, in principle, observable,
these final states cannot interfere.
The finite mass of the $s$ quark introduces small corrections
to the limit of maximum polarization,
but any large mixing induced $\CP$ violation would be a signal for new physics.
Since a single weak phase dominates the $b \to s \gamma$ transition in the 
Standard Model, the cosine term is also expected to be small.

Atwood {\it et al.}~\cite{ref:cp_uta:bsg:aghs} have shown that 
an inclusive analysis with respect to $\KS\pi^0\gamma$ can be performed,
since the properties of the decay amplitudes 
are independent of the angular momentum of the $\KS\pi^0$ system. 
However, if non-dipole operators contribute significantly to the amplitudes, 
then the Standard Model mixing-induced $\CP$ violation could be larger 
than the naive expectation $S ~ -2 (m_s/m_b) \sin \left(2\beta\right)$,
and the $\CP$ parameters may vary over the $\KS\pi^0\gamma$ Dalitz plot, 
for example as a function of the $\KS\pi^0$ invariant mass.

With the above in mind, 
we quote two averages: one for $K^*(892)$ candidates only, 
and the other one for the inclusive $\KS\pi^0\gamma$ decay (including the $K^*(892)$).
If the Standard Model dipole operator is dominant, 
both should give the same quantities 
(the latter naturally with smaller statistical error). 
If not, care needs to be taken in interpretation of the inclusive parameters, 
while the results on the $K^*(892)$ resonance remain relatively clean.
Results from \babar\ and \belle\ are used for both averages;
both experiments use the invariant mass range 
$0.60 \ {\rm GeV}/c^2 < M_{\KS\pi^0} < 1.80 \ {\rm GeV}/c^2$
in the inclusive analysis.

\begin{table}
  \begin{center}
    \caption{
      Averages for $b \to s \gamma$ modes.
    }
    \vspace{0.2cm}
    \setlength{\tabcolsep}{0.0pc}
    \begin{tabular*}{\textwidth}{@{\extracolsep{\fill}}lrcc} \hline 
      \mc{2}{l}{Experiment} & 
      $S_{b \to s \gamma}$ & $C_{b \to s \gamma}$ \\
      \hline
      \mc{4}{c}{$\Kstar(892)\gamma$} \\
      \babar & \cite{ref:cp_uta:bsg:babar:kspi0gamma} & 
      $-0.21 \pm 0.40 \pm 0.05$ & $-0.40 \pm 0.23 \pm 0.04$ \\
      \belle & \cite{ref:cp_uta:bsg:belle:kspi0gamma} & 
      $-0.79 \, ^{+0.63}_{-0.50} \pm 0.10$ & $0.00 \, ^{+0.37}_{-0.38} \pm 0.11$ \\
      \mc{2}{l}{\bf Average} & 
      $-0.38 \pm 0.34$ & $-0.30 \pm 0.20$ \\
      \mc{2}{l}{\small Confidence level} & 
      \mc{2}{c}{\small combined average: $0.50$} \\
      \hline       
      \mc{4}{c}{$\KS \pi^0 \gamma$ (including $\Kstar(892)\gamma$)} \\
      \babar & \cite{ref:cp_uta:bsg:babar:kspi0gamma} & 
      $-0.06 \pm 0.37$ & $-0.48 \pm 0.22$ \\
      \belle & \cite{ref:cp_uta:bsg:belle:kspi0gamma} & 
      $-0.58 \, ^{+0.46}_{-0.38} \pm 0.11$ & $-0.03 \pm 0.34 \pm 0.11$ \\
      \mc{2}{l}{\bf Average} & 
      $-0.26 \pm 0.29$ & $-0.36 \pm 0.19$ \\
      \mc{2}{l}{\small Confidence level} & 
      \mc{2}{c}{\small combined average: $0.38$} \\
      \hline 
    \end{tabular*}
      
    \label{tab:cp_uta:bsg}
  \end{center}
\end{table}

\mysubsection{Time-dependent $\CP$ asymmetries in $b \to u\bar{u}d$ transitions
}
\label{sec:cp_uta:uud}

The $b \to u \bar u d$ transition can be mediated by either 
a $b \to u$ tree amplitude or a $b \to d$ penguin amplitude.
These transitions can be investigated using 
the time dependence of $\Bz$ decays to final states containing light mesons.
Results are available from both \babar\ and \belle\ for the 
$\CP$ eigenstate ($\etacp = +1$) $\pi^+\pi^-$ final state.
\babar\  has also performed an analysis on 
the vector-vector final state $\rho^+\rho^-$, 
which they find to be dominated by the $\CP$ even
longitudinally polarized component
(they measure $f_{\rm long} = 
0.978 \pm 0.014 \, ^{+0.021}_{-0.029}$~\cite{ref:cp_uta:uud:babar:rhorho}).

\begin{figure}
  \begin{center}
    \begin{tabular}{cc}
      \resizebox{0.46\textwidth}{!}{\includegraphics{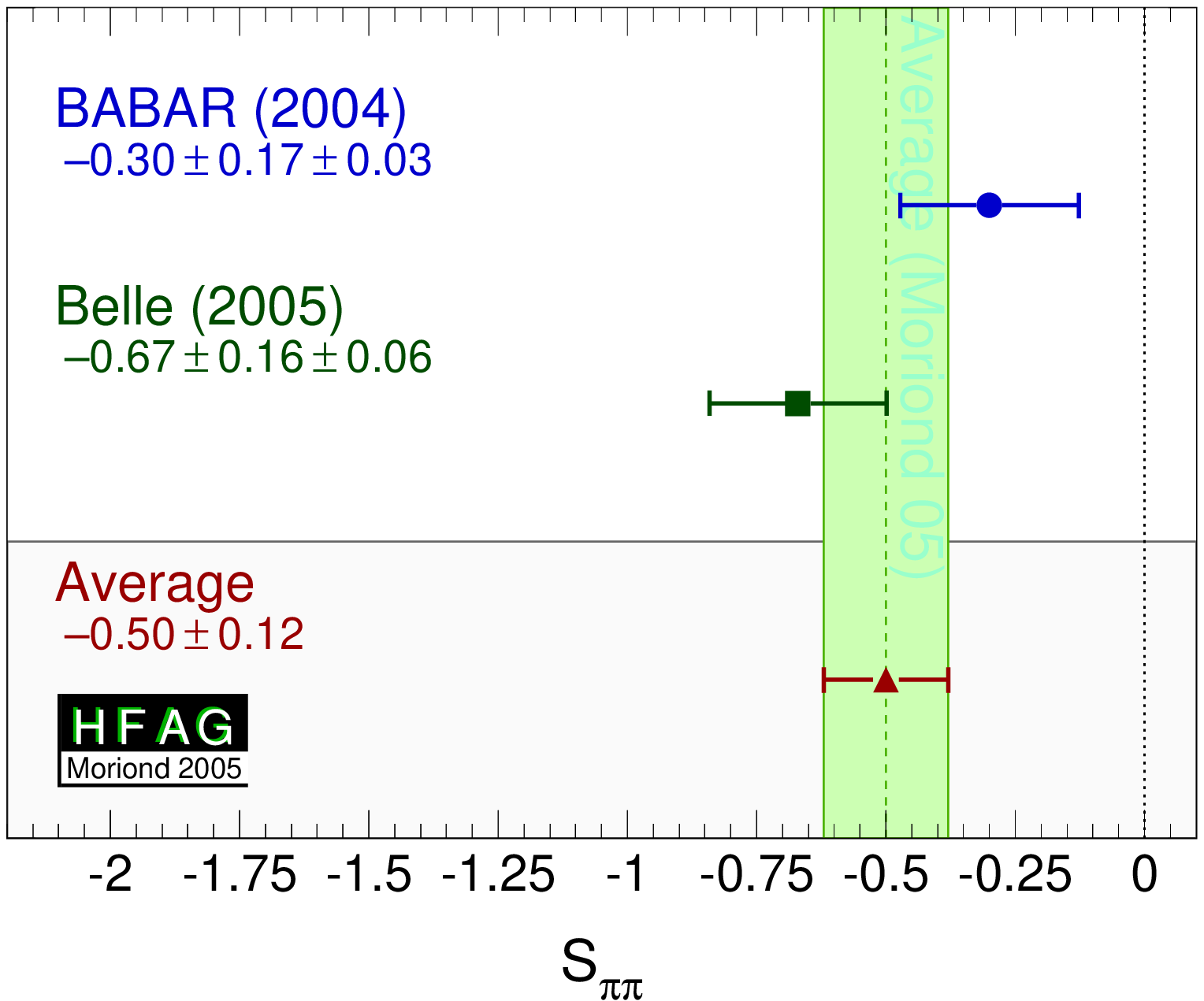}}
      &
      \resizebox{0.46\textwidth}{!}{\includegraphics{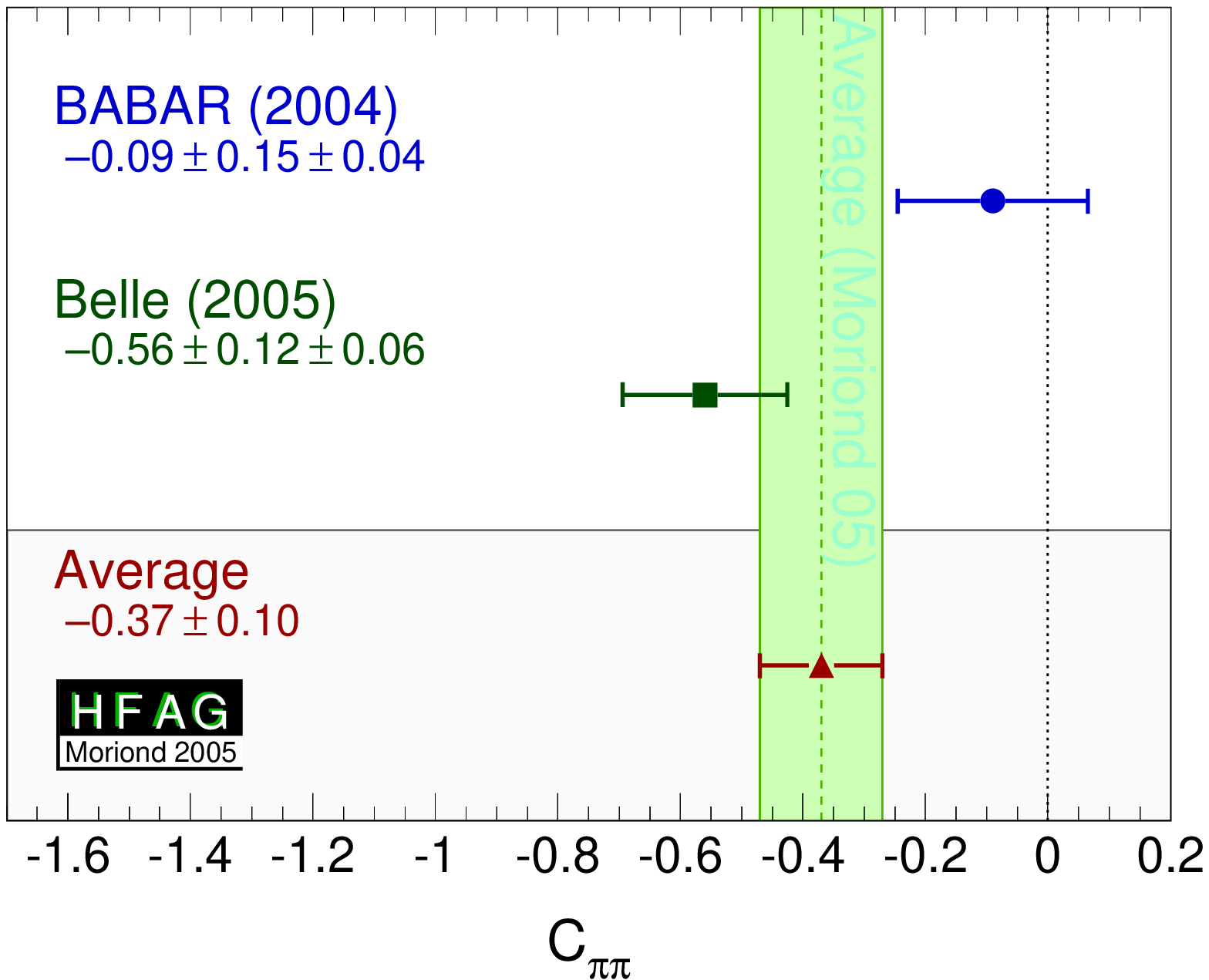}}
    \end{tabular}
  \end{center}
  \vspace{-0.8cm}
  \caption{
    Averages of 
    (left) $S_{b \to u\bar u d}$ and (right) $C_{b \to u\bar u d}$ 
    for the mode $\Bz \to \pi^+\pi^-$.
  }
  \label{fig:cp_uta:uud:pipi}
\end{figure}

For the non-$\CP$ eigenstate $\rho^{\pm}\pi^{\mp}$, 
\belle\ has performed a quasi-two-body analysis,
while  \babar\ performs a time-dependent Dalitz plot (DP) analysis
of the $\pi^+\pi^-\pi^0$ final state~\cite{ref:cp_uta:uud:snyderquinn};
such an analysis allows direct measurements of the phases.
These results, and averages, are listed in Table~\ref{tab:cp_uta:uud}.
The averages for $\pi^+\pi^-$ are shown in Fig.~\ref{fig:cp_uta:uud:pipi}.

\begin{table}
  \begin{center}
    \caption{
      Averages for $b \to u \bar u d$ modes.
    }
    \vspace{0.2cm}
    \setlength{\tabcolsep}{0.0pc}
    \begin{tabular*}{\textwidth}{@{\extracolsep{\fill}}lrcc} \hline 
      \mc{2}{l}{Experiment} & 
      $S_{b \to u\bar u d}$ & $C_{b \to u\bar u d}$ \\
      \hline
      & \mc{3}{c}{$\pi^+\pi^-$} \\
      \babar & ~\cite{ref:cp_uta:uud:babar:pipi} & 
      $-0.30 \pm 0.17 \pm 0.03$ & $-0.09 \pm 0.15 \pm 0.04$ \\
      \belle & ~\cite{ref:cp_uta:uud:belle:pipi} & 
      $-0.67 \pm 0.16 \pm 0.06$ & $-0.56 \pm 0.12 \pm 0.06$ \\
      \mc{2}{l}{\bf Average} & 
      $-0.50 \pm 0.12$ & $-0.37 \pm 0.10$ \\
      \mc{2}{l}{\small Confidence level} & 
      \mc{2}{c}{\small combined average: $0.019~(2.3\sigma)$} \\
      \hline
      & \mc{3}{c}{$\rho^+\rho^-$} \\
      \babar & ~\cite{ref:cp_uta:uud:babar:rhorho} &
      $-0.33 \pm 0.24 \, ^{+0.08}_{-0.14}$ & $-0.03 \pm 0.18 \pm 0.09$ \\
      \hline 
    \end{tabular*}

    \vspace{2ex}

    \resizebox{\textwidth}{!}{
      \setlength{\tabcolsep}{0.0pc}
      \begin{tabular}{@{\extracolsep{2mm}}lrccccc} \hline 
        & \mc{6}{c}{$\rho^{\pm}\pi^{\mp}$ Q2B/DP analysis} \\
        \mc{2}{l}{Experiment} & 
        $S_{\rho\pi}$ & $C_{\rho\pi}$ & $\Delta S_{\rho\pi}$ & $\Delta C_{\rho\pi}$ & ${\cal A}_{CP}^{\rho\pi}$ \\
        \hline
        \babar & ~\cite{ref:cp_uta:uud:babar:rhopi} &
        $-0.10 \pm 0.14 \pm 0.04$ & $ 0.34 \pm 0.11 \pm 0.05$ &
        $\ph{-}0.22 \pm 0.15 \pm 0.03$ & $ 0.15 \pm 0.11 \pm 0.03$ & $-0.088 \pm 0.049 \pm 0.013$ \\
        \belle & ~\cite{ref:cp_uta:uud:belle:rhopi} & 
        $-0.28 \pm 0.23 \, ^{+0.10}_{-0.08}$ & $ 0.25 \pm 0.17 \, ^{+0.02}_{-0.06}$ &
        $-0.30 \pm 0.24 \pm 0.09$ & $ 0.38 \pm 0.18 \, ^{+0.02}_{-0.04}$ & $-0.16 \pm 0.10 \pm 0.02$ \\
        \mc{2}{l}{\bf Average} & 
        $-0.13 \pm 0.13$ & $ 0.31 \pm 0.10$ &
        $ 0.09 \pm 0.13$ & $ 0.22 \pm 0.10$ & $-0.102 \pm 0.045$ \\
        \hline
        & & & \mc{2}{c}{${\cal A}^{+-}_{\rho\pi}$} & \mc{2}{c}{${\cal A}^{-+}_{\rho\pi}$} \\
        \hline
        \babar & \cite{ref:cp_uta:uud:babar:rhopi} &
        & \mc{2}{c}{$\ph{-}0.25 \pm 0.17 \, ^{+0.02}_{-0.06}$} & \mc{2}{c}{$-0.47 ^{\,+0.14}_{\,-0.15} \pm 0.06$} \\         
        \belle & \cite{ref:cp_uta:uud:belle:rhopi} & 
        & \mc{2}{c}{$-0.02 \pm 0.16^{\,+0.05}_{\,-0.02}$} & \mc{2}{c}{$-0.53 \pm 0.29^{\,+0.09}_{\,-0.04}$} \\
        \mc{2}{l}{\bf Average} & 
        & \mc{2}{c}{$-0.15 \pm 0.09$} & \mc{2}{c}{$-0.47^{\,+0.13}_{\,-0.14}$} \\
        \hline 
      \end{tabular}
    }

    \vspace{2ex}

    \setlength{\tabcolsep}{0.0pc}
    \begin{tabular*}{\textwidth}{@{\extracolsep{\fill}}lrcc} \hline 
      \mc{4}{c}{$\rho^{\pm}\pi^{\mp}$ DP analysis} \\
      \mc{2}{l}{Experiment} & $\alpha \ (^\circ)$ & $\delta_{+-} \ (^\circ)$ \\
      \hline
      \babar & \cite{ref:cp_uta:uud:babar:rhopi} &
      $113 \, ^{+27}_{-17} \pm 6$ & $-67 \, ^{+28}_{-31} \pm 7$ \\ 
      \hline 
    \end{tabular*}
      
    \label{tab:cp_uta:uud}
  \end{center}
\end{table}

\begin{figure}
  \begin{center}
    \begin{tabular}{cc}
      \resizebox{0.46\textwidth}{!}{\includegraphics{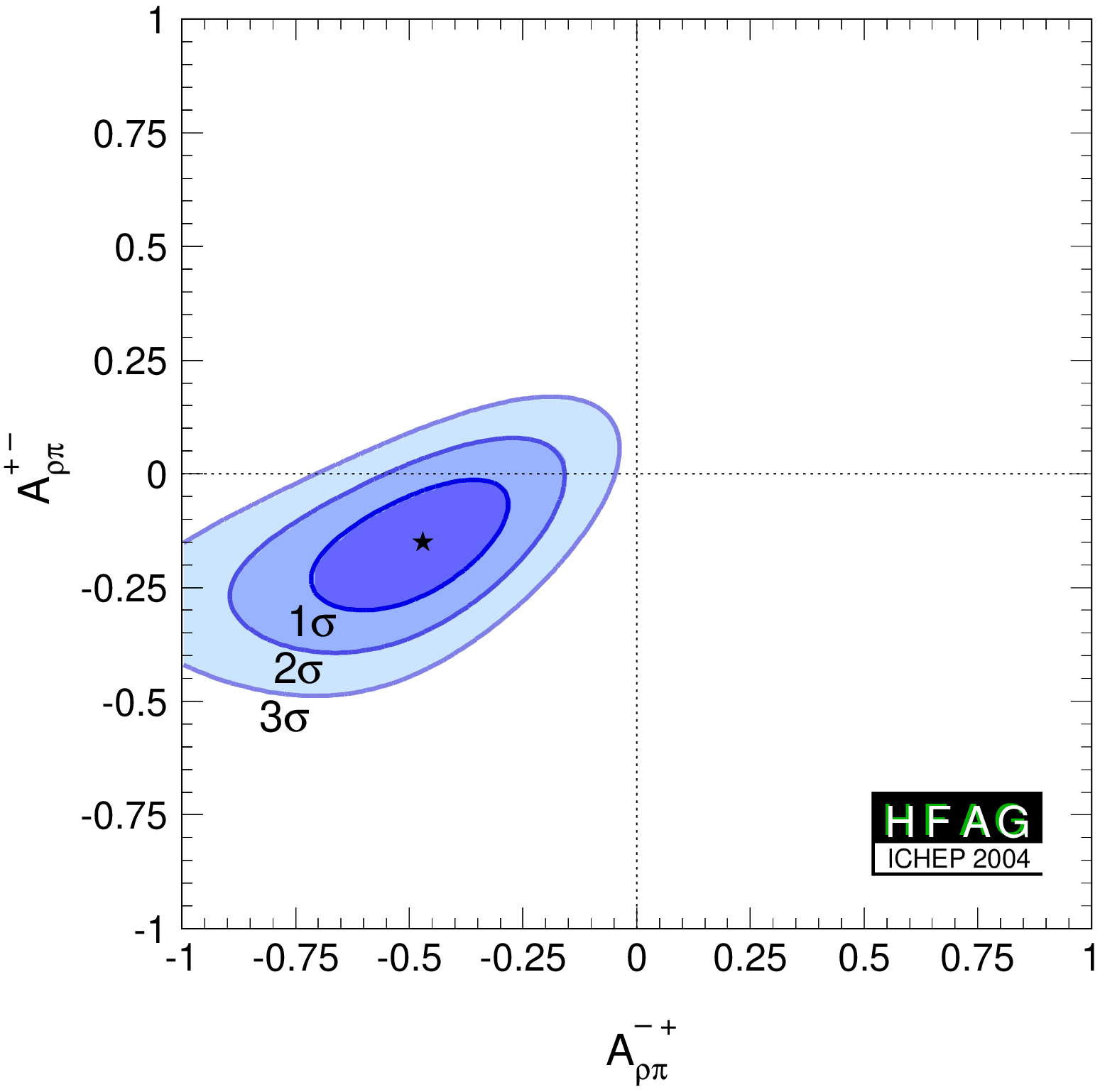}}
    \end{tabular}
  \end{center}
  \vspace{-0.8cm}
  \caption{
    	Direct $\CP$ violation in $\Bz\to\rho^\pm\pi^\mp$. The no-\CP violation
	hypothesis is excluded at the $3.4\sigma$ level.
  }
  \label{fig:cp_uta:uud:rhopi-dircp}
\end{figure}

If the penguin contribution is negligible, 
the time-dependent parameters for $\Bz \to \pi^+\pi^-$ and $\Bz \to \rho^+\rho^-$ 
are given by
$S_{b \to u\bar u d} = \etacp \sin(2\alpha)$ and
$C_{b \to u\bar u d} = 0$.
With the notation described in Sec.~\ref{sec:cp_uta:notations}
(Eq.~(\ref{eq:cp_uta:non-cp-s_and_deltas})), 
the time-dependent parameters for $\Bz \to \rho^\pm\pi^\mp$ are,
neglecting penguin contributions, given by
$S_{\rho\pi}   = \sqrt{1 - (\frac{\Delta C}{2})^2}\sin(2\alpha)\cos(\delta)$,
$\Delta S_{\rho\pi} = \sqrt{1 - (\frac{\Delta C}{2})^2}\cos(2\alpha)\sin(\delta)$ and
$C_{\rho\pi} = {\cal A}_{\CP}^{\rho\pi} = 0$,
where $\delta=\arg(A_{-+}A^*_{+-})$ is the strong phase difference 
between the $\rho^-\pi^+$ and $\rho^+\pi^-$ decay amplitudes.
In the presence of the penguin contribution, there is no straightforward 
interpretation of $\Bz \to \rho^\pm\pi^\mp$ in terms of CKM parameters.
However direct $\CP$ violation may arise,
resulting in either or both of $C_{\rho\pi} \neq 0$ and ${\cal A}_{\CP}^{\rho\pi} \neq 0$.
Equivalently,
direct $\CP$ violation may be seen by either of
the decay-type-specific observables ${\cal A}^{+-}_{\rho\pi}$ 
and ${\cal A}^{-+}_{\rho\pi}$, defined in Eq.~(\ref{eq:cp_uta:non-cp-directcp}), 
deviating from zero.
Results and averages for these parameters
are also given in Table~\ref{tab:cp_uta:uud}.
They exhibit a linear correlation coefficient of $+0.59$.
The significance of observing direct $\CP$ violation 
computed from the difference of the $\chi^2$ obtained in the nominal average, 
compared to setting 
$C_{\rho\pi} = {\cal A}^{\rho\pi}_{\CP} = 0$
is found to be $3.4\sigma$ in this mode. 
The confidence level
contours of ${\cal A}^{+-}_{\rho\pi}$ versus ${\cal A}^{-+}_{\rho\pi}$ 
are shown in Fig.~\ref{fig:cp_uta:uud:rhopi-dircp}.


Some difference is seen between the 
\babar\ and \belle\ measurements in the $\pi^+\pi^-$ system.
The confidence level of the average is $0.019$,
which corresponds to a $2.3\sigma$ discrepancy.  Since there is no
evidence of systematic problems in either analysis,
we do not rescale the errors of the averages.

The precision of the measured $\CP$ violation parameters in
$b \to u\bar{u}d$ transitions allows constraints to be set on the UT angle $\alpha$. 
In addition to the value of $\alpha$ from the \babar\ time-dependent DP analysis,
given in Table~\ref{tab:cp_uta:uud},
constraints have been obtained with various methods:
\begin{itemize}
\item 
  Both \babar~\cite{ref:cp_uta:uud:babar:pipi}
  and  \belle~\cite{ref:cp_uta:uud:belle:pipi} have performed 
  isospin analyses in the $\pipi$ system.
  \babar\ exclude $29^\circ < \alpha < 61^\circ$ at the $90\%$ C.L. while
  \belle\ exclude $19^\circ < \alpha < 71^\circ$ at the $95.4\%$ C.L.
  In both cases, only solutions in $0^\circ$\textemdash$180^\circ$ are considered.
\item Using the measured time-dependent \CP violation parameters 
  in longitudinally polarized $\Bz \to \rho^+\rho^-$ 
  decays~\cite{ref:cp_uta:uud:babar:rhorho},
  in combination with the upper limit for the 
  $\Bz \to \rho^0\rho^0$ branching fraction~\cite{ref:cp_uta:uud:babar:rho0rho0},
  and the measurement of the branching fraction and longitudinal polarization
  of $\Bp \to \rho^+\rho^0$~\cite{ref:cp_uta:uud:babar:rho0rho+,ref:cp_uta:uud:belle:rho0rho+},
  \babar\ performs an isospin analysis~\cite{ref:cp_uta:uud:gronaulondon}
  and obtains $\alpha = (100 \pm 13)^\circ$, using the solution closest to the 
  CKM combined fit average.
  The $90\%$ C.L. allowed region for that solution is $79^\circ < \alpha < 123^\circ$.
\item The CKMfitter group~\cite{ref:cp_uta:ckmfitter} uses the 
  measurements from \belle\ and \babar\ given in Table~\ref{tab:cp_uta:uud},
  with other branching fractions and \CP asymmetries in 
  $\B\to\pi\pi,~\rho\pi$ and $\rho\rho$ modes, 
  to perform isospin analyses for each system.  They then combine the results
  to obtain $\alpha = (97.9 \, ^{+5.0}_{-6.4})^\circ$.
  Similarly, the UTFit group~\cite{ref:cp_uta:utfit} 
  obtain $\alpha = (94.9 \pm 6.6)^\circ$.
\end{itemize}
Note that each method suffers from ambiguities in the solutions.
All the above measurements correspond to the choice
that is in agreement with the global CKM fit.

At present we make no attempt to provide an HFAG average for $\alpha$.
More details on procedures to calculate a best fit value for $\alpha$ 
can be found in Refs.~\cite{ref:cp_uta:ckmfitter,ref:cp_uta:utfit}.


\mysubsection{Time-dependent $\CP$ asymmetries in $b \to c\bar{u}d / u\bar{c}d$ transitions
}
\label{sec:cp_uta:cud}

Non-$\CP$ eigenstates such as $D^\pm\pi^\mp$, $D^{*\pm}\pi^\mp$ and $D^\pm\rho^\mp$ can be produced 
in decays of $\Bz$ mesons either via Cabibbo favoured ($b \to c$) or
doubly Cabibbo suppressed ($b \to u$) tree amplitudes. 
Since no penguin contribution is possible,
these modes are theoretically clean.
The ratio of the magnitudes of the suppressed and favoured amplitudes, $R$,
is sufficiently small (predicted to be about $0.02$),
that terms of ${\cal O}(R^2)$ can be neglected, 
and the sine terms give sensitivity to the combination of UT angles $2\beta+\gamma$.

\begin{figure}
  \begin{center}
    \resizebox{0.60\textwidth}{!}{\includegraphics{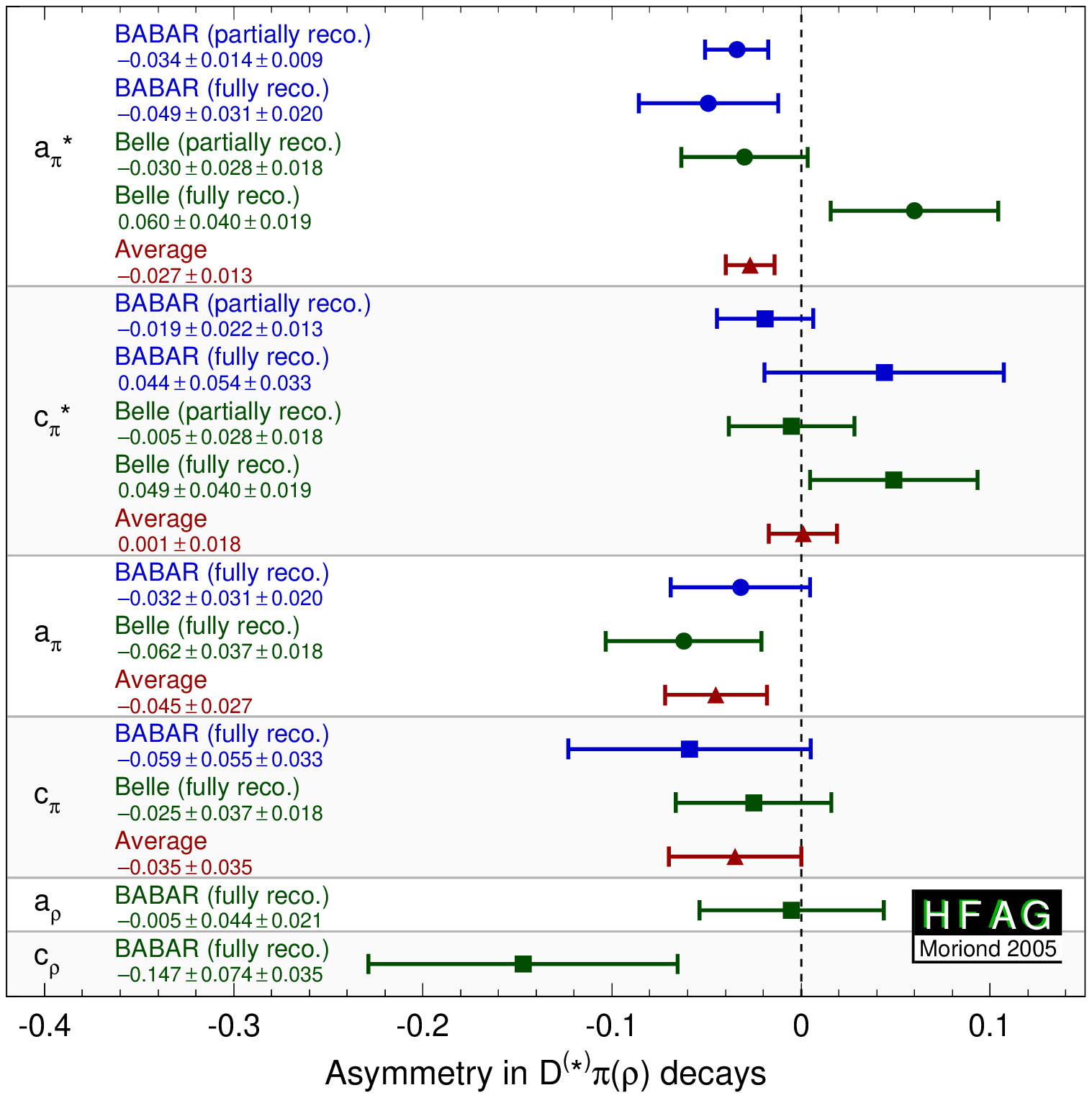}}
  \end{center}
  \vspace{-0.8cm}
  \caption{
    Averages for $b \to c\bar{u}d / u\bar{c}d$ modes.
  }
  \label{fig:cp_uta:cud}
\end{figure}

As described in Sec.~\ref{sec:cp_uta:notations:non_cp:dstarpi},
the averages are given in terms of parameters $a$ and $c$.
$\CP$ violation would appear as $a \neq 0$.
Results are available from both \babar\ and \belle\ in the modes
$D^\pm\pi^\mp$ and $D^{*\pm}\pi^\mp$; for the latter mode both experiments 
have used both full and partial reconstruction techniques.
Results are also available from \babar\ using $D^\pm\rho^\mp$.
These results, and their averages, are listed in Table~\ref{tab:cp_uta:cud},
and are shown in Fig.~\ref{fig:cp_uta:cud}.

\begin{table}
  \begin{center}
    \caption{
      Averages for $b \to c\bar{u}d / u\bar{c}d$ modes.
    }
    \vspace{0.2cm}
    \setlength{\tabcolsep}{0.0pc}
    \begin{tabular*}{\textwidth}{@{\extracolsep{\fill}}lrcc} \hline 
      \mc{2}{l}{Experiment} & $a$ & $c$ \\
      \hline
      \mc{4}{c}{$D^{*\pm}\pi^{\mp}$} \\
      \babar (full rec.) & \cite{ref:cp_uta:cud:babar:full} &
      $-0.049 \pm 0.031 \pm 0.020$ & $\ph{-}0.044 \pm 0.054 \pm 0.033$ \\
      \belle (full rec.) & \cite{ref:cp_uta:cud:belle:full} &
      $\ph{-}0.060 \pm 0.040 \pm 0.019$ & $\ph{-}0.049 \pm 0.040 \pm 0.019$ \\
      \babar (partial rec.) & \cite{ref:cp_uta:cud:babar:partial} &
      $-0.034 \pm 0.014 \pm 0.009$ & $-0.019 \pm 0.022 \pm 0.013$ \\
      \belle (partial rec.) & \cite{ref:cp_uta:cud:belle:partial} &
      $-0.030 \pm 0.028 \pm 0.018$ & $-0.005 \pm 0.028 \pm 0.018$ \\
      \mc{2}{l}{\bf Average} & 
      $-0.027 \pm 0.013$ & $0.001 \pm 0.018 $ \\
      \mc{2}{l}{\small Confidence level} & 
      \small $0.22$ & \small $0.52$ \\
      \hline
      \mc{4}{c}{$D^{\pm}\pi^{\mp}$} \\
      \babar (full rec.) & \cite{ref:cp_uta:cud:babar:full} &
      $-0.032 \pm 0.031 \pm 0.020$ & $-0.059 \pm 0.055 \pm 0.033$ \\
      \belle (full rec.) & \cite{ref:cp_uta:cud:belle:full} &
      $-0.062 \pm 0.037 \pm 0.018$ & $-0.025 \pm 0.037 \pm 0.018$ \\
      \mc{2}{l}{\bf Average} & $-0.045 \pm 0.027$ & $-0.035 \pm 0.035$ \\
      \mc{2}{l}{\small Confidence level} & 
      \small $0.59$ & \small $0.66$ \\
      \hline
      \mc{4}{c}{$D^{\pm}\rho^{\mp}$} \\
      \babar (full rec.) & \cite{ref:cp_uta:cud:babar:full} &
      $-0.005 \pm 0.044 \pm 0.021$ & $-0.147 \pm 0.074 \pm 0.035$ \\
      \hline 
    \end{tabular*}
    \label{tab:cp_uta:cud}
  \end{center}
\end{table}

\mysubsection{Rates and asymmetries in $\Bmp \to \DorDstar K^{(*)\mp}$ decays
}
\label{sec:cp_uta:cus}

As explained in Sec.~\ref{sec:cp_uta:notations:cus},
rates and asymmetries in $\Bmp \to \DorDstar K^{(*)\mp}$ decays
are sensitive to $\gamma$.
Various methods using different $\DorDstar$ final states exist.

Results are available from both \babar\ and \belle\ on GLW analyses
in the decay modes $\Bmp \to D\Kmp$, $\Bmp \to \Dstar\Kmp$ and $\Bmp \to D\Kstarmp$.
Both experiments use the 
$\CP$ even $D$ decay final states $K^+K^-$ and $\pi^+\pi^-$ in all three modes; 
both experiments also use only the $\Dstar \to D\pi^0$ decay, 
which gives $\CP(\Dstar) = \CP(D)$. 
For $\CP$ odd $D$ decay final states, 
\belle\ uses $\KS\pi^0$, $\KS\eta$ and $\KS\phi$ in all three analyses, 
and also use $\KS\omega$ in $D\Kmp$ and $\Dstar\Kmp$ analyses. 
\babar\ uses $\KS\pi^0$ only for $D\Kmp$ analysis; 
for $D\Kstarmp$ analysis they also use $\KS\phi$ and $\KS\omega$
(and assign an asymmetric systematic error due to $\CP$ even pollution 
in these $\CP$ odd channels~\cite{ref:cp_uta:cus:babar:dstarcpk}).
The results and averages are given in Table~\ref{tab:cp_uta:cus:glw}.

\begin{table}
  \begin{center}
    \caption{
      Averages from GLW analyses of $b \to c\bar{u}d / u\bar{c}d$ modes.
    }
    \vspace{0.2cm}
    \resizebox{\textwidth}{!}{
      \setlength{\tabcolsep}{0.0pc}
      \begin{tabular}{@{\extracolsep{2mm}}lrcccc} \hline 
        \mc{2}{l}{Experiment} & 
        $A_{\CP+}$ & $A_{\CP-}$ & $R_{\CP+}$ & $R_{\CP-}$ \\
        \hline
        \mc{6}{c}{$D_{\CP} K^-$} \\
        \babar & \cite{ref:cp_uta:cus:babar:dcpk} &
        $\ph{-} 0.40 \pm 0.15 \pm 0.08$ & $\ph{-} 0.21 \pm 0.17 \pm 0.07$ & 
        $ 0.87 \pm 0.14 \pm 0.06$ & $ 0.80 \pm 0.14 \pm 0.08$ \\
        \belle & \cite{ref:cp_uta:cus:belle:dcpk} &
        $\ph{-} 0.07 \pm 0.14 \pm 0.06$ & $-0.11 \pm 0.14 \pm 0.05$ & 
        $ 0.98 \pm 0.18 \pm 0.10$ & $ 1.29 \pm 0.16 \pm 0.08$ \\
        \mc{2}{l}{\bf Average} & 
        $ 0.22 \pm 0.11$ & $ 0.02 \pm 0.12$ & $ 0.91 \pm 0.12$ & $ 1.02 \pm 0.12$ \\
        \hline
        \mc{6}{c}{$\Dstar_{\CP} K^-$} \\
        \babar & \cite{ref:cp_uta:cus:babar:dstarcpk} &
        $-0.10 \pm 0.23 \, ^{+0.03}_{-0.04}$ & $$ & 
        $ 1.06 \pm 0.26 \, ^{+0.10}_{-0.09}$ & $$ \\
        \belle & \cite{ref:cp_uta:cus:belle:dcpk} &
        $-0.27 \pm 0.25 \pm 0.04$ & $ 0.26 \pm 0.26 \pm 0.03$ & 
        $ 1.43 \pm 0.28 \pm 0.06$ & $ 0.94 \pm 0.28 \pm 0.06$ \\
        \mc{2}{l}{\bf Average} & 
        $-0.18 \pm 0.17$ & $ 0.26 \pm 0.26$ & $ 1.24 \pm 0.20$ & $ 0.94 \pm 0.29$ \\
        \hline
        \mc{6}{c}{$D_{\CP} K^{*-}$} \\
        \babar & \cite{ref:cp_uta:cus:babar:dcpkstar} &
        $-0.09 \pm 0.20 \pm 0.06$ & $-0.33 \pm 0.34 \pm 0.10 \, ^{+0.00}_{-0.06}$ & 
        $ 1.77 \pm 0.37 \pm 0.12$ & $ 0.76 \pm 0.29 \pm 0.06 \, ^{+0.04}_{-0.14}$ \\
        \belle & \cite{ref:cp_uta:cus:belle:dcpkstar} &
        $-0.02 \pm 0.33 \pm 0.07$ & $ 0.19 \pm 0.50 \pm 0.04$ & $$ & $$ \\
        \mc{2}{l}{\bf Average} & 
        $-0.07 \pm 0.18$ & $-0.16 \pm 0.29$ & $$ & $$ \\
        \hline
      \end{tabular}
    }
    \label{tab:cp_uta:cus:glw}
  \end{center}
\end{table}

For ADS analysis, both \babar\ and \belle\ have studied
the mode $\Bmp \to D\Kmp$;
\belle\ has also studied $\Bmp \to \pi^\mp$
and \babar\ has also analyzed the $\Bmp \to \Dstar\Kmp$ mode
($\Dstar \to D\pi^0$ and $\Dstar \to D\gamma$ are studied separately).
In all cases the suppressed decay $D \to K^+\pi^-$ has been used.
The results and averages are given in Table~\ref{tab:cp_uta:cus:ads}.
Note that although no clear signals for these modes have yet been seen,
the central values are given in Table~\ref{tab:cp_uta:cus:ads}.
In $\Bm \to \Dstar\Km$ decays there is an effective shift of $\pi$
in the strong phase difference between the cases that the $\Dstar$ is 
reconstructed as $D\pi^0$ and $D\gamma$~\cite{ref:cp_uta:cus:bg}.
As a consequence, the different $D^*$ decay modes are treated separately.

\begin{table}
  \begin{center} 
    \caption{
      Averages from ADS analyses of $b \to c\bar{u}d / u\bar{c}d$ modes.
    }
    \vspace{0.2cm}
    \setlength{\tabcolsep}{0.0pc}
    \begin{tabular*}{\textwidth}{@{\extracolsep{\fill}}lrcc} \hline 
      \mc{2}{l}{Experiment} & $A_{ADS}$ & $R_{ADS}$ \\
      \hline
      \mc{4}{c}{$D K^-$, $D \to K^+\pi^-$} \\
      \babar & \cite{ref:cp_uta:cus:babar:dk_ads} &
      & $ 0.013 \, ^{+0.011}_{-0.009}$ \\ 
      \belle & \cite{ref:cp_uta:cus:belle:dk_ads} &
      $0.88 \, ^{+0.77}_{-0.62} \pm 0.06$ & $0.023  \, ^{+0.016}_{-0.014} \pm 0.001$ \\
      \mc{2}{l}{\bf Average} & 
      $0.88 \, ^{+0.77}_{-0.62}$ & $0.017 \pm 0.009$ \\
      \hline
      \mc{4}{c}{$\Dstar K^-$, $\Dstar \to D\pi^0$, $D \to K^+\pi^-$} \\
      \babar & \cite{ref:cp_uta:cus:babar:dk_ads} &
      & $-0.001 \, ^{+0.010}_{-0.006}$ \\
      \hline
      \mc{4}{c}{$\Dstar K^-$, $\Dstar \to D\gamma$, $D \to K^+\pi^-$} \\
      \babar & \cite{ref:cp_uta:cus:babar:dk_ads} &
      & $ 0.011 \, ^{+0.019}_{-0.013}$ \\
      \hline 
      \mc{4}{c}{$D \pi^-$, $D \to K^+\pi^-$} \\
      \belle & \cite{ref:cp_uta:cus:belle:dk_ads} &
      $ 0.30 \, ^{+0.29}_{-0.25} \pm 0.06$ & $ 0.0035 \, ^{+0.0010}_{-0.0009} \pm 0.0002$ \\
      \hline 
    \end{tabular*}
    \label{tab:cp_uta:cus:ads}
  \end{center}
\end{table}

For the Dalitz plot analysis, both \babar\ and \belle\ have studied
the mode $\Bmp \to D\Kmp$.
Both have also studied the mode $\Bmp \to \Dstar\Kmp$;
\belle\ has used only $\Dstar \to D\pi^0$,
while \babar\ has used both $\Dstar$ decay modes and 
taken the effective shift in the strong phase difference into account.
\belle\ has also released results using $\Bmp \to D\Kstarmp$;
in this case an additional uncertainty arises from other possible
$\Bmp \to D\KS\pi^\mp$ contributions.
In all cases the decay $D \to \KS\pi^+\pi^-$ has been used.
The results are given in Table~\ref{tab:cp_uta:cus:dalitz}.
Since the measured values of $r_B$ are positive definite, 
and since the error on $\gamma$ depends on the value of $r_B$, 
some statistical treatment is necessary to correct for bias. 
At present, both \babar\ and \belle\ use frequentist treatments.
At present, we make no attempt to average the results.

\begin{table}
  \begin{center} 
    \caption{
      Averages from Dalitz plot analyses of $b \to c\bar{u}d / u\bar{c}d$ modes.
    }
    \vspace{0.2cm}
    \setlength{\tabcolsep}{0.0pc}
    \begin{tabular*}{\textwidth}{@{\extracolsep{\fill}}lrccc} \hline 
      \mc{2}{l}{Experiment} & $\gamma \ (^\circ)$ & $\delta_B \ (^\circ)$ & $r_B$ \\
      \hline
      \mc{5}{c}{$D K^-$, $D \to \KS \pi^+\pi^-$} \\
      \babar & \cite{ref:cp_uta:cus:babar:dk_dalitz} &
      $$ & $104 \pm 45 \, +{+17}_{-21} \, ^{+16}_{-24}$ & $ 0.12 \pm 0.08 \pm 0.03 \pm 0.04$ \\
      \belle & \cite{ref:cp_uta:cus:belle:dk_dalitz} &
      $ 64 \pm 19 \pm 13 \pm 11$ & $ 157 \pm 19 \pm 11 \pm 21$ & $ 0.21 \pm 0.08 \pm 0.03 \pm 0.04$ \\
      \mc{2}{l}{\bf Average} & 
      \mc{3}{c}{\sc in preparation} \\
      \hline
      \mc{5}{c}{$\Dstar K^-$, $\Dstar \to D\pi^0$ or $D\gamma$, $D \to \KS \pi^+\pi^-$} \\
      \babar & \cite{ref:cp_uta:cus:babar:dk_dalitz} &
      $$ & $ 296 \pm 41 \, ^{+14}_{-12} \pm 15$ & $ 0.17 \pm 0.10 \pm 0.03 \pm 0.03$ \\
      \belle & \cite{ref:cp_uta:cus:belle:dk_dalitz} &
      $ 75 \pm 57 \pm 11 \pm 11$ & $ 321 \pm 57 \pm 11 \pm 21$ & $ 0.12 \, ^{+0.16}_{-0.11} \pm 0.02 \pm 0.04$ \\
      \mc{2}{l}{\bf Average} & 
      \mc{3}{c}{\sc in preparation} \\
      \hline
      \mc{5}{c}{$D K^-$ and $\Dstar K^-$ combined} \\
      \babar & \cite{ref:cp_uta:cus:babar:dk_dalitz} &
      $ 70 \pm 31 \, ^{+12}_{-10} \, ^{+14}_{-11}$ \\
      \belle & \cite{ref:cp_uta:cus:belle:dk_dalitz} &
      $ 68 \, ^{+14}_{-15} \pm 13 \pm 11$ \\
      \mc{2}{l}{\bf Average} & 
      \mc{3}{c}{\sc in preparation} \\
      \hline 
      \mc{5}{c}{$D K^{*-}$, $D \to \KS \pi^+\pi^-$} \\
      \belle & \cite{ref:cp_uta:cus:belle:dkstar_dalitz} &
      $112 \pm 35 \pm 9 \pm 11 \pm 8$ & $353 \pm 35 \pm 8 \pm 21 \pm 49$ & $ 0.25 \pm 0.18 \pm 0.09 \pm 0.04 \pm 0.08$ \\
      \hline
    \end{tabular*}
    \label{tab:cp_uta:cus:dalitz}
  \end{center}
\end{table}


\mysection{Charmless \B-decay branching fractions and
               their asymmetries }
\label{sec:rare}

The aim of this section is to provide the branching fractions and
the partial rate asymmetries ($A_{CP}$) of rare $B$ decays. The asymmetry is
defined as $A_{CP} = \frac{N_{\Bbar} -N_B}{N_{\Bbar} +N_B}$, where $N_{\Bbar}$ 
and $N_B$ are number of $\Bzb/\Bm$ and $\Bz/\Bp$, respectively.
Four different $B$ decay categories are     
considered: charmless mesonic, baryonic, radiative and leptonic. Rare mesonic 
decays with charm are not in our scope but results of charmful baryonic decays
are included. Measurements supported with  written documents are accepted in 
our 
the averages; written documents could be journal papers, 
conference contributed papers, preprints or conference proceedings.  
Results from  $A_{CP}$ measurements  obtained from time dependent analyses 
are  listed and described in Sec.~\ref{sec:cp_uta}.

So far all branching fractions assume equal production of charged and
neutral $B$ pairs.  The best measurements to date show that this is
still a good approximation (see Sec.~\ref{sec:bfraction}).
For branching fractions, we provide either averages or the most stringent
90\% confidence level upper limits.  If one or more experiments have
measurements with $>$4$\sigma$ for a decay channel, 
all available central values
for that channel are used in the averaging.  We also give central values
and errors for cases where the significance of the average value is at
least $3 \sigma$, even if no single measurement is above $4 \sigma$.
For $A_{CP}$ we provide averages in all cases.

Our averaging is performed by maximizing the likelihood,
\begin{eqnarray}
    {\mathcal L} = \prod_i {\mathcal P}_i(x) , 
\end{eqnarray}
where ${\mathcal P_i}$ is the probability density function (PDF) of the
$i$th  measurement, and $x$ is the branching fraction or $A_{CP}$.
The PDF is modeled by an asymmetric Gaussian function with the measured
central value as its mean and the quadratic sum of the statistical
and systematic errors as the standard deviations. The experimental
uncertainties are considered to be uncorrelated with each other when the 
averaging is performed. No error scaling is applied when the fit $\chi^2$ is 
greater than 1 since we believe that tends to overestimate the errors
except in cases of extreme disagreement (we have no such cases).

At present, we have measurements of 236 $B$ decay modes including 4 $B_s$
decays newly added in this update cycle and asymmetry
measurements for 43 of these decays.  The averages of polarization 
measurements for $\B\to$ charmless vector meson are also added in this
update cycle.  These results are reported in about 150 separate papers. 
 Because the number of references is so large, we do
not include them with the tables shown here but the full set of
references is available quickly from active gifs at the 
``Winter 2005'' link on 
the rare web page: {\tt http://www.slac.stanford.edu/xorg/hfag/rare/index.html}

\mysubsection{Mesonic charmless decays}

\begin{table}[!htbp]
\begin{center}
\caption{
$B^+$ branching fractions 
(in units of $10^{-6}$). Upper limits are at 90\% CL.
Values in {\red red} ({\blue blue}) are new {\red published} 
({\blue preliminary}) result since PDG2004  [as of April 2, 2005].
}
\scriptsize
\begin{tabular}{|lccccccc|} 
\sgline
RPP\#   & Mode & PDG2004 Avg. & BABAR & Belle & CLEO & CDF & New Avg. \\
\sglinespb
117                                               & 
$K^0\pi^+$                                        & 
$18.8\pm 2.1$                                     & 
\blue{$\err{26.0}{1.3}{1.0}$}                     & 
\red{$\err{22.0}{1.9}{1.1}$}                      & 
$\aerrsy{18.8}{3.7}{3.3}{2.1}{1.8}$               & 
\nodata                                           & 
$24.1 \pm 1.3$                                    \\

118                                               & 
$K^+\pi^0$                                        & 
$12.9\pm 1.2$                                     & 
\blue{$\err{12.0}{0.7}{0.6}$}                     & 
\red{$\berr{12.0}{1.3}{1.3}{0.9}$}                & 
$\aerrsy{12.9}{2.4}{2.2}{1.2}{1.1}$               & 
\nodata                                           & 
$12.1 \pm 0.8$                                    \\

119                                               & 
$\etapr K^+$                                      & 
$78\pm 5$                                         & 
\blue{$\err{68.9}{2.0}{3.2}$}                     & 
\blue{$\err{78}{6}{9}$}                           & 
$\aerr{80}{10}{9}{7}$                             & 
\nodata                                           & 
$70.8 \pm 3.4$                                    \\
\end{tabular}
\end{center}

\vspace{-0.4cm} 
\hspace*{-0.8cm} 
$\dag$Product BF - daughter BF taken to be 100\%; 
~$\ddag$Larger of two solutions taken; 
~\S $M_{\phi\phi}<2.85$ GeV/$c^2$
\end{table}

\large

\begin{table}[!htbp]
\begin{center}
\caption{
$B^0$ branching fractions (in units of $10^{-6}$).
Upper limits are at 90\% CL.
Values in {\red red} ({\blue blue}) are new {\red published} 
({\blue preliminary}) result since PDG2004  [as of April 2, 2005].
}
\scriptsize
\begin{tabular}{|lccccccc|} 
\sgline
RPP\#   & Mode & PDG2004 Avg. & BABAR & Belle & CLEO & CDF & New Avg. \\
\hline
123                                               & 
$K^+\pi^-$                                        & 
$18.5\pm 1.1$                                     & 
$\err{17.9}{0.9}{0.7}$                            & 
\red{$\err{18.5}{1.0}{0.7}$}                      & 
$\aerrsy{18.0}{2.3}{2.1}{1.2}{0.9}$               & 
\nodata                                           & 
$18.2 \pm 0.8$                                    \\

124                                               & 
$K^0\pi^0$                                        & 
$\cerr{9.5}{2.1}{1.9}$                            & 
\blue{$\err{11.4}{0.9}{0.6}$}                     & 
\red{$\berr{11.7}{2.3}{1.2}{1.3}$}                & 
$\aerrsy{12.8}{4.0}{3.3}{1.7}{1.4}$               & 
\nodata                                           & 
$11.5 \pm 1.0$                                    \\

125                                               & 
$\etapr K^0$                                      & 
$63\pm 7$                                         & 
\blue{$\err{67.4}{3.3}{3.2}$}                     & 
\blue{$\berr{68}{10}{9}{8}$}                      & 
$\aerr{89}{18}{16}{9}$                            & 
\nodata                                           & 
$68.6 \pm 4.2$                                    \\
\end{tabular}
\end{center}
\vspace{-0.4cm}
\hspace{1.cm} $\dag$Product BF - daughter BF taken to be 100\%, 
 $\ddag$Relative BF converted to absolute BF
\end{table}

 \clearpage
{\vspace{-2cm}
\mysubsection{Radiative and leptonic decays}}

\begin{table}[!htbp]
\begin{center}
\caption{
Compilation of $B^+$ semileptonic and radiative branching fractions
(in units of $10^{-6}$). Upper limits are at 90\% CL.
Values in {\red red} ({\blue blue}) are new {\red published} 
({\blue preliminary}) result since PDG2004  [as of April 2, 2005].
}
\end{center}
\scriptsize
\begin{tabular}{|lcccccc|} \hline
RPP\# &Mode & PDG2004 Avg. & BABAR  & Belle & CLEO
 & New Avg.  \\
 \sglinespb


160                                               & 
$K^*(892)^+\gamma$                                & 
$38\pm 5$                                         & 
{\red $38.7\pm 2.8 \pm 2.6$}                      & 
{\red $42.5\pm 3.1\pm 2.4$}                       & 
$\aerr{37.6}{8.9}{8.3}{2.8} $                     & 
$40.3 \pm 2.6$                                    \\


161                                               & 
$K_1(1270)^+\gamma$                               & 
$<99$                                             & 
\nodata                                           & 
{\red $\err{43}{9}{9}$}                           & 
\nodata                                           & 
$43 \pm 13$                                       \\


162                                               & 
$K^+\phi\gamma$                                   & 
$3.4\pm 1.0$                                      & 
\nodata                                           & 
$\err{3.4}{0.9}{0.4}$                             & 
\nodata                                           & 
$3.4 \pm 1.0$                                     \\
\hline
\end{tabular}
~\S~$M_{K\pi\pi} < 2.4$ GeV/$c^2$
~\ddag~Central values are not significant.
\end{table}

\begin{table}[!htbp]
\begin{center}
\caption{
Compilation of $B^0$ semileptonic and radiative branching fractions
(in units of $10^{-6}$).  Upper limits are at 90\% CL.
Values in {\red red} ({\blue blue}) are new {\red published} 
({\blue preliminary}) result since PDG2004  [as of April 2, 2005].
}
\end{center}
\hspace*{-0.0cm} 
\scriptsize
\begin{tabular}{|lcccccc|} \hline
RPP\# &Mode & PDG2004 Avg. & BABAR  & Belle & CLEO
 & New Avg.  \\ \sglinespb
162                                               & 
$K^*(892)^0\gamma$                                & 
$43\pm 4$                                         & 
{\red $39.2\pm 2.0\pm 2.4$}                       & 
{\red $40.1\pm 2.1\pm 1.7$}                       & 
$\aerr{45.5}{7.2}{6.8}{3.4}$                      & 
$40.1 \pm 2.0$                                    \\


163                                               & 
$K^0\phi\gamma$                                   & 
$<8.3$                                            & 
\nodata                                           & 
$<8.3$                                            & 
\nodata                                           & 
$<8.3$                                            \\


164                                               & 
$K^+\pi^-\gamma$~\dag                             & 
$4.6\pm 1.4$                                      & 
\nodata                                           & 
$\aerrsy{4.6}{1.3}{1.2}{0.5}{0.7}$                & 
\nodata                                           & 
$4.6 \pm 1.4$                                     \\
\hline
\end{tabular}
\\
~\dag~$1.25$ GeV/$c^2 < M_{K\pi} < 1.6$ GeV/$c^2$
~\ddag~ Central values are not significant.
\end{table}

\begin{table}[!htbp]
\begin{center}
\caption{
Compilation of $B$ semileptonic and radiative branching fractions
(in units of $10^{-6}$).  Upper limits are at 90\% CL.
Values in {\red red} ({\blue blue}) are new {\red published} 
({\blue preliminary}) result since PDG2004  [as of April 2, 2005].
}
\end{center}
\hspace*{-0.0cm} 
\scriptsize
\begin{tabular}{|lcccccc|} \hline
RPP\# &Mode & PDG2004 Avg. & BABAR  & Belle & CLEO & New Avg.  \\ \sglinespb
~60                                               & 
$K_3^*(1780)\gamma $                              & 
$<3000$                                           & 
\nodata                                           & 
{\blue $<2.8$}                                    & 
\nodata                                           & 
{ $<2.8$}                                         \\


~67                                               & 
$s\gamma$                                         & 
$330\pm 40$                                       & 
{\blue $\berr{338}{19}{64}{42}$}                  & 
$\berrsyt{355}{32}{30}{31}{11}{7}$                & 
$\berr{321}{43}{32}{29}$                          & 
$\cerr{339}{30}{27}$                              \\


$~-$                                              & 
$s\gamma$ with baryons                            & 
New                                               & 
\nodata                                           & 
\nodata                                           & 
\red $<38$~\dag                                   & 
 $<38$~\dag                                       \\
\hline
\end{tabular}
\\  
 ~\dag $E_\gamma > 2.0$ GeV; ~\ddag $M(\ell^+\ell^-)>0.2$ GeV/$c^2$
\end{table}

 \clearpage

\begin{table}[!htbp]
\begin{center}
\caption{
Compilation of $B$  leptonic  branching fractions
(in units of $10^{-6}$). Upper limits are at 90\% CL.
Values in {\red red} ({\blue blue}) are new {\red published} 
({\blue preliminary}) result since PDG2004  [as of April 2, 2005].
}
\vspace{0.3cm}
\begin{tabular}{|lcccccccc|} \hline
RPP\# &Mode & PDG2004 Avg. & BABAR  & Belle & CLEO &CDF & D0 & New Avg.  \\ 
\sglinespb
~12                                               & 
$e^+\nu$                                          & 
$<15$                                             & 
\nodata                                           & 
{\blue $<5.4$}                                    & 
$<15$                                             & 
\nodata                                           & 
\nodata                                           & 
{ $<5.4$}                                         \\


~13                                               & 
$\mu^+\nu$                                        & 
$<21$                                             & 
{\red $<6.6$}                                     & 
{\blue $<2.0$}                                    & 
$<21$                                             & 
\nodata                                           & 
\nodata                                           & 
{ $<2.0$}                                         \\


~14                                               & 
$\tau^+\nu$                                       & 
$<570$                                            & 
\blue $<330$                                      & 
\blue $<290$                                      & 
$<840$                                            & 
\nodata                                           & 
\nodata                                           & 
 $<290$                                           \\
\hline
\end{tabular}
\end{center}
\end{table}

\vspace{-1cm}
\mysubsection{Baryonic decays}

\begin{table}[!htbp]
\begin{center}
\caption{
Compilation of $B^+$ baryonic branching fractions (in units of $10^{-6}$).
Upper limits are at 90\% CL.
values in {\red red} ({\blue blue}) are new {\red published} 
({\blue preliminary}) result since PDG2004  [as of April 2, 2005].
}
\scriptsize
\begin{tabular}{|lcccccc|} 
\sgline
RPP\#   & Mode & PDG2004 Avg. & BABAR & Belle & CLEO & New Avg. \\
\sgline
201                                               & 
$p\overline p \pi^+$                              & 
$<3.7$                                            & 
\nodata                                           & 
{\red $\aerr{3.06}{0.73}{0.62}{0.37}$}            & 
$<160$                                            & 
$\cerr{3.06}{0.82}{0.72}$                         \\


204                                               & 
$p\overline p K^+$                                & 
$\cerr{4.3}{1.2}{1.0}$                            & 
{\blue$\err{6.7}{0.9}{0.6}$}                      & 
{\blue$\aerr{5.30}{0.45}{0.39}{0.58}$}            & 
\nodata                                           & 
$\cerr{5.74}{0.61}{0.60}$                         \\

$~$                                               & 
$\Theta^{++} \overline p$ $\dag$                  & 
New                                               & 
\nodata                                           & 
{\blue$<0.091$}                                   & 
\nodata                                           & 
{$<0.091$}                                        \\
\hline
\end{tabular}
\vspace*{0.6cm}

$\dag$ Product BF - daughter BF taken to be 100\%: 
$\Theta(1540)^{++}\to K^+p$ (pentaquark candidate); 
${\mathcal G}(2220)\to p\overline p$ (glueball candidate);
$X_c^0(3350)\to\overline{\Lambda_c}^- p$.
\end{center}
\end{table}

\begin{table}[!htbp]
\begin{center}
\caption{
Compilation of $B^0$ baryonic branching fractions
(in units of $10^{-6}$). Upper limits are at 90\% CL.
values in {\red red} ({\blue blue}) are new {\red published} 
({\blue preliminary}) result since PDG2004  [as of April 2, 2005].
}
\vskip 0.25cm
\scriptsize
\begin{tabular}{|lcccccc|} 
\sgline
RPP\#   & Mode & PDG2004 Avg. & BABAR & Belle & CLEO & New Avg. \\
\sgline
212                                               & 
$p \overline{p}$                                  & 
$< 1.2$                                           & 
{\red $<0.27$}                                    & 
{\blue $<0.41$}                                   & 
$<1.4$                                            & 
{ $<0.27$}                                        \\

214                                               & 
$p \overline{p} K^0$                              & 
$<7.2$                                            & 
\nodata                                           & 
{\blue $\aerr{1.20}{0.32}{0.22}{0.14}$}           & 
\nodata                                           & 
$\cerr{1.20}{0.35}{0.26}$                         \\

$~$                                               & 
$\Theta^+ K^0$ $\dag$                             & 
New                                               & 
\nodata                                           & 
{\blue $<0.23$}                                   & 
\nodata                                           & 
{ $<0.23$}                                        \\


$~-$                                              & 
$p \overline{p} K^{*0}$                           & 
New                                               & 
\nodata                                           & 
{\red $<7.6$}                                     & 
\nodata                                           & 
{ $<7.6$}                                         \\


215                                               & 
$p \overline\Lambda \pi^-$                        & 
$\cerr{4.0}{1.1}{1.0}$                            & 
\nodata                                           & 
\blue{$\aerr{3.27}{0.62}{0.51}{0.39}$}            & 
$<13$                                             & 
$\cerr{3.27}{0.73}{0.64}$                         \\


216                                               & 
$p \overline\Lambda K^-$                          & 
$<0.82$                                           & 
\nodata                                           & 
$< 0.82$                                          & 
\nodata                                           & 
$< 0.82$                                          \\

217                                               & 
$p \overline\Sigma^0 \pi^-$                       & 
$<3.8$                                            & 
\nodata                                           & 
$< 3.8$                                           & 
\nodata                                           & 
$< 3.8$                                           \\


218                                               & 
$\Lambda \overline\Lambda$                        & 
$<1.0$                                            & 
\nodata                                           & 
{\blue $<0.69$}                                   & 
$<1.2$                                            & 
{ $<0.69$}                                        \\


224                                               & 
$\overline\Lambda_c^- p \pi^+\pi^-$               & 
$1300\pm 400$                                     & 
\nodata                                           & 
{\blue $1030\pm 90 \pm 295 $}                     & 
$\berr{1670}{190}{470}{460} $                     & 
$1207 \pm 262$                                    \\


225                                               & 
$\overline\Lambda_c^- p$                          & 
$22\pm 8$                                         & 
\nodata                                           & 
$\aerr{21.9}{5.6}{4.9}{6.5}$                      & 
$<90 $                                            & 
$\cerr{21.9}{8.6}{8.1}$                           \\

%

229                                               & 
$\overline\Sigma_c^{--}(2520)p \pi^+$             & 
$160\pm 70$                                       & 
\nodata                                           & 
{\blue$\err{104}{23}{30}$}                        & 
\nodata                                           & 
$104 \pm 37$                                      \\

%

230                                               & 
$\overline\Sigma_c^0(2520)p \pi^-$                & 
$<121$                                            & 
\nodata                                           & 
{\blue$\err{33}{19}{10}$}                         & 
\nodata                                           & 
$33 \pm 21$                                       \\

%

231                                               & 
$\overline\Sigma_c^0(2455)p \pi^-$                & 
$100\pm 80$                                       & 
\nodata                                           & 
{\blue $\err{97}{21}{30}$}                        & 
$220\pm 60 \pm 64 $                               & 
$115 \pm 33$                                      \\

%

232                                               & 
$\overline\Sigma_c^{--}(2455)p \pi^+$             & 
$280\pm 90$                                       & 
\nodata                                           & 
{\blue $\err{115}{22}{33} $}                      & 
$370\pm 80\pm 113$                                & 
$134 \pm 38$                                      \\

%

233                                               & 
$\overline\Lambda_c^-(2593) p$                    & 
$<110$                                            & 
\nodata                                           & 
\nodata                                           & 
$<110$                                            & 
$<110$                                            \\

            
\hline
\end{tabular}
\end{center}

$\dag$ Product BF - daughter BF taken to be 100\%: 
$\Theta(1540)^+\to p K^0$ (pentaquark candidate).
\end{table}

\mysubsection{$B_s$ decays}
 
%
%

\begin{table}[!htbp]
\caption{
Compilation of $B_s$   
branching fractions (in units of $10^{-6}$). Upper limits are at 90\% CL.
values in {\red red} ({\blue blue}) are new {\red published} 
({\blue preliminary}) result since PDG2004  [as of April 2, 2005].
}
\vskip 0.25cm
\begin{center}
\begin{tabular}{|lccccc|} \hline
RPP\# &Mode & PDG2004 Avg. &  CDF & D0
 & New Avg.  \\ \sglinespb
$15$                                              & 
$\phi\phi$                                        & 
$<1183 $                                          & 
\blue $\aerr{14}{6}{5}{6}$                        & 
\nodata                                           & 
$\cerr{14}{8}{7}$                                 \\

$24$                                              & 
$\mu^+\mu^-$                                      & 
$<2.0 $                                           & 
\red$<0.58 $                                      & 
\red$<0.41$                                       & 
$<0.41$                                           \\

$26$                                              & 
$e^{\pm}\mu^{\mp}$                                & 
$<6.1 $                                           & 
$<6.1 $                                           & 
\nodata                                           & 
$<6.1 $                                           \\

$27$                                              & 
$\phi\mu^+\mu^-$                                  & 
$<47 $                                            & 
$<47 $                                            & 
\nodata                                           & 
$<47 $                                            \\

\hline
\end{tabular}
\end{center}
\end{table}

\mysubsection{Charge asymmetries}

\begin{sidewaystable}[!htbp]
\parbox{8.2in}{\caption{
Compilation of charmless hadronic $CP$ asymmetries for charged $B$ decays.
Values in {\red red} ({\blue blue}) are 
new {\red published} 
({\blue preliminary}) result since PDG2004  [as of April 2, 2005].
}}

\vspace*{ 0.5cm}
\scriptsize
\begin{tabular}{|lccccccc|} 
\hline
RPP\#   & Mode & PDG2004 Avg. & BABAR & Belle & CLEO & CDF & New Avg. \\
\sglinespb

117                                               & 
$K^0\pi^+$                                        & 
$0.03\pm 0.08$                                    & 
{\blue$\err{-0.087}{0.046}{0.010}$}               & 
{\red$\err{0.05}{0.05}{0.01}$}                    & 
$\err{0.18}{0.24}{0.02}$                          & 
\nodata                                           & 
$-0.020 \pm 0.034$                                \\

118                                               & 
$K^+\pi^0$                                        & 
$-0.10\pm 0.08$                                   & 
{\blue $\err{0.06}{0.06}{0.01}$}                  & 
{\red $\err{0.04}{0.05}{0.02}$}                   & 
$\err{-0.29}{0.23}{0.02}$                         & 
\nodata                                           & 
$0.04 \pm 0.04$                                   \\
\end{tabular}
\end{sidewaystable}

\begin{sidewaystable}[!htbp]
\begin{center}
\parbox{7.5in}{\caption{
Compilation of charmless hadronic $CP$ asymmetries for $B^\pm/B^0$ admixtures.
Values in {\red red} ({\blue blue}) are new {\red published} 
({\blue preliminary}) result since PDG2004  [as of April 2, 2005].
}}

\vspace*{0.5cm}   
\scriptsize
\begin{tabular}{|lccccccc|}
\hline
RPP\#    & Mode & PDG2004 Avg. & BABAR & Belle & CLEO & CDF & New Avg. \\
\sglinespb
~56                                               & 
$K^*\gamma$                                       & 
$-0.01\pm 0.07$                                   & 
{\red $-0.013\pm 0.036\pm 0.010$}                 & 
{\red $-0.015\pm 0.044\pm 0.012$}                 & 
$0.08\pm 0.13\pm 0.03$                            & 
\nodata                                           & 
$-0.010 \pm 0.028$                                \\


%

~67                                               & 
$s\gamma$                                         & 
$\err{-0.079}{0.108}{0.022}$                      & 
\red$\err{0.025}{0.050}{0.015}$                   & 
\red$\err{0.002}{0.050}{0.026}$                   & 
$\err{-0.079}{0.108}{0.022}$                      & 
\nodata                                           & 
$0.004 \pm 0.036$                                 \\


103                                               & 
$s\ell\ell$                                       & 
New                                               & 
\red$\err{-0.22}{0.26}{0.02}$                     & 
\nodata                                           & 
\nodata                                           & 
\nodata                                           & 
$-0.22 \pm 0.26$                                  \\


\hline
\end{tabular}
\end{center}

\vskip -1cm
\begin{center}
\parbox{8.5in}{\caption{
Compilation of charmless hadronic $CP$ asymmetries for neutral $B$ decays.
Values in {\red red} ({\blue blue}) are new {\red published} 
({\blue preliminary}) result since PDG2004  [as of April 2, 2005].
}}

\vspace*{0.5cm}
\scriptsize
\begin{tabular}{|lccccccc|} 
\sgline
RPP\#   & Mode & PDG2004 Avg. & BABAR & Belle & CLEO & CDF & New Avg. \\
\sglinespb

123                                               & 
$K^+\pi^-$                                        & 
$-0.09\pm 0.04$                                   & 
{\red $\err{-0.133}{0.030}{0.009}$}               & 
{\red $\err{-0.101}{0.025}{0.005}$}               & 
$\err{-0.04}{0.16}{0.02}$                         & 
{\blue $\err{-0.04}{0.08}{0.006}$}                & 
$-0.109 \pm 0.019$                                \\

124                                               & 
$K^0\pi^0$ $\dag$                                 & 
\nodata                                           & 
\nodata                                           & 
\nodata                                           & 
\nodata                                           & 
\nodata                                           & 
\nodata                                           \\
\end{tabular}
\vskip 0.5cm\end{center}

\large
\hspace{-1.9cm}
\dag~Measurements of time-dependent $CP$ asymmetries are listed on 
the Unitarity Triangle home page. (http://www.slac.stanford.edu/xorg/hfag/triangle/index.html) 

\end{sidewaystable}

 \clearpage
\mysubsection{Polarization measurements}
%

%

\begin{table}[!htbp]
\caption{
Compilation of the longitudinal polarization fraction $f_L$ for $B^+$ decays.
Values in {\red red} ({\blue blue}) are new {\red published} 
({\blue preliminary}) result since PDG2004  [as of April 2, 2005].
\vspace{0.3cm}
}
\begin{center}
\begin{tabular}{|lccccc|} 
\sgline
RPP\#   & Mode & PDG2004 Avg. & BABAR & Belle & New Avg. \\
\sglinespb

%

$~-$                                              & 
$K^{*0}\rho^+$                                    & 
New                                               & 
\blue{$\err{0.79}{0.08}{0.04}$}                   & 
\blue{$\berr{0.43}{0.11}{0.05}{0.02}$}            & 
$0.66 \pm 0.07$                                   \\

138                                               & 
$K^{*+}\rho^0$                                    & 
$\aerr{0.96}{0.04}{0.15}{0.04}$                   & 
$\aerr{0.96}{0.04}{0.15}{0.04}$                   & 
\nodata                                           & 
$\cerr{0.96}{0.06}{0.15}$                         \\

%

156                                               & 
$\phi K^{*+}$                                     & 
$0.46\pm0.12\pm0.03$                              & 
$\err{0.46}{0.12}{0.03}$                          & 
\blue{$\err{0.52}{0.08}{0.03}$}                   & 
$0.50 \pm 0.07$                                   \\

182                                               & 
$\rho^+\rho^0$                                    & 
$0.96\pm0.06$                                     & 
$\aerr{0.97}{0.03}{0.07}{0.04}$                   & 
$\err{0.95}{0.11}{0.02}$                          & 
$\cerr{0.97}{0.05}{0.07}$                         \\

186                                               & 
$\omega\rho^+$                                    & 
New                                               & 
\red{$\aerr{0.88}{0.12}{0.15}{0.03}$}             & 
\nodata                                           & 
$\cerr{0.88}{0.12}{0.15}$                         \\

\sglinespt
\end{tabular}
\end{center}
\end{table}

\begin{table}[!htbp]
\caption{
Compilation of the full angular analysis of $B^+\to\phi K^{*+}$
Values in {\red red} ({\blue blue}) are new {\red published} 
({\blue preliminary}) result since PDG2004  [as of April 2, 2005].
\vspace{0.3cm}
}
\begin{center}
\begin{tabular}{|lccccc|} 
\sgline
\RPP & Parameter & PDG2004 Avg. & BABAR & Belle & New Avg. \\
\sglinespb
\nodata                                           & 
$f_\perp$                                         & 
New                                               & 
\nodata                                           & 
\blue{$\err{0.19}{0.08}{0.02}$}                   & 
$0.19 \pm 0.08$                                   \\

\nodata                                           & 
$\phi_\parallel$                                  & 
New                                               & 
\nodata                                           & 
\blue{$\err{2.10}{0.28}{0.04}$}                   & 
$2.10 \pm 0.28$                                   \\

\nodata                                           & 
$\phi_\perp$                                      & 
New                                               & 
\nodata                                           & 
\blue{$\err{2.31}{0.30}{0.07}$}                   & 
$2.31 \pm 0.31$                                   \\

\sglinespt
\end{tabular}
\end{center}
\hspace*{1.5cm}BR, $f_L$ and $A_{CP}$ are tabulated separately.
\end{table}

\large

\begin{table}[!htbp]
\caption{
Compilation of the longitudinal polarization fraction $f_L$ for $B^0$ decays.
Values in {\red red} ({\blue blue}) are new {\red published} 
({\blue preliminary}) result since PDG2004  [as of April 2, 2005].
\vspace{0.3cm}
}
\begin{center}
\begin{tabular}{|lccccc|} 
\sgline
RPP\#   & Mode & PDG2004 Avg. & BABAR & Belle & New Avg. \\
\hline




154                                               & 
$\phi K^{*0}$                                     & 
$0.57\pm0.11$                                     & 
\red{$\err{0.52}{0.05}{0.02}$}                    & 
\blue{$\err{0.45}{0.05}{0.02}$}                   & 
$0.48 \pm 0.04$                                   \\

%

%

%

%

203                                               & 
$\rho^+\rho^-$                                    & 
New                                               & 
\red{$\berr{0.99}{0.03}{0.04}{0.03}$}             & 
\nodata                                           & 
$\cerr{0.99}{0.05}{0.04}$                         \\

\sglinespt
\end{tabular}
\end{center}
\end{table}

\begin{table}[!htbp]
\caption{
Compilation of the full angular analysis of $B^0 \to \phi K^{*0}$
Values in {\red red} ({\blue blue}) are new {\red published} 
({\blue preliminary}) result since PDG2004  [as of April 2, 2005].
\vspace{0.3cm}
}
\begin{center}
\begin{tabular}{|lccccc|} 
\sgline
\RPP & Parameter & PDG2004 Avg. & BABAR & Belle & New Avg. \\
\sglinespb
\nodata                                           & 
$f_\perp = \Lambda_{\perp\perp}$                  & 
New                                               & 
\red{$\err{0.22}{0.05}{0.02}$}                    & 
\blue{$\aerr{0.31}{0.06}{0.05}{0.02}$}            & 
$0.26 \pm 0.04$                                   \\

\nodata                                           & 
$\phi_\parallel$                                  & 
New                                               & 
\red{$\aerr{2.34}{0.23}{0.20}{0.05}$}             & 
\blue{$\aerr{2.40}{0.28}{0.24}{0.07}$}            & 
$\cerr{2.36}{0.18}{0.16}$                         \\

\nodata                                           & 
$\phi_\perp$                                      & 
New                                               & 
\red{$\err{2.47}{0.25}{0.05}$}                    & 
\blue{$\err{2.51}{0.25}{0.06}$}                   & 
$2.49 \pm 0.18$                                   \\

\nodata                                           & 
$A_{CP}^0$                                        & 
New                                               & 
\red{$\err{-0.06}{0.10}{0.01}$}                   & 
\blue{$\err{0.13}{0.12}{0.04}$}                   & 
$0.01 \pm 0.08$                                   \\

\nodata                                           & 
$A_{CP}^\perp$                                    & 
New                                               & 
\red{$\err{-0.10}{0.24}{0.05}$}                   & 
\blue{$\err{-0.20}{0.18}{0.04}$}                  & 
$-0.16 \pm 0.15$                                  \\

\nodata                                           & 
$\Delta\phi_\parallel$                            & 
New                                               & 
\red{$\aerr{0.27}{0.20}{0.23}{0.05}$}             & 
\blue{$\err{-0.32}{0.27}{0.07}$}                  & 
$0.03 \pm 0.18$                                   \\

\nodata                                           & 
$\Delta\phi_\perp$                                & 
New                                               & 
\red{$\err{0.36}{0.25}{0.05}$}                    & 
\blue{$\err{-0.30}{0.25}{0.06}$}                  & 
$0.03 \pm 0.18$                                   \\

\hline                                            & 
$f_\parallel = \Lambda_{\parallel\,\parallel}$    & 
New                                               & 
\red{$\err{0.26}{0.05}{0.02}$}                    & 
\blue{$\err{0.24}{0.06}{0.02}$}                   & 
$0.25 \pm 0.04$                                   \\

\nodata                                           & 
$\mathcal{A}_T^0 = -0.5\Lambda_{\perp 0}$         & 
New                                               & 
\red{$\err{0.11}{0.05}{0.01}$}                    & 
\blue{$\err{-0.08}{0.08}{0.02}$}                  & 
$0.06 \pm 0.04$                                   \\

\nodata                                           & 
$\mathcal{A}_T^\parallel=-0.5\Lambda_{\perp\parallel}$& 
New                                               & 
\red{$\err{-0.02}{0.04}{0.01}$}                   & 
\blue{$\err{-0.01}{0.05}{0.01}$}                  & 
$-0.02 \pm 0.03$                                  \\

\nodata                                           & 
$\Lambda_{\parallel 0}$                           & 
New                                               & 
\red{$\err{-0.50}{0.12}{0.03}$}                   & 
\blue{$\err{-0.45}{0.11}{0.02}$}                  & 
$-0.47 \pm 0.08$                                  \\

\nodata                                           & 
$\Sigma_{00}$                                     & 
New                                               & 
\red{$\err{0.03}{0.05}{0.01}$}                    & 
\blue{$\err{-0.06}{0.05}{0.01}$}                  & 
$-0.02 \pm 0.04$                                  \\

\nodata                                           & 
$\Sigma_{\parallel\parallel}$                     & 
New                                               & 
\red{$\err{-0.05}{0.06}{0.01}$}                   & 
\blue{$\err{-0.01}{0.06}{0.01}$}                  & 
$-0.03 \pm 0.04$                                  \\

\nodata                                           & 
$\Sigma_{\perp\perp}$                             & 
New                                               & 
\red{$\aerr{0.02}{0.06}{0.05}{0.01}$}             & 
\blue{$\err{0.06}{0.06}{0.01}$}                   & 
$0.04 \pm 0.04$                                   \\

\nodata                                           & 
$\Sigma_{\perp 0}$                                & 
New                                               & 
\red{$\err{-0.41}{0.14}{0.03}$}                   & 
\blue{$\aerr{-0.41}{0.16}{0.14}{0.04}$}           & 
$\cerr{-0.41}{0.11}{0.10}$                        \\

\nodata                                           & 
$\Sigma_{\perp\parallel}$                         & 
New                                               & 
\red{$\aerr{-0.06}{0.09}{0.08}{0.02}$}            & 
\blue{$\err{-0.06}{0.10}{0.01}$}                  & 
$\cerr{-0.06}{0.07}{0.06}$                        \\

\nodata                                           & 
$\Sigma_{\parallel 0}$                            & 
New                                               & 
\red{$\aerr{0.18}{0.11}{0.13}{0.03}$}             & 
\blue{$\err{-0.11}{0.11}{0.02}$}                  & 
$0.01 \pm 0.09$                                   \\

\sglinespt
\end{tabular}
\end{center}
Results below the line have been derived from the primary results.
BR, $f_L$ and $A_{CP}$ are tabulated separately.
\end{table}

\newpage \normalsize

\section{Summary }
\labs{summary}

 This article provides the updated world averages for 
$b$-hadron properties as of winter 2005 conferences (Moriond and CKM05 etc.).
A brief summary of the results described in Secs. 
\ref{sec:life_mix}-\ref{sec:rare} is given in 
Table~\ref{tab_summary}.

\begin{table}
\caption{ Brief summary of the world averages as of winter 2005 conferences.}
\label{tab_summary}
\begin{center}
\begin{tabular}{|l|c|}
\hline
 {\bf\boldmath \b-hadron lifetimes} &   \\
 ~~$\tau(\Bd)$  & \hfagTAUBD \\
 ~~$\tau(\Bu)$  & \hfagTAUBU \\
 ~~$\tau(\Bs\to~\mbox{flavour specific})$  & \hfagTAUBSSL \\
 ~~$\bar{\tau}(\Bs) = 1/\Gs$  & \hfagTAUBSMEANCON \\
 ~~$\tau(\Bc)$  & \hfagTAUBC \\
 ~~$\tau(\Lb)$  & \hfagTAULB \\
\hline
 {\bf\boldmath \b-hadron fractions} &   \\
 ~~$f^{+-}/f^{00}$ in \Ups\ decays  & \hfagFF \\ 
 ~~$\fBd=\fBu$ at high energy & \hfagFBD \\
 ~~\fBs\ at high energy  & \hfagFBS \\
 ~~\fbb\ at high energy  & \hfagFBB \\
\hline
 {\bf\boldmath \Bd\ and \Bs\ mixing parameters} &   \\
 ~~\dmd &  \hfagDMDWU \\
 ~~$|q/p|_{\particle{d}}$ & \hfagQP  \\
 ~~\dms  &  $> \rm \hfagDMSWLIM~at~\CL{95}$ \\
 ~~$\DGGs = (\Gamma_{\rm L} - \Gamma_{\rm H})/\Gs$ & \hfagDGSGSCON \\
\hline
 {\bf\boldmath Semileptonic \B decay parameters} &   \\
 ~~${\cal B}(\BzbDstarlnu)$  & $( 5.33 \pm 0.20)\%$ \\
 ~~${\cal B}(\BzbDplnu)$      & $( 2.13 \pm 0.20)\%$ \\
 ~~${\cal B}(\Bb\to  X\ell\nub)$   & $(10.90 \pm 0.23)\%$ \\
 ~~$|V_{\particle{cb}}|\ (\BzbDstarlnu)$   
       & $ [41.4 \pm 1.0({\rm exp}) \pm 1.8({\rm theo})] \times 10^{-3}$ \\
 ~~$|V_{\particle{cb}}|\ (\BzbDplnu)$   
       & $ [40.4 \pm 3.6({\rm exp}) \pm 2.3({\rm theo}) ] \times 10^{-3}$ \\
 ~~$|V_{\particle{ub}}|$ (inclusive)    & $ (4.70 \pm 0.44 ) \times 10^{-3}$ \\
\hline
 {\bf\boldmath $\CP(t)$ and Unitarity Triangle angles } &   \\
 ~~$\stwob(\phi_1)$ (all charmonium) & $0.726 \pm 0.037$ \\
 ~~$\stwob(\phi_1)_{\rm eff}$ (all $b\to s$ penguin) & $0.43 \pm 0.07$ \\
\hline
 {\bf\boldmath Rare \B decays} &   \\
 ~~$A_{\CP}(\particle{\Bd\to K^+\pi^-})$ & $-0.109 \pm 0.019$ ($5.7\,\sigma$) \\
\hline
\end{tabular}
\end{center}
\end{table}

The HFAG provided the first averages at 2003 winter conferences
and have given updates at summer and winter conferences, as
well as annual PDG averages.

The accuracies of \b-hadron lifetimes and \Bd mixing parameters 
have been considerably improved with the asymmetric \B factory
results compared to the previous LEP working group 
averages~\cite{LEPHFS}. 
In 2004 summer, the \babar and \belle collaborations reported the
simultaneous measurements of \B lifetime and \dmd with 
improved precision using increased data samples.
Because of correlation between lifetime and \dmd which 
complicates the average procedures, these new results were not included
in the last averages~\cite{hfag_hepex}.
In this update, the two-dimensional averaging technique has been developped
taking into account the correlations.  Inclusion of these results changed
the average of \dmd from $0.502\pm 0.006\invps$~\cite{hfag_hepex} 
to \hfagDMDWU.
In the previous update, the average of $|q/p|_d$ was newly added.
In this update, the average of \DGGs is provided
including recent results from CDF and \dzero using $\Bs\to J/\psi\phi$
decay mode with time-dependent angular analysis.

For $|V_{\particle{cb}}|$ and $|V_{\particle{ub}}|$ average values,
various new measurements have been available from CLEO, \babar, and Belle.
Accordingly, the statistical uncertainties of averages are significantly 
reduced from LEP working group averages~\cite{LEPHFS}.
Considerable progress has been also made in theoretical side, but the
reduction of uncertainty is somewhat slower in some cases (\eg\
$|V_{\particle{cb}}|$ in exclusive modes).
The determination of $|V_{\particle{cb}}|$ from the inclusive semileptonic
branching fraction in combination with hadronic mass and lepton energy 
moments is under discussion and foreseen in the future update.
In the previous update, a substantial change occurred with the usage of
the Belle photon spectrum measurement in $b \to s \gamma$ decays.
This is used for the determination of shape function parameters
which enter the $|V_{ub}|$ calculation in inclusive charmless semileptonic
\B\ decays.  The new shape function parameters result in 
substantially smaller error in $|V_{ub}|$ than previous one.
In this update, several preliminary results have become published results.
Only small changes happen in averages from the previous averages due to
above changes and rescaling by updated common input parameters.

Measurements by \babar\ and \belle 
of the time-dependent $\CP$ violation parameter
$S_{b \to c\bar c s}$ in \B decays to charmonium and a neutral kaon
have established $\CP$ violation in $\B$ decays,
and allow a precise extraction of 
the Unitarity Triangle parameter $\stwob / \sin\! 2\phi_1$.
The $\B$ factories have also provided various measurements of
time-dependent $\CP$ asymmetries in hadronic $b \to s$ penguin decays,
establishing $\CP$ violation in these modes.
In this update cycle, the results for $K^0_SK^0_S K^0_S$ and $\omega K^0_S$
from \babar are added in the average.
Intriguingly, the measured parameters exhibit deviations 
from the Standard Model expectation.
The significance of this effect depends on the treatment of the
theoretical error.
Results from time-dependent analyses 
with the decays $\Bz \to \pi^+\pi^-, \rho^\pm\pi^\mp$ and $\rho^+\rho^-$ 
allow, via various methods, 
constraints on the Unitarity Triangle angle $\alpha/\phi_2$.
Constraints on the third Unitarity Triangle angle $\gamma/\phi_3$
have been obtained by \babar\ and \belle,
using $\Bm \to \DorDstar \Km$ decays with Dalitz plot analysis of
the subsequent $D \to \KS\pi^+\pi^-$ decay.

For the rare \B\ decays, branching fractions and charge asymmetries of many 
new decay modes have been measured recently, mostly by \babar\ and Belle.
Since there are several hundred measurements in the tables in \Sec{rare},
we highlight only the measurement of $A_{\CP}(\particle{\Bd\to K^+\pi^-})$,
which provides the observation of direct \CP violation ($5.7\,\sigma$) in 
the \B meson system.  The averages of polarization measurements
for $\B\to$ charmless vector meson pair and branching fractions for 
\Bs meson decays which have been newly available from CDF and \dzero are 
added from this update cycle.

\section*{ Acknowledgements }

We would like to thank collaborators of \babar, Belle, CDF, CLEO, \dzero,
LEP, and SLD experiments who provided fruitful results on \b-hadron
properties and cooperated with the HFAG for averaging.
These results are thanks to the excellent operations of the 
accelerators and collaborations with experimental groups by the 
accelerator groups of PEP-II, KEKB, CESR, Tevatron, LEP, and SLC.
We specially thank J.~Alexander and U.~Langenegger who lead the
averaging work and essential contributions until the previous 
update cycle.
Some of the averages have been obtained based on the discussions 
between theorists for better understanding and improvement on the
theoretical uncertainties.  


\end{thebibliography}


\end{document}